# SOCIAL BIG DATA ANALYTICS OF CONSUMER CHOICES:

# A TWO SIDED ONLINE PLATFORM PERSPECTIVE

by

Meisam Hejazi Nia

APPROVED BY SUPERVISORY COMMITTEE:

____________________________________

Brian T. Ratchford, Chair

____________________________________

Ozalp Ozer, Co-Chair

____________________________________

Dmitri Kuksov

____________________________________

Ahmet Serdar Simsek

Dedicated to my kind parents

SOCIAL BIG DATA ANALYTICS OF CONSUMER CHOICES:

A TWO SIDED ONLINE PLATFORM PERSPECTIVE

by

MEISAM HEJAZI NIA, MS

DISSERTATION

Presented to the Faculty of

The University of Texas at Dallas

in Partial Fulfillment

of the Requirements

for the Degree of

DOCTOR OF PHILOSOPHY IN

MANAGEMENT SCIENCE

THE UNIVERSITY OF TEXAS AT DALLAS

August 2016

ACKNOWLEDGEMENTS

Thanks to my dissertation committee for their support, guidance, and dedication. Dr. Ratchford and Dr. Ozer, thanks for your helpful advice and trust, and for always believing in me; Dr. Simsek thanks for your helpful advice, trust, immense patience, and guidance. Dr. Kuksov, thanks for your support and suggestions.

Thanks Mom, Dad, and siblings, for always believing in me and staying in touch. This work was not possible without sacrificing four years of my life that I could have spent with you. It seems like yesterday when I left my country pursuing a dream. I thank God that today it is a reality.

April 2016

SOCIAL BIG DATA ANALYTICS OF CONSUMER CHOICES:

A TWO SIDED ONLINE PLATFORM PERSPECTIVE

Publication No. ___________________

Meisam Hejazi Nia, PhD
The University of Texas at Dallas, 2016

Supervising Professor: Brian T. Ratchford, Chair
Ozalp Ozer, Co-Chair

This dissertation examines three distinct big data analytics problems related to the social aspects of consumers' choices. The main goal of this line of research is to help two sided platform firms to target their marketing policies given the great heterogeneity among their customers. In three essays, I combined structural modeling and machine learning approaches to first understand customers' responses to intrinsic and extrinsic factors, using unique data sets I scraped from the web, and then explore methods to optimize two sided platforms' firms' reactions accordingly.

The first essay examines "social learning" in the mobile app store context, controlling for intrinsic value of hedonic and utilitarian mobile apps, price, advertising, and number of options available. The proposed model extracted a social influence proxy measure from a macro diffusion model using an unscented Kalman filter, and it incorporated this social influence measure in a mixed logit choice model with hierarchical Dirichlet Process prior. Results suggest significant effects of social influence, which underscores the importance of choosing different marketing policies for pervasive goods. The comparison of mobile app adoption parameters suggests that among several classical goods mobile app adoption pattern is very similar to that of music CDs. The simulation counterfactual analysis suggests that early targeted viral marketing policy might be an optimal strategy for the app-store platforms.

The second essay investigates bidders' anticipated winner and loser regret in the context of the eBay online auction platform. I developed a structural model that accounts for bidders' learning and their anticipation of winner and loser regrets in an auction platform. Winner and loser regrets are defined as regretting for paying too much in case of winning an auction and regretting for not bidding high enough in case of losing it, respectively. Using a large data set from eBay and empirical Bayesian estimation method, I quantify the bidders' anticipation of regret in various product categories, and investigate the role of experience in explaining the bidders' regret and learning behaviors. The counterfactual analyses showed that shutting down the bidder regret via appropriate notification policies can increase eBay's revenue by 24%.

The third essay investigates the effects of Gamification incentive mechanisms in an online platform for user generated content. I use an ensemble method over LDA, mixed normal and k-mean clustering methods to segment users into competitors, collaborators, achievers, explorers and uninterested users. Then, I develop a state-dependent choice model that accommodates the effect of number of badges, the rank in the leaderboard, reputation points, inertia, and reciprocity, and allow for heterogeneity by Dirichlet Process prior. The results suggested that estimating the model on small samples generate biased estimates. Furthermore, they suggest that the effects of Gamification elements are heterogeneous, significantly positive or negative for different users. I found sensitivity patterns that explain importance of certain Gamification elements for users with certain nationalities. These findings help the Gamification platform to target its users. The simulation counterfactual analysis suggests that a two sided platform can increase the number of user contributions, by making earning badges more difficult.

TABLE OF CONTENTS

LIST OF FIGURES

LIST OF TABLES

# CHAPTER 1

# SOCIAL LEARNING AND DIFFUSION OVER THE PERVASIVE GOODS: AN EMPIRICAL STUDY OF AN AFRICAN APP-STORE

Meisam Hejazi Nia

Naveen Jindal School of Management, Department of Marketing, SM32

The University of Texas at Dallas

800 W. Campbell Road

Richardson, TX, 75080-3021

## 1.1. ABSTRACT

I developed a structural model that combines a macro diffusion model with a micro choice model to control for social influence on the mobile app choices of customers over app-stores. Social influence is measured by the density of adopters within the proximity of the customers. Using a large data set from an African app-store and Bayesian estimation methods, I quantify the effect of social influence on customer choices over the app-store, and investigate the effect of ignoring this process in estimating customer choices. I find that customer choices on the app-store are explained better by off-line density rather than online density of adopters, and ignoring social influence in estimation results in biased estimates. Furthermore, my results showed that the mobile app-adoption process is very similar to adoption of music CDs, among all other classical goods. My counterfactual analysis showed that the app-store can increase its revenue by 13.6% through the viral marketing policy (e.g., sharing with friends and family button).

Keywords: mobile app-store, social learning, state space model, structural model, semi parametric Bayesian, MCEM, unscented Kalman filter, hierarchical mixture model, genetic optimization.

## 1.2. INTRODUCTION

Smartphones pervade the global telecommunication market to such an extent that, for example, in the US a consumer has the option to adopt a smartphone handset on a postpaid contract, no matter which mobile operator (e.g., T-Mobile, Verizon, AT&T, or Sprint) the consumer selects. The smartphone handsets and the mobile apps are complements. A mobile app-store (e.g., Google play, Apple and Microsoft’s app stores) acts as a two sided platform that matches consumers to the mobile app publishers/developers. The mobile app platform revenue comes from two sources: selling the paid apps, or advertising on the freemium apps. As a result, for the app-store platform, the consumers' adoption of the mobile apps represents a critical problem.

The app-store platform has a lot of information about the consumers' download behavior, enabling it to customize its marketing actions to target different consumers' based on their different behaviors. For example a mobile app platform should decide between the free trial and the viral referral strategies. A viral referral strategy can be useful if consumers' preferences are interrelated, because of the psychological benefits of social identifications/learning/inclusion or the utilitarian benefits of the network externalities. However, a trial strategy is useful if consumers' have a learning cost or an uncertainty about the mobile apps.

It is not uncommon for customers to have interrelated preference for mobile apps. Online forums are filled with questions about requests for mobile app recommendation, [1] and, in fact, app-stores try to inform users about the popularity of mobile apps. The interdependence of mobile app choices is not only relevant for online world, but also for offline world. It is hard for customers to know what mobile app they want, so they find new mobile apps from family friends and colleagues. App-stores have tried to facilitate this process by creating "Tell a Friend" and "Share This Application"[2]. Therefore, an app-store platform needs a framework to quantify not only the effect of mobile app characteristics, but also the effect of online and offline social influence on customer choices to design policies to affect mobile app choices of its customers.

---

[1] "Mobile Applications Forum." CNET. Accessed April 02, 2016. http://www.cnet.com/forums/mobile-apps/.

[2] WonderHowTo. "How to Share Your Favorite Mobile Apps with Your Friends." Business Insider. June 16, 2011. Accessed April 02, 2016. http://www.businessinsider.com/how-to-share-your-favorite-mobile-apps-with-your-friends-2011-6

Given this context, I asked the following questions: (1) How can I design a targeting approach for an app-store platform? (2) How does the social learning process of the mobile apps' customers differ from that of the classical economy products, such as a color TV? (3) How can an app-store platform capture the heterogeneity of its customers and the variation in the mobile apps to customize its marketing actions? (4) What are the key elements of the consumers' utility of adopting a mobile app that allows an app-store platform to group and target its potential customers?

To answer these questions, I combined a macro social learning diffusion model with a micro choice model. I used a choice model to study the adoption behavior of the consumers. To control for social influence, I applied a filtering technique (i.e., Unscented Kalman Filter) on another aggregated data set to create a time varying measure of social influence. Also, to control for mobile app characteristics, price, and advertising, I used a factor model. I ran the filtering technique on two aggregate adoption data sets for approximately two hundred days. These data sets include, on the one hand, the cumulative number of adopters within a local city in Africa, and on the other hand, the cumulative number of adopters across all thirty cities in which the platform under the current study globally operates. I refer to these two data sets as the aggregate data sets from now on. I ran the choice model and the factor model on a data set of a sample of choices of one hundred forty seven consumers over twenty weeks. I refer to this sample as micro sample.

I used a social learning diffusion model of Van den Bulte and Joshi (2007) to model the simultaneous diffusion of the mobile apps on the app store. Such a modeling approach presents two challenges. The first concerns mobile app consumers' choice sparse data, because the download of a mobile app is a rare event. To address this challenge, I aggregated the data at an app-category level. The second challenge involves dealing with the possible measurement error. For this purpose, I cast the Van den Bulte and Joshi (2007) model into a discrete time state space model. The use of Gaussian Process to filter the measurement error is quite common in online mission critical systems such as robotics as well. In this case, I had to filter two double-degree polynomial differential equations of each mobile app category's diffusion. I used an Unscented Klaman Filter (UKF), an approach introduced to Machine Learning and Robotics to estimate the non-linear diffusion equation up to third order precision (Julier and Uhlman 1997; Wan and van der Merve 2001). This approach is an alternative to Extended Kalman Filter (EKF) which estimates the non-linear diffusion equation only up to first order precision.

I further used a hierarchical prior with a seemingly unrelated regression (SUR) model to use the shared information in the simultaneous diffusion of the mobile app categories, and to avoid the over fitting of the model with three hundred macro parameters. To estimate the macro diffusion model in the short planning time horizon of an app-store platform, I used a Monte Carlo Expectation Maximization (MCEM) approach to optimize the Maximum A Posteriori (MAP) of the parameters, in contrast to a possibly slow convergence Bayesian sampling algorithms, such as Gibbs and Metropolis Hastings. To deal with the problem of the stochastic surface search of the MCEM approach, I used a genetic optimization algorithm, with an initial population that is a perturbed version of the estimates found in Van den Bulte and Joshi (2007) study.

Next, I used the outcome of the macro diffusion model as a measure for social influence in the structural choice model to extract factors of customers' mobile app choices. The choice of a mobile app adoption is very sparse over time. In other words, I expected to observe several zeros in the data. To deal with such sparsity and to filter the possible noise of the data, I aggregated the data on the characteristics of the mobile app categories, and the cumulative number of imitators at a weekly level. Further, to not discard the multi-collinear data on the mobile app-characteristics, I used a factor model to recover the underlying factors of the mobile apps profiles. To name these factors, I merge the factor loading profiles and practitioners' knowledge of customers' mobile app choices.

Given the mobile app-category latent factors, the density of the imitators, and the download history of the app-store platforms, I used a mixture normal multinomial logit model to represent each consumer's choice of mobile app-adoption. I estimated this model by MCMC sampler. The hierarchical modeling and the weighting scheme I used make the approach appropriate for the big data, because the mixture normal prior allows for flexible structure that yet may not over fit. This modeling approach is appropriate for the context of online retailers, in which the distribution of choices follows long tail distribution (Anderson 2006).

I estimated this model over a data set of a newly launched app-store in Africa during May 2013 and a supplementary dataset of network location of the mobile app-store users scraped from web. The sample consists of mobile app choices of approximately 20,000 customers that reside in 30 cities that the app store is available, among which approximately 1,000 resides in a city in Africa. Mobile apps belong to various categories among which I selected 10 categories (presented in table 1.3) that were less sparse. The estimation results show that, social influence significantly affects customer app-adoption choices, and I find that social influence at offline world (within the city) explains the customer choices better than social influence at online world (within the 30 cities that app-store performs). I also find that not controlling for social influence in mobile app choices of customers results in biased customer preference estimates. Furthermore, I find that among many different classical goods, mobile app adoption pattern is very similar to music CDs adoption pattern.

I further used the estimated micro choice model to simulate a counterfactual policy that intervenes in social influence to affect consumers' choices. I find a policy that increases mobile app diffusion by 13.6%. This step is a form of prescriptive analytics that I built over the descriptive and the predictive analytics steps. Furthermore, I find individual specific preference parameters estimated by the choice model, which can help the mobile app-store target its customers.

The current study is mostly related to studies on consumers' peer effect by Yang and Allenby (2003), Stephen and Toubia (2010), Lehmmes and Croux (2006), and Nair et al. (2010). Also, it is related to studies on the global macro diffusion by Van den Bult and Joshi (2007), Putsis et al. (1997), and Dekimpe et al. (1997). Another relevant research stream includes studies on micro diffusion models by Dover et al. (2012), Chaterjee and Eliashberg (1990), and Young (2009). The last stream of relevant studies includes studies on the app store platform by Ghose and Han (2014), Carare (2012), Garg and Telang (2013), Liu et al. (2012), Ghose et al. (2011), Ghose and Han (2011b), Ghose and Han (2011a), and Kim et al. (2008). Although these studies have contributed greatly to the understanding of the phenomenon, none has created a pipeline which combines the macro diffusion modeling and the micro structural choice modeling approaches to allow the app-store platform to target its consumers. The proposed approach allows the app-store to target its customers by applying the descriptive, predictive, and prescriptive analytics over a high volume, high velocity, high variety, and high veracity big data.

Thus, this paper, contributes to the emerging literature on the prescriptive data analytics of the mobile app-store platform in three ways. First, it introduces the combination of macro simultaneous social learning adoption model and micro structural choice modeling approaches to design a method that allows the app-store platforms to target their heterogeneous consumers, using their big data. Second, this paper benchmarks the parameters of social learning mobile app adoption against those of classical economy goods such as the color-TV, personal computer, music CD, and radio-head. Third, this paper shows that social influence at offline (local city level) drives mobile app choices of customers on the app-store, and ignoring social learning process creates biased estimates. Fourth, this paper shows the power of its proposed model for prescriptive analytics over the big data, by finding an optimal viral marketing policy (e.g., share with friends and family) for the app-store that can increase its total expected diffusion by 13.6%. Last but not least, to estimate the proposed social learning model, this paper employs SUR, UKF, MCEM, and genetic algorithm to maximize the MAP estimate of the macro diffusion model. In addition, it uses a hierarchical mixture normal prior over its multinomial logit choice model, estimating it using MCMC sampling method. These approaches that allow for a flexible heterogeneity pattern and for a robust filtering of process and measurement errors, as well as computational feasibility of big data analytics, should be of interest to academia and a number of commercial entities interested in not only the descriptive and predictive, but also the prescriptive analytics of their big data.

## 1.3.LITERATURE REVIEW

This study draws upon several streams investigated, within the literature: (1) the interdependence of consumer preference; (2) mobile app store dynamics; and (3) global macro and micro diffusion and social learning. Given the breadth of these areas across multiple disciplines, the following discussion represents only a brief review of these relevant streams. Table 1.1 presents a summary of the position of this study in the literature.

Table 1.1. Literature Position of this study

| Stream of Study | Interdependence of consumer preference | App Store | Global micro/macro Simultaneous Diffusion |
|---|---|---|---|
| Current study | * | * | * |
| Yang and Allenby (2003); Stephen and Toubia (2010); Lehmmes and Croux (2006); Bell and Song (2007); Aral and Walker (2011); Nair et al. (2010); Bradlow et al. (2005); Hartmann (2010); Yang et al. (2005); Narayan et al. (2011); Kurt et al. (2011); Chung and Rao (2012); Choi et al. (2010). | * | - | - |
| Ghose and Han (2014); Carare (2012); Garg and Telang (2013); Liu et al. (2012); Ghose et al. (2011); Ghose and Han (2011b); Ghose and Han (2011a); Kim et al. (2008). | - | * | - |
| Van den Bulte and Joshi (2007); Yong (2009); Chatterjee and Eliashberg (1990); Putsis et al. (1997); Dekimpe et al. (2000); Neelamegham and Chintagunta (1999); Talukdar et al. (2002); Gatignon et al. (1989); Takada and Jain (1991); Dover et al. (2012). | - | - | * |

### 1.3.1. Interdependence of consumer preference

Quantitative models of consumer purchase behavior often do not recognize that consumers' choices may be driven by the underlying social learning processes within the population. Economic models of choice typically assume that an individual's latent utility is a function of the brand and attribute preferences, rather than the preferences of the other customers. However, for pervasive experience goods, a new model which accounts for these underlying forces and preferences may better explain consumers' choices. Many studies have tried to address this issue, using cross sectional data to model the consumers' preference dependency (Yang and Allenby 2003), online social network seller interaction data to quantify the network value of the consumers (Stephen and Toubia 2010), the customer trials data at Netgroceer.com to determine the importance of its consumers' spatial exposure (Bell and Song 2007), and physician's prescription choices and their self-reported information to demonstrate the significant effect of network influence on consumers' choices (Nair et al. 2010).

Other researchers have also reported on the critical role played by social proximity in shaping consumer preferences. Bradlow et al. (2005) builds on the previous literature to suggest that the demographic and the psychometric proximity measures are important for consumers' choice. Hartmann (2010) uses customer data to show a correlation between social interactions and the equilibrium outcome of an empirical discrete game. Yang et al. (2005) demonstrate the interdependence of spouses' TV viewership to suggest the need for considering choice interdependency. Narayan et al. (2011) employ conjoint experience data to highlight the effects of peer influences, and finally Choi et al. (2010) draw from an internet retailer's dataset to establish the importance of imitation effects in a geographical and a demographical proximity. However, although all studies are significant in suggesting the role of social influence on decision making, none has modeled consumers' mobile apps choices' interdependence.

### 1.3.2. Mobile app store dynamics

Recently, a stream of literature has emerged that pertains to the dynamics of mobile app store. Some studies have addressed Apple and the Google platforms' competition (Ghose and Han 2014), Google play's fermium strategy (Liu et al. 2015), and Apple's app-store's bestseller rank information influence on sales (Carare 2012; Garg and Telang 2013). Other studies consider the relation between the content generation/consumption (Ghose and Han 2011b), the internet usage and mobile internet characteristics (Ghose and Han 2011a), users' browsing behavior on mobile phones and personal computers (Ghose et al. 2011), and voice and short message price elasticity (Kim et al. 2008).Although these studies are represent attempts to teach us more about the nature of the mobile app-market, none has extracted the effect of social dynamics on the consumers' choices in the context of the mobile app-store, at both the macro and micro levels.

### 1.3.3. Global Macro and Micro Diffusion and Social Learning

Two main streams of literature in product diffusion are relevant to this study: the micro diffusion models, and the global diffusion and social learning models. The earliest micro diffusion model considers consumers' Bayesian learning from the signals that follow a Poisson process (Chaterjee and Eliashberg 1990). Later studies emphasize the need for micro-diffusion modeling (Young 2009), and critically review the aggregation and homogeneity of diffusion models (Peres et al. 2010). To remedy the issues, some studies proposed micro network topology approaches (Iyengar and Van den Bulte 2011; Dover et al. 2012). Other studies suggest structural modeling of consumers' dynamic-forward looking adoption choices (Song and Chintagunta 2003), and systematic conditioning to heterogeneous consumer's adoption choices (Trusov et al. 2013). Peres et al. (2010) presents a review of this literature stream.

Parallel with the micro diffusion literature, a stream of studies provide solutions for heterogeneous social learning process (Van den Bulte and Joshi 2007), the mixing (interactions) of adoption process (Putsis et al. 1997), simultaneous diffusion (Dekimpe et al. 1998), supply-side relationship (e.g., production economies) and omitted variables (e.g., income) correlations (Putsis and Srinivasan 2000), and the effect of macro-environmental variables (Talukdar et al. 2002).

This study builds on this literature, by proposing a prescriptive machine learning pipeline that combines the advantages of both macro and micro modeling approaches. The approach that I have proposed recognize that the app-store platform's data may be a noisy measure of the variables of interest. I deal with the sparsity of the choices through a combination of aggregation, filtering, hierarchy, and SUR processes. I suggest a data cleaning and modeling approaches that may be suitable for the big data variety, velocity, veracity, and volume of the app-store platform. To estimate the model, I also suggest a genetic optimization meta-heuristic approach, which enables the stochastic surface optimization.

## 1.4. MODEL

I start the modeling section with the choice of individuals $(i = 1,...,I)$ at the app-store. I am interested to model consumer's mobile app choices. However, to recognize the long tail distribution of mobile app choices (which creates sparsity), I aggregated the choice data at app category level. The customer makes a choice of mobile app category j $(j = 0,1,...,J)$ at a given week $(t = 0,1,...,T)$, where $j = 0$ denotes the outside good option. The model of consumer app choice is different from the prior studies (Carare 2012) in not modeling aggregate purchases, but modeling individual specific choices, through a rich set of mobile app category characteristics.

The model is similar to nested logit model structure of studies such as Ratchford (1982) and Kok and Xu (2011), yet to recognize sparsity of end choices, I aggregated the choices within the nest as the choice of the nests. This model may be useful for mobile app-store owners, because they concern about the diffusion of mobile apps within the categories rather than the diffusion of each instances of mobile app in seclusion.

I specified the utility of consumers' choice of app categories on the app store in the following form:

$$u_{ijt} = \alpha_{ij} + \alpha_{i11} s_{it} + \alpha_{i12} \hat{c}_{jt}^{imm} + \alpha_{i13} F_{1jt} + \alpha_{i14} F_{2jt} + \alpha_{i15} F_{3jt} + \varepsilon_{it}^{j} \quad (1)$$

where $\alpha_{ij}$ denotes the random coefficient of individual i's preference for mobile app category j. $F_{1jt}$, $F_{2jt}$ and $F_{3jt}$ denote factors that control for variation in observable mobile app characteristics/quality, price, and advertising (the structure of factors are explained later). $\hat{c}_{jt}^{imm}$ denotes time varying social influence measure (the structure of the measure is explained later), and $s_{it}$ denotes history of consumers i's app downloads until time t, which controls for state dependence and app-choice interdependence. Particularly, if consumer i downloads an app at t-1, then $s_{it} = s_{it-1} + 1$, otherwise if the consumer selects outside option, then $s_{it}$ remains unchanged: i.e., $s_{it} = s_{it-1}$. This specification induces a first-order Markov process on the choices. Controlling for state dependence and social influence helps to consider the potential correlation between customers' choices across the categories and across the individuals. Table 1.2 presents the definition of the variables and the parameters.

Assuming the random utility term $\varepsilon_{it}^{j}$ has type I extreme value distribution, consumer's i's probability of selecting the app category j at time t is given by a multinomial logit model, based on the deterministic portion $v_{ijt}$ of random utility $u_{ijt}$ as follows:

$$p_{ijt} = \frac{\exp(v_{ijt})}{1 + \sum_{j=1}^{J} \exp(v_{ijt})} \tag{2}$$

where the mean utility of outside good is set to zero, i.e., $v_{it}^{0} = 0$.

Vector of mobile app category characteristic includes average file size of mobile apps (a proxy for the app quality), frequency of featuring, average and variance of mobile app prices, the number of paid or free apps and their ratio, and the average tenure (time since creation) of all the mobile apps within the category. These variables can act as measures of (proxy for) competition. I assumed each of these pieces of the data contains some information that may be important for the consumer, but these pieces are highly correlated. Therefore, to get a better insight, I reduced the variation in these variables into three factors that preserve 85% of the variation. Formally, I used the following factor model process:

$$x_{jt} = bF_{jt} + e_{jt}^{'}, e_{jt}^{k'} \sim N(0, E) \tag{3}$$

To model consumer social learning, I used filtered latent time varying density of imitators $\widehat{c}_{jt}^{imm}$.

This approach is similar to the classical practice of modeling consumer's response to featured and display products, in which the modeler includes an aggregate measure into the choice model to measure the consumers' response. Furthermore, the theoretical interpretation of this modeling approach is that as the number of imitators within the population increases, the possibility that an individual observes another individual who has already adopted the mobile app increases. As a result, the consumer may become more or less likely to adopt a mobile app within mobile app category j. This theory is similar by micro modeling diffusion proposed by Chatterjee and Eliashberg (1990), except that the model does not assume that consumer receives information with a Poisson process, so the process can be a non-homogeneous Poisson process (inter-arrival time is not memory-less anymore). In other word, I endogenize consumers' information receiving process in the choice model. The approach serves as an alternative to the micro-modeling approach used by Yang and Allenby (2003) to incorporate interdependence of awareness and preferences of consumers, but this model is useful when micro spatial structure information is not available. My proposed approach may be relevant to the context of pervasive goods, because these goods are more visible in daily interactions.

There are two approaches to capture the density of imitators in the model. The first approach is to model it as a latent state variable, and recover it from the choice model. Although fancy, this approach may not be the best approach over big data, because it is computationally intractable. The alternative approach is to use the aggregate diffusion data to filter the number of imitators. This approach combines macro aggregate diffusion modeling with micro choice modeling methods, to endognize the number of imitators, an approach that may be more suitable for the big data. In this approach, I can use an aggregate diffusion data, to filter the number of imitators with two degree polynomial linear model. Then, I can use the filtered data in the choice model, to run a nonlinear model on the data set of individual choice of consumers.

To sum up, I used the whole dataset to filter the density of influential and imitators for the mobile app category j within the population at the given time t. I casted the social-learning diffusion differential-equations (Van den Bulte and Joshi 2007) into a discrete state space model.

This model is like a double barrel Bass diffusion model, and allows for heterogeneity in the adopters, by segmenting the observed cumulative number of adopters into the latent number of imitators and influentials. In contrast to classical log likelihood and non-linear least square methods, my filtering approach increases estimation robustness to process and measurement noises (Srinivasan 1999; Xie et al. 1997).

$$y_{jt} = \theta_j c_{jt}^{Inf} + (1-\theta_j) c_{jt}^{imm} + v_{jt}, v_{jt} \sim N(0, V_j)$$
$$\dot{c}_{jt}^{Inf} = (p_j^{\inf} + q_j^{\inf}(\frac{c_{jt-1}^{\inf}}{M_j^{\inf}}))(M_j^{\inf} - c_{jt-1}^{\inf}) + e_{jt}^{\inf}, e_{jt} \sim MVN(0, W)$$
$$\dot{c}_{jt}^{imm} = (p_j^{imm} + q_j^{imm}(w_j(\frac{c_{jt-1}^{\inf}}{M_j^{\inf}}) + (1-w_j)(\frac{c_{jt-1}^{imm}}{M_j^{imm}})))(M_j^{imm} - c_{jt-1}^{imm}) + e_{jt}^{imm} \quad (4)$$

where $y_{jt}$ denotes the observed cumulative number of adopters of mobile apps in the mobile app category j at time (day) t. $c_{jt}^{\inf}$ denotes the latent cumulative number of adopters in influential segment for app category j at time (day) t. $c_{jt}^{imm}$ denotes the latent cumulative number of adopters in imitator segment for app category j at time(day) t. $\theta_j$ denotes the size of the segment of influential adopters, and it is bound between zero and one. $p_j^{\inf}$ denotes independent (random) rate of adoption of influential adopters, and $q_j^{\inf}$ denotes the dependent (influenced by other influential adopters) rate of adoption of influential adopters. $p_j^{imm}$ denotes independent (random) rate of adoption of imitator adopters, and $q_j^{\inf}$ denotes the dependent (influenced by other adopters) rate of adoption of imitator adopters. $w_j$ denotes the degree of influence of influential adopters on the adoption of imitators. $v_{jt}$ denotes the noise of observation equation, and $(e_{jt}^{\inf}, e_{jt}^{imm})$ denotes the vector of noises of state equations.

In summary the first equation denotes the observation equation and the second two the state equations of the state space model. The first equation uses a discrete latent model to integrate over the cumulative number of influential and imitator adopters. The second equation captures the adoption process of influential adopters segment, and the third captures the adoption process of imitator adopters segment. The imitators are different behaviorally from the individuals in influential segment, in that they learn not only from themselves, but also from individuals in influential segment.

This model of social influence measure is more suitable for the context of mobile apps, as it captures more social learning process (Van den Bulte and Joshi 2007) than information cascade process (Bass 1969). Furthermore, it allows for heterogeneity in the adoption process, by segmenting the adopters into influential and imitator segments. Van den Bulte and Joshi (2007) find a closed form solution for this model. I considered that the data may have measured with noise. As a result, to control for this potential measurement error, I use a state space model structure with observation and state noises.

I recognized that there is shared information in the diffusion of various mobile app categories on the app store. As a result, I modeled these differential equations of social learning across mobile app categories jointly and simultaneously. This joint modeling captures shared information at two levels: covariance and prior.

To account for the simultaneity, on the covariance level, I modeled the state variance of the latent measure of cumulative influential and imitator adopters, and the variance of state equation of cumulative influential and imitator adopters through a seemingly unrelated regression (SUR) model. The SUR model is presented formally in (4) as modeling the joint distribution of the state equations in a multivariate normal model structure, rather than modeling the state equation error terms individually.

To jointly model the diffusions, I used a hierarchical model (prior) with conditionally normal distribution constraint on the fixed app-category specific diffusion parameters, which is $\Phi_j = (p_j^{\inf}, p_j^{imm}, q_j^{\inf}, q_j^{imm}, M_j^{\inf}, M_j^{imm}, \theta_j, w_j)$. This Bayesian process shrinks the fixed app-category specific parameters toward the popularity of each mobile app, because it is expected that more popular mobile apps have higher rate of imitator adoptions and market size. Formally, I defined the following structure:

$$\Phi_j = \Delta_o Pop_j + o_j, o_j \sim N(0, \sigma_o^2) \tag{5}$$

where $\Phi_j$ denotes vector of non-state (fixed) parameters of the diffusion. $Pop_j$ denotes the popularity of mobile app category j. $\Delta_o$ denotes the hyper parameter of app category specific parameter shrinkage, and $o_j$ denotes the noise of the hierarchical model, or the unobserved heterogeneity of the mobile app categories.

I accounted for heterogeneity in the individual choice parameters by modeling the choices' parameters random effects. To consider the possibility of misspecification that may result from rigid normal prior, I adopted the flexible semi-parametric approach proposed by Dube et al. (2010). This approach assumes a mixture of multivariate normal distributions over the parameters' prior, to allow for thick tail skewed multimodal distribution. I denote the vector of fixed consumer-level parameters by $\mathrm{A}_i = (\alpha_{i1}, \alpha_{i2}, .., \alpha_{i15})$. I accommodated consumer heterogeneity by assuming that $\mathrm{A}_i$ is drawn from a distribution common across consumers, in two stages. I employed a mixture of normal as the first stage prior, to specify an informative prior that also does not overfit. The first stage consists of a mixture of $K$ multivariate normal distribution, and the second stage consists of prior on the parameters of the mixture of normal density, formally:

$$\begin{aligned} & p(A_i - \Delta z_i \mid \pi, \{\mu_k, \Sigma_k\}) = \sum_{k=1}^{K} \pi_k \phi(A_i - \Delta z_i \mid \mu_k, \Sigma_k) \\ & \pi, \{\mu_k, \Sigma_k\} \mid b \end{aligned} \tag{6}$$

where $b$ denotes the hyper-parameter for the priors on the mixing probabilities and the parameters governing each mixture component. $K$ denotes the number of mixture components. $\{\mu_k, \Sigma_k\}$ denotes mean and covariance matrix of the distribution of individual specific parameter vector $A_i$ for mixture component k. $\pi_k$ denotes the size of the $k'th$ component of mixture model, and $\phi$ denotes the normal density function distribution. $z_i$ denotes information set about customer i, which here only includes only the tenure (the number of days from customer i's registration on the app store). $\Delta$ denotes the parameter of correlation between choice response parameter and information set about customer i.

To obtain a truly non-parametric estimate using the mixture of normal model it is required that the number of mixture components $K$ increase with the sample size. I adopted the approach proposed by Rossi (2014), called non-parametric Bayesian approach. This approach is equivalent to the approach mentioned above when $K$ tends to infinity. In this structure, the parameters of mixture normal model have Dirichlet Process (DP) prior. Dirichlet process is the generalization of Dirichlet distribution for infinite atomic number of partitions. This process represents the distribution of a random measure (i.e., probability). Dirichlet process has two parameters, the first is the base distribution, which is the prior distribution on the parameters of the multivariate

Normal-Inverse Wishart (N-IW) conjugate prior distribution for the distribution for the partitions that the choice parameters are drawn from, and the second parameter is the concentration parameter. Formally, the prior for the individual specific choice parameters has the following structure:

$$\begin{aligned} &\theta_{k1} = (\mu_{k1}, \Sigma_{k1}) \sim DP(G0(\lambda), \alpha^{d}) \\ &G0(\lambda) : \mu_{k1} \mid \Sigma_{k1} \sim N(0, \Sigma_{k1} a^{-1}), \Sigma_{k1} \sim IW(\nu, \nu \times \upsilon \times I) \\ &\lambda(a, \nu, \upsilon) : a \sim Unif(\vec{a}, \overleftarrow{a}), \nu \sim d - 1 + \exp(z), z \sim Unif(d - 1 + \vec{\nu}, \overleftarrow{\nu}), \upsilon \sim Unif(\vec{\upsilon}, \overleftarrow{\upsilon}) \\ &\alpha^{d} \sim (1 - (\alpha - \vec{\alpha}) / (\overleftarrow{\alpha} - \vec{\alpha}))^{power} \end{aligned} \tag{7}$$

where $G0(\lambda)$ denotes the base distribution or measure (i.e., the distribution of hyper-parameters of the prior distribution of the partitions). $\lambda$ denotes the random measure, which represents the probability distribution of $(a, \nu, \upsilon)$. $(a, \nu, \upsilon)$ denotes the hyper-parameters of the prior distribution of the partitions that the choice parameters belong to, which represent the behavior parameters of the latent segments. $d$ denotes the number of choice parameters per customer (in my case d is equal to 15). $\alpha^{d}$ denotes the concentration (also referred to as precision, tightness, or innovation) parameter. The idea is that DP is centered over the base measure $G0(\lambda)$ with N-IW with precision parameter $\alpha^{d}$ (larger value denotes tight distribution). $(\vec{a}, \overleftarrow{a}, \vec{\nu}, \overleftarrow{\nu}, \vec{\upsilon}, \overleftarrow{\upsilon})$ denotes the hyper parameters vector for the second level prior on hyper parameters of prior over the partitions distribution of the choice parameters.

Dirichlet Process Mixture (DPM) is referred to the distribution over the probability measure defined on some sigma-algebra (collection of subsets) of space $\aleph$, such that the distribution for any finite partition of $\aleph$ is Dirichlet distribution (Rossi 2014). In my case, the probability measure over the partitions for mean and variance of random coefficient response parameters of individual choice parameters sigma-algebra has the Normal-Inverse-Wishart conjugate probability. For any subset of customers $C$ of $\aleph$ :

$$E[G(C_\lambda)] = G_0(A_\lambda)$$
$$Var(G(C_\lambda)) = \frac{G_0(A_\lambda)(1-G_0(A_\lambda))}{\alpha^d+1} \tag{8}$$

By De Finetti theorem, integrating (marginalizing) out the random measure $G$ results in the joint distribution for the collection of individual specific mean and covariance of random coefficient choice parameters as follows:

$$p(\mu_.,\Sigma_.) = \int p(\mu_.,\Sigma_. \mid G)p(G)dG \tag{9}$$

This joint distribution can be represented as a sequence of conditional distributions that has exchangeability property:

$$p((\mu_1,\Sigma_1),...,(\mu_I,\Sigma_I)) = p((\mu_{1.},\Sigma_1))p((\mu_2,\Sigma_2) \mid (\mu_1,\Sigma_1))...p((\mu_I,\Sigma_I) \mid (\mu_1,\Sigma_1),...,(\mu_{n-1.},\Sigma_{n-1})) \tag{10}$$

The DP process is similar in nature to Chinese Restaurant Process (CRP) and Polya Urn. In the CRP, there is a restaurant with infinite number of tables (analogous to partitions of mean and variance of the individual choice random coefficients). A customer entering the restaurant selects the tables randomly, but he selects the table with probability proportional to the number of customers that have sat on the table so far (in which case the customer behaves similar to the other customers who are sitting at the selected table). If the customer selects a new table, he will behave based on a parameter that he randomly selects from restaurant customer behavior parameters (so not necessarily identical to the parameters of the other tables).

Table 1.2. Model Variable Definitions

| Variable | Description |
|---|---|
| App Category Daily Download( $y_{jt}$ ) | Cumulative number of consumers who download an app in app category j up until a given day t |
| App Category Weekly Download Latent ( $c_{jt}^{\inf}$ ) | Latent cumulative number of consumers from influential segment, who download an app in app category j up until a given day t. Consumers from influential segment only learn from each other, and not from imitators. |
| App Category Weekly Download Latent ( $c_{jt}^{imm}$ ) | Latent cumulative numbers of consumers from imitator segment, who download an app in app category j up until a given day t. Consumers from imitator segment learn both from each other, and adopters in influential segment. |
| Segment size ( $\theta_j$ ) | A parameter between zero and one that define the size of the influential segment |
| Internal Market Force ( $p_j^{\inf}$ , $p_j^{imm}$ ) | The random Poisson rate of adoption of individuals in influential and imitator segment respectively. |
| External Market Force ( $q_j^{\inf}$ , $q_j^{imm}$ ) | The endogenized imitation rate of adoption of individuals in influential and imitator segment respectively. |
| Learning split ( $w_j$ ) | The degree to which the individuals in imitator segment learn from adopters in the influential segment |
| Market size ( $M_j^{\inf}, M_j^{imm}$ ) | The market size of individual in influential and imitator segments respectively. |
| Category hierarchy parameters $\{\mu_{k2}, \Sigma_{k2}\}$ | Parameter of locally weighted regression parameters of the hierarchical prior of app category diffusion parameters |
| Full covariance matrix of state equation( $W$ ) | Full covariance matrix of state equation of macro diffusion model, which may suggest complementarity or substitution. |
| Variance of observation equation ( $V_j$ ) | Variance of observation equation of macro diffusion model |
| Category data ( $x_{jt}$ ) | Category j characteristic data at day t, including Average file size, total number of adds featured in the category, average price, variance of price, paid app options, free app options, fraction of free to paid apps within the category, average tenure of each app category, total app options within the category |
| Category Factors( $F_{jt}$ ) | Reduced factors explaining the variation in category data |
| Factor loading of Category $(b)$ | Factor loading of data item j of category data vector |
| Consumer utility from app category ( $u_{ijt}$ ) | Consumer i's utility from selecting an app in app category j at week t |
| App category preference ( $\alpha_{ij}$ ) | App category specific preference of consumer i |
| Individual download history state ( $s_{it}^j$ ) | State of individual i's download history in a given category j until week t |
| $\alpha_{i11}...\alpha_{i15}$ | Utility parameters of consumer i |
| $p_{ijt}$ | Probability of selecting an app in category j at time t |
| $\pi_1, \{\mu_{k1}, \Sigma_{k1}\}$ | Parameter of hierarchical mixture of normal components of individual choice parameters |
| $v_{jt}, e_{jt}, e'_{jt}$ | Error terms of observation/state equation and factor model |

The Polya Urn process has also the same structure. In this process, the experimenter starts by drawing balls with different colors from the urn. Any time the experimenter has a ball with a given color drawn from the urn, he will add an additional ball with the same color to the urn, and he also returns the drawn ball. The distribution of number of customers sitting at each table in CRP and number of balls in each color in Polya Urn follow DP.

An alternative way is the approach proposed by Dube et al. (2010) to fit models with successively large numbers of components and to gauge the adequacy of the number of components by examining the fitted density associated with the selected number of components. However, the process of model selection is tedious in this case.

Table 1.2 presents the definition of variable and parameters. To sum up, I used the combination of macro diffusion model and micro choice model that considers the big data nature of the current study: variety, velocity, veracity, and volume. On the variety aspect, I used a flexible semi parametric mixture of normal distribution as prior on the individual choice model. For velocity, I used a simpler linear state space model on the daily data over the full sample, and I aggregated this data at weekly level to use it in the micro individual choice non-linear model. For volume aspect of the data, I considered sparsity nature, so I aggregated both macro diffusion and micro individual choice data for mobile app instances within the mobile app categories, and I used a factor model to summarize the sparse characteristics of the mobile app categories. Finally, for veracity, I casted the social diffusion model into a discrete time state space model to add a layer of robustness to the potential misspecification and process errors. Figure 1.1 presents the box and arrow diagram of the proposed model.

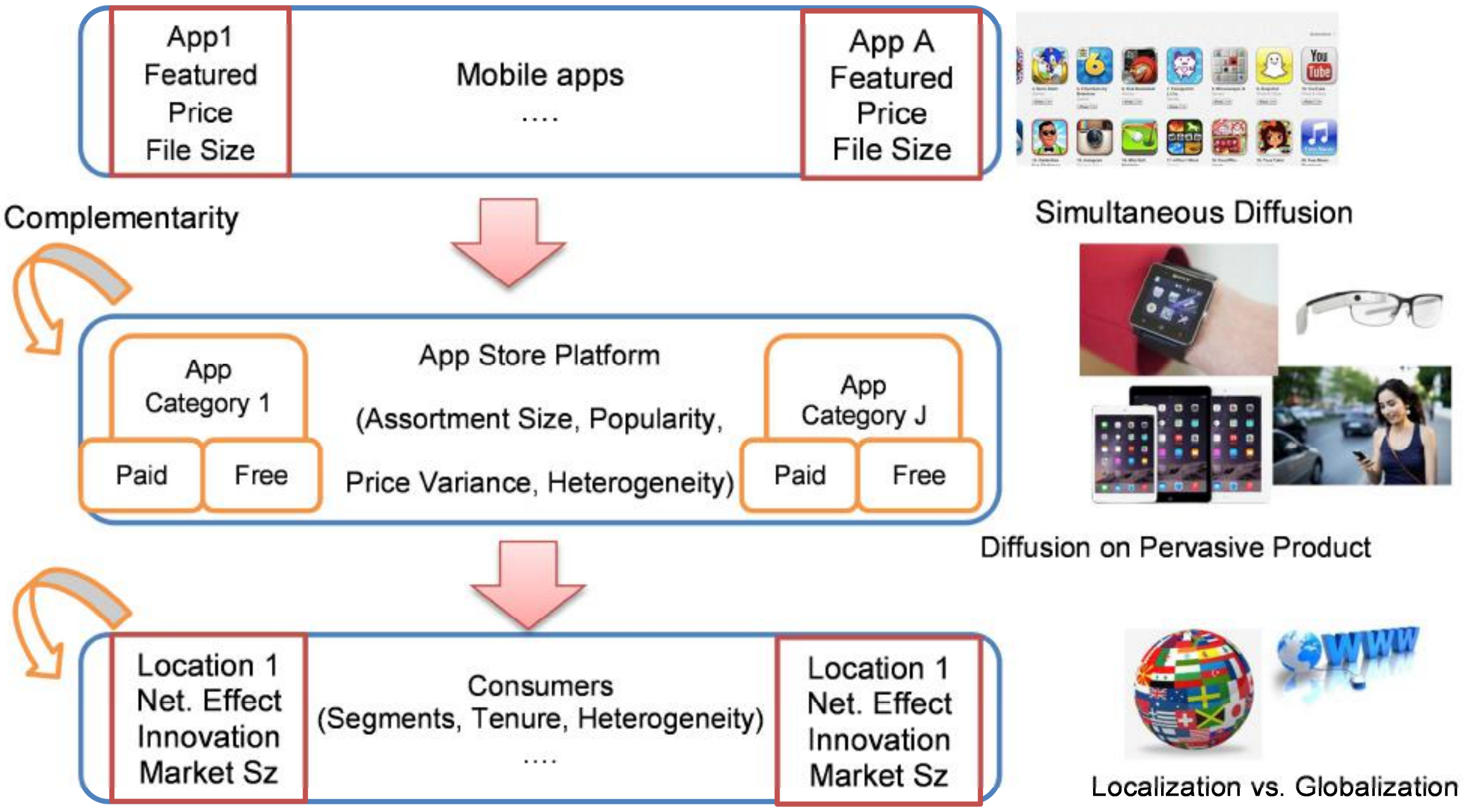

Figure 1.1. Box and Arrow Representation of the Model

**1.5.DATA**

The data set were collected by an African telecom operator on individual choices of downloading mobile apps from the app store platform of its global partner. The app-stores are a type of two sided platform, as they match consumers' and developers/publishers without taking the ownership of the mobile apps. The app-store I studied is launched within around 330 days prior to the current study in 2013 and 2014. I used the aggregate download data for a period of around 190 to 259 days as the macro sample, and the data on download choices of a sample of 1,258 consumers for a period of 124 days as the micro sample. The macro sample therefore includes between 1,900 to 2,590 observations, which might not be considered big, but the small sample includes approximately 160,000 observations, which might be considered big for non-linear models.

A big data set such as ours creates a trade-off in estimation. On one side, I had a big data that can give insight in a short planning horizon, given that I used a linear model. On the other hand, I had computationally intensive methods that can give insights with prescriptive power, given that the data is not big. I wanted to have a method that gives us the advantage of both big data and a computationally intensive method. As a result, I used a second degree polynomial macro-model of social learning diffusion over the macro sample and the non-linear computationally intensive micro choice model over the micro sample.

To deal with the sparsity of the data, which is driven by the long tail distribution of the mobile apps' adoptions, and to reduce the daily noise in the data, I aggregated the data of the micro sample at weekly level before I fed it to the choice model. Second, I aggregated the macro app adoption, and micro app download choice data at app category level to limit the study to the topic of interest for the app-store platform, as well as to handling the data volume. In addition, I used a flexible Bayesian prior to shrink the individual specific choice parameters. I investigated two sources of consumer preference interdependence: local and global. For the local interdependence, I filtered the macro sample data to individual adopters who live in a city under the current study. To do this filtering, I used a data set of mapping IP addresses to cities that I collected by crawling World Wide Web. For the global interdependence, I did not use this filtering, so I used the aggregate information about the mobile app adoptions within all thirty cities from all five continents.

Table 1.3. Categories Basic Statistics

| Index | Category | Total Downloads within local city |
|---|---|---|
| 1 | Dating | 27 |
| 2 | eBook | 414 |
| 3 | Education & Learning | 24 |
| 4 | Health/Diet/Fitness | 42 |
| 5 | Internet & WAP | 52 |
| 6 | Movie/Trailer | 597 |
| 7 | POI/Guides | 22 |
| 8 | Reference/Dictionaries | 55 |
| 9 | TV/Shows | 135 |
| 10 | Video & TV | 105 |

In a nutshell, the data consists of around 20,000 consumers, with around 3,000 consumers in a local African city under the current study. This local city has around 4,000 app downloads for the duration of the current study. Twenty thousand global and three thousand local consumers' who make choices for a course of six month classifies the current data as a big one, for its variety, velocity, veracity, and volume. Table 1.3 illustrates the list of the categories that I selected and their corresponding total downloads within the local city under this study. Each of the 1,258 customers under the study adopts only one of the mobile apps during the course of study, so on all other days she selects outside option. This observation may suggest that a mixed logit choice model might be a suitable model, only if that an inter-temporal dependencies between the choices are controlled.

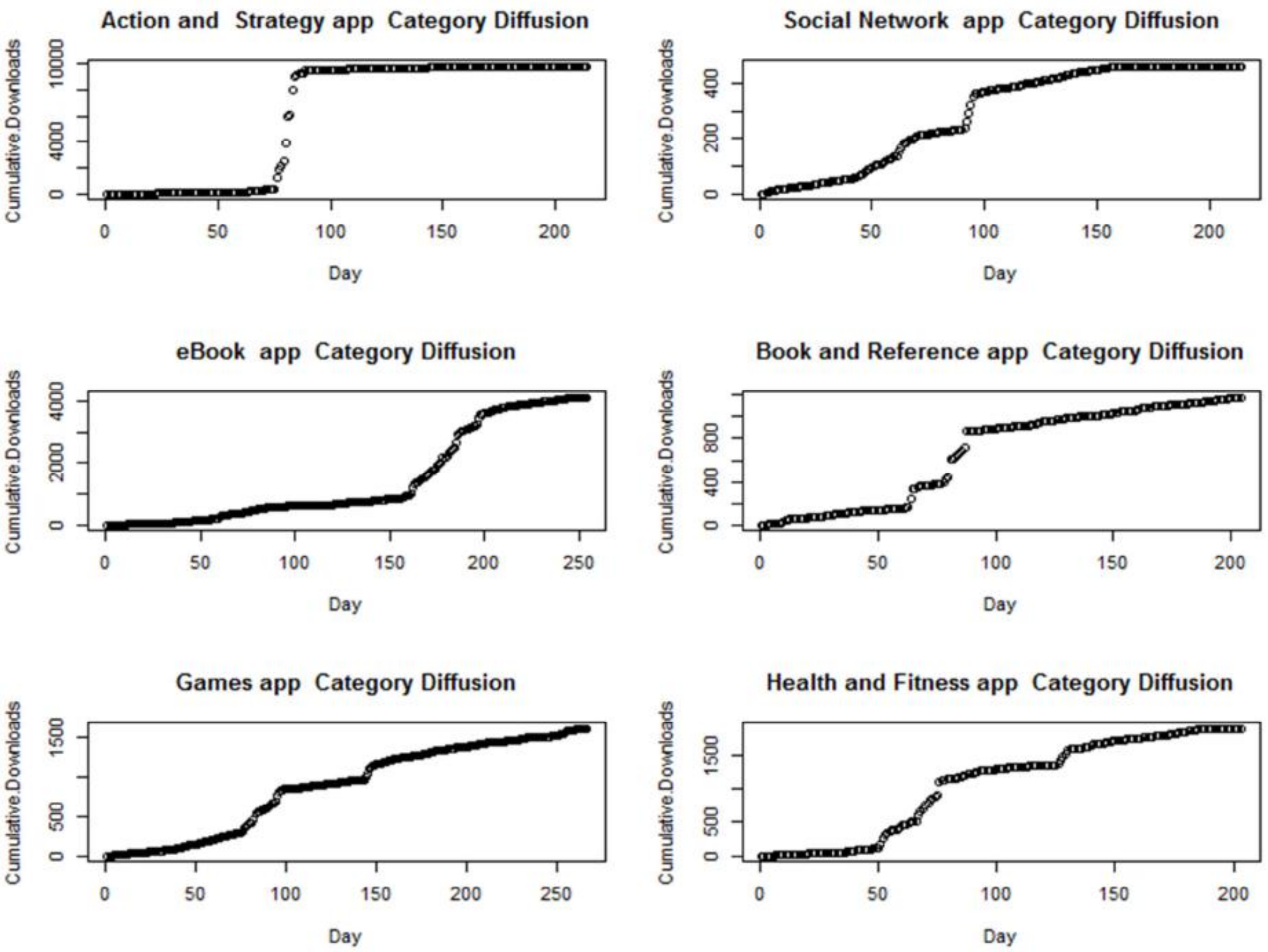

Figure 1.2. Intercontinental (across 30 cities) Diffusion Curves for the mobile apps within the Categories

The dataset also included longitude and latitude of each IP address. However, as mobile phones are usually attached to the customers who might move within the city, aggregating locations at city level might be relevant. Moreover, this assumption is innocuous, because of the social nature of mobile phones and mobile apps (i.e., usage of mobile and mobile phones in social atmosphere). Particularly, mobile phones have become inseparable part of societies, to the point that not only customers use them when they are alone in the bus, when they are to sleep, or even when they are in the class, but also they use them in their parties, in their offices, in their leisure times, and generally in any social events. Mobile phones use in social events makes mobile apps visible, and this visibility can create social learning opportunities. In the figure 1.2, I plotted the diffusion curves of the cumulative adoptions of a sample of six mobile app categories.

For each mobile app category, I had the average file size, the total number of apps featured, the average and the variance of app prices, and the number of paid and free options. Table 1.4 presents the basic statistics of these variables. To explain the heterogeneity in individual responses, I used the data on the tenure of each customer. I defined tenure as the number of days since each customer has subscribed to the app-store. As different types of consumers (i.e., influential and imitators) with different psychological traits adopt the technology at different points in time (Kirton 1976), I used the tenure of consumers as a proxy for the psychological traits that can explain the heterogeneity in consumers' choice responses.

Table 1.4. Mobile app categories basic statistics

| Category Data Summary | Mean | Variance | Min | Max |
|---|---|---|---|---|
| Number of available apps in the Category | 35 | 1250 | 12 | 141 |
| Average tenure of apps in the category (Days) | 316 | 6,386 | 169 | 498 |
| Number of available free apps in the category | 32 | 908 | 7 | 120 |
| Average days that an app is featured in the category | 0.12 | 0.05 | 0.00 | 0.71 |
| Average file size of apps in the category (MB) | 2.00 | 4.00 | 0.50 | 8.00 |
| Variance of prices of apps in the category | 0.51 | 1.09 | 0.00 | 3.75 |

To explain heterogeneity in app store categories, I used the popularity of the mobile app categories on the Apple app store. As the Apple app store is the founder and the leader of the app-store platforms and its consumers are more affluent ones (possibly more influential ones), [3] I expect that the popularity of the mobile app categories on the Apple app store to explain the diffusion of the mobile app categories on the other app store platforms as well. Therefore, I used the mobile app categories' popularity on the Apple app store to explain the heterogeneity in the mobile app category parameters. These popularity statistics is presented in figure 1.3. This figure shows the long tail distribution of the popularity of the mobile apps.

[3] "App Store (iOS)." Wikipedia. Accessed March 23, 2016. http://en.wikipedia.org/wiki/App_Store_(iOS).

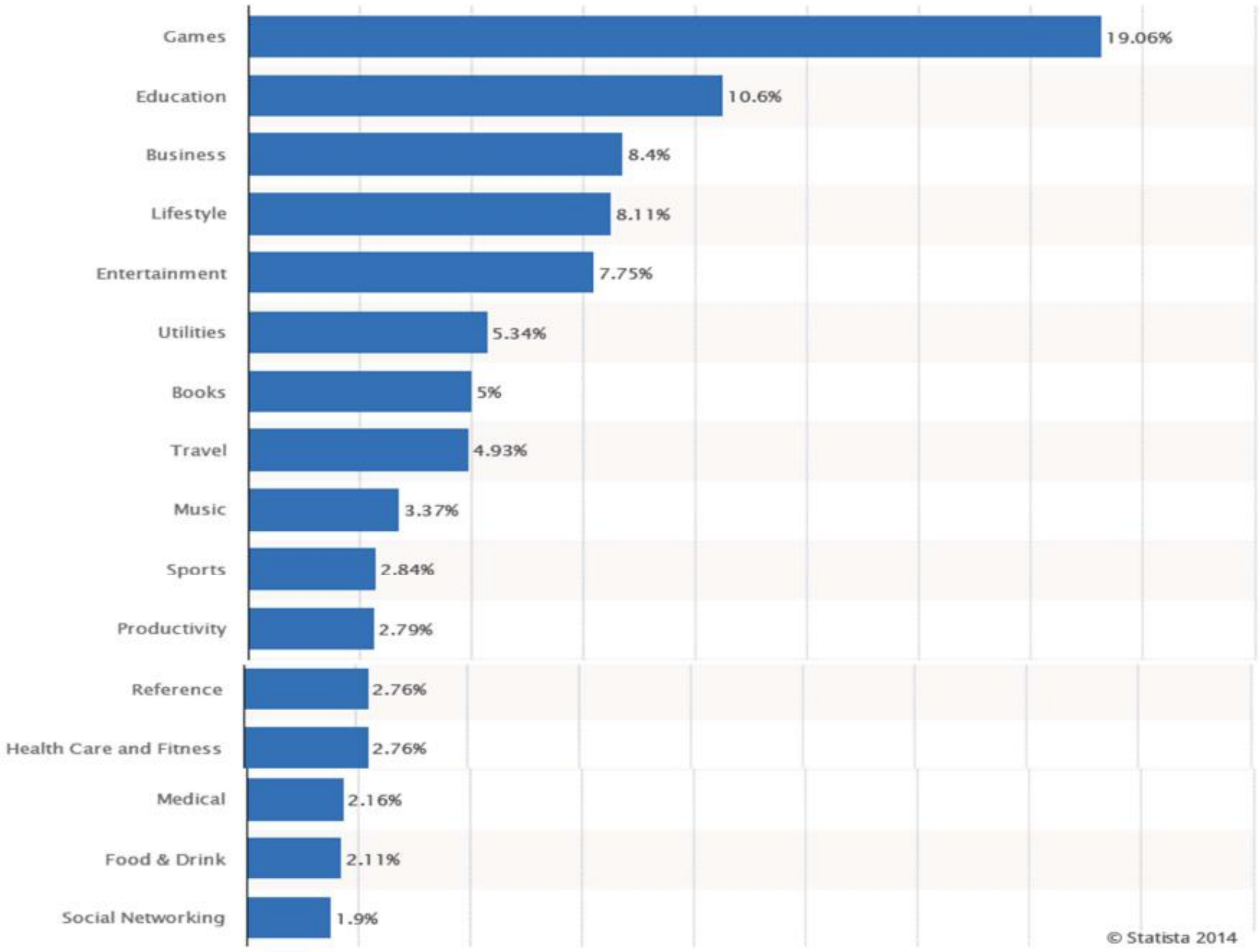

Figure 1.3. Popularity (market share) of App Categories on Apple Inc. App Store

Figure 1.4, from Distimo, a mobile app market research company, suggests that some mobile app categories are more susceptible to be paid, and others are more susceptible to be free. The high share of free mobile apps is an important observation in this figure. The same feature exists in the data sets I used in this study. This feature suggests that the key cost factor that the consumers incur might be the cost of learning about the application, supposedly from others (e.g., their friends, or over internet). This observation suggests that social influence might be an important factor for adoption decision, but a formal model is required to confirm this conjecture.

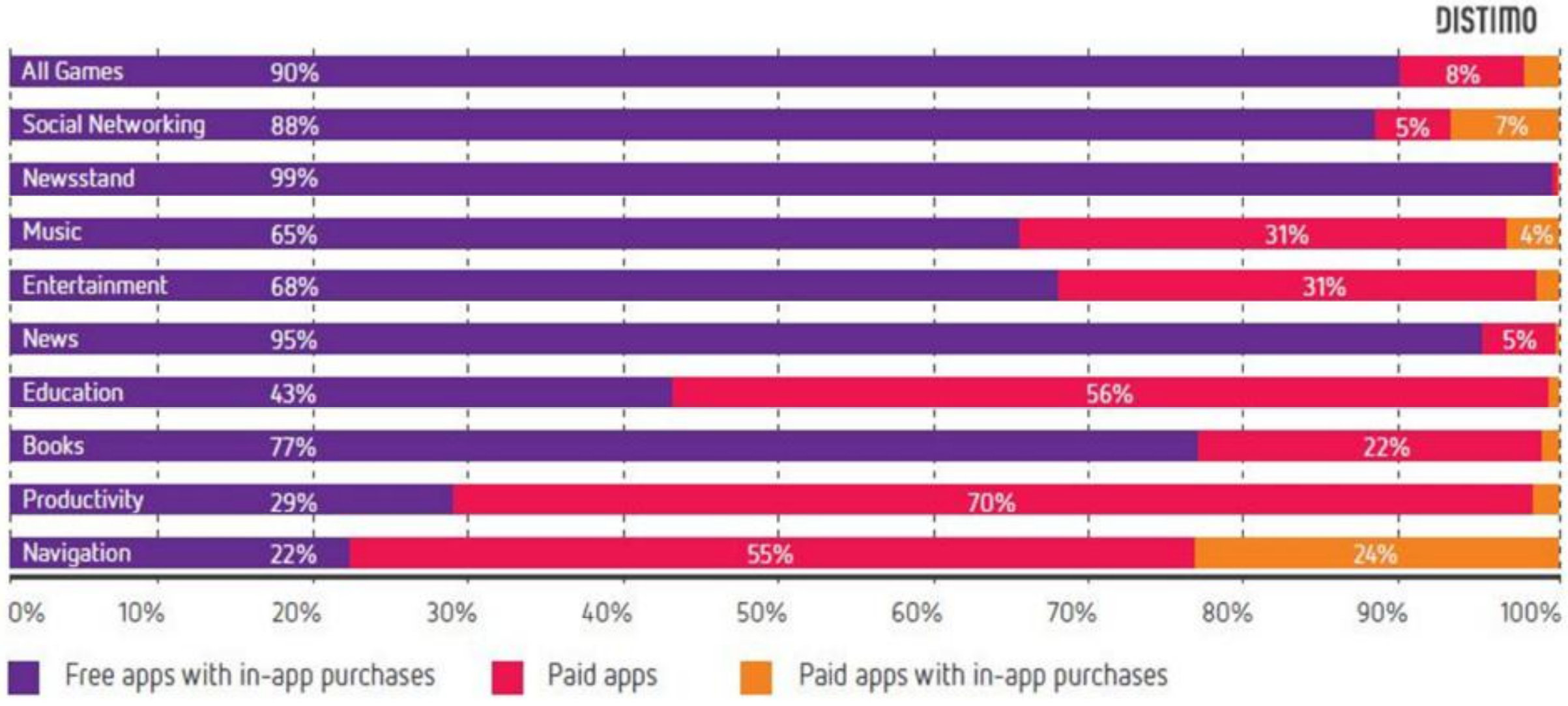

Figure 1.4. Free mobile apps versus paid mobile apps

Anecdotal evidence suggests that from mental accounting perspective consumers perceive the paid mobile apps as investment, for which they are willing to pay money, and the free mobile apps, as entertainment, [4] for which the customer might be less inclined to pay. Guided by the same intuition, I also classified mobile app categories in the sample into two categories: utilitarian and hedonic categories. The utilitarian category includes: device tools, health/diet/fitness, internet/WAP, and reference/dictionary mobile apps. These types of mobile apps might be prominent for their utility rather than their entertainment. The hedonic category includes: ebook, games, humor/jokes, logic/puzzle, and social networks mobile apps. These types of mobile apps might be more relevant for their entertainment features. This categorization might allow to further investigate whether customers really value the utilitarian mobile app categories more than the hedonic mobile app categories, based on the customer choice parameters.

---

[4] Chang, Ryan. "How to Price Your App: Free or Paid - Envato Tuts Code Article." Code Envato Tuts. February 19, 2014. Accessed March 23, 2016. http://code.tutsplus.com/articles/how-to-price-your-app-free-or-paid--mobile-22105.

Finally, the mobile apps are more similar to durable goods than to non-durables. Therefore, the consumers' choice of downloading a mobile app may be sparse in nature. Sparsity here means that several choices of the consumers are no download or outside option choices. A suitable modeling approach that can handle this sparsity might be hierarchical Bayesian approach, which borrows information from other sample items, when the information on an individual is sparse.

## 1.6. IDENTIFICATION AND ESTIMATION

In order to identify the choice model, I used a random coefficient logit specification, which has a fixed diagonal scale. To set the location of the utility, I normalized the utility of outside option to zero. To minimize the concerns about endogeneity (omitted variable), I control for potential correlations between choices by explicitly modeling the inter-temporal choice interdependence in the choice history state variable. I also control for potential confounding effects of price, advertising, and product characteristics by including the latent factors of variation in these variables in the choice model. To control for potential measurement error in the social influence measure, I use Kalman Filter, and I control for potential simultaneity in the social measures through Seemingly Unrelated Regression (SUR) model structure. In addition, by random coefficient structure, the modeling approach also minimizes the concern for Independence from Irrelevant Alternatives (IIA), as it allows for heterogeneity in the individual specific choice behavior parameters.

I identified individual level choice parameters using the micro sample panel of individuals, a sample that consists of twenty week micro choices of a 1,258 customers. Bayesian shrinkage with flexible DP prior helps to identify the large set of individual specific parameters, without over-fitting. The mixture normal distribution is subject to label switching problem (i.e., the permutation of segment assignment returns the same likelihood). However, I immunized myself to this problem by limiting my inference to the joint distribution rather than individual segment assignment. To estimate the micro choice model on the micro sample, I used multinomial logit with DP prior on the individual specific hyper-parameter (Bayesian semi-parametric) estimation code from Bayesm package in R. This method uses Metropolis-Hasting Random-Walk (MH-RW) method to estimate conditional choice probabilities on cross-sectional units (i.e., customers). The limitation of MH-RW is that random walk increments shall be tuned to conform as closely as possible to the curvature in the individual specific conditional posterior, formally defined by:

$$p(A_i \mid y_i, \mu, \Sigma, z_i, \Delta) \propto p(y_i \mid A_i)\, p(A_i \mid \mu, \Sigma, z_i, \Delta) \tag{11}$$

Without prior information on highly probable values of first stage prior (i.e., $p(A_i \mid .)$), tuning the Metropolis chains given limited information of cross-sections (i.e., each customer) by trial is difficult (this problem exacerbates when each customer does not have some of the choice items selected at all in his history). Therefore, to avoid singular hessian, the fractional likelihood approach proposed by Rossi et al. (2005) is implemented in the used approach. Formally rather than using individual specific likelihood, MH-RW approach forms a fractional combination of the unit-level likelihood and the pooled likelihood as follows:

$$l_i^*(A_i) = l_i(A_i)^{(1-w)} \left(\prod_{i=1}^{I} l_i(A_i \mid y_i)\right)^{w\beta}, \beta = \frac{n_i}{N}, N = \sum_{i=1}^{I} n_i \tag{12}$$

where $w$ denotes the small tuning parameter to control the effect of pooled likelihood $\prod_{i=1}^{I} l_i(A_i \mid y_i)$. $\beta$ denotes a parameter chosen to properly scale the pooled likelihood to the same order as the unit likelihood. $n_i$ denotes the number of observations for customer $i$. Using this approach, the MH-RW generates samples conditional on the partition membership indicator for individual $i$ from proposal density $N(0, s^2\Omega)$, so that:

$$\Omega = (H_i + V_A^{-1})^{-1}, H_i = -\frac{\partial^2 \log l_i^*}{\partial A \partial A'}\Big|_{A=\hat{A}_i} \tag{13}$$

where $\hat{A}_i$ denotes the maximum of the modified likelihood $l_i^*(A_i)$, and $V_A$ denotes normal covariance matrix assigned to the partition (i.e., segment) that customer $i$ belongs to.

This approach considers that $A_{i.}$ is sufficient to model the random coefficient distribution. To estimate the infinite mixture of normal prior for choice parameters, a standard data augmentation with the indicator of the normal component is required. Conditional on this indicator, I can identify a normal prior for each customer $i$ parameters. The distribution for this indicator is Multinomial, which is conjugate to Dirichlet distribution, formally:

$$\begin{aligned} &\pi \sim Dirichlet(\alpha^d) \\ &z^i \mid \pi \sim Mult-Nom(\pi) \end{aligned} \tag{14}$$

As a result posterior can be defined by:

$$z^i \sim Mult-Nom(\frac{\alpha^1}{\sum_j \alpha^j},...,\frac{\alpha^K}{\sum_j \alpha^j})$$

$$\pi \mid z^i \sim Dirichlet(\alpha^1 + \delta_1(z^i),...,\alpha^k + \delta_K(z^i)) \tag{15}$$

where $\delta_j(z^i)$ denotes indicators for whether or not $z^i = j$. This result is relevant for DP as any finite subset of customers' choice-behavior parameters' partitions has Dirichlet distribution, and finite sample can only represent finite number of partitions. Exchangeability property of partitions allows the used estimation approach to sequentially draw customer parameters given the indicator value as follows:

$$(\mu_i,\Sigma_i) \mid (\mu_1,\Sigma_1),...,(\mu_{i-1},\Sigma_{i-1}) \sim \frac{\alpha^d G_0 + \sum_{j=1}^{i-1} \delta_{(\mu_j,\Sigma_j)}}{\alpha + i - 1} \tag{16}$$

The next portion of this approach's specification is the definition of the size of the finite clusters over the finite sample that is controlled by $\pi$. Rossi (2014) suggests augmenting Sethuraman's stick breaking notion for draws of $\pi$. In this notion, a unit level stick is iteratively broken from the tail with proportion to the draws with beta distribution with parameter one and $\alpha^d$, and the length of the broken portion defines the $k'th$ element of the probability measure vector $\pi$ (a form of multiplicative process), formally:

$$\pi_k = \beta_k \prod_{i=1}^{k-1}(1-\beta_i), \beta_k \sim Beta(1,\alpha^d) \tag{17}$$

In this notion, $\alpha^d$ determines the probability distribution of the number of unique values for the DP mixture model, formally by:

$$\Pr(I^* = k) = \left\| S_i^{(k)} \right\| \left(\alpha^d\right)^k \frac{\Gamma(\alpha^d)}{\Gamma(i+\alpha^d)}, S_i^{(k)} = \frac{\Gamma(i)}{\Gamma(k)}(\gamma + \ln(i))^{k-1} \tag{18}$$

where $I^*$ denotes the number of unique values of $(\mu, \Sigma)$ in a sequence of $i$ draws from the DP prior. $S_i^{(k)}$ denotes Sterling number of first kind, and $\gamma$ denotes Euler's constant. Furthermore, to facilitate assessment, this approach suggests the following distribution for $\alpha^d$, rather than Gamma distribution:

$$\alpha^d \propto \left(1 - \frac{\alpha^d - \breve{\alpha}}{\vec{\alpha} - \breve{\alpha}}\right)^{\phi} \tag{19}$$

where $\vec{\alpha}$ and $\breve{\alpha}$ can be assessed by inspecting the mode of $I^* | \alpha^d$. $\phi$ denotes the tunable power parameter to spread prior mass appropriately. An alternative to Gibbs sampler employed by this approach might be collapsed Gibbs sampler that integrates out the indicator variable for partition (segment) membership of each customer, but Rossi (2014) argues that such an approach does not improve the estimation procedure. Appendix 1.C presents the series of conditional distribution that this approach employs in its Gibbs sampling to recover individual specific choice parameters.

I identified the latent cumulative number of influential and imitators of mobile app categories with observed cumulative number of adopters in the complete dataset. Also to avoid over fitting, I used a normal prior on the fixed social learning macro diffusion model to regularize the likelihood of the model. Although the local and global aggregate data sets only have two thousands observations, Bayesian shrinkage of parameters allows identification of parameters. To estimate the latent cumulative number of influential and imitators, I used the maximum a posteriori (MAP) method, a popular method in machine learning, as an alternative method to Markov Chain Monte Carlo (MCMC) sampling methods. This approach uses an optimization method to maximize the a posteriori of the model parameters. I used genetic algorithm for the optimization, as the number of parameters that I estimated for the social learning diffusion model is around 300: 210 covariance elements of state covariance matrix, 10 elements of observation covariance matrix, and 80 elements of fixed parameters of the diffusion differential equations. Gradient descend optimization method has complexity of $O(P)$ per iteration, but requires tuning learning parameter, and the quasi newton optimization method has the complexity of $O(P^2)$ per iteration, where P is the number of parameters to estimate. This complexity translates to long run time over big data, in which the number of parameters increases with the variety, and volume of the data. As a result, I adopted the genetic algorithm approach that Venkatesan et al. (2004) finds comparable to the classical gradient descend or the Quasi Newton approach. In addition, genetic algorithm is known as global optimization method, in contrast to local optimization of Quasi Newton method. Given that a latent state space model like mixture models has multiple local maxima, genetic algorithm might be more prone to find the global maxima than a Quasi Newton method.

In order to estimate the macro social learning diffusion model, I used Unscented Kalman Filter (UKF) nested within a Monte Carlo Expectation Maximization (MCEM) method. Unscented Kalman Filter (UKF) is an approach proposed in robotics literature (Wan and Merwe 2001, Julier and Uhlmann 1997), which achieves third order accuracy in estimating the latent state in a state space model, as opposed to the Extended Kalman Filter (EKF) that only achieves the first order accuracy, with the same order of computational complexity, i.e., $O(T)$. The basic idea behind UKF is that rather than using the closed form first order tailor expansion term, for the measurement updating of the latent state, by computing Jacobean vector, it uses an Unscented Transformation (UT) to transform sigma vector of points around the mean, and the mean of the latent state prior of a nonlinear state equation, to estimate the transformed normal distribution posterior parameters. I explained UKF algorithm in appendix 1.B.

The MCEM approach starts with an initial vector of parameters. Then it uses MCMC, UKF in this case, to recover the latent state distribution, and a set of samples. Given the latent state samples, it computes the expected log likelihood, and it searches for the parameters that maximize this expected log likelihood (de Valpine 2012). MCEM is appealing for its speed, compared with the full MCMC sampling method. However, MCMC approach is notorious for slow convergence, and both approaches may suffer from finding only the local maxima. The exercise of global optimization genetic algorithm stochastic search may be a remedy to this stochastic surface search problem. In the optimization, I used transformation to make sure that the market sizes of the social learning diffusion model are positive, and parameters of effect of learning from imitator and influential in the imitator state equation ($w$), and the segment size of influential and imitators ($\theta$) are between zero and one. I used just in time compiler in R to speed up the estimation process.

In summary, I used MCEM, UKF, MAP, GA, and SUR methods to estimate the social learning model on the aggregate sample data, which consists of the aggregate number of adoption of twenty thousand adopters of mobile apps in ten mobile app categories for two hundred days, and I used MCMC sampling to estimate the mixture normal multinomial logit model of the micro mobile app choices of a hundred forty seven customers in ten app categories over twenty weeks.

## 1.7. RESULTS

Table1.5. presents the log-likelihood of the proposed models. Model 1 and 2 represent social learning aggregate diffusion models over local adoption (only adopters within one city) and global adoption (total number of adopters across 30 cities). Local social learning model dominates global social learning model by the likelihood. This result might suggest that mobile app adoption process is more locally rather than global coordinated. Model 3 and 4 use the filtered number of imitators as a measure for social influence, and model 5 and 6 use the observed number of adopters as a measure for social influence. Domination of model 5 and 6 over model 3 and 4 by log likelihood might suggest that not only number of imitators but other social factors might be the driving factor for mobile app adoptions. This other factor might include the social force of differentiators (in fact the result of micro analysis reconfirms the existence of such potential). The dominance of model 5 and 6 over model 7 (the model with no social learning) suggests that in fact social learning is an important force that drives individual mobile app adoption choices (the bias in the parameter estimates when social learning is ignored is discussed later).

Table 1.5. MODEL COMPARISON

| Model | Description | Number of obs. | Log Lik. |
|---|---|---|---|
| 1 | Local Adoption (aggregate sample) | 2,000 | -20,724.16 |
| 2 | Global Adoption (aggregate sample) | 2,000 | -21,649.32 |
| 3 | Choice Explained by Local Imitators Signals (micro sample) | 22,644 | -25,921.92 |
| 4 | Choice Explained by Global Imitators Signals (micro sample) | 22,644 | -38,310.49 |
| 5 | Choice Explained by Local Adopters Signals (micro sample)* | 22,644 | -12,252.85 |
| 6 | Choice Explained by Global Adopters Signals (micro sample) | 22,644 | -15,275.20 |
| 7 | Choice with No social influence measure (micro sample) | 22,644 | -15,977.04 |

* dominant model

Finally, dominance of model 5 over model 6 reconfirms the result from aggregate model that social learning at local level (within the city) rather than global level (for example over the world wide web) drives the adoption choices of the customers. This finding for mobile apps (as a form of pervasive good) contrasts with findings about the adoption of traditional goods that emphasize the importance of learning over World Wide Web (Putsis et al. 1998).

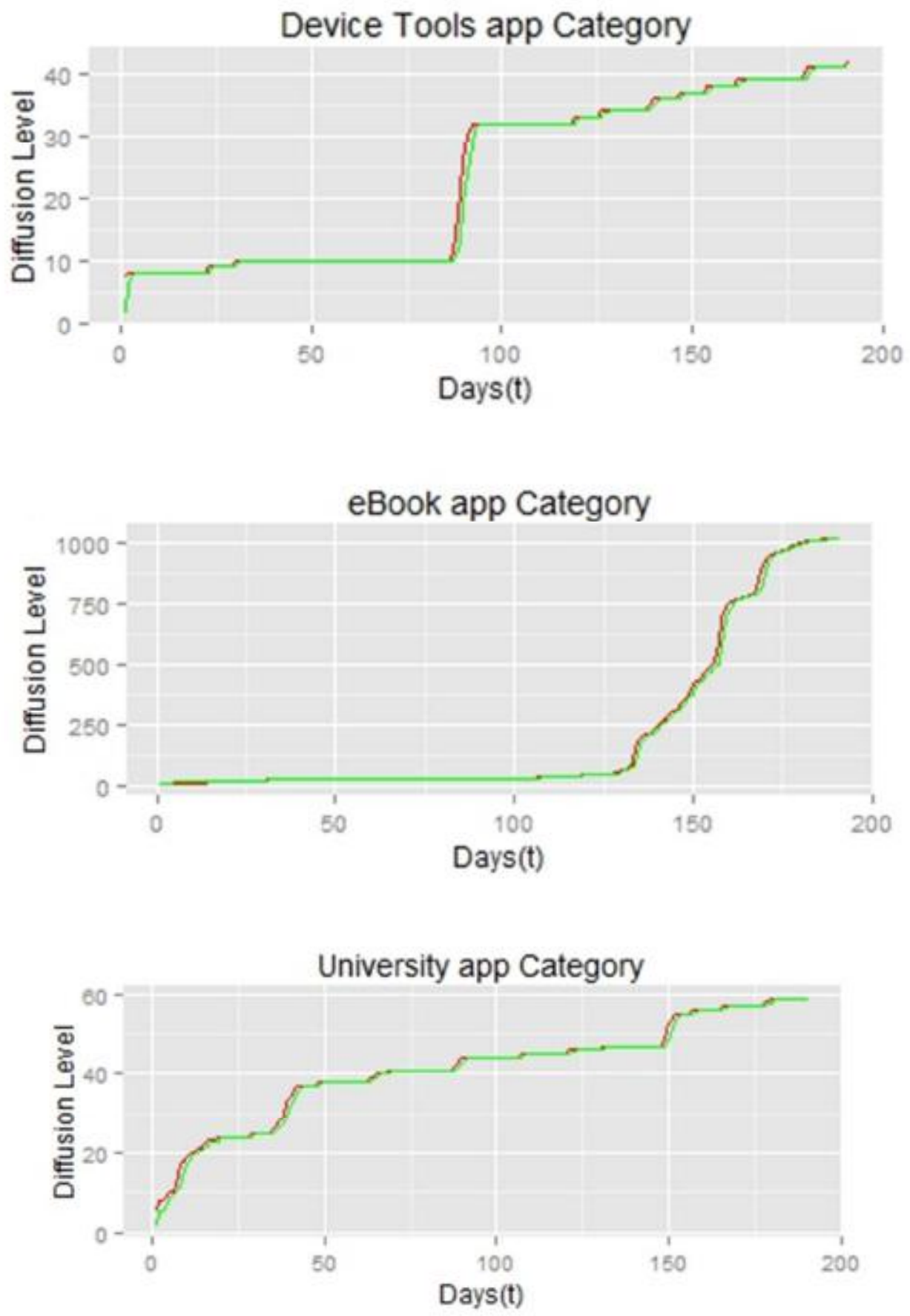

Figure 1.5. 1-Step-ahead Forecast for Local Diffusion (Green Line: a step ahead; Red line: the actual)

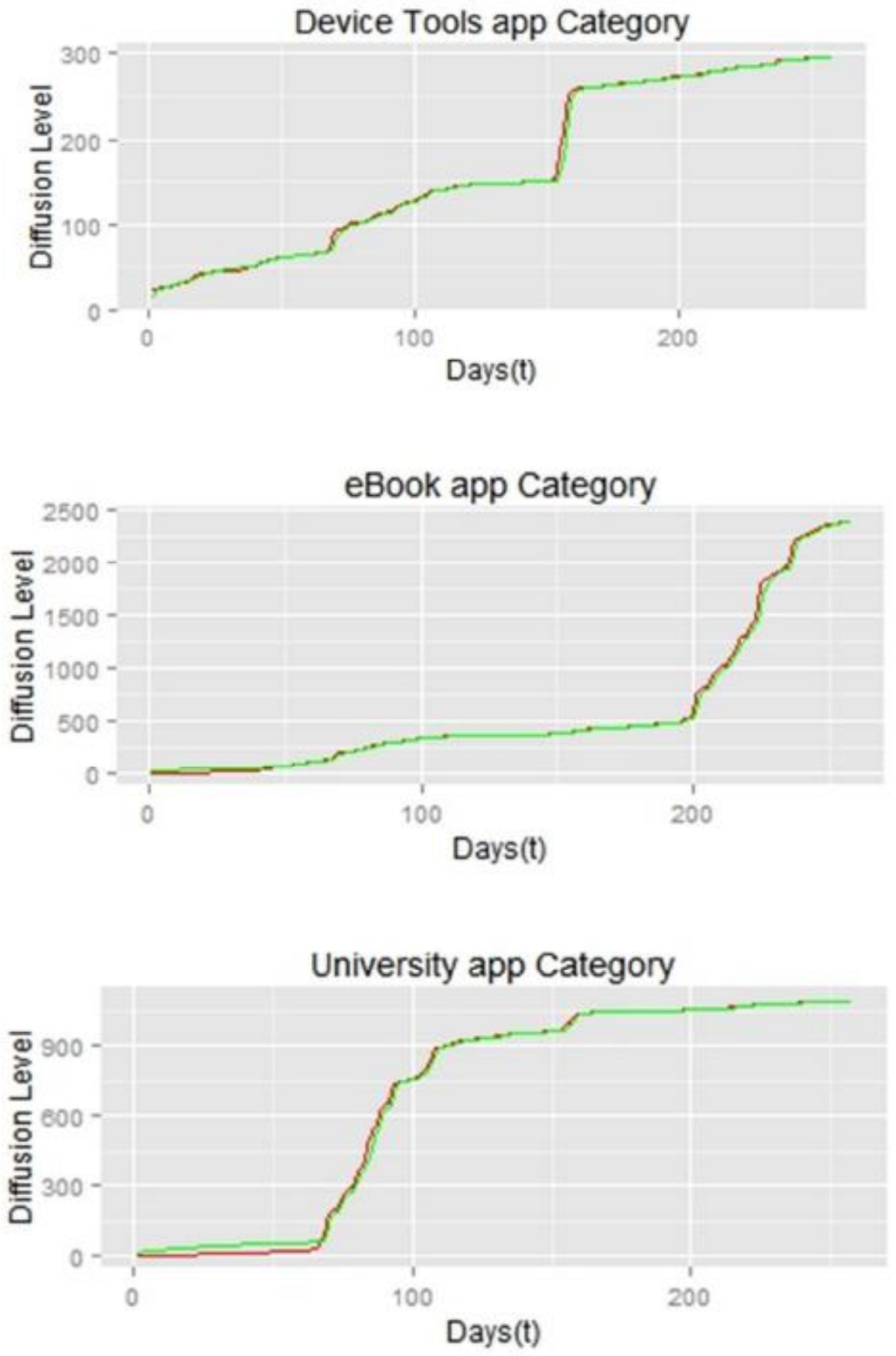

Figure 1.6. 1-Step-ahead Forecast for Global Diffusion level (Green line: a step ahead, Red line: the actual)

Table 1.6. Performance of the Proposed Model for local and international category adoption

| **Description** | **MAD** | **MSE** |
|---|---|---|
| Local Category Adoption | 0.64 | 1.48 |
| International Category Adoption | 0.03 | 0.12 |

Figure 1.5 and 1.6 present a step ahead forecast versus the observed cumulative number of adopters at both the local and the global level. This visualization together with the Mean Absolute Deviation (MAD) and the Mean Square Error (MSE) presented in table 1.6, suggest that social learning macro diffusion model fits the app-store platforms' macro app diffusion data reasonably well.

I benchmarked the estimates with Van den Bulte and Joshi (2007) paper, and relative to the market size the MSE is in reasonably good range.

Table 1.7. Factor Loading Matrix (Varimax rotation)

| **Loadings/Components** | **C1** | **C2** | **C3** |
|---|---|---|---|
| Average File size (a proxy for app quality) | 0.77 | -0.07 | -0.09 |
| Dummy variable of Is Featured | 0.82 | 0.3 | 0.01 |
| Average Price | -0.06 | 0.94 | -0.28 |
| Variance of Price | -0.05 | 0.94 | -0.19 |
| Number of Paid app Options | 0.97 | -0.09 | -0.09 |
| Number of Free app Options | 0.96 | -0.15 | -0.03 |
| Fraction of Free apps to Paid Apps | -0.09 | -0.25 | 0.87 |
| Average Tenure (time from creation) | -0.08 | 0.67 | 0.48 |
| Total number of app Options | 0.96 | -0.14 | -0.03 |

Table 1.7 presents the result of the factor analysis to extract the latent factor of mobile app characteristics. I used Varimax rotation to be able to interpret the factors. I named the factors both from supply side and the demand side in table 1.8.

Table 1.8. Factor Names

| **Factor** | **Supply side Name** | **Demand Side Name** |
|---|---|---|
| C1 | Red Ocean app categories | Popular Apps |
| C2 | Paid app categories | Investment Apps |
| C3 | Free app categories | Freemiums |

I limited the factor/principle components to three, as it captures already 0.85% of the variation in the data. Number of paid mobile apps and free mobile apps load highly into the first factor, so I expected that there is high demand for these mobile apps that has brought app developer/publishers to develop many mobile apps. Further, Ghose and Hann (2014) use the average file size of a mobile app as a proxy for the quality of the mobile apps, and this mobile app category feature also loads highly into the first factor, so I may be able to call the first factor as popular mobile apps. The average and variance of the prices load highly into the second factor. I called the second factor investment mobile apps, guided by discussion about figure 1.4 (in the data section), and I refer to the third factor by Freemiums, because the fraction of free mobile apps is much higher than paid mobile apps.

Table 1.9. PARAMETER ESTIMATES: Global Adoption

| | $p^{\inf}$ | $q^{\inf}$ | $p^{imm}$ | $q^{imm}$ | $M^{\inf}$ | $M^{imm}$ | $w$ | $\theta$ |
|---|---|---|---|---|---|---|---|---|
| Mobile App Categories: | | | | | | | | |
| Device Tools | 0.024 | 0.000 | 0.278* | 0.192* | 50* | 580* | 0.010* | 0.039* |
| eBooks | 0.024 | 0.000 | 0.274* | 0.189* | 260* | 4540* | 0.007* | 0.044* |
| Games | 0.026 | 0.000 | 0.293* | 0.202* | 80* | 1150* | 0.009* | 0.046* |
| Health/Diet/Fitness | 0.025 | 0.000 | 0.288* | 0.199* | 100* | 1600* | 0.008* | 0.048* |
| Humor/Jokes | 0.026 | 0.000 | 0.297* | 0.205* | 100* | 1410* | 0.008* | 0.043* |
| Internet/WAP | 0.026 | 0.000 | 0.296* | 0.204* | 100* | 1580* | 0.010* | 0.039* |
| Logic/Puzzle/Trivia | 0.024 | 0.000 | 0.275* | 0.190* | 90* | 1440* | 0.009* | 0.039* |
| Reference/Dictionaries | 0.026 | 0.000 | 0.296* | 0.204* | 90* | 1440* | 0.007* | 0.048* |
| Social Networks | 0.026 | 0.000 | 0.297* | 0.205* | 40* | 390* | 0.008* | 0.044* |
| University | 0.025 | 0.000 | 0.281* | 0.193* | 130* | 2080* | 0.009* | 0.046* |

***** $p \leq 0.05$

Tables 1.9 and 1.10 present the parameter estimates of the social learning diffusion models studies over global (across the cities within the app store) and local (within the city of interest). Over both local and global diffusion data, the independent random adoption rate for individuals in influential segment is not significant statistically across different mobile app categories, except for eBooks (which might be driven by its assortment size). However, this rate is significantly higher than dependent adoption rate for this segment, which might suggest that the model is properly identifying the behavior of the segment of influentials. For influential segment, the rate of independent adoption is very close to the same rate for Everclear music CD that Van den Bulte and Joshi (2007) find. However, for this segment, the dependent rate of adoption is similar to the same rate for foreign language CD adoption in the mentioned study. This result might be driven by the low search cost of influential segment on the app-store, which in turn drives their learning less from others.

Table 1.10. PARAMETER ESTIMATES: Local Adoption

| Mobile App Categories: | $p^{\inf}$ | $q^{\inf}$ | $p^{imm}$ | $q^{imm}$ | $M^{\inf}$ | $M^{imm}$ | $w$ | $\theta$ |
|---|---|---|---|---|---|---|---|---|
| Device Tools | 0.025 | 0.000 | 0.282* | 0.194* | 5* | 80* | 0.100* | 0.046* |
| eBooks | 0.024* | 0.000 | 0.278* | 0.191* | 103* | 1952* | 0.032* | 0.044* |
| Games | 0.024 | 0.000 | 0.275* | 0.189* | 3* | 56* | 0.246* | 0.038* |
| Health/Diet/Fitness | 0.025 | 0.000 | 0.282* | 0.194* | 7* | 120* | 0.782* | 0.038* |
| Humor/Jokes | 0.026 | 0.000 | 0.299* | 0.206* | 6* | 99* | 0.506* | 0.041* |
| Internet/WAP | 0.025 | 0.000 | 0.285* | 0.197* | 11* | 200* | 0.738* | 0.041* |
| Logic/Puzzle/Trivia | 0.025 | 0.000 | 0.282* | 0.194* | 6* | 113* | 0.344* | 0.043* |
| Reference/Dictionaries | 0.026 | 0.000 | 0.299* | 0.206* | 12* | 225* | 0.940* | 0.042* |
| Social Networks | 0.025 | 0.000 | 0.281* | 0.193* | 3* | 48* | 0.658* | 0.040* |
| University | 0.025 | 0.000 | 0.281* | 0.194* | 6* | 113* | 0.555* | 0.042* |

***** $p \leq 0.05$

For imitator segment, in almost all the categories rate of independent adoption (mean of 0.288) is greater than rate of dependent adoption (mean of 0.198). For this segment, the independent rate of adoption is significantly more than the same rate for goods proposed in classical economy that Van den Bulte and Joshi (2007) report. This difference can be driven by the low search cost of mobile apps for imitators. The dependent rate of adoptions is similar to the same rate for Everclear music CD that Van den Bulte and Joshi (2007) report. For global adoption, across the mobile app categories the weight of influential in driving imitators' dependent choice of adoption is 0.009 which is similar to the same parameter for Everclear music CD. However, this rate is 0.50 for local adoption, which is similar to the same rate for John Hiatt music CD (Van den Bulte and Joshi 2007). The size of influential segment in the observed sample for global adopter data is 0.044 and for local adopter data is 0.042 which is very similar to the same rate for Everclear music CD (Van den Bulte and Joshi 2007). To sum up, these results might suggest that customers adoption behavior for mobile apps is very similar to music CD adoptions, except that the independent rate of adoptions for imitators are higher, but the dependent rate of adoptions for influentials is less, driven by the lower search cost.

Table 1.11 summarizes the individual parameters distribution for the choice model that uses local number of adopters (unfiltered density) as a proxy for social influence. The negative mean for the preference parameter for each mobile app categories indicates higher preference of outside options for customers. In the city under the study, the customers prefer mobile apps in Health/Diet/Fitness, Games, Internet/WAP, and device tools relatively more than mobile apps in social network, ebooks, and Humar/Jokes categories. Relative to the apple app-store popularity statistics (presented in figure 1.3 in data section), the surprising result is high preference of the customers for Health/Diet/Fitness mobile apps. This information can help this app-store to target its marketing communication message by highlighting this mobile app category.

The mean for the distribution of download history state parameter is negative and significant. This negative effect of history suggests that this app-store is not doing well in retaining the customers, perhaps for its appearance and its nonoptimal shopping shelf. However, the effect of social influence is positive and significant, which suggests that there is positive spill-over (possibly because of awareness effect) of adoption within the population.

Table 1.11. PARAMETER ESTIMATES: Individual Choice effect (Local Adopters)

| | **Estimate** | **Std. Dev.** | **$2.5^{th}$** | **$97.5^{th}$** |
|---|---|---|---|---|
| Category specific preference: | | | | |
| Device Tools $\alpha_1$ | -6.22* | 5.04 | -14.327 | -2.669 |
| eBooks $\alpha_2$ | -11.34* | 3.14 | -15.290 | -6.381 |
| Games $\alpha_3$ | -4.35* | 3.76 | -11.222 | -2.296 |
| Health/Diet/Fitness $\alpha_4$ | -4.1 | 2.18 | -5.939 | 2.982 |
| Humor/Jokes $\alpha_5$ | -16.32* | 5.85 | -22.097 | -9.715 |
| Internet/WAP $\alpha_6$ | -5.41* | 2.29 | -8.021 | -3.021 |
| Logic/Puzzle/Trivia $\alpha_7$ | -14.2* | 3.49 | -18.122 | -8.332 |
| Reference/Dictionaries $\alpha_8$ | -8.48* | 1.92 | -11.092 | -4.547 |
| Social Networks $\alpha_9$ | -10.54 | 3.47 | -15.530 | 0.076 |
| University $\alpha_{10}$ | -5.78* | 1.39 | -7.791 | -2.916 |
| States: | | | | |
| Individual download history State $\alpha_{11}$ | -27.27* | 5.46 | -34.350 | -13.821 |
| Latent imitation level $\alpha_{12}$ | 0.02* | 0.01 | 0.011 | 0.035 |
| App category characteristics (factors): | | | | |
| Popularity of app category $\alpha_{13}$ | 1.32 | 0.63 | -0.830 | 1.767 |
| Investment apps category $\alpha_{14}$ | 5.34 | 1.75 | -0.922 | 7.230 |
| Hedonic apps category $\alpha_{15}$ | 7.13 | 4.28 | -6.606 | 10.330 |

* $p<0.05$

Appendix 1.D presents the same result table for choice model with local imitators, models with global imitators/adopters, and model with no social influence. The model with no social influence underestimates the preference for mobile apps almost in all the categories except for eBook, Humor/Jokes, Reference/Dictionary, and University. In addition, this model underestimates the effect of popularity, investment, and free characteristics of the mobile apps.

In summary, a model that does not account for social influence returns bias estimates for the parameters.

Table 1.12. PARAMETER ESTIMATES: Individual Choice Hierarchical Model (Local Adopters): CustomerTenure (number of days since registeration on the app-store) explanation of the effects

| Parameter explained by Tenure | **Estimate** | **Std. Dev.** | **2.5th** | **97.5th** |
|---|---|---|---|---|
| Category specific preference: | | | | |
| Device Tools $\alpha_1$ | -0.00044* | 1.01E-04 | -0.00058 | -0.00023 |
| eBooks $\alpha_2$ | -0.00048* | 2.63E-04 | -0.00087 | -0.00006 |
| Games $\alpha_3$ | -0.00041* | 4.46E-05 | -0.00049 | -0.00032 |
| Health/Diet/Fitness $\alpha_4$ | -0.0008* | 7.30E-05 | -0.00092 | -0.00061 |
| Humor/Jokes $\alpha_5$ | -0.00091* | 2.49E-04 | -0.00126 | -0.00046 |
| Internet/WAP $\alpha_6$ | 0.00011 | 7.58E-05 | -0.00002 | 0.00025 |
| Logic/Puzzle/Trivia $\alpha_7$ | -0.00056* | 1.29E-04 | -0.00081 | -0.00035 |
| Reference/Dictionaries $\alpha_8$ | -0.00028 | 1.50E-04 | -0.00046 | 0.00002 |
| Social Networks $\alpha_9$ | -0.00001 | 9.45E-05 | -0.00016 | 0.00020 |
| University $\alpha_{10}$ | 0.00018 | 1.36E-04 | -0.00007 | 0.00034 |
| States: | | | | |
| Individual download history State $\alpha_{11}$ | -0.00136* | 3.33E-04 | -0.00193 | -0.00081 |
| Latent imitation level $\alpha_{12}$ | 0.00006* | 7.64E-06 | 0.00004 | 0.00007 |
| App category characteristics (factors): | | | | |
| Popularity of app category $\alpha_{13}$ | 0.00001 | 2.06E-05 | -0.00003 | 0.00005 |
| Investment apps category $\alpha_{14}$ | -0.00006 | 6.77E-05 | -0.00016 | 0.00007 |
| Hedonic apps category $\alpha_{15}$ | 0.00021 | 1.35E-04 | -0.00003 | 0.00043 |

* $p<0.05$

Table 1.12 presents correlation between customer tenure (number of days since registeration on the app-store) and choice parameters of customers. Those who register early to the app-store (potentially with innovator personality) have higher preference for mobile apps in Internet/WAP and university mobile app categories. This correlation might be relevant as mobile innovators might be more interested in improving their performance oriented apps. In addition, these customers are more sensitive to download history and social influence. This result is aligned with the chasm on the product life cycle theories that argue that if the product does not pass the acceptance of early adopters it will fall into the chasm, leading to early failiure.

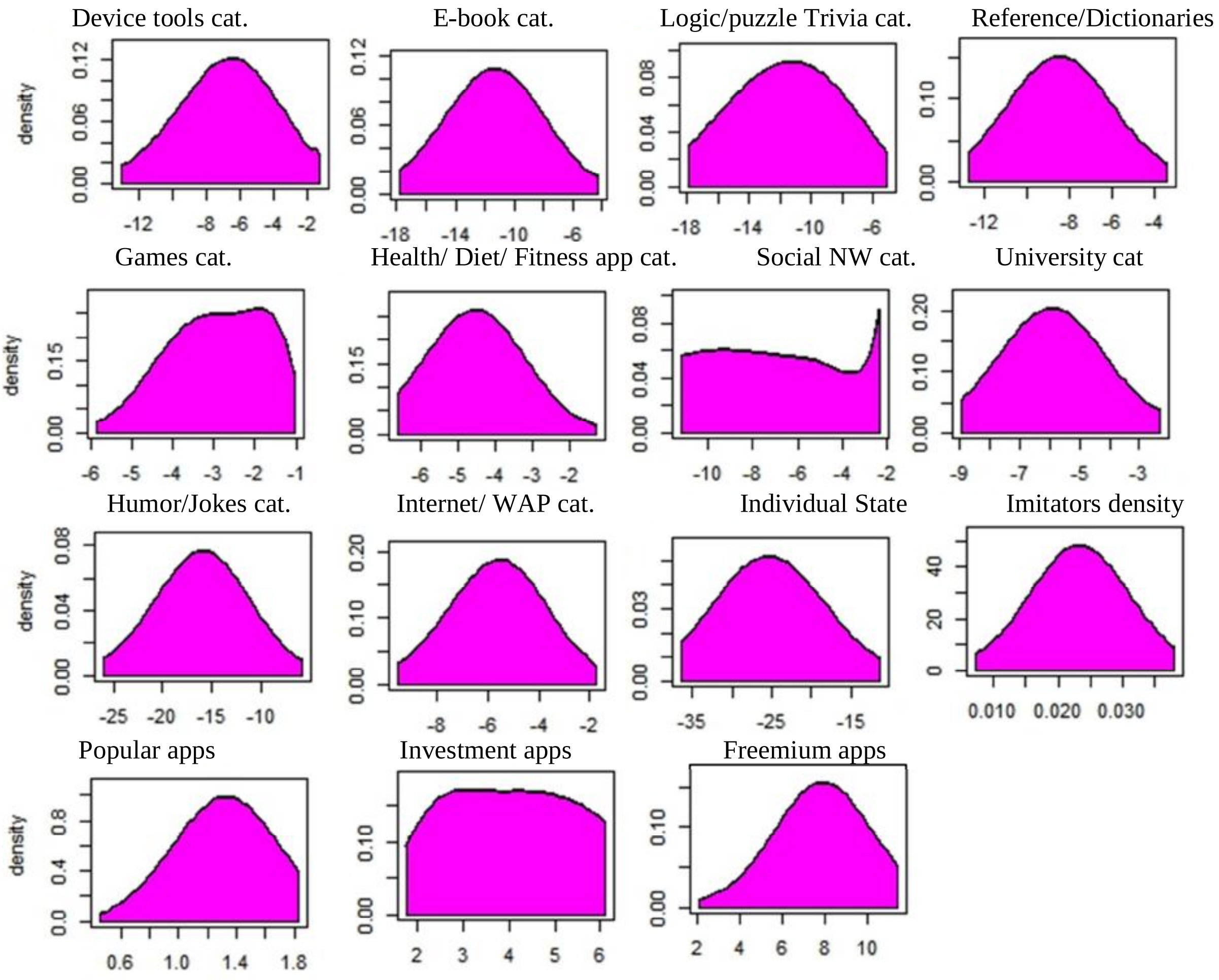

Figure 1.7. PARAMETER DISTRIBUTION: Heterogeneity in Individual Choice (Local Adopters)

Figure 1.7 presents the distribution of choice parameters. This distribution has heavy tail, which highlights the importance of allowing for flexible heterogeneity distribution for the choice parameters.

Table 1.13. PARAMETER ESTIMATES: Individual Choice effect
(Local Adopters)

| Total number of users: 1258 | **Positive Significant** | **Negative Significant** |
|---|---|---|
| Category specific preference: | | |
| Device Tools $\alpha_1$ | 0 | 1253 |
| eBooks $\alpha_2$ | 0 | 1258 |
| Games $\alpha_3$ | 0 | 1258 |
| Health/Diet/Fitness $\alpha_4$ | 53 | 1205 |
| Humor/Jokes $\alpha_5$ | 0 | 1258 |
| Internet/WAP $\alpha_6$ | 0 | 1258 |
| Logic/Puzzle/Trivia $\alpha_7$ | 0 | 1258 |
| Reference/Dictionaries $\alpha_8$ | 0 | 1258 |
| Social Networks $\alpha_9$ | 53 | 1205 |
| University $\alpha_{10}$ | 0 | 1258 |
| States: | | |
| Individual download history State $\alpha_{11}$ | 0 | 1258 |
| Latent imitation level $\alpha_{12}$ | 1257 | 0 |
| App category characteristics (factors): | | |
| Popularity of app category $\alpha_{13}$ | 1205 | 53 |
| Investment apps category $\alpha_{14}$ | 1205 | 53 |
| Hedonic apps category $\alpha_{15}$ | 1205 | 53 |

Targeting is a relevant application of micro choice modeling for app-stores. Table 1.13 presents the distribution of significance and sign of each of the choice parameters at individual customer level. Knowing the distribution of negative and positive response helps the app-store to target 53 customers that do not prefer health/diet/fitness or social network mobile apps. This correct targeting might help improving the usability of the app store.

## 1.8. COUNTERFACTUAL ANALYSIS

The advantage of the individual specific choice model for the app-store platforms is that it allows estimating the implications of the social influence policy for total expected adoption by simulation. I ran three counterfactual scenarios using the estimated choice model by modifying the level of social influence. Furthermore, I use the estimated model to find the optimal dynamic level of social influence to maximize the diffusion over the app-store platform. Formally, I solve the following optimization problem:

$$\max_{\{c_{jt}^{imm}\}} \sum_{t=1}^{T} \sum_{j=2}^{J} \sum_{i=1}^{I} \frac{\exp(u_{ijt})}{1+\sum_{j=1}^{J} \exp(u_{ijt})}$$

$$c_{jt-1}^{imm} \leq c_{jt}^{imm}, \forall j,t \qquad (20)$$

Table 1.14 presents the implications of each of these four policies. Surprisingly shutting down the social influence improves the total expected adoptions of mobile apps on this app-store platform. This further confirms that this platform does not have enough quality to retain its customers. However, an optimal social influence policy shows 13.6% increase in total expected adoptions of the platform. This optimal policy decreases adoption of mobile apps in Reference/Dictionary category, but increases the expected adoption of mobile app categories in Logic/Puzzle/Trivia, device tools, and Games the most. A common characteristic of these three categories is their popularity, so I tried to explain these improvements by optimal policy with the popularity of mobile apps in each of the categories over the Apple's app-store.

Table 1.14. COUNTERFACTUAL ANALYSIS: Change in the adoption level by intervening social influence

| Category specific counterfactual results: | original expected adoption | shut down social influence | 1% more social influence | 1%less social influence | An optimal social influence |
|---|---|---|---|---|---|
| Device Tools | 875.83 | -57% | 0.8% | -0.7% | 55.8% |
| eBooks | 189.45 | -1% | 0.0% | 0.0% | 2.5% |
| Games | 187.51 | 19% | -0.3% | 0.3% | 58.6% |
| Health/Diet/Fitness | 22.21 | 0% | 0.0% | 0.0% | 6.1% |
| Humor/Jokes | 255.09 | 0% | 0.0% | 0.0% | 0.6% |
| Internet/WAP | 1042.20 | 23% | -0.4% | 0.5% | 14.1% |
| Logic/Puzzle/Trivia | 249.12 | 25% | -0.6% | -0.2% | 109.1% |
| Reference/Dictionaries | 1262.09 | 16% | -0.4% | 0.3% | -36.2% |
| Social Networks | 21.66 | 0% | 0.0% | 0.0% | 0.3% |
| University | 18.08 | -1% | 0.0% | 0.0% | 1.5% |
| Total improvement | 4123.25 | 1% | -0.1% | 0.1% | 13.6% |

Table 1.15 presents the result of regressing the improvement under optimal social influence policy on popularity rank of the mobile app category. The correlation between mobile app category popularity and the improvement under optimal policy is positive and significant. This result suggests that more popular mobile app categories have more improvement under optimal policy. This result indicates that this app-store can improves its adoption by 13.6% if it can use social influence to increase the adoption of more popular mobile app categories.

Table 1.15. COUNTERFACTUAL ANALYSIS: Explain optimal social influence improvement with popularity rank of the app category on the app-store

| | *Coefficients* | *Standard Error* | *p-value* | *t Stat* |
|---|---|---|---|---|
| Intercept | -0.185 | 0.147 | 0.245 | -1.254 |
| Category popularity rank | 0.050* | 0.015 | 0.010 | 3.388 |

* p<0.01

Finally figure 1.8 presents social influence level for this optimal policy. This policy suggests early increase in the social influence by potentially a viral marketing campaign.

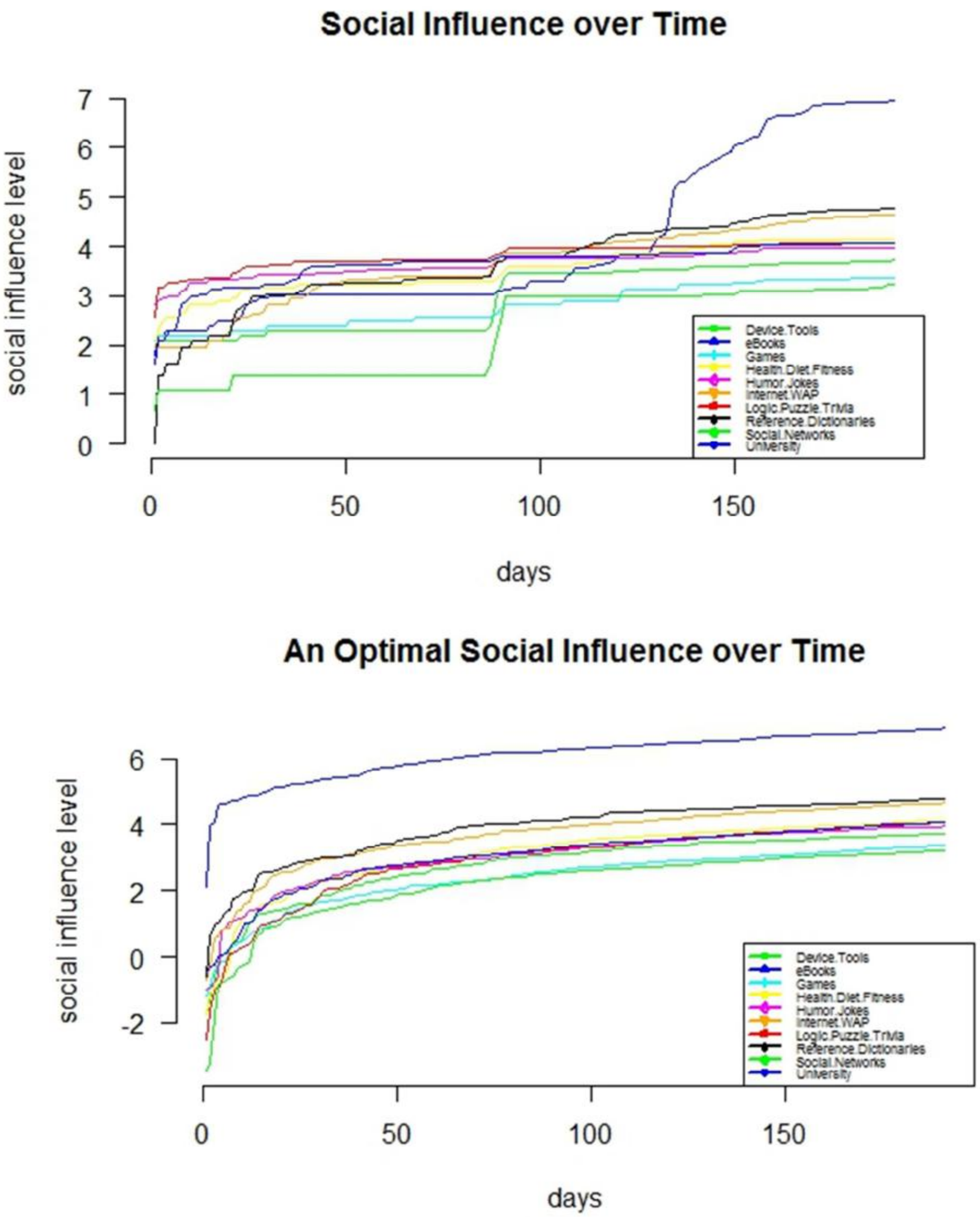

Figure 1.8. COUNTERFACTUAL ANALYSIS: an optimal social influence strategy to increase expected adoption level by 14% (log scale)

## 1.9. CONCLUSION

In this paper, I developed an approach that combines macro diffusion model with micro choice model to allow app-stores to target their customers and proposed Dirichlet Process to model customers' heterogeneity, and Unscented Kalman Filter to estimate social influence measure. Then, using a large data set from an African app-store, I showed that social influence is an important factor in determining adoption choice of customers. My results demonstrate that ignoring social influence in modeling customers' adoption can bias the choice parameters' estimates. Furthermore, my results indicate that social influence on mobile app adoption choices is effective locally (within the city of the study) rather than globally (over the internet). I benchmarked the mobile app adoption process against the same process for classical economy goods, and I find that mobile app adoption process is similar to the same process for music CDs.

I further illustrated how estimated model can be used to analyze counterfactual scenario where the app-store platform optimizes its intervening social influence. This counterfactual analysis showed that, if this app store runs viral marketing campaign focusing on more popular mobile app categories, it can increase its total adoption by 13.6%. I believe that my modeling approach, proposed estimation method, and derived empirical insights in this paper can be of interest to both practitioners and scholars in academia.

# CHAPTER 2

# DO BIDDERS ANTICIPATE REGRET DURING AUCTIONS? AN EMPIRICAL STUDY OF AN AFRICAN APP-STORE

Meisam Hejazi Nia

Naveen Jindal School of Management, Department of Marketing, SM32

The University of Texas at Dallas

800 W. Campbell Road

Richardson, TX, 75080-3021

## 2.1.ABSTRACT

I developed a structural model that accounts for bidders' learning and their anticipation of winner and loser regrets in an auction platform. Winner and loser regrets are defined as regretting for paying too much in case of winning an auction and regretting for not bidding high enough in case of losing it, respectively. Using a large data set from eBay and empirical Bayesian estimation method, I quantify the bidders' anticipation of regret in various product categories, and investigate the role of experience in explaining the bidders' regret and learning behaviors. I also showed how the results can be used to increase eBay's revenue significantly. The counterfactual analyses showed that shutting down the bidder regret via appropriate notification policies can increase eBay's revenue by 24%.

Keywords: winner and loser regret in auctions, affiliated value auction, emotionally rational bidders, Bayesian updating structural model

## 2.2. INTRODUCTION

It is not uncommon to regret one's own bidding decision at the end of an online auction. Whether this is about regretting for giving up too easily on a bidding war or regretting for losing self-control and bidding too high, bidders more than rarely feel discomfort about their final bids. eBay forums are filled with questions like "I won an auction but regret: What can I do?", and, in fact, eBay tries to educate its users for bidding without regrets[5]. Winning an auction on eBay is a contract to complete the sale, and not honoring this contract has serious consequences, including being banned from any transactions on its website. Therefore, the desire to avoid these consequences, along with the bad experiences one can have about the product, seller, or one's own bidding behavior or even the common sense lead to an anticipation of end-of-auction regret during the bidding period.

[5] Bertolucci, Jeff. "Big Data Analytics: Descriptive Vs. Predictive Vs. Prescriptive - InformationWeek." InformationWeek. December 31, 2013. Accessed March 23, 2016. http://www.informationweek.com/big-data/big-data-analytics/big-data-analytics-descriptive-vs-predictive-vs-prescriptive/d/d-id/1113279.

I consider two types of regret that are studied in the auction literature: Bidders might feel *winner regret* when they win an auction but feel they pay too much, since their winning depends on them being the most optimistic among the auction participants about the market value of the auction item and/or the honesty of the seller (Bajari and Hortacsu 2003b). On the other hand, a bidder might feel *loser regret* when she loses an auction in which the winning bid turns out to be less than her valuation of the item. Clearly, the latter type of regret realizes when the bidders bid naively (or strategically) instead of bidding their true valuations of the items. Intuitively, anticipation of winner (resp. loser) regret should make the bidders lower (resp. increase) their bids.

There are many studies in the auction literature that show that the anticipation of both types of regret significantly affects the bidding behaviors of auction participants in various settings. Experimental studies such as Filiz Ozbay and Ozbay (2007) and Engelbrecht-Wiggans and Katok (2008) study these effects under the first-price sealed-bid auction setting and show that they have significant implications on the bidding behavior. Although, in theory, the second price nature of eBay auctions implies that bidders should not experience winner regret (see, for example, Ariely and Simonson 2003), Bajari and Hortacsu (2003a) and Yin (2006) investigate the ``winner's curse[6]" in eBay auctions and suggest that eBay bidders anticipate it too, and hence act strategically[7]. I also focus on eBay markets in this paper.

Auction platforms, such as eBay, act as a two sided market by connecting sellers and bidders without taking ownership of the auction item. Auctions in these platforms involve a vast amount of different sellers, bidders, and products in different categories, and hence they exhibit a high level of heterogeneity in behavior. Although many descriptive and predictive tools are studied in the auction literature to deal with this heterogeneity and large data sets (e.g., Park and Bradlow 2005; Bradlow and Park 2007; Zeithammer and Adams 2010), the prescriptive analyses remain limited[8]. However, counterfactual analyses relying on structural models that control for consumers' decision processes can have remarkable contributions. One such contribution is being able to investigate the effects of notification policies (such as notifying bidders about the similar auctions in the past) on the platform revenue. To work towards filling this gap, in this paper, I developed a structural model to explain the bidder behaviors in an online auction platform.

Considering all the requirements for a viable explanation of the auction platforms, I asked the following questions: Can I design a computationally tractable system to estimate bidders' bidding behaviors in an online auction platform? To what extent do bidders anticipate winner and loser regret and how do they vary in bidders' experience and learning behavior? What is the effect of intervening bidders' regret by notification policies on the auction platform's revenue?

---

[6] I use the term "winner regret" to refer to the explained phenomenon, but "winner's curse" is also used in the literature.

[7] Zeithammer and Adams (2010) suggest that sealed-bid second price auction is not a good abstraction for eBay auctions. Bidders' naivety can also result in this inconsistency. I comment more on this issue in Results section.

[8] Bertolucci, Jeff. "Big Data Analytics: Descriptive Vs. Predictive Vs. Prescriptive - InformationWeek." InformationWeek. December 31, 2013. Accessed March 23, 2016. http://www.informationweek.com/big-data/big-data-analytics/big-data-analytics-descriptive-vs-predictive-vs-prescriptive/d/d-id/1113279.

To answer these questions, I developed my model considering many important aspects of bidders' behaviors. In particular, I account for the emotionally laden context of auctions where, in addition to the regret anticipation, the bidders do not know the item's market value and learn about the value of the product during the bidding process, for example, by gaining additional information about the auction item or resolving some of the uncertainties about the seller or about their own needs (Hossain 2008, Zeithammer and Adams 2010, Okenfels and Roth 2002). I also consider the fact that bidders tend to bid incrementally in online auctions (Chakarvarti et al. 2002; Zeithammer and Adams 2010), and have different levels of experience which affects their bidding behaviors (Ariely et al. 2005; Wilcox 2000; Srinivasan and Wang 2010). I further take into account both common and private value components of the auction item (i.e., affiliated value), and bidders' learning from the current highest bid during the bidding process. Due to the emotionally laden context of eBay auctions, I assume that bidders might lose their global focus, as Ariely and Simonson (2003) suggest, so they show inertia and generally do not search across auctions (Haruvy and Leszcyc 2010).

I used the following estimation strategy to identify the parameters of my model: First, at each discrete time period during the bidding process, I modeled the utility of a bidder consisting of her expected profit, i.e., the difference between her valuation and bid, and anticipated winner and loser regrets. Second, assuming the observed bid at each period is the one that maximizes the bidder's utility for a given valuation level, I derived –using the first order condition for the utility function— the bidder's revealed latent valuation for the auction item at that time period. Finally,

I assumed that this derived latent valuation (which is now a function of the bidder regret, among others) consists of a common value, a private value, and a component consisting of bidder's learning the value of the auction item from the highest observed bid. This approach allows me to identify the regret parameters.

More specifically, since there is a common value component of the auction item, the number of bidders also matter in this process. To account for the bounded rationality and incomplete information in the model, I considered that the bidders perceive the observed bid and number of bidders as a noisy measure of the latent bid and the latent number of bidders. I modeled the bidders' Bayesian learning of the latent bid and the latent number of bidders using Kalman Filter theory, which Jap and Naik (2008) introduced to the auction literature. In this structure, I assumed that bidders' beliefs about the latent bid and latent number of bidders follow a first order Markov process. Similarly, I assumed that the common value element of valuation follows a first order Markov process as well, with a drift and a common time varying signal.

To account for heterogeneity and to avoid over-fitting the data with a large number of parameters, I clustered the bidders and utilized the eBay-specified auction clusters, and shrunk the bidder specific (regret and valuation) and auction specific (evolution of bids and the number of bidders) parameters within these bidder and auction clusters. I used a mixture normal distribution model to cluster the bidders using their observed characteristics, which are used as proxies for bidders' experience level. Finally, I optimized the Maximum a Posteriori (MAP) of the model given the segment and cluster membership of each of the bidders and auction items over a large eBay data set. To optimize MAP, I used simulated annealing method, which is a metaheuristic global optimization method used both in robotics and portfolio optimization in finance (Crama and Schyns 2003; Zhuang et al 1994)[9].

I estimated this model over an eBay data set that I crawled and scraped from the web in May 2014. This sample consists of around 58,000 bids of around 12,000 bidders in around 1,600 auctions that offered items for sale in 19 different categories presented in Table 1. The estimation results show that, in all auction categories, both winner and loser regrets are significant and I find a positive relationship between winner and loser regret. I also find that those who are more regretful stick to status quo, i.e., they update their valuations less frequently and learn less from others. Furthermore, I find that experience can explain the heterogeneity in the bidders' learning, updating, and regretting behavior.

I further used the estimated model to analyze a counterfactual scenario where the auction platform shuts down the bidders' winner regret. This analysis shows that, if an auction platform can shut down winner regret of bidders by its notification policies, it can increase its revenue by 24%. I also observed that shutting down winner regret can cause the highest bid to increase two to four folds in some auctions.

Using notification policies to affect the bidders' behaviors is not uncommon in eBay. For instance, my personal interview with an eBay scholar suggested that eBay is concerned about bidders' loser regret that might lead to a potential churn effect. Therefore, they use notifications to inform the bidders who might lose the auction if they do not change their bids. Empirical evidence of significant (winner and/or loser) regret of bidders might invoke using similar notification policies in online auction platforms[10]. Such policies are studied in the experimental literature as well, and shown to be effective in influencing the regret levels of bidders (see, for example, Engelbrecht-Wiggans and Katok 2008).

---

[9] I used this optimization approach because the Quasi Newton, Broyden–Fletcher–Goldfarb–Shanno (BFGS) optimization, or Bayesian sampling methods are computationally intractable over large data sets.

I believe the contributions of my paper can be of interest to both practitioners and scholars in academia. My contributions are threefold: First, I consider bidders' anticipation of winner and loser regret in the affiliated value setting, and propose a tractable empirical Bayesian method to estimate a structural model of bidder demand in an online auction platform. This model allows the auction platforms to run counterfactual scenarios. In this way, I contribute to the line of descriptive and predictive auction models for auction customer relationship management (Bradlow and Park 2007, Park and Bradlow 2005, Jap and Naik 2008, and Zeithammer and Adams 2010).

Second, I model the learning and affiliated value of bidders, and, by allowing for incremental valuation revelation in the proposed model, I allow for the incremental naïve bidding behavior. The importance of these features in a model are emphasized by Okenfels and Roth (2002), Hossain (2008), and Zeithammer and Adams (2010). This aspect of my model contributes to the stream of papers that model common and private value auctions structurally (e.g., Laffont et al. 1995, Bajari and Hortacsu 2003, Haile et al. 2003, and Haile and Tamer 2003). Unlike these papers, I consider the auctions as emotionally laden social contexts. To the best of my knowledge, I am the first to model the bidders' Bayesian learning and affiliated value updating processes to account for bidders' updating their uncertain valuations.

---

[10]Notifications providing information about similar auction items, such as the highest bids, paid amounts, and number of bidders in those auctions can help to influence the winner regret of bidders. Similarly, more granular information about the sellers might help with the trust issues, which again can affect the winner regret level.

Third, I contribute to the auction regret literature by proposing a method that identifies regret parameters structurally using field data. Importance of bidders' anticipation of winner and loser regret are emphasized in the literature, for example, by Ariely and Simonson (2003), Filiz Ozbay and Ozbay (2007), and Engelbrecht Wiggans and Katok (2008). The latter two studies use experiments to show that notification policy can affect the bidders' feeling regret and potential over and under bidding. I used real company data in this paper, and my structural modeling approach allows the auction platforms to quantify the impacts of new policies, target different bidders, and customize their operations conditional on bidder behaviors.

The rest of the paper is organized as follows: In Section 2, I review the relevant literature. Section 3 provides a detailed description of data. Section 4 describes my structural model and how the empirical Bayesian method can be used to estimate the parameters of the model using eBay data. I interpret the estimation results in Section 5, and explain how I can use the estimated parameters in testing a counterfactual scenario where the auction platform shuts down the bidder regret in Section 6. Next, I test the robustness of some of my assumptions and methods in Section 7. Finally, Section 8 presents my concluding remarks and discussion for future research directions.

### 2.3. LITERATURE REVIEW

My paper resides at the intersection of four streams of literature: (1) customer relationship management using auction big data; (2) bounded rationality, trembling hand, learning, and the affiliated value of bidders; (3) the emotionally rational or regretful bidders; (4) the theoretical, experimental, and empirical studies of auctions. I explore each one of them in the following sections.

### 2.3.1. Customer Relationship Management of Auction Platforms

Numerous studies investigate the behavior of bidders on the auction platform to estimate the demand or to extract information that can be used in customer relationship management and targeting. For example, Park and Bradlow (2005) propose a stochastic model to identify the bidders, the conditions in which they bid, and the amount of their bids, which are useful for customer relationship management. Bradlow and Park (2007), further extend their research by proposing a record breaking stochastic approach to recover the latent number of bidders in the context of the first price auction. Both studies acknowledge that empirical literature has demonstrated flaws in the theoretical prediction of auctions, but they argue that these flaws can be corrected by a model which accounts for the behavioral aspect of the bidders' decision making. More recently, another predictive study conducted by Jap and Naik (2008) uses the Kalman Filter theory to develop a ``Bid Analyzer" that allows one to estimate the distribution of auction participants' latent bids. All of these papers call for new studies to model the structural aspect of bidders' decisions for policy experimentation.

To model the bidder behavior structurally, Bajari and Hortacsu (2003) make the simplifying assumption that the eBay auction with proxy bidding approximates a second price auction. Their study employs a data set from a 1998 coin auction on eBay to estimate a reduced form common value model (for tractability), but it calls for studies that model affiliated value (i.e., existence of both common and private value elements). Although their proposed model relies on the assumption that bidders are fully rational, they try to recover winner curse from measuring the amount shed by a bidder when a new bidder enters the auction. Zeithammer and Adams (2010) carry out a series of statistical tests to cast doubt on the assumptions that the proxy bidding mechanism is equivalent to the second price sealed bid auction, and that bidders' bids are equal to their valuations. They recommend employing a reduced form modeling approach. In discussing Zeithammer and Adam's study (2010), Hortacsu and Nielsen (2010) and Srinivasan and Wang (2010) note that, although some of its tests are questionable, its main hypothesis is strongly supported. Both of these commentaries call for a structural model based on Zeithammer and Adam's new findings, particularly those indicating that both naïve and sophisticated bidders might exist on eBay, and bidders' experience plays an important role on their bidding behaviors.

Yao and Mela (2010) take the auction platform as a two sided market, and jointly model the choices of bidders and sellers structurally to extract the value of the customer lifetime and the impact of the commission policy on the auction platform revenue. They consider only one auction category and model bidders' disutility in the form of historical cost function, rather than in the form of winner and loser regret. In parallel with their paper, Haruvy and Leszczyc (2010) also model the disutility of bidding in the form of the inertia of bidders within an auction. They attribute this inertia to search cost.

All of the above studies unanimously suggest that future policy experiments is possible only by a structural model of bidder learning in which heterogeneity is explained by experience measures. They also agree that such a study should model common and private values jointly, in the form of an affiliated value model. Built over the above studies, I model the bidders' anticipation of winner and loser regret structurally. Understanding such consumer behaviors can benefit the auction platforms by allowing it to target its policies toward helping naïve consumers learn, if such learning is predicted to improve the revenues of auction platforms.

### 2.3.2. Bounded Rationality, Learning, and Affiliated Value of Bidders

There are many papers discussing that consumers are bounded rational, and their action is subject to flaws (see, for example, Simon 1972; Selten 1975; Kahneman and Tversky 1979; Tversky and Kahneman 1992; Camerer and Weber 1992; Hey and Orme 1994; Camerer and Ho 1994; Kahneman 2003). Various theories explain why consumers behave bounded rationally, from decision making and psychology perspectives (Simon 1972; Ellison 2006; Salant 2011; Kaufman 1999). Bounded rationality is referred to as naïve bidding in the auction context.

Naive bidders are known to bid in an ad-hoc manner or by matching their bids with others bids. In particular, Ely and Hossain (2008) define the naive bidder as the one who acts as if the amount she pays conditional on winning equals to her bid in eBay auctions. Additionally, Okenfels and Roth (2002) also define naïve and inexperienced bidder as a bidder who mistakenly treats the eBay auction as an English first-price auctions in which the winner pays the maximum bid.

Furthermore, Kagel et al. (1987) posit that the dominant strategy equilibrium does not organize second-price auction outcomes, as bids consistently exceed private values. Other studies posit that experience and learning can reduce bidders' bounded rationality, fostering more rational behavior. For example, Wilcox (2000) finds that experience leads to behavior which is more consistent with auction theory although the proportion of experienced bidders who behave in a manner inconsistent with the theory is quite large.

Ariely et al. (2005) find that experience reduces but does not eliminate considerable incremental bidding. In this respect, Ockenfels and Roth (2002) examine the multiple-bid phenomena to consider how bidders get information from other's bids, and then revise their willingness to pay in an auction with independent values. To describe bidders' learning, Hossain (2008) suggests that bidders do not always know their exact private valuation for a good and so learn spontaneously from the posted price. He concludes that bidders obtain information about their own and others' preferences as they participate in the auction.

These experimental studies are particularly relevant to my research in the sense that they emphasized the role of naïve bidder, learning, and experience. However, built over these studies, my paper integrates these processes in a structural model to help the auction platform manage and target its bidders. Moreover, in contrast to these studies, my paper accounts for behavioral regularities that stem from bidders' anticipation of winner and loser regret, another form of bounded rationality.

### 2.3.3. Emotionally Rational or Regretful Bidders

Several studies explain the bounded rationality of auction bidders by referring to bidders' uncertainty about the value of the commodity, which suggests that bidders might anticipate winner and loser regret in their decision. In particular, Holt and Sherman's (1994) theoretical study describes the acceptance of a bid as an informative event because it signals an overestimation of unknown value. They mention that winning/losing might result in regret, so the bidder might anticipate winner/loser regret in her decisions. In testing a regret theory in a first-price sealed-bid auction setting, Filiz-Ozbay and Ozbay (2007) find that the anticipation of winner and loser regret can be modified by a notification policy. Also, Engelbrecht-Wiggans and Katok (2008) find a similar phenomenon, and they conclude that the policy of revealing losing bids may decrease the auction holders' revenue. Although experiments are helpful for making causal inferences, an auction site might need a structural model to run counterfactual analysis and to target its bidders.

Regret construct has been the subject of many studies in the consumer behavior, psychology, decision science, behavioral economics, and marketing literature. A stream of literature in psychology and consumer behavior defines regret as a negative psychological response which occurs when an individual believes that a present situation would have been better if only she had decided differently (Peluso 2011; Gilovich and Medvec 1995; Van Dijk and Zeelenberg 2005; Simonson 1992; Zeelenberg et al. 2000; Inman and Zeelenberg 2002; Roes 1994). This regret can affect the consumers' decision-making through counterfactual thinking (Roes 1994). In particular, the consumer might consider the possible negative outcome of a previous choice in her future decisions and so might regulate her behavior to decide differently ex ante, by being regret averse (Peluso 2011; Zeelenberg and Pieters 2007; Boles and Messik 1995; Tsiros and Mittaal 2000).

Many studies in psychology literature classify the different types of regrets according to action and inaction regret categories. The first category refers to consumers' feelings of sorrow for what they have done, and the second refers to consumers' feelings of sorrow for what they have not done. The former is analogous to winner and the latter to loser regret discussed in the auction literature (Filiz-Ozbay and Ozbay 2007; Engelbrecht-Wiggans and Katok 2008). Furthermore, action regret has short term effect and evokes intense feeling, and inaction regret has long term effect and evokes wistful feelings (Gilovich et al. 1998; Gilovich and Medvec 1995; Keinan and Kivetz 2008).

Bell (1982) argues that, after making a decision under uncertainty, the decision maker may discover the relevant outcomes by learning that another alternative would have been preferable. This learning creates a sense of loss or regret that, if incorporated explicitly into the expected utility framework, better predicts individuals' decisions. According to Loomes and Sugden (1986), the violation of the conventional expected utility suggests that important influential choice factors are overlooked, perhaps because of the misspecification of the conventional theories. They propose an alternative approach, formulating a theory of expected modified utility to account for the individual's capacity to anticipate feelings of regret and rejoice. Such theory rests on two fundamental assumptions: First, many people experience the sensations called regret and rejoice; and, second, they try to anticipate and take into account those sensations in making decisions under uncertainty. Guided by the mentioned study's observation and suggestions, many theoretical studies incorporated regret to explain how the optimum pricing strategy might change in a new setting that incorporates regret (Popescu and Wu 2007; Nasiry and Popescu 2011; Heidhues and Koszegi 2008; Su and Zhang 2009; Diecidue et al. 2012; Nasiry and Popescu 2012; Ozer and Zheng 2012). All of the above mentioned studies are useful in expanding the domain of knowledge about consumers' regret and the effect of such phenomenon on the consumers' decisions. However, none of them has modeled both the rational and emotional aspects of bidders' decision making in the context of the online auctions, where bidders' values are affiliated. In this context, the bidders learn the value of the commodities by observing others bids as well.

### 2.3.4. Theoretical, Experimental, and Empirical Auctions studies

I classify the papers in this section into three categories based on the modeling assumption about the bidders' valuations: independent private value, common value, and affiliated value models. The model of independent private value assumes each bidder has a different private value known only to him (Laffont et al. 1995; Guerrere et al. 2000; Haile and Tamer 2003). In respect to the common value assumption, Haile et al. (2003) proposes a non-parametric test for first-price sealed-bid auctions based on the fact that winners curse might exist in such auctions. Bajari and Hortacsu (2003) also assume that eBay's auction can be approximated with second-price sealed-bid auction to estimate a structural model of common value to recover winner curse. However, they acknowledge that a better option might be assuming that the eBay auction is affiliated value. Finally, affiliated value is a form of valuation that is drawn from a joint distribution of valuations, consisting of both private and common value component (Li et al. 2002; Campo et al. 2003). Although these studies expand the domain of knowledge about the implication of various assumptions, they do not consider bidders' emotional response and bounded rationality. Chakravarti et al. (2002) call for future studies of this issue by emphasizing that the learning process might alter the valuations of bidders by an "information cascade". They suggest that such learning and value affiliation might induce strategic emulation of preceding bidders without considering private signals. In this paper, I incorporate learning, value affiliation, and emotion in a structural model.

Structural models can consider either consumers' learning, in the form of an adaptive Bayesian learning model, or the consumers' expectation, in the form of a forward-looking approach. Zeithammer (2006) argues that buyers can benefit from forward-looking strategies if they take into account the information provided by the announcements of upcoming auctions. He implicitly states that a forward-looking model for bidders in online auction is intractable, so in developing such a model, he uses several simplifying assumptions. Given all these simplifying assumptions, it is not clear that a forward looking approach has much more merit than the Bayesian adaptive-learning approach. Further, it is not clear how an emotionally laden environment of an auction might foster the forward-looking behavior of the bidders.

Smith (1989) notes that auction contexts are often emotion-laden and suggests that the outcomes reflect communal legitimization of both price and allocation given uncertainty about value, preferences and fairness. Chakravarti et al. (2002) suggest that the individual and social nature of the value determination processes is a fertile area for future research. Furthermore, whether bidders experience regret or not when bidding aggressively and winning may depend on their cognitive skills for counterfactual reasoning and their facilities with motivational processes (e.g., dissonance and attribution) for managing the emotions of victory and defeat, according to Tsiros and Mittal (2000).

The Filiz-Ozbay and Ozbay (2007) and Engelbrecht-Wiggans and Katok (2007) studies focus on experimentally attributing underbidding and overbidding to regret theory; Astor et al. (2011) finds that aforementioned studies' theoretical predictions for the effect of regret holds, by employing an approach that combines auction experiment with psychological measures that indicate emotional involvement. Furthermore, Greenleaf (2004) shows that the auction sellers also anticipate regret and rejoice when they set the reserve price, which is the lowest auction price that the seller will accept. Engelbrecht-Wiggans (1989) proposes a utility theory that depends not only on the profit, but also on the regret of the outcomes (e.g., money left on the table). Further, Engelbrecht-Wiggans and Katok (2007) points out that in the case of independent private value first-price sealed-bid auctions, bidders bid above risk neutral Nash equilibrium, which can be explained only by regret theory.

Overall, built on the aforementioned studies, I develop a general model that nests all learning, experience, value affiliation, and bidders' anticipation of winner and loser regret in a structural form that allows an auction platform to target its customers and run counterfactual policies. Without a model that controls for all these mechanisms, the revenue implication of bidders' regret for the auction platform may not be clear.

### 2.4. DATA

I acquired the data set by crawling and scraping a sample of auctions from eBay website, during May 2014. It consists of 58,285 bids of 12,247 bidders on 1,647 auction items within various auction categories. eBay's revenue is based on a complex system of fees for services, including listing product features ($0.10 to $2) and a Final Value Fee for each sale (10% of the total amount of the sale, i.e., price of the item plus shipping charges), and it exceeded $17.90 billion in 2014. Millions of collectibles, décor, appliances, computers, furnishings, equipment, domain names, vehicle, and other miscellaneous items are sold on eBay daily. Generally, sellers can auction anything on the site as long as it is not illegal and it does not violate the eBay prohibited and restricted item policy.

eBay uses a bidding mechanism called proxy bidding. This mechanism asks the bidder to submit the maximum amount she is willing to pay for the item, which is called a proxy bid. Then, eBay's software bidding agent (called proxy engine) bids incrementally on the bidder's behalf up to this maximum value, which remains hidden from other bidders until someone outbids it. As new proxy bids enter, the proxy engine sets the current winning bid to the second highest bidder’s maximum value plus the minimum increment specified by eBay. The current winning bid is displayed on the auction board throughout the auction. At the end of the auction, the bidder with the highest proxy bid wins the item and pays a price equal to the second highest bidder's maximum bid plus the increment. This process makes eBay auctions a hybrid of the English and second-price sealed bid auctions. Table 2.1 presents a possible path for the proxy and observed bids in an auction where the starting price is $25 and the minimum increment is $1.

Table 2.1. A sample bid sequence on an eBay auction with $25 reservation bid and $1 minimum increment

| Bid number | Max. bid (unobservable) | Bid on the auction board (observable) |
|---:|---:|---:|
| 1 | $50 | $25 |
| 2 | $40 | $41 |
| 3 | $70 | $51 |
| 4 | $65 | $66 |

In eBay’s website, I observe both the amount that each bidder puts in as her proxy bid and the bids automatically generated by eBay’s proxy engine. I filtered the automatic bids out to be able to work with the actual bids of the bidders. Note that, in eBay’s system, if someone puts a bid between the displayed (automatic) bid and the highest proxy bid, this action will not reveal the highest proxy bid – the highest proxy bid will only be revealed after someone outbids it. Hence, even though the displayed bids always increase over time, the proxy bids may not be in increasing order (see the example in Table 2.1). I sorted the bids before using it in the model estimation to overcome this issue. Figure 2.1 presents the evolution of observed bids across six sample items.

For this study, I randomly selected 19 auction categories. The selected categories have both luxury and widely available goods, so the sample allows to test whether the regret levels are different across these two categories. For instance, a bidder might regret more for losing a luxury item auction than losing a necessity item auction. Table 2.2 shows the categories that I use in this study along with the number of auctions in each category. I classified the first nine categories as luxury good categories, and the next ten categories as the widely available goods.

Table 2.2. Auction categories in the eBay data

| Auction category | Number of auction Items |
|---|---|
| Jewelry and Watches | 149 |
| Collectibles | 103 |
| Crafts | 78 |
| Pottery and Glass | 74 |
| Antiques | 68 |
| Art | 70 |
| Entertainment Memorabilia | 88 |
| Tickets and Experiences | 91 |
| Stamps | 72 |
| Toys and Hobbies | 93 |
| Books | 84 |
| Clothing, Shoes and Accessories | 84 |
| Gift Cards and Coupons | 85 |
| Music | 86 |
| Consumer Electronics | 83 |
| DVDs and Movies | 87 |
| Dolls and Bears | 84 |
| Health and Beauty | 74 |
| Video Games and Consoles | 93 |

Table 2.3 presents a sample of auction items. For each auction item, I know its title, category, number of bidders, number of bids, and the duration of the auction. I call this auction specific information. In my data set, the average number of bidders and bids in each auction are 9.52 and 49.19 with standard deviations of 4.58 and 19.40, respectively. The average duration of auctions is 4.74 days with a standard deviation of 1.67 days.

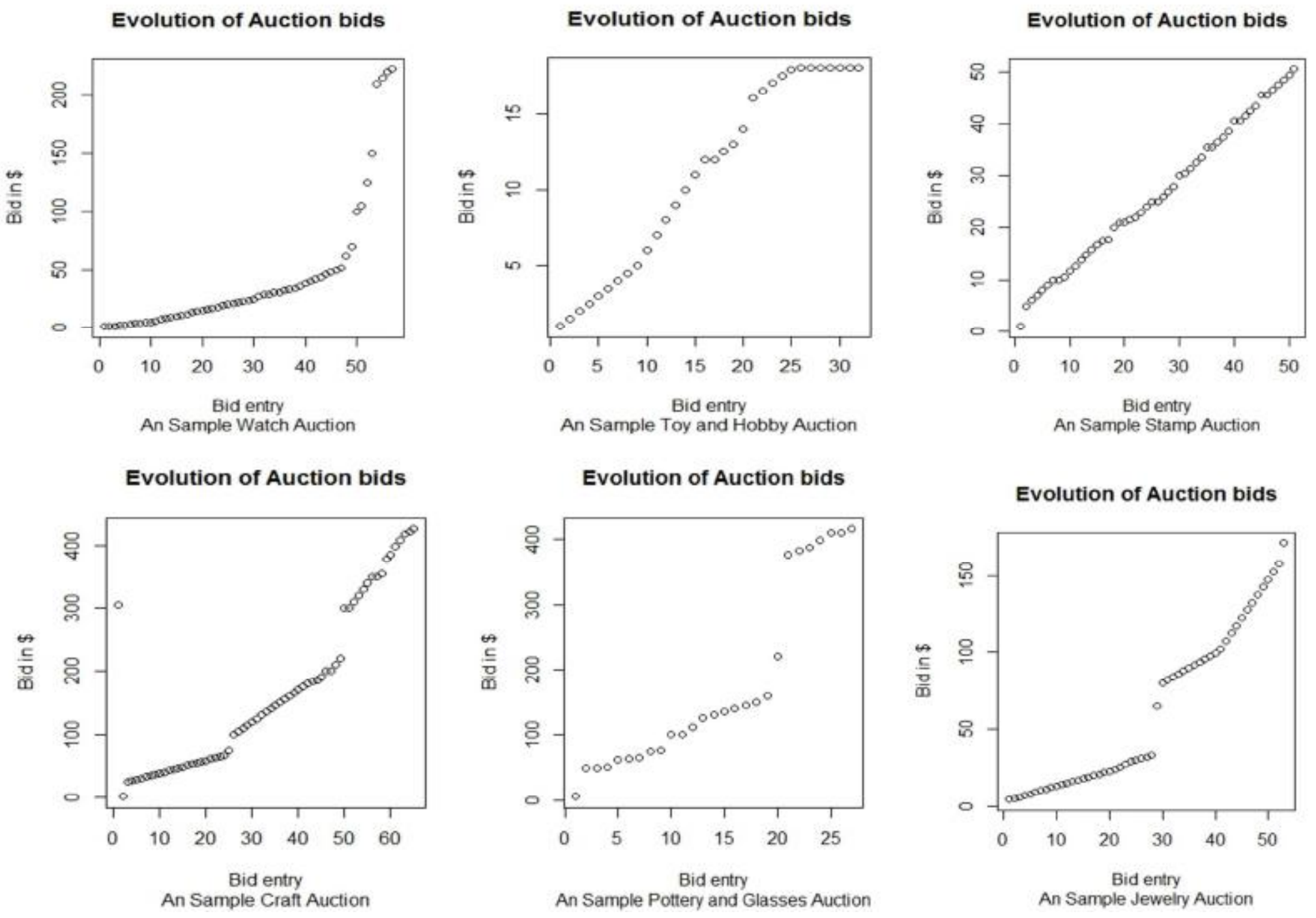

Figure 2.1. Evolution of Bids in six sample auctions

Table 2.3. Sample auctions in the eBay data

| Auction Item Title | auction category | winning bid | number of bids | number of bidders | Ended |
|---|---|---|---|---|---|
| Vintage Original Co-op porcelain sign | Collectibles | $1,000.00 | 92 | 12 | May 18, 2014 , 2:15PM |
| $3/1 Pantene Product Coupons Shampoo Conditioner Styler | Gift Cards & Coupons | $17.50 | 30 | 5 | May 19, 2014 , 6:30PM |
| Genesis Breyer P-Orridge "Naked Eye" Autographed Camera w/ Original Negatives | Entertainment Memorabilia | $900.00 | 75 | 9 | May 22, 2014 , 2:00AM |

Figure 2.2 presents the evolution of the number of bidders for a sample of six auctions in different auction categories. An interesting observation in these auctions is a spike at the rate of entrance at the last minutes. This behavior is known as sniping in the auction literature.

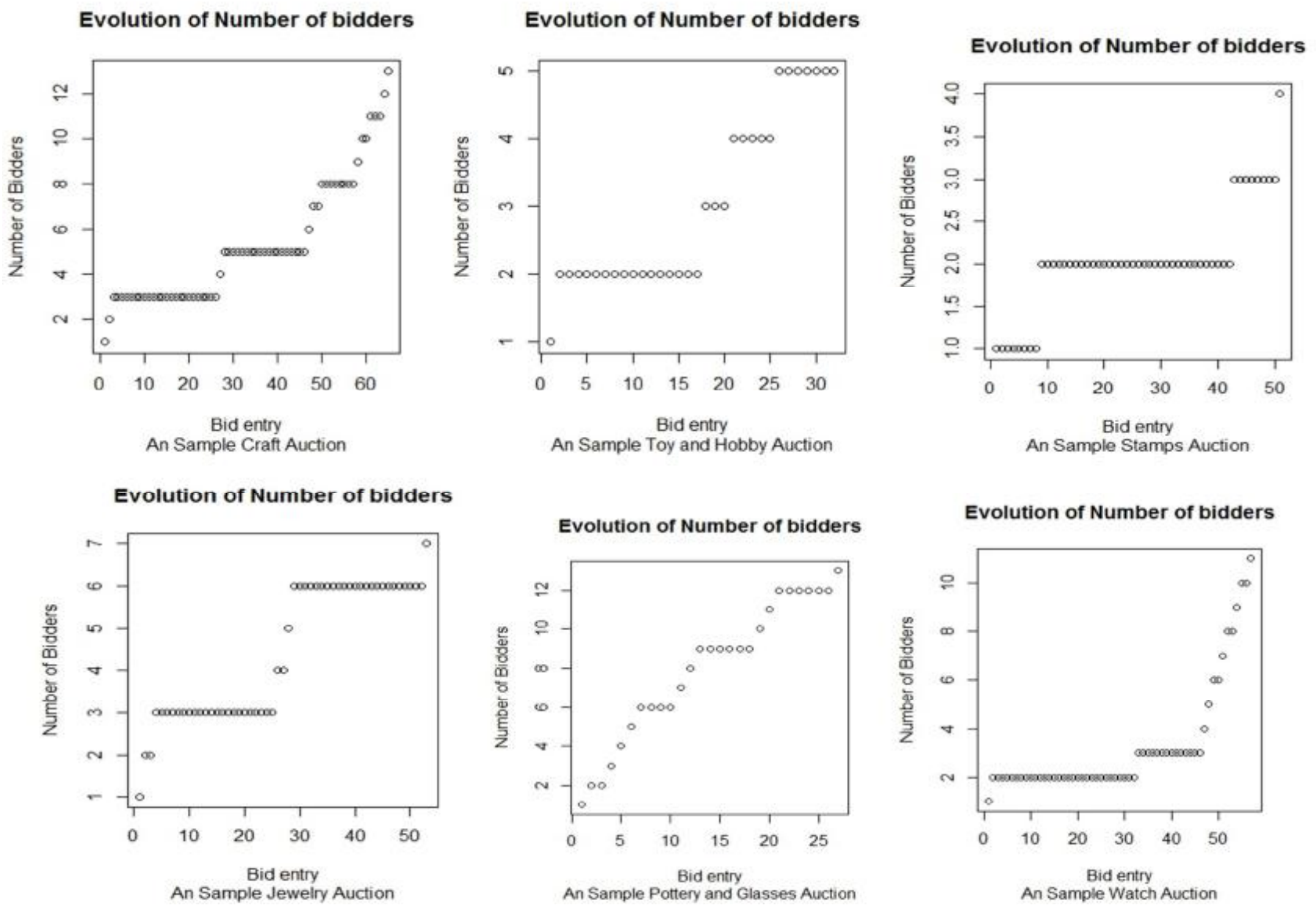

Figure 2.2. Evolution of number of participating bidders in six sample auctions

eBay employs feedback as a reputation mechanism for its members to decrease their uncertainty about bidder and seller characteristics. While buyers can leave sellers negative, neutral or positive feedback, sellers can leave buyers positive feedback or choose not to leave feedback. Over time, eBay members develop a feedback profile, or reputation, based on these ratings. This information appears next to the members’ name and on the members’ profile.

In my data set, each bidder attends only one of the auctions. I observe each bidder’s feedback score, number of bids on the item in question, total number of bids within the last 30 days, total number of items / categories bid on within the last 30 days, and number of bid activity with the current seller. This bidder specific information capture different types of proxies for the experience of the bidders, and, hence, I utilize them in my analyses. For example, while bidding on one auction category might show high level of concentration, bidding on three categories might show high level of differentiation.

Table 2.4 presents the summary statistics of average bidder characteristics within each auction category. It shows significant heterogeneity among bidders in different auction categories. Another important observation from Table 2.4 is that, on average, bidders in each auction category have bid at least three times for the same item, which suggests a multiple (incremental) bidding behavior. This behavior, which is also reported in the auction literature for similar settings (see, for example, Zeithammer and Adams 2010) suggests that bidders might not enter their valuations as proxy bids, as second-price sealed bid auction theory suggests.

Table 2.4. Summary statistics of the average bidder characteristics within each of 19 auction categories

| Characteristic | Mean | SD | min | max |
|---|---|---|---|---|
| size | 644.53 | 238.21 | 453 | 1550 |
| avg. feedback score | 714.53 | 260.39 | 342 | 1301 |
| sd feedback score | 2763.37 | 1897.84 | 745 | 8033 |
| avg. Number of bids on this item | 4.84 | 0.74 | 3 | 7 |
| sd Number of bids on this item | 8.05 | 1.93 | 4 | 13 |
| avg. total number of bids in 30 days | 195.16 | 100.57 | 56 | 504 |
| sd total number of bids in 30 days | 493.74 | 260.94 | 110 | 1065 |
| avg. Number of items bidded on in 30 days | 93.63 | 57.41 | 30 | 264 |
| sd Number of items bidded on in 30 days | 251.42 | 197.83 | 53 | 1001 |
| avg. Bidding Activity with current Seller | 28.63 | 8.51 | 17 | 50 |
| sd Bidding Activity with current Seller | 31.11 | 4.05 | 24 | 40 |
| avg. Number of categories bided on | 2.05 | 0.39 | 1 | 3 |
| sd Number of categories bided on | 1 | 0 | 1 | 1 |

## 2.5.MODEL DEVELOPMENT

I developed an agent-based structural model to predict the revenue implications of possible auction platform policies, such as the notification policy, by counterfactual analyses. I model bidders' actions in a Bayesian adaptive-learning structure. The term adaptive-learning refers to the bidders' updating beliefs about the value of the item, the distribution of the bids, and the number of bidders conditioned on observing noisy signals (Jap and Naik 2008). Bayesian learning approach is appropriate in my setting, because auctions are emotionally laden settings, in which users' preferences are correlated (Chakravarti et al. 2002). In this environment, majority of bidders are naïve, so they learn the value of the auction items by observing others' bids (Hossain 2008; Zeithammer and Adams 2010; Okenfels and Roth 2002). Furthermore, another advantage of the Bayesian-learning approach is that it is computationally tractable in the auction context[11].

### 2.5.1 Modeling the valuation of auction items

I identify the anticipated loser and winner regret of the bidders by first modeling the bidder's valuation with a dynamic adaptive utility maximization approach. This approach incorporates the anticipated regret of the bidder in the utility specification. Then --using the first order condition for the utility function--I derive the latent valuation of the bidders and embed it into another valuation specification combining the affiliated values of bidders and learning from others in the bidding process. This method provides the required identifying equations for the anticipated loser and winner regrets for each bidder.

---

[11] Forward-looking approaches are intractable in this setting; see, for example, Zeithammer (2006), which proposes many restrictive simplifying assumptions to deal with the intractability of a forward-looking model.

#### 2.5.1.1. Dynamic adaptive utility maximization approach:

I first specify the utility of an emotionally rational bidder. In a second price auction setting, the term "emotionally rational bidder" refers to a bidder that, rather than bidding her private value as suggested by the auction theory, acts naively by comparing her bid with the bids of the population. The reason of this comparison might be a lack of information about the value of the auction item and/or about the seller or bidder's past bad experiences. In this way, the bidder anticipates a possible regret for potentially winning or losing with his current bid, so ex-ante the bidder compares her bid with the highest bid of the population. To quantify this phenomenon, I adopt the utility function format that Engelbrecht-Wiggans and Katok (2008) and Filiz Ozbay and Ozbay (2007) specify for emotionally rational bidders. Formally, the utility is defined as

$$u_{it} = \underbrace{(v_{it} - b_{it})G_{t-1}(b_{it})}_{\text{Rational gain from}} - \underbrace{\int_{z_t \le b_{it}} \alpha_i (b_{it} - z_t) dG_{t-1}(z_t)}_{\text{Winner regret}} - \underbrace{\int_{b_{it} \le z_t \le v_{it}} \beta_i (v_{it} - z_t) dG_{t-1}(z_t)}_{\text{Loser regret}} \quad (1)$$

where $u_{it}$ denotes the utility of bidder $i = 1..I$ at the time of bidding the t'th bid, $t = 1..T$, in a particular auction (I suppressed the auction subscripts for ease of exposition). $v_{it}$ denotes the time varying value of bidder i at time t, and $b_{it}$ denotes the bid that bidder i raises at time t. $z_t$ denotes the maximum bid among all other participating bidders and $G_t$ is the cdf of $z_t$. Note that $G_{t-1}(b_{it})$ is the probability of winning the auction (based on the beliefs at time *t-1*) after bidding $b_{it}$ at time *t*. $\alpha_i$ and $\beta_i$ denote the winner and loser regret parameters of bidder *i*, respectively. All the notation I use in this paper is summarized in Table 2.5.

Table 2. 5. Notation

| Notation | Description |
|---|---|
| $u_{it}$ | The utility of bidder $i = 1..I$ at time of bidding the tth bid, $t = 1..T$ , in auction j, which is suppressed for ease of presentation |
| $v_{it}$ | The time varying value of bidder i at time of raising tth bid in auction j; The measure of the valuation of bidder i at time of raising tth bid in auction j |
| $b_{it}$ | The bid that bidder i raises at time t |
| $G_{t-1}$ | The time varying belief of bidders about the distribution of maximum bid $z_t$ response of all other participating bidders |
| $\alpha_i$ | The winner regret parameter of bidder i |
| $\beta_i$ | The loser regret parameter of bidder i |
| $g_{t-1}(.)$ | The density function, or derivative of cumulative distribution function of bids $G_{t-1}(.)$ |
| $b_{jt}$ | The tth bid in auction $j = 1 \ldots J$ |
| $\theta_{jt}$ | The latent tth bid in auction $j = 1 \ldots J$ |
| $\varepsilon_{jt}$ | The Normally distributed noise of entering the bid into the system, or observation noise, which has auction specific variance of $\sigma_{jv}$ |
| $\tau_j$, $\gamma_j$ | The evolution and drift factors of the latent bid, in system equation |
| $\omega_{jt}$ | the noise of evolution of the latent bids within the auction, which has auction specific variance of $\sigma_{jw}$ |
| $F_t(.)$ | The time varying cumulative distribution function of bids; The auction specific subscript j is suppressed |
| $f_t(.)$ | The time varying density function of bids, assuming that the distribution is normal; The auction specific subscript j is suppressed |
| $n_t$ | The time varying number auction participants; The auction specific subscript j is suppressed |
| $n_{jt}$ | The observed number of bidders at time of tth bid in auction j |
| $\kappa_{jt}$ | The latent number of bidders at time of tth bid in auction j |
| $\tau_{jt}$ | The time trend, or the count of bids that have entered so far |
| $\iota_j$ | The average rate of entrance parameter |
| $\eta_j$ | The rate of sniper entrance |
| $\xi^1{}_{jt}$ | The normally distributed system noise, which has the variance of $\sigma^1{}_{j\xi}$ |
| $\zeta^1{}_{jt}$ | The normally distributed observation noise, which has the variance of $\sigma^1{}_{j\zeta}$ |
| $b_{-it}$ | The maximum bid that others have raised until the tth bid |
| $\varphi_{it}$ | The affiliated value of bidder i when tth bid is raised |
| $\vartheta_{jt}$ | The auction specific time varying common value element of this valuation |
| $\delta_i$ | The parameter of revelation of the value |
| $\zeta^2{}_{it}$ | The private signal error term, which has the auction specific variance of $\sigma^2{}_{j\zeta}$ |
| $\xi^2{}_{jt}$ | The common signals that bidders receive, and it has auction specific variance of $\sigma^2{}_{j\xi}$ |
| $d_i$ | The vector of measures for experience of bidder i |
| $m_c$ | The mean of this vector across members of cluster c |
| $\pi_c$ | The propensity of population membership in segment c |
| $\Theta_i = (\alpha_i, \beta_i, \delta_i, \rho_i)$ | The vector of regret, valuation and learning parameter of bidder i |
| $Ind_i$ | The segment that individual i is its member |
| $f^j$ | The information vector of the auction item j, with n information items (i.e., columns) |
| $\theta'$ | The latent prior of membership of an auction in an auction cluster |
| $z_n$ | The latent cluster index of feature j |
| $f^j{}_n$ | The nth observed information item of auction js information vector |
| $\alpha'$ and $\beta'$ | The parameters of the LDA model to estimate |
| $\psi_j = (\gamma_j, \tau_j, \iota_j, \eta_j)$ | The auction specific parameters of the evolution of belief about the bids and number of bidders |
| $clus_j$ | The cluster membership index for auction j |

This utility specification has three components[12]: The first component is the expected profit of the bidder from winning the auction. The second component is the anticipated winner regret for paying higher: Winner regret is defined as a multiplier of the difference between the bidder's bid and highest bid of others in case the bidder wins the auction, and it depends on the distribution of maximum bids of other bidders. The third component is anticipated loser regret, which occurs when the bidder loses an auction even though the winning bid is lower than her valuation. In this case, loser regret is defined as a multiplier of the difference between the bidder's valuation and the winning bid. The underlying assumption for this specification is that bidding is a noisy process, so bidders form a belief about the distribution of the latent bids, which I describe next.

[12] A criticism to the proposed utility specification might be that bidders might search across different auctions. However, studies such as Haruvy and Leszcyc (2010) show that bidders have inertia, and unless there is an incentive, they do not search across auctions. Ariely and Simonson (2003) also posit that when bidders are emotionally involved with the auction, they lose their global view of all the options that are available to them (i.e., search), so they act bounded rationally, and only focus on selecting the bid amount.

Consistent with the suggestion that bidders learn during the auction[13] (Hossain 2008, Zeithammer and Adams 2010, Okenfels and Roth 2002) and bidding is a noisy process (Jap and Naik 2008), I assume that the mean of bids follows a first order Markov process. Formally, I define

$$b_{jt} = \theta_{jt} + \varepsilon_{jt}, \qquad \varepsilon_{jt} \sim N(0, \sigma_{jv}) \tag{4}$$

$$\theta_{jt} = \tau_j \theta_{jt-1} + \gamma_j + \omega_{jt}, \qquad \omega_{jt} \sim N(0, \sigma_{jw}) \tag{5}$$

where $b_{jt}$ denotes the t'th bid in auction $j = 1 \ldots J$ , $\theta_{jt}$ denotes the latent bid, and $\varepsilon_{jt}$ denotes the normally distributed noise of entering the bid into the system (i.e., trembling hand of Selten 1975) or observation noise (i.e. bounded rationality of Simon 1972), which has auction specific variance, $\sigma_{jv}$. $\tau_j$ and $\gamma_j$ denote evolution and drift factors of the latent bid, respectively, and $\omega_{jt}$ denotes the noise of evolution (or the unobserved evolution factor) of the latent bids within the auction, which has auction specific variance, $\sigma_{jw}$.

[13] As mentioned in the Introduction, bidders can learn via various processes, such as gaining additional information about the auction item or resolving some of the uncertainties about the seller or about their own needs, etc. This type of learning is different than learning the value of the item from the bidders of the other auction participants.

Let $B_i$ be the random variable denoting the latent bid of bidder *i*, $F_t(.)$ and $f_t(.)$ be the time varying cumulative distribution and density functions of latent bids (uniform across bidders), respectively (assuming the density function exists), and $n_t$ be the time varying number of auction participants. Cumulative distribution and density functions for maximum bid of other $n_t - 1$ bidders at time *t* are formally defined as (I suppress the auction index *j* for ease of exposition):

$$\begin{aligned} G_t(\theta_t) &= P_t(\max(B_1, ..., B_{n_t - 1}) \le \theta_t) = P_t(B_1 \le \theta_t, ..., B_{n_t - 1} \le \theta_t) \\ &= P_t(B_1 \le \theta_t)...P_t(B_{n_t - 1} \le \theta_t) = F_t(\theta_t)...F_t(\theta_t) = F_t(\theta_t)^{n_t - 1} \end{aligned} \quad (6)$$

$$g_t(\theta_t) = (n_t - 1) F_t(\theta_t)^{n_t - 2} f_t(\theta_t) \quad (7)$$

Many studies such as Park and Bradlow (2005) and Bradlow and Park (2007) propose methods to recover the latent number of bidders. In this study, for the purpose of parsimony and simplicity, I assume that customers use the same Bayesian updating structure for the evolution of both the bids and the number of bidders. This assumption is reasonable since there is potential observation noise for the number of bidders[14].

---

[14] Usually bidders only skim through the bids to get a high level understanding of number of bidders, and since each bidder bids multiple times, double counting or missing one bidder might be completely natural for bounded rational bidder. In addition, bidders do not know if any bidder has left the auction or not at the time of consideration, so the cumulative number is a noisy signal.

Formally, I use the following first order Markov process to specify the evolution of the actual number of bidders[15]:

$$n_{jt} = \kappa_{jt} + \zeta^1{}_{jt}, \qquad \zeta^1{}_{jt} \sim N(0, \sigma^1{}_{j\zeta}) \tag{8}$$

$$\kappa_{jt} = \kappa_{jt-1} + \iota_j + \eta_j \tau_{jt} + \xi^1{}_{jt}, \qquad \xi^1{}_{jt} \sim N(0, \sigma^1{}_{j\xi}) \tag{9}$$

where $n_{jt}$ and $\kappa_{jt}$ denote the observed and latent number of bidders at time of t'th bid in auction *j*, respectively. $\iota_j$ denotes the average rate of entrance between bidding times and $\eta_j$ is the change in that entrance rate, which is multiplied by $\tau_{jt}$, which denotes the time trend in the auction. This specification allows me to model the sniping behavior explicitly (see, for example, Roth and Ockenfels 2000 for further discussion of sniping). $\zeta^1{}_{jt}$ is the observation noise, normally distributed with mean zero and variance $\sigma^1{}_{j\zeta}$, and $\xi^1{}_{jt}$ is the system noise in the rate of entrance and exit, also normally distributed with mean zero and variance $\sigma^1{}_{j\xi}$. Therefore, bidders update their expectations about the latent number of bidders at each point in time by observing the cumulative number of distinct bidders, who have bid up until that moment.

The last step of the model development in this approach is deriving the expression for valuation. I assume that, at each time *t*, bidders optimize their utility by selecting the optimal bid, $b_{it}$, given the valuation that they decide to reveal at the time. As a result, the bid, $b_{it}$, satisfies the following first order condition:

---

[15] Given that theory and many empirical studies suggest that bidders are bounded rational for various reasons, it is reasonable to assume that bidders follow a simpler parsimonious approximation, such as my model, rather than a complex one.

$$\frac{\partial u_{it}}{\partial b_{it}} = -G_{t-1}(b_{it}) + (v_{it} - b_{it})g_{t-1}(b_{it}) - \alpha G_{t-1}(b_{it}) - \alpha b_{it} g_{t-1}(b_{it}) + \alpha b_{it} g_{t-1}(b_{it}) + \beta(v_{it} - b_{it})g_{t-1}(b_{it}) = 0 \qquad (3)$$

Inverting equation (3) gives a measure for bidders' valuation. Hence, valuation is specified as

$$v_{it} = \frac{G_{t-1}(b_{it}) + b_{it}g_{t-1}(b_{it}) + \alpha_i G_{t-1}(b_{it}) - \alpha_i b_{it} g_{t-1}(b_{it}) + (\alpha_i + \beta_i) b_{it} g_{t-1}(b_{it})}{g_{t-1}(b_{it}) + \beta_i g_{t-1}(b_{it})} \qquad (10)$$

However, bids $b_{it}$ are noisy, so a better measure of valuation consists of the expectation of the right hand side of equation (10) over the distribution $F_{t-1}(.)$ of latent bids $\theta_{it}$. Therefore, valuation takes the following form:

$$v_{it} = E_\theta \left[ \frac{G_{t-1}(\theta_{it}) + \theta_{it} g_{t-1}(\theta_{it}) + \alpha_i G_{t-1}(\theta_{it}) - \alpha_i \theta_{it} g_{t-1}(\theta_{it}) + (\alpha_i + \beta_i)\theta_{it} g_{t-1}(\theta_{it})}{g_{t-1}(\theta_{it}) + \beta_i g_{t-1}(\theta_{it})} \right] \qquad (11)$$

The right hand side of the equation (11) is fully specified, but in order to estimate the unknown regret parameters, another specification of $v_{it}$ is required. To derive such a specification, the affiliated valuation and learning theory provides an appropriate ground, which I analyze next.

#### 2.5.1.2. Affiliated valuation and learning approach:

In this approach, I model a bidder's valuation of an auction item as a combination of three components: a common value, a private value, and a component consisting of bidder's learning the value of the auction item from the bids of other participants (see Hossain 2008; Zeithammer and Adams 2010; and Okenfels and Roth 2002 for further justification of this specification). As a result, the time varying valuation has the following specification:

$$v_{it} = \rho_i b_{-i(t-1)} + \varphi_{it} \qquad (12)$$

where $b_{-i(t-1)}$ denotes the bid that the bidder sees in the auction board at t-1, and $\varphi_{it}$ denotes the affiliated value of bidder i when tth bid is raised. The affiliated value consists of a private signal that only the bidder receives, and a common signal that all the bidders receive. To control for both types of these unobserved signals, I model the affiliated value evolution in the state space format, where the private signal is the error of the observation equation, and the common signal is the error of the valuation state equation. In addition, I assume that these signals affect the valuation higher at higher value items, and lower when the value of the item is lower (i.e., the signals are heteroscedastic). Therefore, consistent with Zeithammer and Adams (2010), I consider a log-log model of affiliated valuation evolution, which has the following form:

$$\begin{aligned} \log(\varphi_{it}) &= \log(\vartheta_{jt}) + \log(\delta_i) + \zeta^2{}_{it}, \qquad \zeta^2{}_{it} \sim N(0, \sigma^2{}_{j\zeta}) \\ \log(\vartheta_{jt}) &= \log(\vartheta_{jt-1}) + \xi^2{}_{jt}, \qquad \xi^2{}_{jt} \sim N(0, \sigma^2{}_{j\xi}) \end{aligned} \tag{13}$$

where $\varphi_{it}$ denotes the affiliated value of bidder i that raises t'th bid, and $\vartheta_{jt}$ denotes auction specific time varying common value element of this valuation. $\delta_i$ denotes the parameter of revelation of the value, and $\zeta^2{}_{it}$ denotes the private signal error term, which has the auction specific variance of $\sigma^2{}_{j\zeta}$[16]. $\xi^2{}_{jt}$ denotes the common signals that bidders receive between time *t-1* and *t*, and it has auction specific variance of $\sigma^2{}_{j\xi}$. This specification allows the bidders to reveal their private values gradually when they bid multiple times. I incorporated this hiding process to allow for the later bids to be systematically higher, consistent with the Zeithammer and Adams (2010), and Okenfels and Roth (2002).

[16] In an ideal scenario, each bidder would have different distribution for their private values. However, this assumption significantly complicates the model and makes it impossible to estimate the bidder-specific parameters using the available data set. My data set is sparse in the sense that many bidders do not raise their bids more than three or four bids within each auction.

#### 2.5.1.3. Identification of loser and winner regret:

Assuming that the two approaches discussed above –dynamic utility maximization and affiliated valuation and learning approaches-- give the same valuation for a particular auction item, I can combine equations (11) - (13) in the following way:

$$\log\left( E_\theta\left[\frac{G_{t-1}(\theta_{it})+\theta_{it}g_{t-1}(\theta_{it})+\alpha_i G_{t-1}(\theta_{it})-\alpha_i\theta_{it}g_{t-1}(\theta_{it})+(\alpha_i+\beta_i)\theta_{it}g_{t-1}(\theta_{it})}{g_{t-1}(\theta_{it})+\beta_i g_{t-1}(\theta_{it})}\right]-\rho_i b_{-it}\right)$$
$$=\log(\vartheta_{jt})+\log(\delta_i)+\zeta^2{}_{it},\zeta^2{}_{it}\sim N(0,\sigma^2{}_{j\zeta})$$
$$\log(\vartheta_{jt})=\log(\vartheta_{jt-1})+\xi^2{}_{jt},\xi^2{}_{jt}\sim N(0,\sigma^2{}_{j\xi})$$

(14)

For each bidding time t, the set of equations (4)-(9) and (14) provides the required identifying equations for the anticipated loser and winner regrets for each bidder.

### 2.5.2. Accounting for heterogeneity

eBay auctions and their participants' behaviors show high level of heterogeneity, as also mentioned in the Data section. Although bidders bid multiple times (incrementally), the number of observations is not enough to identify each bidder's parameters, so I use a level of shrinkage through Bayesian prior on the auction specific and individual specific parameters. Many studies including Srinivasan and Wang (2010), Wilcox (2000), and Ariely et al (2005) emphasize the influence of experience on the behaviors of the bidders. I use the bidders' information as a proxy for their experience at the hierarchical level in order to shrink the parameters of bidders with similar experience. I ran the estimation procedure in two steps: In the first step, I segment the bidders into *K* clusters and in the second step, I condition on the segment index of bidders while running the estimation procedure[17]. This two-step approach helps speeding up the estimation procedure.

In the same manner, to account for heterogeneity in the auction specific parameters, I shrink auction parameters given the eBay-specified auction clusters. Similar to bidder specific parameters, I conditioned on the cluster membership in the estimation procedure to shrink the auction specific parameters. In Section 2.7, I tested applying a clustering technique on the auctions too, rather than using the eBay-specified clusters. My main model results turned out to be robust to this method, which suggests that eBay auction clusters were indeed informative.

### 2.5.3. Estimation Procedure

The total number of parameters of the model is large, mainly to account for heterogeneity: There are four parameters, $\Theta_i = (\alpha_i, \beta_i, \delta_i, \rho_i)$, for each of the 12,603 bidders, and 10 parameters, $\psi_j = (\gamma_j, \tau_j, \iota_j, \eta_j)$ and the variances of six state-space equation error terms $\Sigma_j = (\sigma_{vj}, \sigma_{wj}, \sigma^1{}_{\zeta j}, \sigma^1{}_{\xi j}, \sigma^2{}_{\zeta j}, \sigma^2{}_{\xi j})$, for each of the 1646 auctions. This makes a total of 66,872 parameters. This large number of parameters over a data set of 58,285 bids most likely causes over-fitting problem. However, Bayesian shrinkage of parameters across clusters allows me to identify the model. Additionally, I put constraints on the evolution of bids and the number of bidders to be able to identify the model more efficiently. These constraints assure that the evolution parameters of both the bids and the numbers of bidders are non-negative. I also put constraints on the valuation growth and learning from others' bids, consistent with the theory which suggests that valuation is positive, and the bidder can either learn from the highest bid to increase her valuation or not learn at all. Finally, I rounded up the latent number of bidders recovered from the state space model, as the number of the bidders should be an integer number. I explain my estimation algorithm in this section and provide a pseudocode of it in Appendix C.

[17] It is not clear whether or not the model would over-fit and learn noise rather than the actual behavior of bidders if I had incorporated unobserved parameters of bidder responses in a form of a mixture normal model embedded within the estimation procedure.

#### 2.5.3.1 Clustering estimation:

In the first step of the estimation procedure, I clustered the bidders based on the observed data. To cluster bidders with similar experiences, I assume, in each segment, the experience of members is a noisy measure of the segment's mean experience, so I use a mixture normal fuzzy clustering[18]. Formally, the likelihood of the mixture normal clustering approach has the following structure:

$$\prod_{i=1}^{I} P_{Norm}(d_i \mid m, v, \pi_i) = \prod_{i=1}^{I}\sum_{c=1}^{K} \pi_{ic} P_{Norm}(d_i \mid m_c, v_c) \tag{15}$$

where $d_i$ denotes the vector of measures for experience of bidder *i*. $m=(m_1,\ldots,m_K)$, $v=(v_1,\ldots,v_K)$ where $m_c$ and $v_c$ denote the mean and variance of $d_i$ across members of cluster c, respectively, and $\pi_i = (\pi_{i1},\ldots \pi_{iK})$ where $\pi_{ic}$ is the propensity of population membership in segment c. I maximize this likelihood function with respect to ($m$, $v$, $\pi$) using an Expectation Maximization (EM) algorithm, which, for each bidder, provides a probability distribution for segment memberships. Finally, I assign each bidder to the segment with the highest probability. Therefore, for the shrinkage parameters, I formally have:

---

[18] The term "fuzzy" is used for methods which estimate a distribution for the cluster memberships, rather than assigning the bidders to certain clusters.

$$\Theta_i \sim MVN(.\,|\,Ind_i, d_i) \tag{16}$$

where $Ind_i$ denotes bidder i's segment. This specification provides flexible patterns of bidders' responses.

I used Mclust package in R to perform this clustering. This package uses the Bayesian information criterion (BIC) for model selection.

To cluster similar auctions, I utilized the eBay-specified auction clusters. Formally, I obtain the following structure:

$$\Psi_j \sim MVN(.\,|\,clus_j, D_j) \tag{18}$$

where $clus_j$ denotes cluster membership index of auction j.

In summary, the model uses the hierarchical multivariate normal prior for both bidder specific and auction specific parameters, conditional on their segment and cluster membership (I use $clus_j$ to denote the cluster membership index of auction j). This procedure accounts for heterogeneity in these entities, and prevents over-fitting to the data and learning noise.

**2.5.3.2.Maximum A Posteriori (MAP) optimization:**

In the second step, given the segment and cluster membership information of each bidder and auction, I optimized Maximum A Posteriori (MAP) of the model parameters over the data[19]. Considering the assumption that bidders update their belief about the distribution of the bids sequentially, I use Kalman Filter theory (Kalman and Bucy 1961). Introduced by Jap and Naik (2008) to auction literature, Kalman Filter starts with a prior on the distribution of latent measures, and it updates the posterior distribution of the latent measures sequentially, using Kalman gain factor (the variance of signals proportion) to weight the observed signal and the prior in a Bayesian updating process. The advantage of Kalman Filter to other filters is its closed form, which significantly improves the estimation speed. To estimate the latent state space model with Kalman Filter, I used a Monte Carlo Expectation Maximization (MCEM) approach suggested by De Valpine (2012). This approach embeds Kalman Filter in the optimization method. In other words, given the non-state parameters $\Phi = (\Theta, \Psi, \Sigma)$, the procedure estimates the mean and variance of the latent space, and then uses Monte Carlo simulation to estimate the full joint likelihood of the observation $y_t$ and the latent state $\Delta_t = (\theta_t, \kappa_t, \vartheta_t)$, which has the following form:

$$l(y_1,..., y_T \mid \Phi) = \pi(\Delta_0 \mid \Phi)\prod_{t=1}^{T} p(y_t \mid \Delta_t, \Phi)\pi(\Delta_t \mid \Delta_{t-1}, \Phi) \tag{19}$$

[19] Alternatives such as Gibbs and Metropolis Hasting sampling methods are computationally intractable over large data sets.

The posterior of the model has the following form:

$$\begin{aligned} &P(\tau_j, \gamma_j, \iota_j, \eta_j, \alpha_i, \beta_i, \rho_i, \delta_i, \Sigma \mid b_{jt}, b_{-jt}, n_{jt}, d_i, D_j) = \\ &\prod_{j=1}^{J}\prod_{i=1}^{I}\prod_{t=1}^{T_j}\left[\int P_{Norm}(b_{jt} \mid \theta_{jt}, \sigma_{jv}) \times P_{Norm}(\theta_{jt} \mid \theta_{jt-1}, \sigma_{jv}, \tau_j, \gamma_j) d\theta_j\right] \\ &\times\left[\int P_{Norm}(n_{jt} \mid \kappa_{jt}, \sigma^1{}_{j\zeta}) \times P_{Norm}(\kappa_{jt} \mid \kappa_{jt-1}, \sigma^1{}_{j\xi}, \iota_j, \eta_j) d\kappa_j\right] \\ &\times\left[\int P_{Norm}(b_{jt} \mid b_{-jt}, \alpha_i, \beta_i, \rho_i, \delta_i, \sigma^2{}_{j\zeta}, \vartheta_{jt}, \kappa_{jt}) \times P_{Norm}(\vartheta_{jt} \mid \vartheta_{jt-1}, \sigma^2{}_{j\xi}) d\vartheta_j\right] \\ &\times P_{Norm}(\alpha_i, \beta_i, \rho_i, \delta_i \mid \mu_{ind(i)}, \sigma_{ind(i)}, d_i) \\ &\times P_{Norm}(\tau_j, \gamma_j, \iota_j, \eta_j \mid \mu_{clust(j)}, \sigma_{clust(j)}, D_j) \\ &\times \delta_{\hat{\Sigma}}(\Sigma) \times \delta_{\hat{\mu}_{clust(j)}}(\mu_{clust(j)}) \times \delta_{\hat{\sigma}_{clust(j)}}(\sigma_{clust(j)}) \times \delta_{\hat{\mu}_{ind(i)}}(\mu_{ind(i)}) \times \delta_{\hat{\mu}_{ind(i)}}(\mu_{ind(i)}) \end{aligned} \tag{20}$$

where $\theta_j = (\theta_{j1},...,\theta_{jT_j})$, $\kappa_j = (\kappa_{j1},..,\kappa_{jT_j})$, $\vartheta_j = (\vartheta_{j1},...,\vartheta_{jT_j})$, and $\delta_{?}(.)$ denotes the Dirac delta function[20]. The first, second, and third lines denote likelihood of error terms of the state space equations (4) and (5) --specified for the belief of bidders about the bid distribution-- equations (8) and (9) --specified for the belief of bidders about the number of bidder distribution--, and equation (14) --specified for the evolution of affiliated valuations, respectively. The fourth and fifth lines denote the likelihood of error terms of equation (16) and (18), respectively, specified as hierarchy over individual and auction specific parameters. The sixth line specifies prior on the variance of the three state space equations, and the mean and variance parameters of the hierarchy. I can rewrite the model parameters' posterior based directly on the error terms of the state space equations and hierarchy over individual and auction specific parameters as follows:

$$\begin{aligned}
\prod_{j=1}^{J}\prod_{i=1}^{I}\prod_{t=1}^{T_j} & \left[\int P_{Norm}(\varepsilon_{jt} \mid b_{jt},\theta_{jt},\sigma_{jv}) \times P_{Norm}(\omega_{jt} \mid \theta_{jt},\theta_{jt-1},\sigma_{jv},\tau_j,\gamma_j) d\theta_j\right] \\
& \times\left[\int P_{Norm}({\zeta^1}_{jt} \mid n_{jt},\kappa_{jt},{\sigma^1}_{j\zeta}) \times P_{Norm}({\xi^1}_{jt} \mid \kappa_{jt},\kappa_{jt-1},{\sigma^1}_{j\xi},\iota_j,\eta_j) d\kappa_j\right] \\
& \times\left[\int P_{Norm}({\zeta^2}_{it} \mid b_{jt},b_{-jt},\alpha_i,\beta_i,\rho_i,\delta_i,{\sigma^2}_{j\zeta},\vartheta_{jt},\kappa_{jt}) \times P_{Norm}({\xi^2}_{jt} \mid \vartheta_{jt},\vartheta_{jt-1},{\sigma^2}_{j\xi}) d\vartheta_j\right] \\
& \times P_{Norm}(\alpha_i,\beta_i,\rho_i,\delta_i \mid \mu_{ind(i)},\sigma_{ind(i)},d_i) \\
& \times P_{Norm}(\tau_j,\gamma_j,\iota_j,\eta_j \mid \mu_{clust(i)},\sigma_{clust(i)},D_j) \\
& \times \delta_{\hat{\Sigma}}(\Sigma) \times \delta_{\hat{\mu}_{clust(j)}}(\mu_{clust(j)}) \times \delta_{\hat{\sigma}_{clust(j)}}(\sigma_{clust(j)}) \times \delta_{\hat{\mu}_{ind(i)}}(\mu_{ind(i)}) \times \delta_{\hat{\mu}_{ind(i)}}(\mu_{ind(i)})
\end{aligned} \tag{21}$$

I use an MCEM approach to compute the maximizing parameters of the posterior of the model in (20). This iterative method starts with an initial set of parameter estimations and alternates between an Expectation (E-) step and a Maximization (M-)[21] step until convergence.

---

[20] Dirac delta function is a generalized distribution that is zero everywhere except at the point that its subscript specifies. It represents a normal distribution at the limit when the variance equates to zero.

[21] I actually applied the *Generalized* EM algorithm where, in the M-step, rather than computing the maximizing parameters, I settled with a point that improves the objective. This algorithm has similar properties with EM algorithm (see, McLachlan and Krishnan 2008, for further details).

For the E-step of each iteration, I first perform a Weighted Least Squares (WLS) to project the bidder (resp., auction) specific information to the bidder (resp., auction) specific parameters within each segment as follows:

$$\begin{aligned} \Theta_i &= \mu'_{Ind_i} d_i + \lambda_i, \qquad \lambda_i \sim N(0, \sigma_{Ind_i}) \\ \Psi_j &= \mu'_{clus_j} D_j + \chi_j, \qquad \chi_j \sim N(0, \sigma_{clus_j}) \end{aligned} \tag{22}$$

The estimated parameters of these WLS's, $(\hat{\mu}_{ind(i)}, \hat{\sigma}_{ind(i)}, \hat{\mu}_{clust(j)}, \hat{\sigma}_{clust(j)})$, are then used to compute the prior probabilities of the bidder-specific and auction-specific parameters (fourth and fifth lines in equation (20), respectively). I computed the likelihood contribution of the belief of bidders about the bids and the number of bidders, and the evolution of the valuations (first, second, and the third lines of equation (20)) using Kalman filtering and backward smoothing methods to derive the evolution of the state parameters in each bidding time. I used Monte Carlo sampling method to integrate out the latent state variables. The details of these methods are explained in Appendix C. I used DLM package in R to run the Kalman filtering and backward smoothing.

For the optimization problem in the M-step of each iteration, since a closed form solution for the gradient of the maximum a posteriori of the model is not available, methods such as gradient descent, quasi Newton, and conjugate gradient are computationally intractable. Calculating the gradient numerically will also increase the run-time of the estimation algorithm cubically in the number of the parameters, i.e. $O(P^3TJ)$. Therefore, I used simulated annealing method, which is a generic probabilistic heuristic method for global optimization.

Simulated annealing method uses only function values, so it is relatively slow. It starts with an initial value and, at each iteration, a new point is randomly generated. The algorithm accepts all new points that improves the objective, but also, with a certain probability that gradually decreases, it might accept points that worsen the objective. By accepting the latter type of points, the algorithm avoids being trapped in local minima in early iterations and is able to explore globally for better solutions. (See Belisle 1992 for further discussion of this algorithm.) To the best of my knowledge, I am the first to use simulated annealing in the marketing/OM fields, but it is used in other fields (for example, Crama and Schyns 2003 use this method for complex portfolio optimization, and Zhuang et al 1994 use it for robotics calibration).

I terminated the MCEM algorithm when the Euclidian difference between the parameter estimations of two consecutive iterations became smaller than a pre-specified tolerance or after a maximum number of iterations (I used 1e-8 as the tolerance and 2,000 as the maximum iteration number in this study).

## 2.6. RESULTS

I start presenting the results with the bidder segments that I estimated in the first step. The BIC criterion suggests clustering the bidders into 47 segments. Table 2.6 presents the summary statistics of average bidder characteristics within each bidder segment. As expected, there is considerable heterogeneity between segments.

The optimal MAP is estimated to be -94,280,085. Given that this model is estimated on approximately 60,000 bids of 12,000 bidders in 1,600 auctions, this value is in the expected range. Table 2.7 presents the summary statistics for the bidder-specific parameter estimations: columns 2-5 are across 19 auction categories and columns 6-9 are across estimated 47 bidder segments. The estimated parameters show significant heterogeneity in bidders' parameters across bidder segments. On the other hand, there is not much heterogeneity across auction categories, which suggests that regret and valuation/learning characteristics are more individual specific than category specific.

Table 2.6. Summary statistics of the average bidder characteristics within each of 47 bidder segments

| Characteristic | Mean | SD | min | max |
|---|---|---|---|---|
| Size | 260.55 | 275.02 | 3 | 992 |
| avg. feedback score | 3471.21 | 12399.80 | 48 | 84027 |
| sd feedback score | 3635.04 | 8480.39 | 5 | 45365 |
| avg. Number of bids on this item | 8.57 | 7.67 | 1 | 35 |
| sd Number of bids on this item | 6.79 | 6.77 | 0 | 26 |
| avg. total number of bids in 30 days | 680.32 | 1072.73 | 3 | 4530 |
| sd total number of bids in 30 days | 630.98 | 1066.22 | 0 | 5814 |
| avg. Number of items bid on in 30 days | 257.57 | 412.59 | 1 | 1631 |
| sd Number of items bid on in 30 days | 266.91 | 449.57 | 1 | 2099 |
| avg. Bidding Activity with current Seller | 24.47 | 20.19 | 1 | 100 |
| sd Bidding Activity with current Seller | 19.34 | 11.72 | 0 | 40 |
| avg. Number of categories bid on | 2.19 | 0.57 | 1 | 4 |
| sd Number of categories bid on | 1.13 | 0.64 | 0 | 3 |

Winner (resp. loser) regret parameter is significant in 44 (resp. 45) out of the 47 bidder segments at $p<0.01$ and it is significant in two (resp. one) categories at $p<0.05$. This significance is consistent with the findings of Bajari and Hortacsu (2003) suggesting when there is an element of common value in the valuation, there is potential winner regret anticipation in eBay auctions. Table 2.7 also indicates that the mean of average loser regret is slightly higher (in magnitude) than winner regret, but the means are fairly close to each other.

The significance of both winner and loser regrets and their close magnitudes, on average, are not consistent with the suggestions of Ariely and Simonson (2003), which state that the second price systems used by online auctions like eBay decrease the probability of winner regret, while maximizing loser regret. Therefore, my results indeed support the results of Zeithammer and Adams (2010) suggesting sealed-bid second price auction is not a good abstraction for eBay auctions. Another possible explanation for this inconsistency is bidders' naïve bidding behaviors which do not conform to second price auction theory. The magnitudes of the regret values do not support the claim of Gilovich et al. (1998) either: They suggest that action regret (analogous to winner regret) incites more intensive feeling than inaction regret (analogous to loser regret), which incites wistful feeling.

Table 2.7. Summary statistics for the bidder specific parameter estimations

| | within each auction category (19) | | | | within each bidder segment (47) | | | |
|---|---|---|---|---|---|---|---|---|
| Parameter | min | max | Mean | SD | min | max | Mean | SD |
| avg. winner regret | -1.38 | -1.24 | -1.31 | 0.04 | -1.67 | -0.52 | -1.28 | 0.19 |
| se winner regret | 0.02 | 0.04 | 0.04 | 0.01 | 0.03 | 0.41 | 0.10 | 0.08 |
| avg. loser regret | -1.4 | -1.28 | -1.33 | 0.03 | -1.7 | -0.79 | -1.34 | 0.13 |
| se loser regret | 0.02 | 0.04 | 0.04 | 0.006 | 0.03 | 0.49 | 0.10 | 0.09 |
| avg. valuation param. | 1.17 | 1.28 | 1.23 | 0.03 | 0.79 | 1.42 | 1.22 | 0.10 |
| se valuation param. | 0.02 | 0.04 | 0.03 | 0.004 | 0.02 | 0.27 | 0.08 | 0.06 |
| avg. learning param. | 0.18 | 0.32 | 0.25 | 0.03 | 0 | 0.81 | 0.27 | 0.12 |
| se learning param. | 0.02 | 0.04 | 0.03 | 0.01 | 0.03 | 0.26 | 0.09 | 0.06 |

Looking at the estimation results from the auction category perspective shows that both winner and loser regret are significant in all the categories at $p<0.0001$. To test the hypothesis that luxury and widely available goods convey different levels of regret, I ran pairwise t-test between the parameters of regret in widely available and luxury goods categories. However, the results, which are presented in Table 2.8, did not show any significant difference between luxury and widely available good auctions in terms of regret levels.

Table 2.8. t-Test: Paired Two Sample for Means

| | For Winner Regret | | For Loser Regret | |
|---|---|---|---|---|
| | *Widely Available* | *Luxury* | *Widely Available* | *Luxury* |
| Mean | -1.334 | -1.317 | -1.319 | -1.330 |
| Variance | 0.934 | 0.982 | 0.946 | 0.957 |
| Observations | 6024 | 6024 | 6024 | 6024 |
| Pearson Correlation | 0.008 | | 0.002 | |
| Hypothesized Mean Difference | 0 | | 0 | |
| Df | 6023 | | 6023 | |
| t Stat | -1.002 | | 0.609 | |
| P(T<=t) one-tail | 0.158 | | 0.271 | |
| t Critical one-tail | 1.645 | | 1.645 | |
| P(T<=t) two-tail | 0.316 | | 0.542 | |
| t Critical two-tail | 1.960 | | 1.960 | |

Table 2.7 shows that the mean of the average valuation revelation parameters across bidder segments is 1.22, i.e., at each increment, on average, the bidders reveal 22% more than their previously revealed valuation. The mean of average learning parameters across bidder segments is 0.27, which implies that, on average, the bidders weigh the highest observed bid 27% while updating their valuations by learning from the highest bid. Comparison of the estimated valuation and learning parameters suggests that bidders put more weight on their own valuation than learning from the highest bid. A possible explanation for the low learning level is shill bidding, as Boze and Daripa (2011) suggest. In shill bidding, the seller bids on the auction by herself or through one of her affiliate to cause others to bid higher. It might be possible that the bidders consider such shill bidding, so they discount their learning from the highest bid.

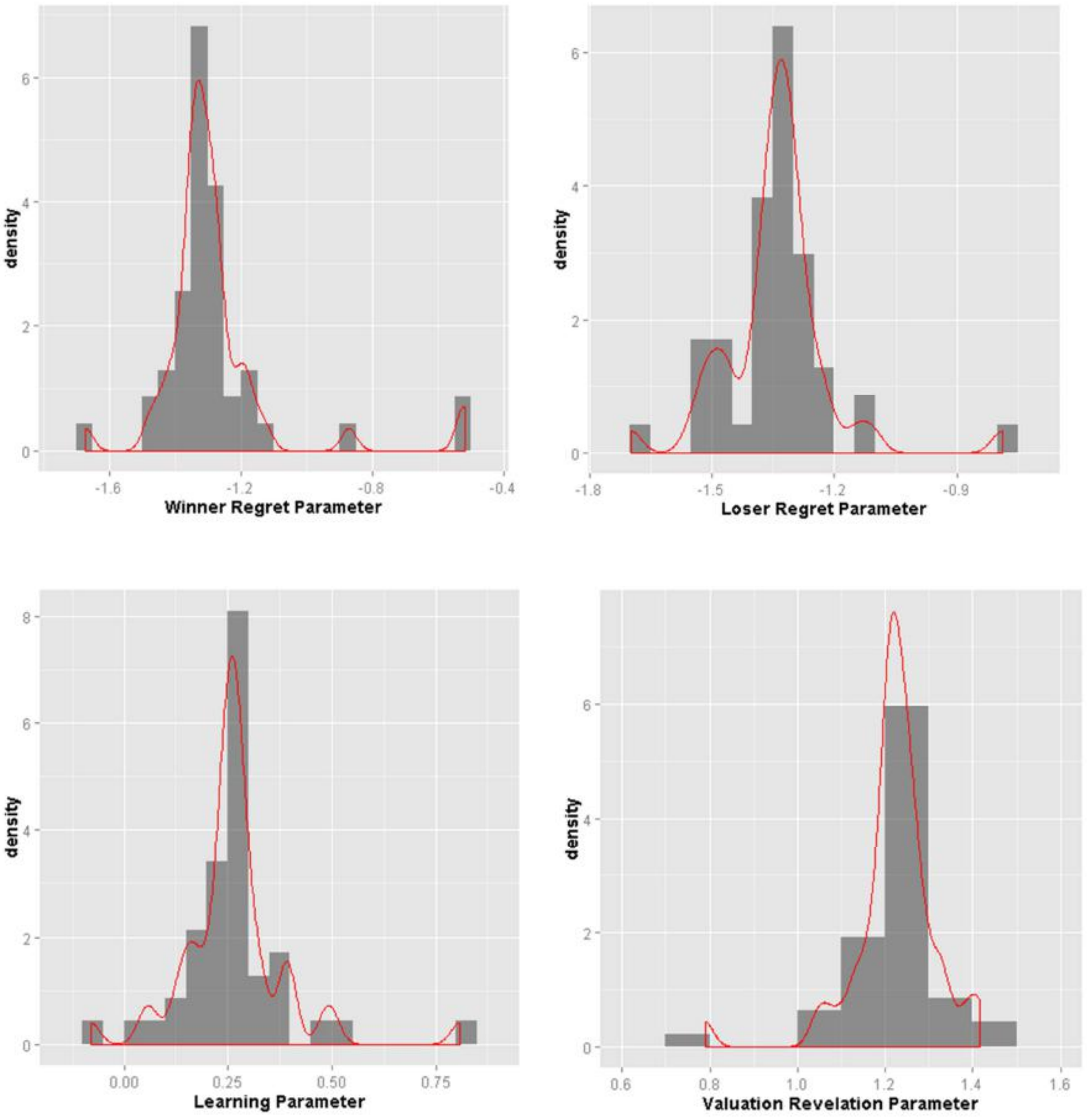

Figure 2.3. Histogram of regret and valuation evolution parameters across bidder segments

Figure 2.3 shows the distribution of regret, learning, and valuation revelation parameters across bidder segments. These distributions have long tails, and, indeed, Shapiro-Wilk normality tests reject normality for these distributions at $p<0.01$, so Gaussian distribution does not represent them well. This observation lends support to the importance of allowing flexible response patterns by clustering the data and shrink different bidders' parameters across their corresponding segment parameter means.

Table 2.9. Relation between the winner regret $\alpha_i$, the loser regret $\beta_i$, the update of valuation parameters $\delta_i$ and learning parameter $\rho_i$ estimates across forty seven bidder segments

| | *winner regret* | *loser regret* | *valuation revelation* | *learning* |
|---|---|---|---|---|
| winner regret | 1 | | | |
| loser regret | 0.427 | 1 | | |
| valuation revelation | 0.662 | 0.589 | 1 | |
| Learning | 0.135 | 0.474 | 0.613 | 1 |

Table 2.10. Relation between the winner regret $\alpha_i$, the loser regret $\beta_i$, the update of valuation parameters $\delta_i$ and learning parameter $\rho_i$ estimates across forty seven bidder segments

| Regressand | Regressor | Estimate | SE | t-stat | p-value |
|---|---|---|---|---|---|
| Winner regret | | | | | |
| | Intercept | -0.43 | 0.27 | -1.61 | 0.11 |
| | loser regret | 0.63** | 0.20 | 3.16 | 0.00 |
| Winner Regret | | | | | |
| | Intercept | -0.78** | 0.20 | -3.901 | 0.000 |
| | Learning | 0.94** | 0.13 | 7.305 | 0.000 |
| | valuation revelation | -0.59** | 0.17 | -3.418 | 0.001 |
| Loser Regret | | | | | |
| | Intercept | -1.64** | 0.16 | -10.118 | 0.000 |
| | Learning | 0.33** | 0.10 | 3.149 | 0.003 |
| | valuation revelation | 0.17 | 0.14 | 1.193 | 0.239 |

** Two tail 0.95% confidence interval significance

Table 2.11. Explaining winner regret $\alpha_i$, the loser regret $\beta_i$, the update of valuation parameters $\delta_i$ and the learning parameter $\rho_i$ estimates across 47 bidder segments

| Regressand | Regressor | Estimate | SE | t-stat | p-value |
|---|---|---|---|---|---|
| Winner Regret $(Adjusted\text{-}R^2 = 0.64)$ | | | | | |
| | Intercept | -1.278* | 0.017 | -75.449 | 0.000 |
| | Segment Size | 0.000 | 0.000 | 0.564 | 0.576 |
| | Bidders Feedback mean | 0.001* | 0.000 | 6.344 | 0.000 |
| | Number of Bids on This item | -0.004 | 0.003 | -1.570 | 0.125 |
| | total number of bids in 30 days | 0.013* | 0.004 | 3.574 | 0.001 |
| | Number of items bid on in 30 days | 0.000 | 0.000 | -1.456 | 0.153 |
| | Bid activity with current Seller | 0.003* | 0.001 | 2.430 | 0.020 |
| | Number of categories Bid on Mean | 0.044 | 0.046 | 0.964 | 0.341 |
| Loser Regret $(Adjusted\text{-}R^2 = 0.30)$ | | | | | |
| | Intercept | -1.341* | 0.016 | -83.916 | 0.000 |
| | Segment Size | 0.000 | 0.000 | -0.737 | 0.466 |
| | Bidders Feedback mean | 0.001* | 0.000 | 4.261 | 0.000 |
| | Number of Bids on This item | -0.004 | 0.003 | -1.734 | 0.091 |
| | total number of bids in 30 days | -0.001 | 0.003 | -0.311 | 0.758 |
| | Number of items bid on in 30 days | 0.000 | 0.000 | -1.025 | 0.312 |
| | Bid activity with current Seller | 0.000 | 0.001 | -0.455 | 0.651 |
| | Number of categories Bid on Mean | -0.027 | 0.043 | -0.635 | 0.529 |
| Learning value from bids $(Adjusted\text{-}R^2 = 0.64)$ | | | | | |
| | Intercept | 0.271* | 0.017 | 16.326 | 0.000 |
| | Segment Size | 0.000 | 0.000 | 1.168 | 0.250 |
| | Bidders Feedback mean | 0.001* | 0.000 | 9.030 | 0.000 |
| | Number of Bids on This item | 0.003 | 0.003 | 1.216 | 0.231 |
| | total number of bids in 30 days | 0.002 | 0.004 | 0.601 | 0.551 |
| | Number of items bid on in 30 days | 0.000 | 0.000 | -0.208 | 0.836 |
| | Bid activity with current Seller | 0.001 | 0.001 | 0.590 | 0.559 |
| | Number of categories Bid on Mean | -0.023 | 0.045 | -0.507 | 0.615 |
| Valuation update $(Adjusted\text{-}R^2 = 0.41)$ | | | | | |
| | Intercept | 1.269* | 0.016 | 79.126 | 0.000 |
| | Segment Size | 0.000 | 0.000 | -0.215 | 0.831 |
| | Bidders Feedback mean | 0.001* | 0.000 | 4.782 | 0.000 |
| | Number of Bids on This item | 0.004 | 0.003 | 1.521 | 0.136 |
| | total number of bids in 30 days | -0.011* | 0.003 | -3.148 | 0.003 |
| | Number of items bid on in 30 days | 0.000 | 0.000 | 1.859 | 0.071 |
| | Bid activity with current Seller | 0.000 | 0.001 | -0.213 | 0.833 |
| | Number of categories Bid on Mean | -0.010 | 0.043 | -0.242 | 0.810 |

* Two tail 0.95% confidence interval significance

I evaluated the correlation between bidder-specific parameters as well. Table 2.9 shows the correlation matrix for regret and learning parameters across 47 bidder segments and Table 2.10 shows the regression analysis between these parameters. The results show that winner regret is positively correlated with loser regret. I explain this result by the type of bidders: Some bidders might be emotional, so they account for both winner and loser regret emotions, and the others might be less emotional so they generally regret less. I also find a negative relationship between learning less from others (status quo tendency) and feeling winner regret, consistent with Inman and Zeelenberg (2002) findings. In other words, I find that bidders who update their valuations based on the new auction board bid less, anticipate more winner regret than others.

I also explain the estimated regret and learning parameters based on the observed characteristics of the bidder segments. These characteristics are good proxies for the bidders' experience and important factors on bidders' behaviors, as I discussed earlier. Table 2.11 presents the result of this analysis. The results show that bidders with more feedback score (i.e., more experience) are less regretful, and learn more from the bids on the auction board. I also find that bidding in several categories correlates with more loser regret and the valuation update correlates positively with the bidders' feedback score. The latter result suggests that bidders with more experience reveal their value more, which is consistent with the dominant strategy of rational bidders in the auction literature.

Table 2.12 presents the summary statistics for the auction-specific parameter estimations. The average of the parameter for the belief about the growth of bids is 1.77 across all auction items, which suggests that bidders believe that bids will exponentially grow as new bids enter. This exponential growth is consistent with the form of the evolution of bids in figure 2.1. The average drift parameter is 5.58 across all auction items, which suggests the dollar value that bidders expect a new bid will increment the previous bid after the growth. The average rate of entrance between two bids is 1.05 across auction items, suggesting bidders expect 1 new bidder watch the auction and ready to bid between two consecutive bids. The last minute rush rate is 2.09 across auction items, which suggests that bidders expect the rate of entrance to triple at the end of the auction. This is consistent with the sniping behavior.

Table 2.12. Summary statistics for the auction specific parameter estimations

| | within each auction category (19) | | | |
|---|---|---|---|---|
| Parameter | min | max | Mean | SD |
| avg. growth of bids | 1.5 | 2 | 1.77 | 0.11 |
| se growth of bids | 0.07 | 0.24 | 0.14 | 0.04 |
| avg. drift of bids | 5.36 | 5.81 | 5.58 | 0.12 |
| se drift of bids | 0.08 | 0.2 | 0.12 | 0.03 |
| avg. last minute flood | 0.9 | 1.26 | 1.05 | 0.09 |
| se last minute flood | 0.06 | 0.24 | 0.13 | 0.05 |
| avg. mean entrance rate | 1.84 | 2.26 | 2.09 | 0.11 |
| se mean entrance rate | 0.08 | 0.24 | 0.14 | 0.04 |

## 2.7. COUNTERFACTUAL ANALYSIS

Notification policies are shown to be effective in influencing the bidder's feeling and anticipating regret (see, for example, Filiz-Ozbay and Ozbay 2007). Such a notification policy for an auction platform such as eBay might suggest sending emails of auction winning bid and amount paid statistics to the users, or to present such information on the website. Notification policies can be conditioned on the bidder behavior to target only naïve bidders. However, to implement a policy change, the auction platform should be able to predict the revenue implications of such an action accurately. From this aspect, in addition to allowing for targeting bidders, the key advantage of modeling the bidders' decision structurally is the capability to study counterfactuals. My empirical results show that the bidders experience significant winner regret on eBay auctions. Therefore, I studied a counterfactual scenario where an auction platform shuts down the winner regret using its notification policy (I first assumed that loser regret is still in effect).

To run this counterfactual scenario, I set the winner regret parameters to zero while keeping all the other parameters in their estimated values. Given this new setting, I started from the first bid, and, at each point in time, I computed the optimal bid of a given bidder by running a Broyden–Fletcher–Goldfarb–Shanno (BFGS) optimization algorithm on the utility function of the bidder, presented in equation (1). Given this new optimal bid, I then updated the time varying parameters of belief about the distribution of the latent bids, by running Kalman Filter, and I computed the optimal bid of the next given bidder. In this way, I simulated the bids of all the bidders at each point in time and determined the winning bid and the amount paid for each auction item in the new environment with no winner regret.

Figure 2.4 presents the results of this analysis on six auction samples. The results show that shutting down the winner regret can increase the winning bid two to four times in some auctions. The results resemble a step function in some proximity, because, as a result of this shut down, some bidders bid so much higher than others that the other bidders' bids became irrelevant, as they are prone to raise a bid with a lower value.

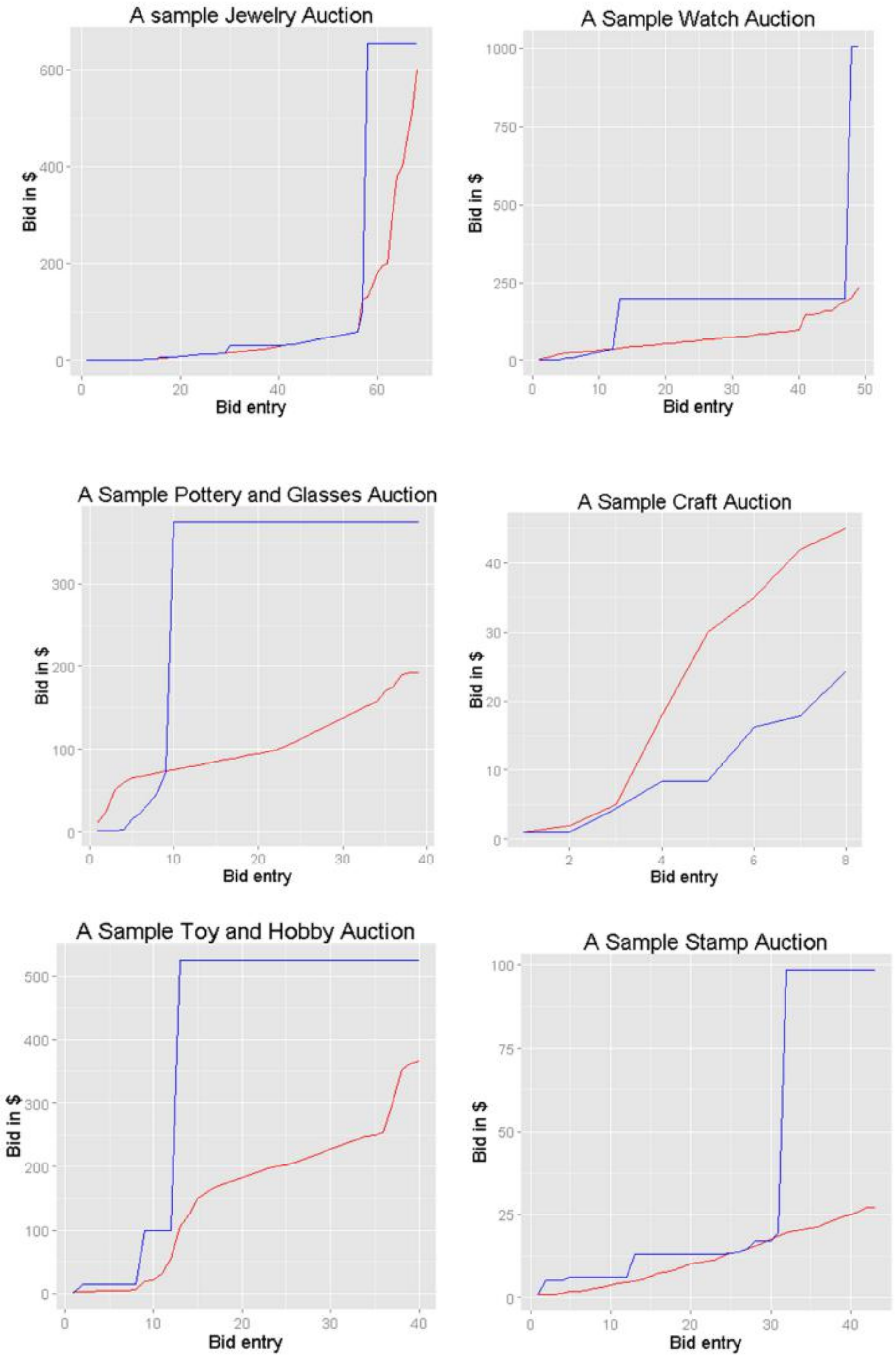

Figure 2.4. Counterfactual analysis of shutting down winner regret (blue line the optimal bidding when regret is shut down, and red line the observed)

Table 2.13. Counterfactual analysis of shutting down only winner and both winner/loser regret

| Auction Category | Number of Auctions | Average improvement of shutting down winner regret | Average improvement of shutting down both winner and loser |
|---|---|---|---|
| Jewelry and Watches | 149 | 28% | 28% |
| Collectibles | 103 | 36% | 32% |
| Clothing, Shoes and Accessories | 84 | 25% | 16% |
| Crafts | 78 | 28% | 31% |
| Pottery and Glass | 74 | 27% | 22% |
| Antiques | 68 | 40% | 49% |
| Toys and Hobbies | 93 | 29% | 30% |
| Stamps | 72 | 61% | 43% |
| Books | 84 | 28% | 30% |
| Tickets and Experiences | 91 | 18% | 5% |
| Art | 70 | 25% | 21% |
| Gift Cards and Coupons | 85 | 40% | 38% |
| Music | 86 | 44% | 27% |
| Consumer Electronics | 83 | 19% | 17% |
| DVDs and Movies | 87 | 53% | 39% |
| Dolls and Bears | 84 | 27% | 39% |
| Entertainment Memorabilia | 88 | 23% | 13% |
| Health and Beauty | 74 | 37% | 40% |
| Video Games and Consoles | 93 | 39% | 38% |
| Total improvement | | 24% | 24% |
| Average improvement across all auctions | | 32% | 29% |

I further studied the revenue implications at auction platform level. Table 2.13 presents the results of this counterfactual analysis within each auction category, and across all the auctions. By shutting down the bidders' winner regret through a notification policy, on average, the auction platform can improve its revenue of each item by 32%, and its total revenue by 24%. Considering category-based improvements indicates that "Stamps" and "DVD and Movies" have the highest improvement, when I shut down the winner regret. Significant improvements by shutting down winner regret occur in widely available good categories such as DVD's and movies, music, health and beauty, books, and gift cards and coupons, in which usually non-expert bidders bid on (stamps and antiques are exceptions). However, smaller improvements occur in tickets and experiences, entertainment memorabilia, and art, which most likely attract more expert bidders. To test this hypothesis I regressed the counterfactual revenue improvement of the auction items on the characteristics of bidders and price. Top part of Table 2.14 presents the results. They suggest that the number of bid on a specific auction item and the total number of items bid on are positively correlated with the improvement in revenue. This can be explained by incremental bidders, i.e., those who bid a lot are naive incremental bidders, and shutting down winner regret improves the revenue more, when bidders are naïve. Furthermore, a high number of auction items bid on is a signal of the bidder's not concentrating on one auction item to win, which is another proxy for less experience of the bidder.

As notification policy might have the potential to remove both types of regrets together, I also experimented the effect of shutting down both winner and loser regret. Since winner and loser regrets affect the bids in the opposite directions, intuitively, such a shutdown should decrease the amount of revenue improvement of shutting down only the winner regret, and, indeed, on average, it improves each auction's highest bid by 29% and the total revenue by 24%, slightly less than shutting down only the winner regret. Regressing the counterfactual revenue improvement on the characteristics of bidders and price did not provide us statistically significant relations in this case (estimation results are presented in bottom part of Table 2.13).

Table 2.14. Counterfactual revenue improvements explained by the characteristics of bidder on each auction bidder category

**Shutting down winner regret**

| | *Coefficients* | *Standard Error* | *t Stat* | *P-value* |
|---|---|---|---|---|
| Intercept | 0.1559 | 0.0921 | 1.6935 | 0.0905 |
| Feedback | 0.0000 | 0.0000 | -0.9519 | 0.3413 |
| Bids on this item | 0.0035 | 0.0020 | 1.7814 | 0.0750 |
| Total bids in 30 days | 0.0000 | 0.0001 | -0.3022 | 0.7625 |
| Number of items bided on | 0.0009 | 0.0004 | 2.5731 | 0.0102 |
| Activity with the Seller | 0.0006 | 0.0011 | 0.5675 | 0.5704 |
| Number of Categories bid on | 0.0231 | 0.0303 | 0.7613 | 0.4466 |

**Shutting down both types of regret**

| | *Coefficients* | *Standard Error* | *t Stat* | *P-value* |
|---|---|---|---|---|
| Intercept | 0.166 | 0.084 | 1.987 | 0.047 |
| Feedback | 0.000 | 0.000 | -0.611 | 0.542 |
| Bids on this item | 0.002 | 0.002 | 1.358 | 0.175 |
| Total bids in 30 days | 0.000 | 0.000 | 0.404 | 0.686 |
| Number of items bided on | 0.000 | 0.000 | 0.756 | 0.450 |
| Activity with the Seller | 0.000 | 0.001 | 0.008 | 0.994 |
| Number of Categories bid on | 0.035 | 0.027 | 1.266 | 0.206 |

## 2.8. ROBUSTNESS CHECKS

I first checked the robustness of my results to some of the modeling assumptions. One assumption in my model is that bidders use their own bids as a proxy for how much they are going to pay in case they win the auction[22]. Therefore, they compare their bids with their beliefs about the maximum bid of other bidders (since the maximum bid of others is not directly observable). I tested an alternative utility specification assuming that when a bidder bids $b_{it}$ at time t, she considers the fact that she would pay an amount between her bid and the currently displayed maximum bid, in case she wins the auction. I model this situation by replacing $b_{it}$ with $\lambda_i b_{it} + (1-\lambda_i)\max_{it}$ where $\max_{it}$ is the maximum bid that is shown on the auction board and $\lambda_i$ is the parameter (to be estimated from data) of expected proportion of the current bid difference that can be possibly filled by new bids.

[22] Even if bidders might be aware of the second price nature of eBay auctions, it still make sense to make this assumption, since the bidders do not know whether there will be new bids between the one on the auction board and their bids before the auction ends. Therefore, it is possible that they will pay a price very close to their own bids even if they win.

I found that this new specification does not change the inference significantly. In the new specification, winner and loser regrets are still significant at $p<0.05$ across all bidder segments except one, and their magnitudes, on average, do not change significantly. Other insights derived from the main model did not change either, so I concluded that my utility specification is robust to this assumption.

I further checked the robustness of my estimation algorithm to the clustering approaches. First, I tested applying a different clustering method on the auctions, rather than using the eBay-specified clusters. In particular, the descriptions in the item titles in Table 2.3 suggest that the auction categories may not be the best way to classify auctions, since the keywords in the product titles also provide useful information for classification. For example the words “Shampoo”, “Conditioner”, and “Styler” in the title of the auction item which is classified as “Gift Card and Coupon”, might suggest that this item is actually closely related to items in “Health and Beauty” category. Or the word “Original” appearing in the descriptions of items that are in “Collectible” and “Entertainment and Memorabilia” categories might suggest that bidders behave similarly for these two items, as a response to this signal. This observation led me to use a model that incorporates not only the auctions' structured information, but also unstructured text description of the auction items.

To be able to incorporate text data, I used frequency matrix of words, and after augmenting it with the auction item information, I used a Latent Dirichlet Allocation (LDA) model to cluster the auction items based on their observed characteristics. Similar to bidder specific parameters, I used a two-step approach, [23] i.e., I first clustered the auctions using the LDA model and then I conditioned on the segment membership in the estimation procedure to shrink the auction specific parameters.

I used a topic modeling package available in R, to run LDA using Variational Bayesian Expectation Maximization (VBEM) method, over a data set of key word frequency and auction information. To create auction key word frequency matrix, I used WordNet python interface to lemmatize the keywords after parsing them, and I only kept the keywords that are available in the dictionary. I explain the details of the LDA estimation procedure in the online companion.

This procedure grouped the auctions into 50 clusters[24]. The average number of bidders (resp. bids) in each cluster ranges from 6 to 13 (resp. 25 to 63) and the average auction duration ranges from 4 to 6 days[25]. The optimal MAP in this case is estimated to be -99,276,228 (5.3% less than the estimated main model MAP). The estimation and counterfactual results can be found in the online companion. Overall, my insights derived from the main model do not change significantly.

---

[23] I did not use LDA in the estimation procedure, mainly for its detrimental effect on the run time.

[24] If I use k-means algorithm, I observe that within group sum of square uniformly decreases as the number of clusters increases (see Figure B.2 in Appendix). However, I used 50 clusters, since some clusters become highly sparse if I use more than 50 clusters.

[25] Except one (outlier) cluster that, on average, lasts one day with one average number of bids and bidders.

Second, to check the robustness of my results to the clustering method used for the bidders, I tested using a k-means algorithm instead of mixture normal fuzzy clustering. Appendix B presents the change in within group sum of square based on the number of clusters using k-means algorithm, [26] and compares the summary statistics of clusters derived by both approaches. These analyses show that both clustering methods provide similar clusters for bidders. The optimal MAP in the case of k-means algorithm is estimated to be -103,428,091, which is 9.7% less than the estimated main model MAP. Furthermore, my main insights are not affected significantly in this case either[27].

## 2.9. CONCLUSION

In this paper, I developed a structural model that accounts for bidders' learning and their anticipation of winner and loser regrets in an auction platform and proposed an empirical Bayesian estimation method to calibrate the parameters of this model. Then, using a large data set from eBay, I showed that bidders anticipate significant levels of regret in various product categories. My results also demonstrate that experience can explain the heterogeneity in the bidders' learning, updating, and regretting behavior.

I further illustrated how the estimated model can be used to analyze a counterfactual scenario where the auction platform shuts down the bidders' winner regret. This counterfactual analysis shows that, if eBay can shut down winner regret of bidders by appropriate notification policies, it can increase its revenue by 24%. I believe that my modeling approach, proposed estimation method, and derived empirical insights in this paper can be of interest to both practitioners and scholars in academia.

[26] The elbow of the curve in Figure B.1 is around 50, which suggests that the range of the optimal number of clusters that BIC criterion suggests in mixture normal clustering (47 clusters) is robust.

[27] Estimation results of this model is available upon request from author.

# CHAPTER 3

# MEASURING GAMIFICATION ELEMENTS' EFFECTS ON USER CONTENT GENERATION: AN EMPIRICAL STUDY OF STACKOVERFLOW'S TWO SIDED PLATFORM'S BIG DATA

Meisam Hejazi Nia

Naveen Jindal School of Management, Department of Marketing, SM32

The University of Texas at Dallas

800 W. Campbell Road

Richardson, TX, 75080-302

## 3.1. ABSTRACT

The cornerstone of the new marketing era consists of user generated content. This information is useful for reducing consumer uncertainty, generating new ideas for new products, and managing the customer relationship. To motivate users to generate content, practitioners use video game elements such as badges, leaderboard, and reputation points for user achievements, in an approach called Gamification. To allow Gamification platforms to target their users, I profile user segments by an ensemble method over LDA, mixed-normal and k-mean clustering, and then I develop a model of state-dependent choices of content generating users. This model captures long tail distribution of user heterogeneity by Dirichlet Process, and investigates the effects of fun and social elements of Gamification, reputation points, rank in the leaderboard, and badges (i.e. gold, silver, bronze) on the users' probabilities to contribute content. I used a big data set of approximately 11,000,000 choices made by 36,000 users across 250 days on Stackoverflow to estimate the mixed binary logit model of users' content contribution choices. I show that estimating the model on smaller random samples generate biased results. The estimation results demonstrate that users show heterogeneous significant positive and negative inertia, reciprocity, intrinsic motivation, and responses to badges, reputation points, and leaderboard ranks. I found interesting sensitivity patterns to Gamification elements for users with different nationality, which allows the Gamification platform to create targeted messages. The counterfactual analysis suggests that the Gamification platform can increase the number of contributions by making earning badges more difficult.

Keywords: Gamification, user generated content, mixed logit with DP prior, semi parametric Bayesian, ensemble segmentation, targeting

## 3.2. INTRODUCTION

An underutilized marketing resource is user generated content, a type of online contents that customers generate and use. User generated content is an important tool for generating word of mouth buzz, collecting new product development ideas, decreasing consumers' uncertainty about an experience goods, engaging brands, and managing customer relationships. However, to use this resource effectively, marketers might need to know how to motivate users to generate more favorable and high quality content. To motivate the users, marketing practitioners have started to use the video games concepts such as badges and points for user achievements, and leaderboards, for user popularity, in a method called Gamification.

According to Gabe Zicherman, the author of "game based marketing", Gamification is the use of game play mechanics for non-game applications (Zichermann and Linder 2010). In other words, Gamification is the process of using game thinking and mechanics to engage an audience and solve problems (Van Grove 2011). Studies show that game play itself stimulates the human brain (releasing dopamine), so Gamification aims to bring the proven mechanics from gaming into marketing (Bosomworth 2011). Gartner predicts by 2016, Gamification will be a vital tool for brands' and retailers' customer loyalty and marketing. However, this report highlights that firms are skeptical about the longevity and the real efficiency of Gamification as a tool to motivate customers[28].

As a result, given the interest in and skepticism about the effects of Gamification mechanics on motivating users, this study asks the following questions: How to model the choices of consumers in response to Gamification mechanics? How to weigh emotional elements such as fun in relation to the mechanical elements, such as badges and leaderboard? How to design a scalable and flexible targeting approach that is feasible on massive streaming Gamification platform data? Are the social aspects of public good contributions, such as reciprocity and reputation, important in motivating users to provide content in a gamified context?

Answering each of these questions helps the Gamification platform to form a different targeted policy to increase the users' content contributions. For example, depending on whether badges are good or bad motivators, the Gamification platform might modify the thresholds of earning them. As points sum up to build the users' reputation, depending on whether different users respond positively or negatively to their reputation changes, the Gamification platform can send a customized list of tasks with different difficulty level to users. In the customized list, the Gamification platform might prioritize tasks to make sure that the community replies to the request of target users who are positively reciprocal. In addition, given that Gamification is about user empowerment, the Gamification platform might want to send positive empowering messages to failed users who have high inertia.

---

[28] Gartner's Gamification predictions for 2020. Growth Engineering website. http://www.growthengineering.co.uk /future-of-gamification-gartner/. Accessed June 7, 2015.

To respond to these questions and take into account the emotional nature of the motivation process, the current study builds its model in the light of the state-dependent utility model in the consumer choice literature. In particular, I included in the state-dependent utility model the elements that might define the observed motivation state of users. A user decides whether or not to contribute, based on this utility. First, I included in the model heterogeneous stimulation level in a form of user specific random effects, guided by the studies in the consumer behavior literature (Mittelstaedt 1976; Joachimsthaler and Lastovicka 1984; Steenkamp and Baumgartner 1992). Second, I considered the number of badges in different categories (i.e. gold, silver, bronze) to have different effects, guided by the Gamification literature (Wei et al. 2015; Li et al. 2015; Deterding 2012; Antin and Churchil 2011; etc.). Furthermore, I allowed the users of different segments to respond differently to the same type of badges.

I considered the social aspect of users' decisions at two levels: first, the reciprocity and the reputation points at state of user utility level; and second, the reach of users at hierarchical level, guided by the literature on behavioral aspect of decision making (Bolton et al. 2013; Bolton et al. 2004; Yoganarasimhan 2013; Lee and Bell 2013; Toubia and Stephen 2013). I also considered that the effects of badges and reputation points in motivating users might be different in the short and long term, similar to the effect of loyalty program rewards and promotions in the marketing literature (Liu 2007; Jedidi et al. 1999, Mela et al. 1997; Lewis 2004).

To estimate the model I use a data set I scraped from StackOverflow by my Python crawler. The data set includes approximately 11,000,000 contribution choices of 36,000 users over a course of approximately 230 days. StackOverflow is a question and answer website, where registered users can post their programming questions, and the other community members can respond. The StackOverflow business model is based on the traditional job listing, Curriculum Vitae search, and unobtrusive advertising. It uses Gamification concepts such as reputation points, badges, and a leaderboard to motivate its users. Community members can up-vote or down-vote a question or an answer, and StackOverflow keeps track of the votes a user receives as reputation points. The platform (i.e. StackOverflow) uses these votes later as a measure to define who receives badges at gold, silver, and bronze levels in different knowledge domains. These domains are specified by tags that a user attaches to the question. In addition, these reputation points define the rank of each user on the leaderboard. I selected StackOverflow as a source of the data for this study, because it implements a successful Gamification mechanics on its question and answer platform (e.g., Antin and Churchill 2011; Wei et al. 2015; Li et al. 2015). To use this data, I wrote a web crawler, and I synthesized the data from various web pages based on the user identity.

Estimating the model over the big data set allows the Gamification platform to target its policies effectively, if the estimates capture the heavy tail of user-heterogeneity parameter-distribution. I employed a mixed binary logit model with hierarchical Dirichlet process, which allows the number of response parameters to increase with sample size. Allowing the number of parameters to increase with the sample size allows the estimation procedure not only to learn the tail more effectively, but also to learn more about the infinitely complex real phenomena as more data becomes available.

To the best of my knowledge no studies in marketing have estimated a choice model over such a big data set. Instead, marketing scholars resort to a linear-probability data-fitting approach to estimate consumers' parameters (Goldfarb and Tucker 2011). An alternative approach is to sample from the data, but throwing away data might not be a relevant strategy for targeting. I showed the estimates for the model using samples with different sizes. The results showed that estimating the model over a smaller sample sizes results in biased estimates. As a result, importance of a quick, flexible, and scalable method is highlighted.

The results show that users can be segmented into competitors (20%), collaborators (21%), achievers (25%), explorers (11%), and uninterested (22%) users. The users show heterogeneous significant positive and negative response to the badges, leaderboard ranks, and reputation points. In addition, users show heterogeneous significant positive and negative inertia, intrinsic motivation, and reciprocity. These results suggest that the Gamification platform can condition its targeted message on the users' responses to increase their content contributions. Particularly, my results identify that certain nationalities are sensitive to certain Gamification elements. For example, American users show significant inertia, increase their contribution when earning silver badges, but decrease their contribution when their reputation is greater. However, European users increase their contribution when their reputation is greater, but they decrease their contribution when they earn Gold badges.

Given the estimated parameters, and the two sided sword effects of the badges, I used a counterfactual analysis to study the effect of modifying the threshold of badges on the response of users. The results suggested that the Gamification platform may want to increase the thresholds of earning the badges rather than decreasing them, to make badges harder to achieve. This recommendation parallels the recommendation in studies on loyalty program effectiveness in the marketing literature that suggests increasing the reward threshold is a good choice. In the Gamification context, this decision is important because badges are once-in-a-lifetime elements, without expiry date.

In summary the current study contributes to the literature in marketing in the following ways: First, although a stream of literature in marketing focuses on various factors that affect the valence of user generated content, and its impact in reducing customer uncertainty (e.g., Weiss et al. 2008; Moe and Schweidel 2012; Godes and Silva 2012; Mallupraganda et al. 2012), the user motivation to contribute content is understudied. The current study tries to narrow this gap by determining which Gamification elements can drive motivation of users to contribute content.

Second, although many practitioners and social psychologists emphasize the role of Gamification as a motivator (Wu 2011; Deterding 2012; Conejo 2014), quantitative measures of Gamification elements such as badges and leaderboard to help the Gamification platform to target its policies are understudied. Two studies in progress by Wei et al. (2015) and Li et al. (2015) use a difference in difference and a hidden Markov model to identify such effects. However, both of these studies assume that the users select the number of contributions, rather than whether to contribute or not. Also these studies do not account for heterogeneity in users' responses to the Gamification elements. Therefore, in the current study I modeled the binary choice of the users while allowing for state-dependency and heterogeneity. Finally, I use ensemble method over LDA, mixed normal, and k-mean clustering methods to profile user segment behaviors, and mixed binary logit model with hierarchical Dirichlet Process prior to recover user specific parameters. These contributions should be of interest to both practitioners and scholars.

## 3.3. LITERATURE REVIEW

This study draws upon several streams within the literature that have investigated, including: (1) User Generated Content (UGC); (2) Gamification mechanisms and rewards in loyalty programs; (3) Optimal stimuli level and state dependent choice models; (4) Behavioral aspects of decision making (altruism, reciprocity, endowment effect, etc.). Given the breadth of these areas across multiple disciplines, what follows is only a brief review of these relevant streams. Table 3.1 presents a list of relevant studies in each literature stream.

Table 3.1. The relevant streams of litrature in five clusters

| Research Area | References |
|---|---|
| User Generated Content and free rider problem | Mallapragada et al. (2012); Godes and Silva (2012); Moe and Schweidel (2012); Chevalier and Mayzlin (2006); Chaudhuri (2011); Chen (2008); Weiss et al. (2008). |
| Gamification elements ,Mechanism, and Loyalty | Li et al. (2015); Wei and Zhu (2015); Conejo (2014); Bittner and Shipper (2014); Salcu and Acatrinei (2013); Roth and Schneckenberg (2012); Kopalle et al. (2012); Zhang and Breugelmans (2012); Wu (2011); Zichermann and Cunningham (2011); Pink (2009); Liu (2007); Shugan (2005); Kivetz and Simonson (2002); Bolton et al. (2000). |
| State dependent choice model, and optimal stimuli | Dubé et al (2008); Seetharaman (2004); Seetharaman (1999); Guadagni and Little (1983); Steenkamp and Baumgartner (1992); Joachimsthaler and Lastovicka (1984); McAlister (1982); Mittelstaedt et al. (1976); Lewis(2004); Jedidi (1999). |
| Behavioral aspect of Decision Making, Altruism, reciprocity | Toubia and Stephen (2013); Lee and Bell (2013); Yoganarasimhan (2013); Bolton et al. (2004); Andreoni(1990); Cornes and Sandler (1994); Bolton et al. (2013); Churchill (2011); Chen et al. (2010); Raban (2009); Chiu et al. (2006); Ren and Kraut (2011); Tedjamulia et al. (2005). |
| Big Data Estimation Methods | McMahan et al. (2013); McMahan (2011); Genkin et al. (2007); Le Cessie and Van Houwelingen, (1992); Murphy (2012). |

### 3.3.1. User Generated Content (UGC)

User generated content in marketing refers to the contents that are both produced and consumed by the same consumers, for example question and answers, blogs, Twitter, social networks, and YouTube videos (Mallapraganda et al. 2012). UGC can also be considered as a form of public goods, because one cannot exclude others from using it after and during usage. Marketing and economics scholars have studied UGC from two perspectives: consumption and production. From the consumption perspective, Chevalier and Mayzlin (2006) find that UGC can have a positive effect on sales. In addition, Weiss et al. (2008) find that the consumers' goal and the social history of the producer affect how consumers perceive the value of UGC. From the production perspective, Godes and Silva (2012), Moe and Schweidel (2012), and Mallapraganda et al. (2012) suggest that the UGC creation process is subject to selection bias due to social influence and the heterogeneity in preferences of the product adopters who enter with different order and at different times. Although these studies are useful, none of them discuss how the firm can affect the UGC creation process by motivating users.

The public good literature in experimental economics fills the gap by studying these incentive compatible mechanisms (Chen 2003). By relaxing strong rationality assumptions, these studies find that punishing altruistically and monetarily, grouping likeminded individuals, and passing advice across generations can motivate the users (Chaudhuri 2011). Although these studies are helpful, they neglect that users' emotion can also be relevant. In particular, psychological studies emphasize that having fun, earning virtual rewards, setting goals, and empowering can also motivate users to contribute UGC. In the current study I focused on quantifying the effect of such psychological factors on the users' choice to generate content, in a gamified context.

### 3.3.2. Gamification mechanisms and rewards in loyalty programs

To motivate users in a non-gaming environment, Gamification uses elements from video games (Bittner and Shipper 2014). Gamification elements can be classified according to three categories: dynamics, mechanics, and components (Zichermann and Cunningham 2011). Game dynamics involve personal-psychological elements of the sense of progression, emotions, relationships, and narratives. Game mechanics involve social-psychological elements, including feedback, rewards, competition, cooperation, and transactions. Game components include achievements, levels, points, badges, leaderboards, and virtual goods. Studies relevant to Gamification can be classified into two groups: quantitative and qualitative studies of general Gamification, and the Gamification role in marketing. Only two studies quantify the effect of Gamification elements. First, Li et al. (2015) use difference-in-difference reduced-form to identify the aggregate level effect of badges on users' number of content contributions. Second, Wei et al. (2015) aim to quantify these effects structurally by a Hidden Markov Model (HMM), but they consider that users plan how many contents to contribute, rather than plan at each point whether to contribute or not (i.e. binary choice). Although helpful, both of the studies fail to control for the users' heterogeneity and effects of users' inertia.

However, the qualitative studies of Gamification emphasize the role of user inertia. These studies emphasize the psychological need of consumers to experience pleasure, fun and empowerment, based on self-determination theory (Wu 2011). Unlike monetary rewards, fun and pleasure are process-focused motivators, rather than outcome-focused motivators (Shen et al. 2015). In Gamification, outcome-focused motivators include social status and reputation. Gamification captures these outcome-focused motivators in points, leaderboard, and badges (Deterding 2012). These studies are helpful, but for marketing purposes, a Gamification platform needs a formal model to quantify the effect of Gamification elements to target users based on their behavior.

Marketing scholars have studied Gamification elements qualitatively. Bitter & Shipper (2014) shows in a case study that Gamification is useful for advertising. Salcu and Acatriney (2013) discuss a case study in which Gamification has worked in the affiliated marketing program. Roth and Schnechenberg (2012) find that Gamification is useful for innovation and creativity. Conejo (2014) posits that Gamification can revolutionize loyalty programs. In particular, Gamification differs from conventional loyalty programs given its emphasis on fun, meaningfulness, and empowerment, in addition to point rewards. The loyalty programs' point rewards have been the subject of many marketing studies (Bolton et al. 2000; Shugan 2005; Kivetz and Simonson 2002; Kopalle et al. 2012, Zhang and Bregelman 2012; Liu 2007; Jedidi et al. 1999; Mela et al. 1997; Lewis 2004). These studies support a model of Gamification elements to control for heterogeneity and short and long term effects, but they do not quantify the effects of Gamification elements on users' content contribution.

### 3.3.3. Optimal stimuli level and state dependent choice models

Inertia and state dependence that qualitative Gamification studies suggest is the subject of two groups of studies in consumer research and marketing science. Consumer research scholars emphasize that, to engage in exploratory behavior, consumers need to be emotionally motivated until they reach their heterogeneous optimal estimation level (Steenkamp and Baumgartner 1992; Joachimsthaler and Lastovicka 1984; Mittelstaedt et al. 1976; Seetharaman et al. 1999; McAlister 1982). Marketing science scholars use fixed effects to model this optimal stimulation level, and they use lagged instant or cumulative choices to control for heterogeneous users' state dependence (Guadagni and Little 1983; Dube et al. 2008). These studies further emphasize that the modeler should allow for a flexible heterogeneity structure to avoid confounding state dependence with heterogeneity. Built on the above research, this study adopted a heterogeneous agent-based state-dependent choice-model, rather than the rule-based simulation approach that Ren and Kraut (2011) adopt to run counterfactual Gamification policies.

### 3.3.4. Behavioral aspects of decision making

Gamification policies aim to motivate users to behave in certain ways, for example to create content. Online content can be considered as a form of public good, because upon publishing its consumption cannot be controlled. Behavioral economics literature concludes that the impure altruism model can explain the public good creation (Cornes and Sandler 1992; Andreoni 1990). In particular, the impure altruism model considers that users might have both private and public incentives to contribute. In marketing literature, Toubia and Stephen (2013) classify user's heterogeneous factors of utility to contribute content into two groups: intrinsic and extrinsic (or image and prestige related). Tedjmulia et al. (2005) argues that, under specific circumstances, the extrinsic factors can affect intrinsic factors, either positively by internalization or negatively by over-justification. Chiu and Want (2006) argue from social psychological perspective that, content creation is more influenced by intrinsic factors such as fun and playfulness than extrinsic factors such as reciprocity and reputation. However, many other studies emphasize the importance of social capital and reputation, as a substitute for monetary incentives, in content creation (Raban 2008; Chen et al. 2010; Toubia and Stephen 2013). Furthermore, Antin and Churchill (2011) discuss that Gamification badges and leaderboards can influence extrinsic factors such as reputation, status affirmation, and group identification, and intrinsic factors such as goal setting, and instructions.

All in all, many marketing modeling studies have emphasized the role of social factors such as reputation and reciprocity on users' choices in various contexts (Bolton et al. 2004; Bolton et al. 2013; Banks et al. 2002; Yoganarasimhan 2013; Lee and Bell 2013). However, no study has yet modeled individual users' content creation choices in terms of Gamification policies that aim to motivate users both intrinsically and extrinsically. The current study aims to narrow this gap.

## 3.4. DATA

I collected the data for this study from StackOverflow, because many studies find that it provides a successful Gamification application (e.g., Antin and Churchill 2011; Wei et al. 2015; Li et al. 2015). Stackoverflow is an open online platform for professional and enthusiast programmers. It was founded in 2008, by a firm which later established Stack Exchange, a network of question and answer websites focused on diverse topics (ranging from physics to writing) and modeled after StackOverflow. In 2014, StackOverflow had 4 million users, and among these users, 77% asked and 65% answered questions. During this period, they generated 11 million questions and displayed an exception level of heterogeneity in their content creation. For example, only 8% of the users answered more than 5 questions. StackOverflow raised $6 million venture capital money in 2010, and its business model is based on three key activities: job listing (like traditional classified advertising), Curriculum Vitae search, and unobtrusive display advertising. The platform is rigid in its focus, requesting the users to ask only questions relevant to its topic and refrain from raising questions that are opinion based or lead to open ended chat. The moderators monitor the violation of this rule through the community members' reports.

The community of content creator users plays a key role in managing StackOverflow's day to day activities, but the community notifies moderators in exceptional cases, for example when the etiquette is not preserved. StackOverflow selects lifetime moderators through a democratic voting procedure, but moderators can resign. According to StackOverflow, moderators act as liaisons between the user community and Stack Exchange. StackOverflow asks users to register and log in before asking a question, but to answer a question, a user must either sign up for an account or post as a guest, in which case the user must register her name and email address.

On Stackoverflow, a user who asks a question has the option to review the answers and to accept an answer as correct. In addition, everyone can vote up or down on either each question or answer. To be able to vote, the user must first register. The sum of all the up-votes and down-votes that a user receives for contributing contents (i.e. questions and answers) is called the user's reputation points. Web surfers can observe a user's reputation level on her profile page and the leaderboard. According to StackOverflow, reputation is a rough measure of how much the community trusts the user's expertise, communication skill, and content quality and relevance. In addition to the weekly reputation information, the leaderboard presents the previous week's information about user's rank and rank change. The leaderboard provides an informal way of tracking reputation within the community. It acts like a leaderboard of a league, and it only tracks the users' points above a threshold of 200 points.

Contributing users can also earn badges as hallmarks of their achievements. Badges have three categories or levels: gold, silver, and bronze. In addition, badges are specific to knowledge domains. In this study, I refer to this detail by using a domain knowledge tag because it is directly relevant to the tag that user attaches to her question. In particular, when a user posts a question, the StackOverflow platform requires her to attach a relevant tag to make the question appear to relevant audience, or sub-community. The total up-votes a user earns by answering the questions relevant to the tag nominate a user for tag badges. The threshold for gold, silver, and bronze badges, are 1000, 400, and 100 respectively, fixed across the tags (or knowledge domain). Table 3.2 presents a sample of these badges.

Table 3.2. Sample set of Bandges in different knowledge domains (tags)

| Tag Badge Name | Type | Definition |
|---|---|---|
| vcl | Bronze | Earn at least 100 total score for at least 20 non-community wiki answers in the vcl tag. |
| entity-framework | Silver | Earn at least 400 total score for at least 80 non-community wiki answers in the entity-framework tag. |
| r | Gold | Earn at least 1000 total score for at least 200 non-community wiki answers in the r tag. These users can single-handedly mark r questions as duplicates. |
| ggplot2 | Silver | Earn at least 400 total score for at least 80 non-community wiki answers in the ggplot2 tag. |
| statistics | Bronze | Earn at least 100 total score for at least 20 non-community wiki answers in the statistics tag. |
| regex | Bronze | Earn at least 100 total score for at least 20 non-community wiki answers in the regex tag. |
| linux-kernel | Silver | Earn at least 400 total score for at least 80 non-community wiki answers in the linux-kernel tag. |
| sql-server-2008 | Gold | Earn at least 1000 total score for at least 200 non-community wiki answers in the sql-server-2008 tag. These users can single-handedly mark sql-server-2008 questions as duplicates. |
| html-helper | Silver | Earn at least 400 total score for at least 80 non-community wiki answers in the html-helper tag. |
| google-app-engine | Silver | Earn at least 400 total score for at least 80 non-community wiki answers in the google-app-engine tag. |

I collected the data for this study, by automatically scraping the Stack Overflow website. The sample includes the 36,915 users who appeared in the leaderboard during the first week of June 2014, and identified themselves by an English name that can be captured by the python crawler and scraper code. I observed users' choices including choices to comment, review, revise, accept, and post-answers. Table 3.3 presents the definition of each of the activities that I observed from users. The activity that I am interested to model is aggregate number of users' contribution, which includes commenting for clarification, answering a question, and revising an answer. To control for potential reciprocity of users, I include the total number of accepted posts, reviews, and asking activities in the reciprocity variable.

Table 3.3. Type of activity, description and inclusion in the dependent or independent variable

| Activity Name | Observations in data set | Type (Proxy) | Definition |
|---|---|---|---|
| Comment | 1,995,665 | Dependent (Choice) | Includes the activity of asking for clarification, suggesting correction, providing meta information about the post (so that not confuse with answer), it is short (600 character), only limited Markup, URL, disposable, and it does not have revision history, and it can be deleted without warning to the author by the moderator. |
| Accepted | 80,446 | Independent (Reciprocity) | Includes the activity of the questioner to review the answers and only accept the one that it finds suitable. |
| Post Answered | 671,772 | Dependent (Choice) | Includes the activity of answering the question that is raised on the platform. |
| Review | 1,017,029 | Independent (Reciprocity) | Includes the activity of the questioner to review the answers that is posted to her question. |
| Post Asked | 129,526 | Independent (Reciprocity) | Includes the activity of the questioner to ask a question. |
| Revision | 812,992 | Dependent (Choice) | Includes the activity of revising the answering post that is raised on the platform. |

I collected these choices' data for 238 days of the sample period, namely from June 2014 to January 2015. In the sample, I observed 11,276,186 users' choices. Table 3.4 shows the frequency of users' declaration of their website and nationality. Given the identifier of the users, I also collected the leadership board information, namely the total reputation points per week, the weekly reputation, the rank, and the rank change, to synthesize with the main data. In addition, I collected the history of each user's badge earnings.

Table 3.4. Sample Observations' statistics

| Observations | Frequency |
|---|---|
| Users | 36,915 |
| website | 13,194 |
| USA | 9,434 |
| UK | 2,362 |
| Australia | 1,133 |
| India | 2,142 |
| Europe | 7,142 |
| Asia | 482 |
| South America | 659 |
| China | 208 |
| Middle East | 892 |

Figure 3.1 presents the total number of content contributions taken from a sample of four users over time. As can be seen, considerable heterogeneity exists in the users' content contribution levels, and on some days, users do not contribute at all. The number of zero contributions of different users, and the heterogeneity in the number of the contributions might offer evidence to explain why a simple regression and a homogeneous response model might not return unbiased estimates. In addition, a great number of non-contribution choices might suggest that the user thinks more about whether to contribute than about how much to contribute.

For each of the users, I also scraped the profile information: tenure, last seen date, the number of profile views, reputation, the number of gold, silver and bronze badges earned, the number of answers, the number of questions, the total amount of reach (i.e. approximate total number of people who viewed the user's posts), user's website, and user's country. Table 3.5 shows the basic statistics of these variables before and after the observation period. The average number of reputation points, badges, questions and answered increased between 10% and 30%.

To better understand the heterogeneity in users' behavior, I segmented users by clustering their observable cross sectional information profile for pre-study period. These profiles are consisted of binary indicator and count data. There are various methods to cluster this data. K-means partition based clustering might be relevant for its employing similarity index based on Euclidean distance. Mixture Normal Fuzzy model based clustering might be relevant for its assumption that observations of a given clusters are noisy measures of cluster centers. Finally, Latent Dirichlet Allocation (LDA) model based clustering method might be relevant for its assumption that each profile attribute might be relevant contingent on the cluster that the observation belongs to. Hubert and Arabie (1985) suggests adjusted random index for comparing clustering results. This method compares the pair assignment of two clustering results and by assuming generalized hyper geometric distribution creates an index that bounds to 1 under perfect agreement, and 0 under random partition (Yeung and Ruzzo 2001).

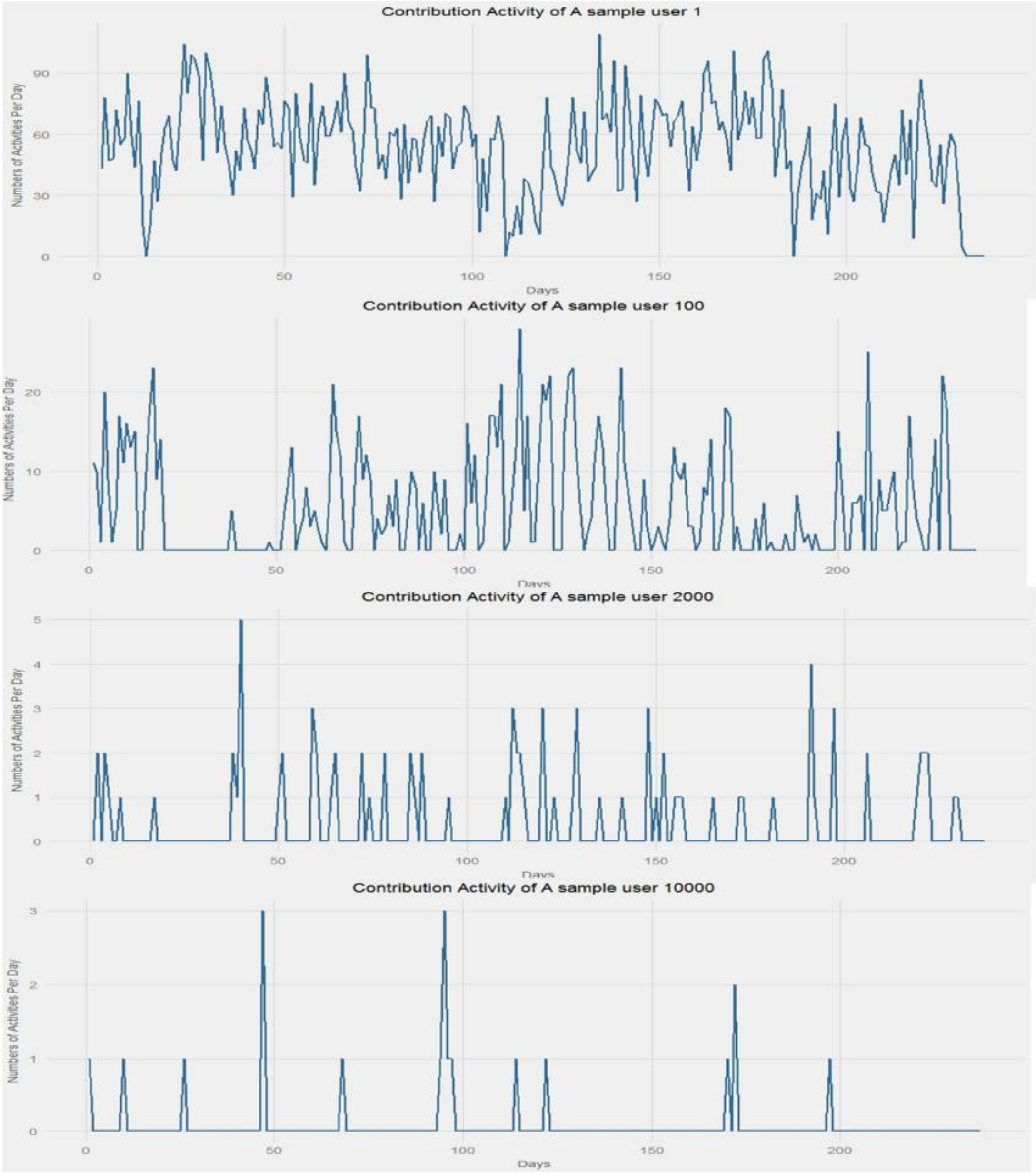

Figure 3.1. Contribution of a sample of four users over time

To have computational tractability, I randomly sampled 20,000 from 36,000 users for clustering, and ran three clustering methods to segment these users. Elbow measure of the within cluster sum of square suggested that k equal to 30 is optimal for k-means clustering method. Bayesian Information Criteria (BIC) measure suggested that five clusters are enough for mixture normal fuzzy clustering method, and Log Likelihood measure suggested that 13 clusters in LDA method represents the data better. I used random adjusted measure to compare the partition membership result of these three methods. Table 3.6 presents the result of this comparison. Mixture Normal and K-means clustering generate more similar result than other couples.

Table 3.5. Sample Observations' basic statistics

| | Pre | | | | Post | | | |
|---|---|---|---|---|---|---|---|---|
| Variable | AVG | SD | Min | Max | AVG | SD | Min | Max |
| Reputation Points | 6,213.99 | 17,650.15 | 1 | 685,463 | 8,066.61 | 21,221.38 | 1 | 773,020 |
| Number of Gold Badges | 5.51 | 8.04 | 2 | 301 | 6.70 | 9.19 | 2 | 344 |
| Number of Silver Badges | 20.98 | 44.88 | 2 | 4,597 | 24.83 | 50.68 | 2 | 5,233 |
| Number of Bronze Badges | 40.86 | 66.08 | 2 | 5,951 | 46.30 | 71.74 | 2 | 6,507 |
| Number of Answers | 188.86 | 527.00 | 0 | 29,950 | 207.56 | 574.53 | 0 | 31,537 |
| Number of Questions | 36.22 | 73.04 | 0 | 1,688 | 40.30 | 78.05 | 0 | 1,737 |

Table 3.6. Adjusted Random Index measure for clustering agreements

| | LDA | Mixture Normal | K-means |
|---|---|---|---|
| LDA | 1.0000 | | |
| Mixture Normal | 0.0005 | 1.0000 | |
| K-means | 0.0000 | 0.3420 | 1.0000 |

As it is not clear which method to choose, I used ensemble method (similar to Strehl and Ghosh 2003) to create results that are more robust to the type of clustering method, rather than using each of these methods In this approach. I combine the results of cluster membership of observation pairs using a hierarchical agglomerative clustering method. For ease of exposition and interpretability, I cut the tree at the level with five clusters. To interpret the segments, this study adopts the terminology of Gamification to classify the users into four groups based on whether the focus is on action versus reaction, or on context versus players (Wu 2012): collaborators (focus on interaction and players), competitors (focus on action and players), explorers (focus on context and interaction), and achievers (focus on context and action). Table 3.7 presents the definition of each of the segments and the proxy variable relevant to the context.

Table 3.7. Gamification Segment Names

| User Segment Name | Focus | proxy variables | Definition |
|---|---|---|---|
| Competitors | Player & Action | High level of reputation and badges | Users will go to great lengths to achieve rewards that confer them little or no gameplay benefit simply for the prestige of having it. |
| Collaborators | Player & Interaction | High number of answers | Users gain the most enjoyment from a game by interacting with other users, and on some occasions, computer-controlled characters with personality |
| Explore | Context & Interaction | High number of Questions | Users find great joy in discovering an unknown glitch or a hidden Easter egg. |
| Achievers | Context & Action | Personal Site in the Profile | Users thrive on competition with other users, and prefer fighting them to scripted computer-controlled opponents |

Table 3.8 presents the behavioral segmentation result based on ensemble method. Users in collaborator segment consist 20% of the sample and show high level of answering activity, although they have not earned significantly more reputation points and badges. Users in achiever segment consist 25% of the sample and declare their website more than users in other segments, and all of them are American. Users in Explorer segment consist 11% of the sample and ask significantly more questions than others. Users in competitor segment consist 20% of the sample and have earned more badges and reputation points than others. Finally, I identified users in uninterested segment that consist 22% of the sample and do not declare their nationality and behave poorly relative to other users with respect to all measures. In fact this information can help the Gamification platform to target its customers.

Table 3.8. Gamification Segment Names (heat map configured at row level)

| Segment Name | Collaborators | Uninterested Users | Achievers | Explorers | Competitors | Whole Sample |
|---|---|---|---|---|---|---|
| Segment Size | 0.21 | 0.22 | 0.25 | 0.11 | 0.20 | 1.00 |
| Website | 0.44 | 0.00 | 0.51 | 0.38 | 0.46 | 0.36 |
| USA | 0.00 | 0.00 | 1.00 | 0.02 | 0.01 | 0.26 |
| UK | 0.29 | 0.00 | 0.00 | 0.00 | 0.00 | 0.06 |
| Austrailia | 0.14 | 0.00 | 0.00 | 0.00 | 0.00 | 0.03 |
| India | 0.27 | 0.00 | 0.00 | 0.00 | 0.00 | 0.06 |
| Euroupe | 0.00 | 0.00 | 0.00 | 0.01 | 0.97 | 0.19 |
| Asia | 0.06 | 0.00 | 0.00 | 0.00 | 0.00 | 0.01 |
| South America | 0.09 | 0.00 | 0.00 | 0.00 | 0.00 | 0.02 |
| China | 0.03 | 0.00 | 0.00 | 0.00 | 0.00 | 0.01 |
| Middle East | 0.11 | 0.00 | 0.00 | 0.00 | 0.00 | 0.02 |
| Tenure | 1,584.87 | 1,333.64 | 1,788.74 | 1,691.32 | 1,623.44 | 1,601.01 |
| Seen since | 30.26 | 22.65 | 33.75 | 366.40 | 25.60 | 66.70 |
| profile Views | 1,301.65 | 172.23 | 846.11 | 886.31 | 1,095.88 | 848.84 |
| Reputation Points | 7,638.74 | 1,906.79 | 6,664.90 | 7,084.18 | 8,237.01 | 6,183.55 |
| Gold Badges | 3.10 | 0.66 | 2.92 | 3.82 | 3.55 | 2.69 |
| Silver Badges | 23.78 | 7.61 | 22.85 | 25.10 | 25.94 | 20.56 |
| Bronz Badges | 46.31 | 18.68 | 43.90 | 48.57 | 49.88 | 40.57 |
| Answers | 237.29 | 54.21 | 196.26 | 212.68 | 251.88 | 186.61 |
| Questions | 33.78 | 19.30 | 30.29 | 60.24 | 36.46 | 33.24 |
| Reach | 5,690,175.70 | 220,400.60 | 4,689,612.80 | 4,582,561.20 | 5,917,558.80 | 4,149,459.60 |

It is important to note that users fall within the continuum of these classifications, and the noted assignments only discriminates based on the strength of each of the proxy signals (i.e. number of questions, number of answers, level of reputation, declaration of a personal webpage), for ease of interpretation. Figure 3.2 presents the median and the quantiles variations of the total number of answers that each user reviews or accepts (i.e. contribution received), the total reputation points of the user, her weekly reputation points and rank evolution across 238 days of the study. The variation in these variables further suggests that an aggregate non-agent-based model might miss the underlying dynamics in the data.

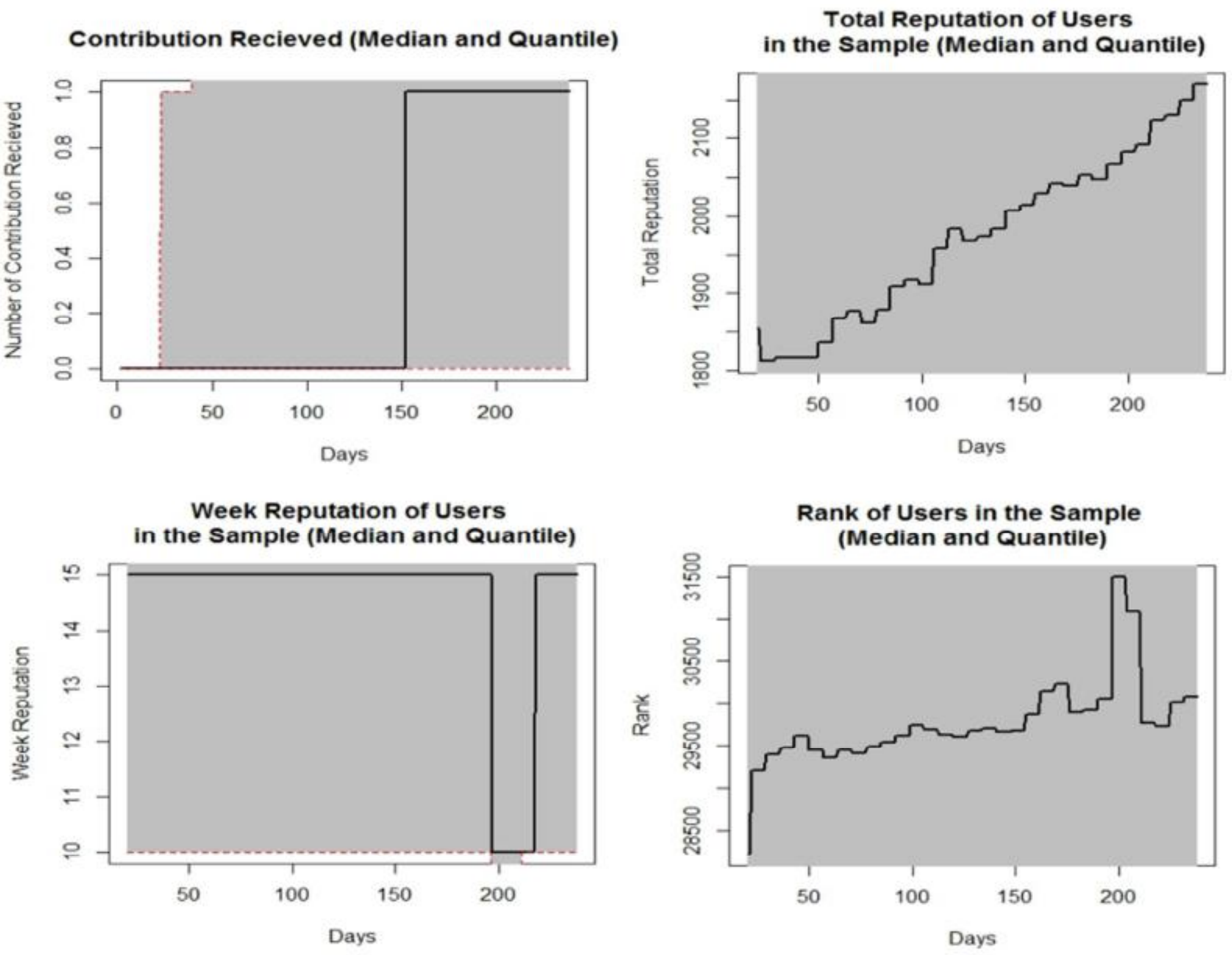

Figure 3.2. Evolution of Explanatory Variables Over time (Median shown in Black, and the interval between 25% and 75% interval is colored gray)

Finally, figure 3.3 illustrates the evolution of the total number of gold, silver and bronze badges. The data reveals peaks in the number of badges earned. I did not find the exact explanation for the peaks from the platform change of the thresholds perspective, but, as the peak is not far from the beginning of the sample, it might be relevant to the seasonal summer period when the programmers have more free time to contribute. Another seasonal peak takes place in September, an occurrence which again might reflect another demand shock because Google shows the same type of StackOverflow search trend peak in both July and September. There is also a periodic structure in the evolution of the badges. The programming nature of the questions explains this cyclical pattern. Generally, the users on StackOverflow are more active during workdays than during weekends.

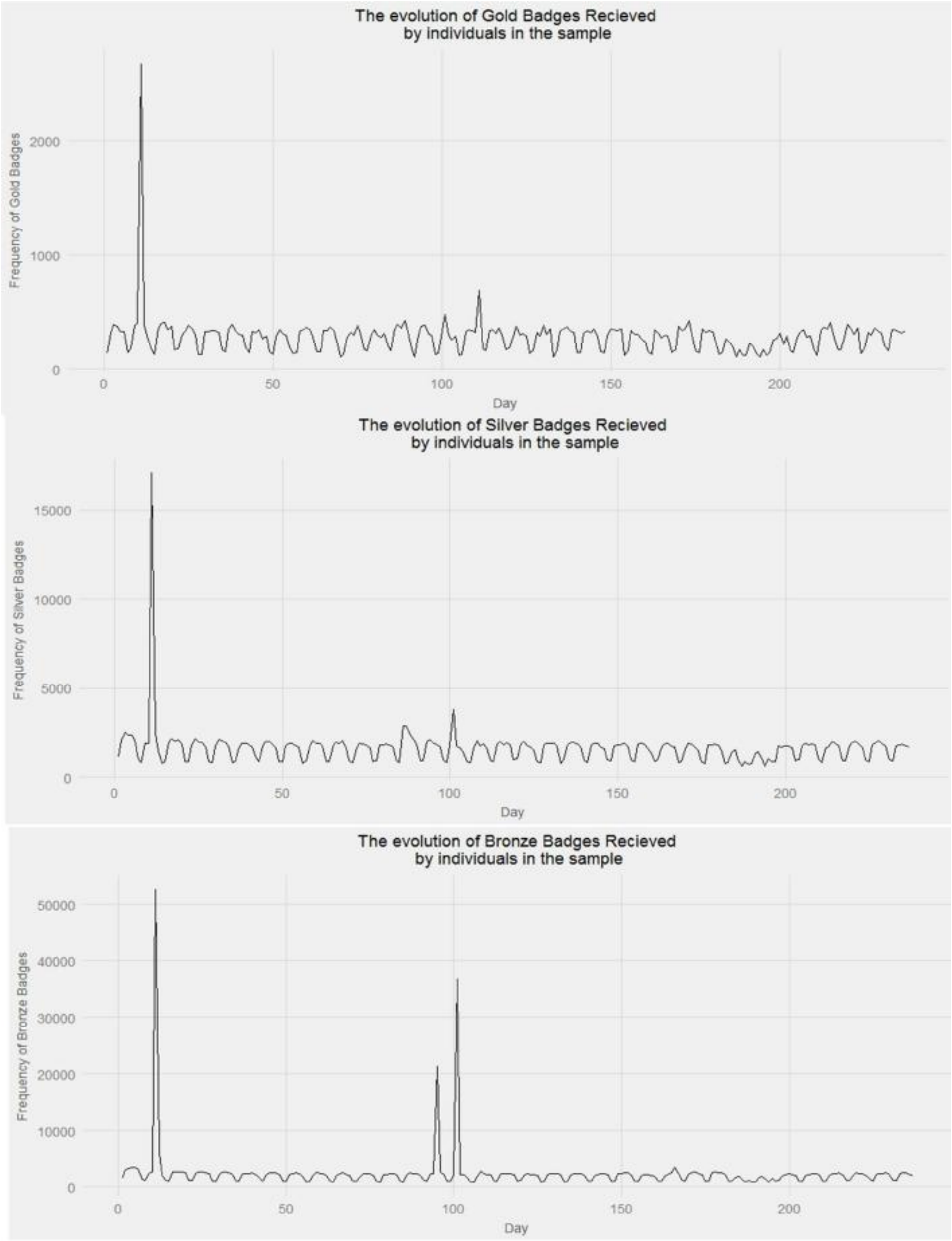

Figure 3.3. Evolution of Total number of Gold, Silver, and Bronze Badges Granted to Users in the Sample

Before discussing the model, it might be worthwhile to note that, to preserve the data of 11,000,000 contribution choices, with approximately 42,000 heterogeneity parameters, the old tabular data structure is not a viable option for the commodity computing devices. Therefore, I used a big data tool, called Apache Spark, which is built on the Hadoop map-reduce structure for data cleaning. The map-reduce structure simply creates a data flow of map and reduce operations. Map operations create assignment outcomes for each of the data points in parallel, across multiple machines (i.e. like a function with each data point). The reduce operation aggregates the outcomes of map operations, based on the group flags in parallel, across multiple machines (i.e. like aggregate functions). The data structure of map-reduce consists of pairs of key (for grouping purposes) and value (the actual data). I used the same structural of key-value pairs of map-reduce model, rather than the tabular structure. In particular, I kept each of the variables in a separate file with a key for user and time indices on separate lines. Furthermore, I used a sparse matrix structure to reduce the size of the badges' explanatory variables data. Next, I present my proposed model.

## 3.5. MODEL

I start this section with explaining the choices of the Gamification platform. In particular the Gamification elements that I considered include: fun element, badges, leaderboard, and reputation points. For example, a Gamification platform might work on the positive environment of social interaction between content producers and consumers, by putting emphasis on different contents (e.g., easy vs. hard, polite vs. not polite questions), to make the engagement more fun. It can also manipulate the threshold of earning badges, to make earning badges harder or simpler. In addition, a Gamification platform can send empowering messages to users whose rank fall on the leaderboard. To find the effect of each of these policies, the Gamification platform should measure the response of the users to the Gamification incentives.

In the context of this study, the choice of users to create content can be in the following forms: to post an answer, to revise, or to comment on a question or an answer. As a result, I considered the outcome of the user choice positive if the user makes any of these choices, and negative if the user selects none. I recognize that based on their unobserved state dependent utility, the users contribute content to the platform by answering questions and commenting and revising the answers. This state includes information about the number of contributions that users have had and the number of Gamification assets and recognitions (e.g., badges, leaderboard rank, and reputation points) they have earned recently and cumulatively. Formally, I define users' state-dependent utility of user $i$ at day $t$ in week $w$ for contributing content, in the following form:

$$\begin{aligned} U_{it} &= \alpha_i + \gamma_{i1} cont_{it-1} + \gamma_{i2} rcv_{it-1} + \gamma_{i3} crep_{iw-1} + \gamma_{i4} rep_{iw-1} + \\ &\gamma_{i5} rnk_{iw-1} + \gamma_{i6} \Delta rnk_{iw-1} + \gamma_{i7} bdg_{it-1} + \gamma_{i8} cbdg_{it-1} + \varepsilon_{it} \end{aligned} \qquad (1)$$

where $\alpha_i$ denotes the fixed stimulation threshold parameter for user $i$ for contributing content. $cont_{it-1}$ denotes total number of contents that user $i$ has contributed until day $t$. $rcv_{it-1}$ denotes total number of answers and comments that user $i$ has received until day $t$ for the question she has raised. $crep_{iw-1}$ denotes the cumulative number of reputation points user $i$ has earned until week $w$. $rep_{iw-1}$ denotes the number of reputation points user $i$ has earned at week $w-1$. $rnk_{iw-1}$ denotes the rank of user $i$ in leaderboard published at the end of week $w-1$. $\Delta rnk_{iw-1}$ denotes the first order rank difference of user $i$ between week $w-1$ and week $w-2$. $bdg_{it-1}$ denotes the vector of number of gold, silver, and bronze badges user $i$ has earned at day $t-1$. $cbdg_{it-1}$ denotes the vector of cumulative number of gold, silver, and bronze badges user $i$ has earned until day $t$. $\varepsilon_{it}$ denotes idiosyncratic unobserved utility error term for user $i$ at day $t$. $\Lambda_i = (\alpha_i, \gamma_{i1}..\gamma_{i8})$ denotes the vector of individual specific choice parameters to estimate. Assuming that a contributor has a random state dependent utility, and that the distribution of the random error term is extreme value, a logit function can model the probability of observing a user contribution. As a result, the likelihood of a user's multiple contributions in a day, follow binomial distribution. Similarly, a mixed logit function can model the choice of multiple users with heterogeneous choice parameters within the population. Next I explain the rationale behind choosing the variable that might explain the observed states of the users, in terms of the Gamification components, inertia, and reciprocity.

The proposed utility model includes user fixed effect $\alpha_i$ to capture users' heterogeneous optimal stimulation level, to represent that users require motivation to contribute content (Salcu and Actrinei 2013; Mittelstaedt 1976; Joachimsthaler and Lastovicka 1984; Steenkamp and Baumgartner 1992).

The total cumulative number of contributions $cont_{it-1}$ acts as proxy for the fun that a user experiences. As a result, a lag cumulative number of contributions might be a state variable to capture the effect of the fun elements of the Gamification platform. Furthermore, the number of content received (i.e. answer to the posted question) acts as the proxy for the social utility of the user. As a result, I included the lagged total number of asked questions, answers reviewed and answers accepted by a user $rcv_{it-1}$ as a proxy for the users' reciprocity state. Another proxy for the social utility $crep_{iw-1}$ of users to contribute content is the level of reputation points, i.e. the number of up-votes a user has received for comment and answers (Bolton et al. 2013; Bolton et al. 2004; Yoganarasimhan 2013; Lee and Bell 2013; Toubia and Stephen 2013). As the reputation point might have both instant and long term effects, the utility of the user incorporates both the weekly level $rep_{iw-1}$, and the cumulative level of user reputation $crep_{iw-1}$ (Wei et al. 2015; Li et al. 2015). Another Gamification element that is proxy signal for social status of user is the lagged leaderboard absolute rank $rnk_{iw-1}$ and rank change $\Delta rnk_{iw-1}$. The latter one might be relevant for potential endowment effect. In other words, an individual might be regretful for losing the last week rank or forgone social status.

Last but not least, badges might also affect users' motivations to contribute content, for both intrinsic (empowerment effect), or extrinsic (social status function) motivations (e.g., Antin and Churchill 2011; Wei et al. 2015; Li et al. 2015). The total number of badges earned at each badge category (i.e. Gold, Silver, Bronze) $cbdg_{it-1}$ and the number of badges earned in previous days $bdg_{it-1}$ might both affect the choice of user to contribute as they show the progress of users in the Gamified environment to their social surroundings. In addition, like the effect of any marketing policy, short term and long term effect of earning the badges might be different (Liu 2007;Jedidi et al. 1999, Mela et al. 1997; Lewis 2004). As a result, consistent with Wei et al. (2015) and Li et al. (2015), my consumer utility model includes both lagged cumulative $cbdg_{it-1}$ and instant number of each of the badges $bdg_{it-1}$. Figure 3.4 shows box and arrow diagram of the components of the state-dependent utility of users to contribute content. Table 3.9 summarizes the definition of variable and parameters.

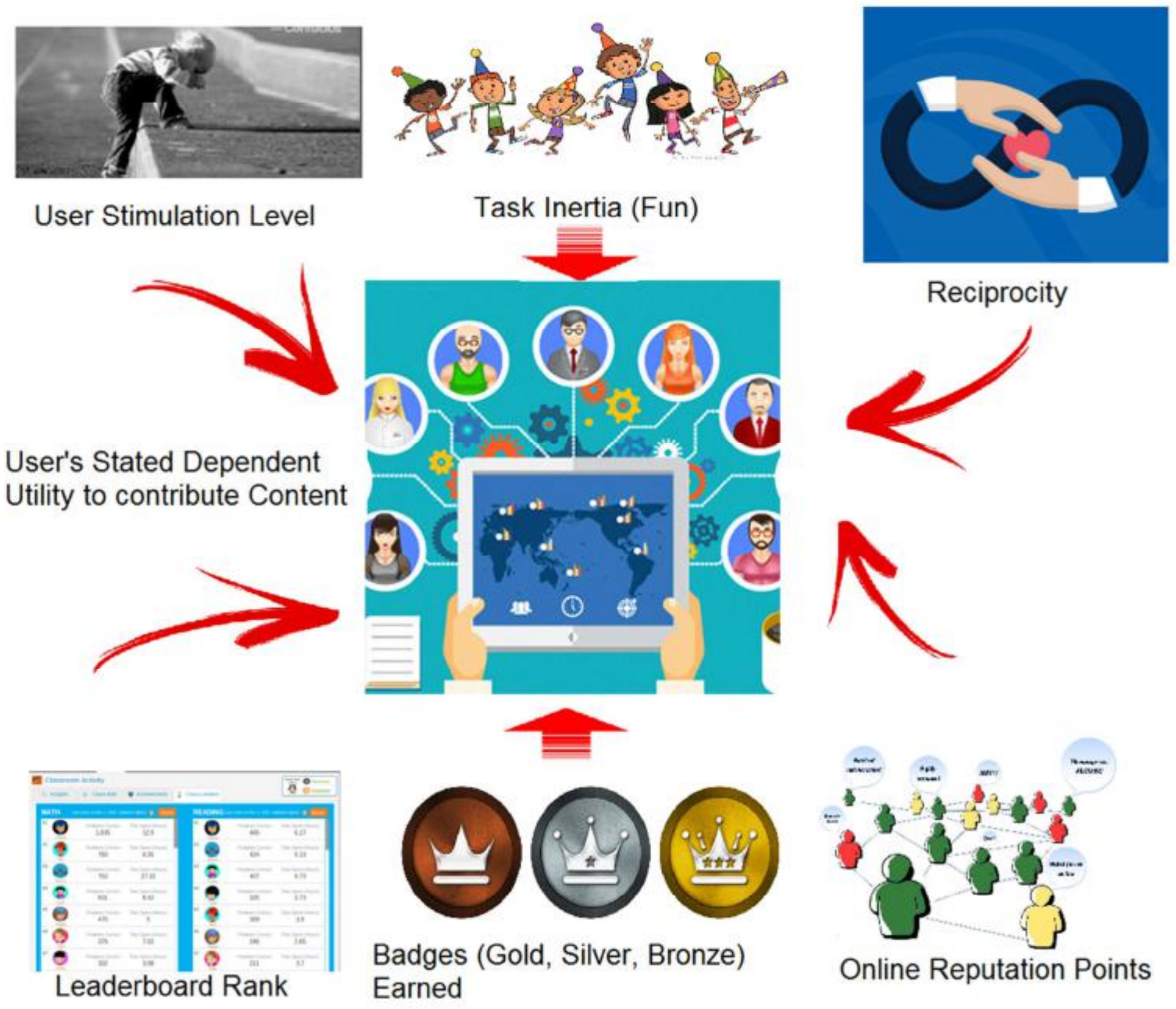

Figure 3.4. Box and Arrow Model of State Dependent Utility of a user to contribute

Last but not least, to control for the unobserved heterogeneity in users' responses to each of the Gamification elements, the model of the users' state-dependent utility allows for flexible patterns of response, through a random coefficient model of users choices, by putting hierarchical Bayesian Dirichlet Process (DP) prior on $\Lambda_i$. This approach allows the number of parameters to increase with the size of the sample, increasing learning as new data is observed.

This approach assumes a mixture of multivariate normal distributions over the parameters' prior, to allow for thick tail skewed multimodal distribution. I accommodated user heterogeneity by assuming that $\Lambda_i$ is drawn from a distribution common across users, in two stages. I employed a mixture of normal as the first stage prior, to specify an informative prior that also does not overfit. The first stage consists of a mixture of $K$ multivariate normal distribution, and the second stage consists of prior on the parameters of the mixture of normal density, formally:

$$p(\Lambda_i - \Delta z_i \mid \pi, \{\mu_k, \Sigma_k\}) = \sum_{k=1}^{K} \pi_k \phi(\Lambda_i - \Delta z_i \mid \mu_k, \Sigma_k) \qquad (2)$$
$$\pi, \{\mu_k, \Sigma_k\} \mid b$$

where $b$ denotes the hyper-parameter for the priors on the mixing probabilities and the parameters governing each mixture component. $K$ denotes the number of mixture components. $\{\mu_k, \Sigma_k\}$ denotes mean and covariance matrix of the distribution of individual specific parameter vector $\Lambda_i$ for mixture component k. $\pi_k$ denotes the size of the $k'th$ component of mixture model, and $\phi$ denotes the normal density function distribution. $z_i$ denotes information set about user i, which here can include indicator of publishing personal website on the profile, indicator of stated nationality from USA, UK, Australia, India, Europe, Asia, South America, China, Middle east, Tenure (the number of days since registration on the Gamification platform), Seen (the number of days since last date that user logged in), number of profile views from internet browsers, total number of reputation points and badges accumulated, number of answered and asked questions, and the total number of internet browser reached by contributing to the platform, until the start of sample period. $\Delta$ denotes the parameter of correlation between choice response parameter and information set about user $i$.

To obtain a truly non-parametric estimate using the mixture of normal model it is required that the number of mixture components $K$ increase with the sample size. I adopted the approach proposed by Rossi (2014), called non-parametric Bayesian approach. This approach is equivalent to the approach mentioned above when $K$ tends to infinity. In this structure, the parameters of mixture normal model have Dirichlet Process (DP) prior. Dirichlet process is the generalization of Dirichlet distribution for infinite atomic number of partitions. This process represents the distribution of a random measure (i.e. probability). Dirichlet process has two parameters, the first is the base distribution, which is the prior distribution on the parameters of the multivariate Normal-Inverse Wishart (N-IW) conjugate prior distribution for the distribution for the partitions that the choice parameters are drawn from, and the second parameter is the concentration parameter. Formally, the prior for the individual specific choice parameters has the following structure:

$$
\begin{aligned}
&\theta_{k1} = (\mu_{k1}, \Sigma_{k1}) \sim DP(G0(\lambda), \alpha^{d}) \\
&G0(\lambda): \mu_{k1} \mid \Sigma_{k1} \sim N(0, \Sigma_{k1} a^{-1}), \Sigma_{k1} \sim IW(\nu, \nu \times \upsilon \times I) \\
&\lambda(a, \nu, \upsilon): a \sim Unif(\vec{a}, \overleftarrow{a}), \nu \sim d - 1 + \exp(z), z \sim Unif(d - 1 + \vec{\nu}, \overleftarrow{\nu}), \upsilon \sim Unif(\vec{\upsilon}, \overleftarrow{\upsilon}) \\
&\alpha^{d} \sim (1 - (\alpha - \vec{\alpha}) / (\overleftarrow{\alpha} - \vec{\alpha}))^{power}
\end{aligned} \tag{3}
$$

where $G0(\lambda)$ denotes the base distribution or measure (i.e. the distribution of hyper-parameters of the prior distribution of the partitions). $\lambda$ denotes the random measure, which represents the probability distribution of $(a, \nu, \upsilon)$. $(a, \nu, \upsilon)$ denotes the hyper-parameters of the prior distribution of the partitions that the choice parameters belong to, which represent the behavior parameters of the latent segments. $d$ denotes the number of choice parameters per user (in my case d is equal to 15). $\alpha^{d}$ denotes the concentration (also referred to as precision, tightness, or innovation) parameter. The idea is that DP is centered over the base measure $G0(\lambda)$ with N-IW with precision parameter $\alpha^{d}$ (larger value denotes tight distribution). $(\vec{a}, \overleftarrow{a}, \vec{\nu}, \overleftarrow{\nu}, \vec{\upsilon}, \overleftarrow{\upsilon})$ denotes the hyper parameters vector for the second level prior on hyper parameters of prior over the partitions distribution of the choice parameters.

Dirichlet Process Mixture (DPM) is referred to the distribution over the probability measure defined on some sigma-algebra (collection of subsets) of space $\aleph$ , such that the distribution for any finite partition of $\aleph$ is Dirichlet distribution (Rossi 2014). In my case, the probability measure over the partitions for mean and variance of random coefficient response parameters of individual choice parameters sigma-algebra has the Normal-Inverse-Wishart conjugate probability. For any subset of users $U$ of $\aleph$ :

$$E[G(U_{\lambda})] = G_0(\Lambda_{\lambda})$$
$$Var(G(U_{\lambda})) = \frac{G_0(\Lambda_{\lambda})(1 - G_0(\Lambda_{\lambda}))}{\alpha^d + 1} \quad (4)$$

By De Finetti theorem, integrating (marginalizing) out the random measure $G$ results in the joint distribution for the collection of user specific mean and covariance of random coefficient choice parameters as follows:

$$p(\mu_., \Sigma_.) = \int p(\mu_., \Sigma_. \mid G) p(G) dG \quad (5)$$

This joint distribution can be represented as a sequence of conditional distributions that has exchangeability property:

$$p((\mu_1, \Sigma_1), ..., (\mu_I, \Sigma_I)) = p((\mu_{1.}, \Sigma_1)) p((\mu_2, \Sigma_2) \mid (\mu_1, \Sigma_1)) ... p((\mu_I, \Sigma_I) \mid (\mu_1, \Sigma_1), ..., (\mu_{n-1.}, \Sigma_{n-1})) \quad (6)$$

The DP process is similar in nature to Chinese Restaurant Process (CRP) and Polya Urn. In the CRP, there is a restaurant with infinite number of tables (analogous to partitions of mean and variance of the individual choice random coefficients). A user entering the restaurant selects the tables randomly, but he selects the table with probability proportional to the number of users that have sat on the table so far (in which case the user behaves similar to the other users who are sitting at the selected table). If the user selects a new table, he will behave based on a parameter that he randomly selects from restaurant user behavior parameters (so not necessarily identical to the parameters of the other tables). The Polya Urn process has also the same structure. In this process, the experimenter starts by drawing balls (representing the parameter of response for each user) with different colors from the urn. Any time the experimenter has a ball with a given color drawn from the urn, he will add an additional ball with the same color to the urn, and he also returns the drawn ball. The distribution of number of customers sitting at each table in CRP and number of balls in each color in Polya Urn follow DP.

Table 3.9. Utility model Variables Definition

| Variable | Description |
|---|---|
| State Dependent Utility( $U_{it}$ ) | State dependent utility of user $i$ at day $t$ in week $w$ |
| Individual Specific Fixed Effect $(\alpha_i)$ | Fixed effect, or fixed optimal threshold level of user $i$ |
| Contribution State ( $cont_{it-1}$ ) | Total contribution level of user $i$, up until the current contribution point in day $t$, demeaned and then normalized by one hundred |
| Reciprocity State ( $rcv_{it-1}$ ) | Total number of contribution received (answers received for her question) by user $i$, up until day $t$, divided by a hundred |
| Reputation State ( $crep_{iw-1}$ ) | Total number of reputation points received by user $i$, up until week $w$ |
| Weekly Reputation ( $rep_{iw-1}$ ) | Total number of reputation point received by user $i$, at the previous week (i.e. week $w-1$) |
| Leaderboard rank ( $rnk_{iw-1}$ ) | Rank of user $i$, in the leaderboard at previous week (i.e. week $w-1$) |
| Leaderboard rank change ( $\Delta rnk_{iw-1}$ ) | First order rank difference for user $i$'s in the leaderboard from the other week to the previous week (i.e. week $w-2$ to week $w-1$) |
| Instant Badge category ( $bdg_{it-1}$ ) | A vector of number of gold, silver, and bronze badges user $i$ earned at the previous day (i.e. day $t-1$) |
| Cumulative Badge Category ( $cbdg_{it-1}$ ) | A vector of total cumulative number of gold, silver, and bronze badges user $i$ earned until the previous day (i.e. day $t-1$) |
| $\gamma_{i1}..\gamma_{i8}$ | User $i$ specific parameters of state dependent utility of user $i$ |
| $\varepsilon_{it}$ | User $i$ and day $t$ specific type one extreme value error |

An alternative way is the approach proposed by Dube et al. (2010) to fit models with successively large numbers of components and to gauge the adequacy of the number of components by examining the fitted density associated with the selected number of components. However, the process of model selection is tedious for big data sets in this case.

To sum up, in this section I modeled the state dependent utility of users to contribute content to the gamified platform. I control for potential self-selection and unobserved heterogeneity of users by defining Dirichlet Process prior on the mixed logit choice model parameters. I also controlled for potential reciprocity and inertia (potential fun) by including the number of contributions sent and received by user until a given choice occasion.

## 3.6. ESTIMATION

In order to identify the choice model, I used a random coefficient (mixed) binary logit specification, which has a fixed scale. To set the location of the utility, I normalized the utility of no contribution option to zero. To minimize the concerns about self-selection, I use different fixed effect (stimulation level to contribute content) for different users. To minimize concerns about endogeneity (omitted variable), I control for potential correlations between choices of various users and unobserved heterogeneity by incorporating multi-modal mixture normal prior on the users' choices parameters (in a form of DP prior). I also control for potential confounding effects of inertia and reciprocity by incorporating the number of send and received contributions. In addition, by random coefficient structure, the modeling approach also minimizes the concern for Independence from Irrelevant Alternatives (IIA), as it allows for heterogeneity in the individual specific choice behavior parameters.

Estimation of the proposed model over a big data set consisting of approximately 11,000,000 million choices of approximately 37,000 users involves various computational and statistical issues, including over-fitting and computational tractability. First, the large number of parameters may cause over-fitting the sample, and this over-fitting may reduce generalizability of results. Bayesian shrinkage with flexible DP prior helps to identify the large set of individual specific parameters, without over-fitting. Second, optimization approaches that use sum of gradient, like the Newton Ralfphson and batch gradient-descent methods, are expensive over this kind of enormous data-set like the one used in this study. To deal with the same type of computational tractability issue in estimating a logit model, Goldfarb and Tucker (2011) resort to a linear probability model for a data set of only 2.5 million observations, with a much lower number of parameters. An alternative approach is to sample a subset of data and estimate the parameters. However, throwing away data by taking small sample might not be a relevant approach for targeting users.

To avoid the sample selection issue and show the effect of sample size, I take random samples of 1K, 5K, and 10K users from approximately 37K users by stratified sampling from strata that are generated from k-mean, mixture normal fuzzy clustering, and Latent Dirichlet Allocation clustering. In addition, I separate cross sectional variable of information set about users before the sample period into fixed (e.g., nationality, webpage declaration) and dynamic (number of question, answers, badges, and reputation points) items. I estimated both the models that incorporate only fixed information set and complete information set (both fixed and dynamic variables) at hierarchy level.

The mixture normal distribution is subject to label switching problem (i.e. the permutation of segment assignment returns the same likelihood). However, I immunized myself to this problem by limiting my inference to the joint distribution rather than user segment assignment. To estimate the content contribution choice model, I used multinomial logit with DP prior on the user specific hyper-parameter (Bayesian semi-parametric) estimation code from Bayesm package in R. This method uses Metropolis-Hasting Random-Walk (MH-RW) method to estimate conditional choice probabilities on cross-sectional units (i.e. users). The limitation of MH-RW is that random walk increments shall be tuned to conform as closely as possible to the curvature in the individual specific conditional posterior, formally defined by:

$$p(\Lambda_i \mid y_i, \mu, \Sigma, z_i, \Delta) \propto p(y_i \mid \Lambda_i) p(\Lambda_i \mid \mu, \Sigma, z_i, \Delta) \tag{7}$$

Without prior information on highly probable values of first stage prior (i.e. $p(\Lambda_i \mid .)$), tuning the Metropolis chains given limited information of cross-sections (i.e. each user) by trial is difficult. Therefore, to avoid singular hessian, the fractional likelihood approach proposed by Rossi et al. (2005) is implemented in the used approach. Formally rather than using individual specific likelihood, MH-RW approach forms a fractional combination of the unit-level likelihood and the pooled likelihood as follows:

$$l_i * (\Lambda_i) = l_i(\Lambda_i)^{(1-w)} \left( \prod_{i=1}^{I} l_i(\Lambda_i \mid y_i) \right)^{w\beta}, \beta = \frac{n_i}{N}, N = \sum_{i=1}^{I} n_i \tag{8}$$

where $w$ denotes the small tuning parameter to control the effect of pooled likelihood $\prod_{i=1}^{I} l_i(\Lambda_i \mid y_i)$. $\beta$ denotes a parameter chosen to properly scale the pooled likelihood to the same order as the unit likelihood. $n_i$ denotes the number of observations for user $i$. Using this approach, the MH-RW generates samples conditional on the partition membership indicator for user $i$ from proposal density $N(0, s^2\Omega)$, so that:

$$\Omega = (H_i + V_\Lambda^{-1})^{-1}, H_i = -\frac{\partial^2 \log l_i *}{\partial\Lambda\partial\Lambda'}\Big|_{\Lambda=\hat{\Lambda}_i} \tag{9}$$

where $\hat{\Lambda}_i$ denotes the maximum of the modified likelihood $l_i*(\Lambda_i)$, and $V_\Lambda$ denotes normal covariance matrix assigned to the partition (i.e. segment) that customer $i$ belongs to.

This approach considers that $\Lambda_{i.}$ is sufficient to model the random coefficient distribution. To estimate the infinite mixture of normal prior for choice parameters, a standard data augmentation with the indicator of the normal component is required. Conditional on this indicator, I can identify a normal prior for each customer i parameters. The distribution for this indicator is Multinomial, which is conjugate to Dirichlet distribution, formally:

$$\begin{aligned} &\pi \sim Dirichlet(\alpha^d) \\ &z^i \mid \pi \sim Mult-Nom(\pi) \end{aligned} \tag{10}$$

As a result posterior can be defined by:

$$\begin{aligned} &z^i \sim Mult-Nom\left(\frac{\alpha^1}{\sum_j \alpha^j},\dots,\frac{\alpha^K}{\sum_j \alpha^j}\right) \\ &\pi \mid z^i \sim Dirichlet(\alpha^1+\delta_1(z^i),\dots,\alpha^k+\delta_K(z^i)) \end{aligned} \tag{11}$$

where $\delta_j(z^i)$ denotes indicators for whether or not $z^i = j$. This result is relevant for DP as any finite subset of user choice-behavior parameters' partitions has Dirichlet distribution, and finite sample can only represent finite number of partitions. Exchangeability property of partitions allows the used estimation approach to sequentially draw customer parameters given the indicator value as follows:

$$(\mu_i,\Sigma_i)\mid(\mu_1,\Sigma_1),\dots,(\mu_{i-1},\Sigma_{i-1}) \sim \frac{\alpha^d G_0+\sum_{j=1}^{i-1}\delta_{(\mu_j,\Sigma_j)}}{\alpha+i-1} \tag{12}$$

The next portion of this approach's specification is the definition of the size of the finite clusters over the finite sample that is controlled by $\pi$. Rossi (2014) suggests augmenting Sethuraman's stick breaking notion for draws of $\pi$. In this notion, a unit level stick is iteratively broken from the tail with proportion to the draws with beta distribution with parameter one and $\alpha^d$, and the length of the broken portion defines the $k'th$ element of the probability measure vector $\pi$ (a form of multiplicative process), formally:

$$\pi_k = \beta_k \prod_{i=1}^{k-1}(1-\beta_i), \beta_k \sim Beta(1,\alpha^d) \tag{13}$$

In this notion, $\alpha^d$ determines the probability distribution of the number of unique values for the DP mixture model, formally by:

$$\Pr(I^* = k) = \left\|S_i^{(k)}\right\| \left(\alpha^d\right)^k \frac{\Gamma(\alpha^d)}{\Gamma(i+\alpha^d)}, S_i^{(k)} = \frac{\Gamma(i)}{\Gamma(k)}(\gamma + \ln(i))^{k-1} \tag{14}$$

where $I^*$ denotes the number of unique values of $(\mu, \Sigma)$ in a sequence of $i$ draws from the DP prior. $S_i^{(k)}$ denotes Sterling number of first kind, and $\gamma$ denotes Euler's constant. Furthermore, to facilitate assessment, this approach suggests the following distribution for $\alpha^d$, rather than Gamma distribution:

$$\alpha^d \propto (1 - \frac{\alpha^d - \underline{\alpha}}{\bar{\alpha} - \underline{\alpha}})^{\phi} \tag{15}$$

where $\bar{\alpha}$ and $\underline{\alpha}$ can be assessed by inspecting the mode of $I^* \mid \alpha^d$. $\phi$ denotes the tunable power parameter to spread prior mass appropriately. An alternative to Gibbs sampler employed by this approach might be collapsed Gibbs sampler that integrates out the indicator variable for partition (segment) membership of each user, but Rossi (2014) argues that such an approach does not improve the estimation procedure. Appendix 3.A presents the series of conditional distribution that this approach employs in its Gibbs sampling to recover individual specific choice parameters. In summary, I used MCMC sampling to estimate the mixture normal multinomial logit model of the content contribution choices of samples of 1K, 5K, and 10K users over two hundred thirty seven days, in a Gamification environment.

## 3.7. RESULTS AND MANAGERIAL IMPLICATIONS

I begin this section by discussing the importance of big data. Table 3.10 presents the log likelihood of twelve models I have tested. Among samples with 1K users, stratified random sample from Latent Dirichlet Allocation clustering has a better fit. In addition, the models that use both fixed and dynamic information set of users at hierarchical level explains users' choice better. However, estimating the model over sample with 5K size suggests that potentially the LDA stratified sample might not have represented the population because the log likelihood does not increase proportionally. In addition, estimate of the model that uses whole information set about user at hierarchical level over the sample with 10K size returns relatively better likelihood than the same model estimated over sample with 5K size. I announce this model dominant, because it uses more information and returns a better likelihood relative to the model estimated over its adjacent sample size (i.e. sample with 5K size).

Table 3.11 presents the distributions of the parameter estimates for the individual content contribution choice model that explains choice parameter with whole information set of users estimated over a sample with 10K random users. These distributions are visualized in figure 3.5. Although I used a flexible mixture normal model, yet the parameter of response has a normal bell shape.

Table 3.10. MODEL COMPARISON

| Model | Description | Number of obs. | Log Lik. |
|---|---|---|---|
| 1 | Uniform Sample Choice explained by all variables | 237,000 | -65,379.08 |
| 2 | LDA stratified Sample Choice Explained by all variables | 237,000 | -61,868.84 |
| 3 | K-mean stratified Sample Choice Explained by all variables | 237,000 | -62,554.44 |
| 4 | Mixture Normal stratified Sample Choice Explained by all variables | 237,000 | -65,164.39 |
| 5 | Mixture Normal stratified Sample Choice Explained by Static HB variables | 237,000 | -66,374.15 |
| 6 | Uniform Sample Choice explained by Static HB variables | 237,000 | -65,943.30 |
| 7 | LDA stratified Sample Choice Explained by Static HB variables | 237,000 | -65,028.38 |
| 8 | K-mean stratified Sample Choice Explained by static HB variables | 237,000 | -63,548.60 |
| 9 | Sample of 5K explained by static HB variables | 1,185,000 | -327,701.60 |
| 10 | Sample of 5K explained by all variables | 1,185,000 | -327,765.30 |
| 11 | Sample of 10K explained by static HB variables | 2,370,000 | -656,838.80 |
| 12 | Sample of 10K explained by all variables* | 2,370,000 | -653,301.00 |

* Dominant model

The distribution of the parameter estimates for this and the other model over samples with sizes of 1K, 5K, and 10K is presented in appendix 3.B. Comparison of these estimates suggest that model that uses only fixed information set of users at hierarchical level overestimates fixed effect (or stimulation level), and it underestimates the effect of leaderboard and badges (long term effect of silver and bronze badges) elements. In addition, estimating the same model over a sample with 5K random users results in underestimation of fixed effect, inertia, rank, and badges (except gold), and it overestimates the effect of reputation points and reciprocity. These results highlight the importance of employing a bigger data set to get a better estimate of Gamification elements.

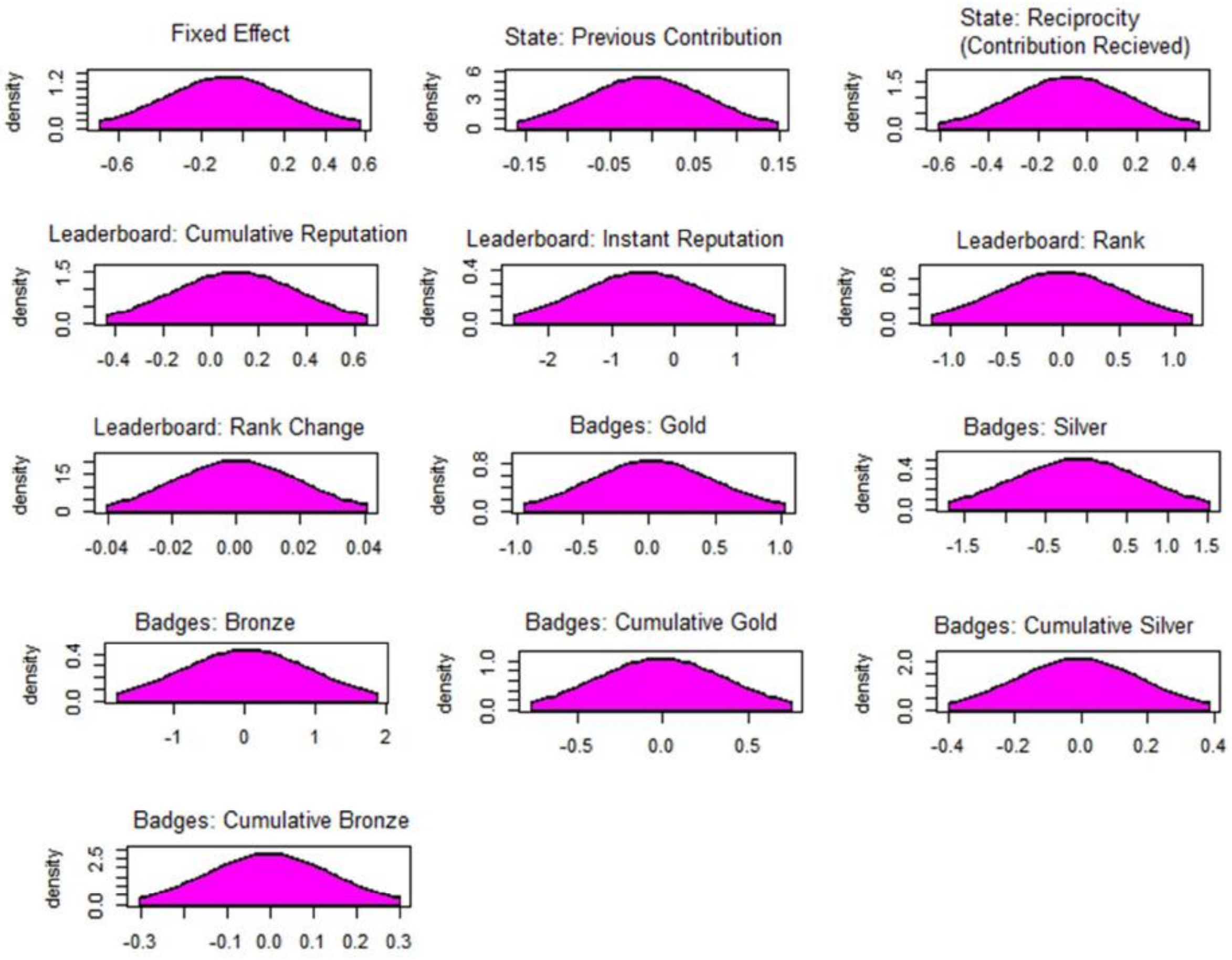

Figure 3.5. HISTOGRAM OF PARAMETER ESTIMATES: Individual Choice parameters

As hierarchical Bayesian method allows recovering individual specific parameters, I can use individual specific parameters to recover significance of parameters. In fact this significance information is useful for the Gamification platform to target its users. Table 3.15 presents the statistics of significance of parameters across the population. It is interesting to note that individual level fixed effect, which I interpret as the stimuli level required to contribute content, is positive significant across 2% and negative significant across 2% of all the users. This finding suggests that, when users contribute content, either an intrinsic (for 2% of users) or an extrinsic motivation (for 98% of users) exists.

Table 3.11. PARAMETER ESTIMATES: Individual Content Contribution Choice explained by whole information set (Sample with 10K explained)

| | **Estimate** | **Std. Dev.** | **2.5th** | **97.5th** |
|---|---|---|---|---|
| Fixed Effect | -0.039 | 0.301 | -0.589 | 0.599 |
| States: | | | | |
| Previous contribution | -0.004 | 0.070 | -0.099 | 0.105 |
| Reciprocity (contribution received) | -0.091 | 0.411 | -0.528 | 0.229 |
| Leader Board: | | | | |
| Cum Reputation | 0.108 | 0.184 | -0.320 | 0.435 |
| Reputation | -0.500 | 0.689 | -1.636 | 1.051 |
| Rank | -0.012 | 0.347 | -0.795 | 0.726 |
| Rank Change | 0.000 | 0.001 | -0.002 | 0.002 |
| Badges | | | | |
| Gold Badge | 0.079 | 0.488 | -0.780 | 0.965 |
| Silver Bade | -0.105 | 0.324 | -0.678 | 0.414 |
| Bronze Badge | 0.011 | 0.405 | -0.764 | 0.738 |
| Cum Gold Badge | -0.014 | 0.228 | -0.399 | 0.359 |
| Cum Silver Badge | -0.006 | 0.188 | -0.295 | 0.297 |
| Cum Bronze Badge | -0.004 | 0.134 | -0.224 | 0.237 |

The number of contents contributed has significant positive effects for 8% of users and significant negative effects for 9% of users on their probability of contributing. This result is also relevant for the Gamification platforms targeting. The Gamification platform can investigate the journey of customers who show inertia (positive effect of previous contribution) and try to generate the similar journey for those who show resistance (negative effect of previous contribution), through its messaging policy. In addition, the Gamification platform can also send customized positive messages to the users with inertia, to keep them as loyal customers. It can also send promotional incentivizing messages to those who show resistance, to motivate them to contribute more (similar to the promotions that are sent to churned customers). In fact, given the domain knowledge of Gamification practitioners, the message for the users with resistance shall emphasize the fun aspect of answering other users' questions (Brittner and Shipper 2014; Wu 2012; Deterding 2012).

Table 3.12. PARAMETER ESTIMATES: Individual Choice effect significance

| | Positive Significant | Negative Significant | % positive | % Negative |
|---|---|---|---|---|
| Fixed Effect | 171 | 239 | 2% | 2% |
| States: | | | | |
| Previous contribution | 819 | 949 | 8% | 9% |
| Reciprocity (contribution received) | 450 | 1228 | 5% | 12% |
| Leader Board: | | | | |
| Cum Reputation | 1041 | 564 | 10% | 6% |
| Reputation | 273 | 673 | 3% | 7% |
| Rank | 619 | 716 | 6% | 7% |
| Rank Change | 1349 | 1444 | 13% | 14% |
| Badges | | | | |
| Gold Badge | 258 | 147 | 3% | 1% |
| Silver Bade | 76 | 127 | 1% | 1% |
| Bronze Badge | 263 | 184 | 3% | 2% |
| Cum Gold Badge | 296 | 320 | 3% | 3% |
| Cum Silver Badge | 842 | 930 | 8% | 9% |
| Cum Bronze Badge | 1037 | 906 | 10% | 9% |

The effect of contribution received (or reciprocity) is positive significant for 5% of users and negative significant for 12% of users. In other words, when the users' questions are answered more often than others, 5% of users are more likely and 12% of them are less likely to answer the community members' questions. The reciprocity result for 5% of users is consistent with the result of studies in information system research of the knowledge market and in the economics of impure altruism that emphasize the importance of reciprocity in users' decisions (Ruben 2009; Chen et al. 2010; Chiu and Wang 2009; Bolton et al. 2013; Andreoni 1990; Cornes and Sandler 1994). However, the negative response to receiving answer by 12% of users might be explained by users' shift of focus on their daily life as opposed to participating on the Gamification platform. This finding suggests that the Gamification platform owner can employ a prioritizing strategy. Such a prioritizing strategy can put higher priority on the questions of users who contribute more when the community answers their questions.

The effect of weekly reputation (instant) is positive significant for 3% and negative significant for 7% of users, and the effect of cumulative reputation is positive significant for 10% of users and negative significant for 7% of users. This result is relevant for the targeting, and I explain the relevance later in this section, but first I explain why the effect can be positive and negative for different users. The negative effect may be explained by moral licensing (Wei et al. 2015), or reversion to the mean. Moral licensing refers to the process by which a user reduces her pro-social activity after being nominated as pro-social. The moral licensing might be more relevant here, because none of the studies by practitioners and academia has yet defended the mean reversion of users in the Gamification context. The positive effect of reputation might be explained by the empowerment effect of Gamification that social psychologists emphasize (Wu 2012). In other words, the reputation points might act as a signal to the user to recognize the potential of helping others.

The effect of rank is positive significant for 6% and negative significant for 7% of users, and the effect of rank change, second order lagged effect, is positive significant for 13% of users, and negative significant for 14% of users. Recognizing these two different effects is useful for targeting in the context of the Gamification platform. The negative second order lagged effect resembles the mean reversion behavior. This behavior might be relevant to the anchorage effect of the rank on the leaderboard. The positive second order lagged effect resembles inertia. In other words, when falling in the leaderboard the user gives up and when rising in the leaderboard, the user works harder. Although the Gamification platform may not have control over the mean reversion of the user, it may be able to affect the users' negative inertia (i.e. giving up when falling in the leaderboard) by positive empowering messages.

The results for the effects of the gold, silver, and bronze badge categories might also be of interest of the Gamification platform, because it can modify the badges' requirements (threshold of points to earn badges) to motivate users. Instant effects of earning Gold badges are positive significant for 3% and negative significant for 1% of users. In addition, the long term effects of earning Gold badges are positive significant for 3% of users and negative significant for 3% of users. Instant effects of earning Silver badges are positive significant for 1% and negative significant for 1% of users. In addition, the long term effects of earning Silver badges are positive significant for 8% of users and negative significant for 9% of users. Instant effects of earning Bronze badges are positive significant for 3% and negative significant for 2% of users. In addition, the long term effects of earning Bronze badges are positive significant for 10% of users and negative significant for 9% of users. Again, these results are useful for targeting as I explain later in this section.

Potential explanations for observing both positive and negative effects of badges across segments are similar to the explanations for observing both positive and negative effects of the number of reputation points (i.e. moral licensing vs. empowerment). However, the means of long term effects of badges across population is negative. These negative long term effects can be explained by the goal setting aspect of badges in a Gamification setting. In other words, users might have set the goal to win badges as hallmark of Gamification, so as they earn these badges, they reduce their content contribution. Given these negative effects, the fact that badges are once-in-a-life- time effect might suggest the Gamification platform a higher point threshold requirement to grant the badges. The counterfactual section quantifies the effect of such policy at aggregate level.

To discuss the targeting aspect of these results, table 3.13 presents the hierarchical parameters' estimates. These results suggest that certain nationalities are more likely to be sensitive to certain aspects of Gamification platform. For the sake of brevity, I only review more interesting patterns. First, American users show more inertia in contribution. They reduce their contribution level, if they have more reputation, but increase it, if they earn silver badges. Second, European users increase their effort, if they have more reputation, but English users decrease their contribution if they earn gold badges. Third, south Americans are more motivated to contribute to StackOverflow. Fourth, Asian users are more reciprocal. They increase their contribution when they earn Silver badges, but decrease it when they earn Gold badges. Fifth, Middle Eastern users are also more reciprocal.

These patterns might be relevant for targeting because, if certain nationality responds positively for example to gold badges in long term, it might be relevant to guide the users with this nationality to earn gold badges easier. However, if users of certain nationality decrease their content contribution when they earn gold badges, then it might be relevant to send messages to these users that confuses them, making earning gold badges difficult. Similar approaches can be designed conditioning on response to silver and bronze badges, and reputation points. In addition, if the platform finds users of certain nationality more reciprocal, it can put answering the questions of these users top priority for other users. All these findings can better guide the Gamification platform toward increasing their users' content contributions.

Table 3.13. PARAMETER ESTIMATES: Individual Choice Hierarchical Model

| | **Estimate** | **Std. Dev.** | **2.5$^{th}$** | **97.5$^{th}$** |
|---|---|---|---|---|
| Fixed Effect | | | | |
| website | -0.043 | 0.072 | -0.140 | 0.137 |
| USA | -0.002 | 0.003 | -0.008 | 0.003 |
| UK | -0.009 | 0.010 | -0.027 | 0.010 |
| Australia | -0.014 | 0.011 | -0.036 | 0.008 |
| India | -0.083 | 0.063 | -0.201 | 0.048 |
| Europe | -0.005 | 0.021 | -0.044 | 0.035 |
| Asia | 0.000 | 0.000 | -0.001 | 0.001 |
| South America | 0.186* | 0.051 | 0.082 | 0.300 |
| China | 0.134 | 0.107 | -0.043 | 0.336 |
| Middle East | -0.044 | 0.085 | -0.200 | 0.101 |
| Tenure | -0.020 | 0.021 | -0.060 | 0.021 |
| Seen | 0.011 | 0.008 | -0.003 | 0.027 |
| Profile Views | -0.002 | 0.006 | -0.013 | 0.009 |
| Reputation | 0.351* | 0.056 | 0.249 | 0.460 |
| Gold Badges | 0.004 | 0.004 | -0.003 | 0.012 |
| Silver Badges | -0.015 | 0.012 | -0.038 | 0.009 |
| Bronze Badges | 0.016 | 0.013 | -0.009 | 0.042 |
| Number of Answers | -0.016 | 0.084 | -0.167 | 0.140 |
| Number of Questions | -0.025 | 0.026 | -0.074 | 0.030 |
| Reach | 0.000 | 0.001 | -0.001 | 0.001 |
| States: | | | | |
| Previous contribution | | | | |
| website | 0.003 | 0.134 | -0.203 | 0.260 |
| USA | 0.167* | 0.093 | 0.007 | 0.365 |
| UK | -0.079 | 0.088 | -0.245 | 0.104 |
| Australia | -0.024 | 0.025 | -0.071 | 0.021 |
| India | -0.010 | 0.008 | -0.026 | 0.008 |
| Europe | 0.000 | 0.007 | -0.013 | 0.013 |
| Asia | -0.106 | 0.079 | -0.284 | 0.016 |
| South America | -0.005 | 0.006 | -0.017 | 0.005 |
| China | 0.011 | 0.019 | -0.024 | 0.052 |
| Middle East | 0.003 | 0.018 | -0.032 | 0.037 |
| Tenure | 0.039 | 0.119 | -0.188 | 0.291 |
| Seen | 0.011 | 0.042 | -0.071 | 0.093 |
| Profile Views | 0.000 | 0.001 | -0.002 | 0.002 |
| Reputation | -0.003 | 0.150 | -0.304 | 0.243 |
| Gold Badges | 0.091 | 0.197 | -0.337 | 0.500 |
| Silver Badges | -0.090 | 0.148 | -0.390 | 0.193 |
| Bronze Badges | 0.048 | 0.039 | -0.036 | 0.123 |
| Number of Answers | -0.009 | 0.013 | -0.034 | 0.016 |
| Number of Questions | -0.003 | 0.010 | -0.022 | 0.017 |
| Reach | 0.397* | 0.183 | 0.029 | 0.708 |
| Reciprocity (contribution received) | | | | |
| website | -0.016* | 0.007 | -0.030 | -0.001 |
| USA | 0.004 | 0.026 | -0.046 | 0.057 |
| UK | 0.009 | 0.026 | -0.045 | 0.059 |
| Australia | -0.221 | 0.199 | -0.648 | 0.104 |
| India | -0.048 | 0.062 | -0.166 | 0.072 |
| Europe | 0.000 | 0.001 | -0.002 | 0.002 |
| Asia | 0.554* | 0.200 | 0.248 | 0.961 |
| South America | -0.118 | 0.228 | -0.499 | 0.368 |
| China | 0.046 | 0.195 | -0.283 | 0.447 |
| Middle East | 0.107* | 0.054 | 0.005 | 0.210 |
| Tenure | -0.009 | 0.016 | -0.041 | 0.024 |
| Seen | 0.014 | 0.014 | -0.012 | 0.039 |
| Profile Views | 0.149 | 0.132 | -0.156 | 0.375 |
| Reputation | -0.004 | 0.006 | -0.015 | 0.007 |
| Gold Badges | 0.029 | 0.020 | -0.007 | 0.067 |
| Silver Badges | -0.004 | 0.019 | -0.040 | 0.036 |
| Bronze Badges | 0.272* | 0.121 | 0.029 | 0.497 |
| Number of Answers | -0.033 | 0.044 | -0.112 | 0.054 |
| Number of Questions | 0.000 | 0.001 | -0.002 | 0.002 |
| Reach | -0.337* | 0.107 | -0.512 | -0.082 |
| Leader Board: | | | | |
| Cum Reputation | | | | |
| website | -0.148 | 0.171 | -0.455 | 0.149 |
| USA | -0.263* | 0.117 | -0.500 | -0.055 |
| UK | -0.031 | 0.041 | -0.109 | 0.049 |
| Australia | 0.012 | 0.013 | -0.014 | 0.038 |
| India | -0.001 | 0.009 | -0.021 | 0.018 |
| Europe | 0.210* | 0.058 | 0.101 | 0.307 |
| Asia | -0.002 | 0.004 | -0.009 | 0.005 |
| South America | 0.010 | 0.012 | -0.013 | 0.034 |
| China | 0.011 | 0.012 | -0.011 | 0.035 |
| Middle East | 0.059 | 0.070 | -0.077 | 0.197 |
| Tenure | -0.020 | 0.026 | -0.069 | 0.032 |
| Seen | 0.000 | 0.001 | -0.001 | 0.001 |
| Profile Views | 0.146 | 0.118 | -0.062 | 0.331 |
| Reputation | -0.063 | 0.098 | -0.237 | 0.117 |
| Gold Badges | -0.279* | 0.077 | -0.435 | -0.127 |
| Silver Badges | 0.026 | 0.030 | -0.029 | 0.092 |
| Bronze Badges | 0.000 | 0.008 | -0.016 | 0.017 |
| Number of Answers | 0.001 | 0.007 | -0.014 | 0.014 |
| Number of Questions | 0.153 | 0.199 | -0.262 | 0.588 |
| Reach | -0.012 | 0.011 | -0.033 | 0.012 |
| Reputation | | | | |
| website | -0.013 | 0.039 | -0.091 | 0.062 |
| USA | -0.049 | 0.033 | -0.113 | 0.014 |
| UK | 0.400 | 0.219 | -0.052 | 0.776 |
| Australia | -0.103 | 0.082 | -0.271 | 0.052 |
| India | 0.000 | 0.002 | -0.003 | 0.004 |
| Europe | -0.399 | 0.284 | -0.920 | 0.090 |
| Asia | -0.401 | 0.247 | -0.925 | 0.068 |
| South America | -0.219 | 0.230 | -0.621 | 0.225 |
| China | 0.208* | 0.079 | 0.069 | 0.376 |
| Middle East | 0.032 | 0.025 | -0.017 | 0.080 |
| Tenure | -0.007 | 0.019 | -0.048 | 0.028 |
| Seen | 0.484* | 0.231 | 0.118 | 0.848 |
| Profile Views | -0.003 | 0.009 | -0.021 | 0.016 |
| Reputation | -0.005 | 0.038 | -0.072 | 0.073 |
| Gold Badges | 0.075 | 0.039 | -0.004 | 0.150 |
| Silver Badges | 0.415* | 0.196 | 0.023 | 0.764 |
| Bronze Badges | 0.029 | 0.075 | -0.121 | 0.180 |
| Number of Answers | 0.000 | 0.002 | -0.003 | 0.003 |
| Number of Questions | -1.080* | 0.274 | -1.500 | -0.596 |
| Reach | -0.184 | 0.248 | -0.643 | 0.321 |
| Rank | | | | |
| website | -0.255 | 0.211 | -0.658 | 0.132 |
| USA | -0.013 | 0.066 | -0.144 | 0.111 |
| UK | -0.005 | 0.023 | -0.050 | 0.041 |
| Australia | 0.008 | 0.018 | -0.027 | 0.043 |
| India | 0.470 | 0.457 | -0.141 | 1.520 |
| Europe | 0.008 | 0.017 | -0.024 | 0.041 |
| Asia | -0.017 | 0.056 | -0.124 | 0.094 |
| South America | -0.044 | 0.058 | -0.158 | 0.066 |
| China | 1.017* | 0.283 | 0.487 | 1.578 |
| Middle East | -0.059 | 0.128 | -0.301 | 0.179 |
| Tenure | 0.000 | 0.003 | -0.005 | 0.006 |
| Seen | -0.691 | 0.711 | -1.877 | 0.524 |
| Profile Views | -1.450* | 0.498 | -2.429 | -0.675 |
| Reputation | -0.125 | 0.371 | -0.773 | 0.603 |
| Gold Badges | -0.162 | 0.149 | -0.405 | 0.131 |
| Silver Badges | -0.057 | 0.040 | -0.133 | 0.018 |
| Bronze Badges | 0.005 | 0.029 | -0.050 | 0.061 |
| Number of Answers | 0.859* | 0.189 | 0.381 | 1.097 |
| Number of Questions | 0.000 | 0.008 | -0.015 | 0.014 |
| Reach | 0.038 | 0.029 | -0.018 | 0.090 |
| Rank Change | | | | |
| website | 0.018 | 0.028 | -0.037 | 0.070 |
| USA | 0.064 | 0.139 | -0.187 | 0.349 |
| UK | -0.054 | 0.061 | -0.168 | 0.074 |
| Australia | 0.000 | 0.001 | -0.003 | 0.003 |
| India | -0.520* | 0.272 | -0.963 | -0.004 |
| Europe | 0.242 | 0.284 | -0.449 | 0.649 |
| Asia | 0.029 | 0.193 | -0.361 | 0.397 |
| South America | 0.058 | 0.069 | -0.094 | 0.162 |
| China | 0.009 | 0.018 | -0.024 | 0.048 |
| Middle East | -0.009 | 0.015 | -0.038 | 0.018 |
| Tenure | 0.000 | 0.000 | 0.000 | 0.000 |
| Seen | 0.000 | 0.000 | 0.000 | 0.000 |
| Profile Views | 0.000 | 0.000 | 0.000 | 0.000 |
| Reputation | 0.000 | 0.000 | 0.000 | 0.000 |
| Gold Badges | 0.000 | 0.000 | 0.000 | 0.000 |
| Silver Badges | 0.000 | 0.000 | 0.000 | 0.000 |
| Bronze Badges | 0.000 | 0.000 | 0.000 | 0.000 |
| Number of Answers | 0.000 | 0.000 | 0.000 | 0.000 |
| Number of Questions | 0.000 | 0.000 | 0.000 | 0.000 |
| Reach | 0.000 | 0.000 | 0.000 | 0.000 |
| Badges | | | | |
| Gold Badge | | | | |
| website | 0.00000 | 0.00002 | -0.00004 | 0.00005 |
| USA | 0.00001* | 0.00001 | 0.00000 | 0.00003 |
| UK | 0.00000 | 0.00000 | -0.00001 | 0.00001 |
| Australia | -0.00028 | 0.00007 | -0.00042 | -0.00016 |
| India | 0.00000 | 0.00001 | -0.00001 | 0.00001 |
| Europe | 0.00001 | 0.00002 | -0.00003 | 0.00004 |
| Asia | 0.00003 | 0.00002 | -0.00001 | 0.00006 |
| South America | -0.0004* | 0.00012 | -0.00063 | -0.00015 |
| China | 0.00000 | 0.00004 | -0.00009 | 0.00008 |
| Middle East | 0.00000 | 0.00000 | 0.00000 | 0.00000 |
| Tenure | 0.0008* | 0.00018 | 0.00048 | 0.00112 |
| Seen | 0.00022 | 0.00015 | -0.00010 | 0.00047 |
| Profile Views | 0.00008 | 0.00011 | -0.00012 | 0.00030 |
| Reputation | -0.00001 | 0.00005 | -0.00009 | 0.00009 |
| Gold Badges | 0.00002 | 0.00001 | -0.00001 | 0.00005 |
| Silver Badges | 0.00001 | 0.00001 | -0.00001 | 0.00002 |
| Bronze Badges | 0.00000 | 0.00001 | -0.00001 | 0.00001 |
| Number of Answers | 0.00000 | 0.00000 | 0.00000 | 0.00000 |
| Number of Questions | 0.00000 | 0.00000 | 0.00000 | 0.00000 |
| Reach | 0.00000 | 0.00000 | 0.00000 | 0.00000 |
| Silver Bade | | | | |
| website | 0.00000 | 0.00001 | -0.00001 | 0.00002 |
| USA | 0.00000 | 0.00000 | -0.00001 | 0.00000 |
| UK | 0.00000 | 0.00000 | 0.00000 | 0.00000 |
| Australia | -0.00004* | 0.00001 | -0.00005 | -0.00002 |
| India | 0.00001 | 0.00001 | -0.00001 | 0.00002 |
| Europe | -0.00002 | 0.00002 | -0.00005 | 0.00001 |
| Asia | 0.00002* | 0.00001 | 0.00001 | 0.00003 |
| South America | 0.00000 | 0.00000 | 0.00000 | 0.00000 |
| China | 0.00000 | 0.00000 | 0.00000 | 0.00000 |
| Middle East | -0.00001* | 0.00000 | -0.00001 | 0.00000 |
| Tenure | 0.00000 | 0.00000 | 0.00000 | 0.00000 |
| Seen | 0.00000 | 0.00000 | 0.00000 | 0.00000 |
| Profile Views | 0.00000 | 0.00000 | 0.00000 | 0.00000 |
| Reputation | 0.00000 | 0.00000 | -0.00001 | 0.00000 |
| Gold Badges | 0.00000 | 0.00000 | 0.00000 | 0.00000 |
| Silver Badges | 0.00000 | 0.00000 | 0.00000 | 0.00000 |
| Bronze Badges | 0.00000 | 0.00000 | 0.00000 | 0.00001 |
| Number of Answers | 0.00000 | 0.00000 | -0.00001 | 0.00000 |
| Number of Questions | 0.00000 | 0.00000 | 0.00000 | 0.00001 |
| Reach | 0.00000 | 0.00000 | 0.00000 | 0.00000 |
| Bronze Badge | | | | |
| website | 0.00000 | 0.00000 | 0.00000 | 0.00000 |
| USA | 0.00000 | 0.00000 | 0.00000 | 0.00000 |
| UK | -0.01716* | 0.00725 | -0.03297 | -0.00624 |
| Australia | 0.00050 | 0.00043 | -0.00038 | 0.00127 |
| India | -0.00228 | 0.00139 | -0.00501 | 0.00036 |
| Europe | 0.00058 | 0.00132 | -0.00209 | 0.00303 |
| Asia | 0.00847 | 0.00809 | -0.01185 | 0.02379 |
| South America | 0.00511 | 0.00327 | -0.00076 | 0.01180 |
| China | 0.00000 | 0.00007 | -0.00013 | 0.00013 |
| Middle East | 0.01705 | 0.01298 | -0.01084 | 0.03890 |
| Tenure | -0.03556* | 0.00920 | -0.05325 | -0.01680 |
| Seen | -0.02181* | 0.01065 | -0.04569 | -0.00412 |
| Profile Views | 0.00002 | 0.00352 | -0.00689 | 0.00666 |
| Reputation | -0.00215* | 0.00105 | -0.00430 | -0.00027 |
| Gold Badges | 0.00013 | 0.00072 | -0.00129 | 0.00150 |
| Silver Badges | 0.00747 | 0.00362 | 0.00147 | 0.01536 |
| Bronze Badges | 0.00011 | 0.00013 | -0.00014 | 0.00035 |
| Number of Answers | -0.00019 | 0.00044 | -0.00105 | 0.00065 |
| Number of Questions | -0.00109* | 0.00048 | -0.00207 | -0.00009 |
| Reach | 0.00532* | 0.00233 | 0.00127 | 0.01005 |
| Cum Gold Badge | | | | |
| website | -0.00036 | 0.00097 | -0.00239 | 0.00152 |
| USA | 0.00000 | 0.00002 | -0.00005 | 0.00005 |
| UK | -0.01033* | 0.00231 | -0.01565 | -0.00638 |
| Australia | 0.00124 | 0.00241 | -0.00301 | 0.00601 |
| India | -0.00685* | 0.00322 | -0.01287 | -0.00047 |
| Europe | 0.00146 | 0.00110 | -0.00076 | 0.00348 |
| Asia | 0.00011 | 0.00033 | -0.00052 | 0.00077 |
| South America | 0.00001 | 0.00024 | -0.00046 | 0.00050 |
| China | -0.00640* | 0.00402 | -0.01528 | -0.00074 |
| Middle East | -0.00011 | 0.00010 | -0.00030 | 0.00010 |
| Tenure | 0.00034 | 0.00034 | -0.00034 | 0.00100 |
| Seen | 0.00066 | 0.00034 | -0.00007 | 0.00133 |
| Profile Views | -0.00568* | 0.00216 | -0.00947 | -0.00058 |
| Reputation | 0.00064 | 0.00070 | -0.00073 | 0.00201 |
| Gold Badges | 0.00000 | 0.00002 | -0.00003 | 0.00004 |
| Silver Badges | 0.00758* | 0.00139 | 0.00492 | 0.01065 |
| Bronze Badges | 0.00253 | 0.00222 | -0.00187 | 0.00703 |
| Number of Answers | 0.00520 | 0.00313 | -0.00062 | 0.01142 |
| Number of Questions | -0.00074 | 0.00072 | -0.00207 | 0.00082 |
| Reach | 0.00011 | 0.00024 | -0.00036 | 0.00055 |
| Cum Silver Badge | | | | |
| website | 0.00001 | 0.00019 | -0.00037 | 0.00038 |
| USA | 0.00055* | 0.00018 | 0.00015 | 0.00076 |
| UK | 0.00001 | 0.00000 | 0.00000 | 0.00001 |
| Australia | -0.00002 | 0.00002 | -0.00006 | 0.00001 |
| India | -0.00003 | 0.00002 | -0.00006 | 0.00000 |
| Europe | 0.00001 | 0.00010 | -0.00015 | 0.00022 |
| Asia | -0.00002 | 0.00003 | -0.00008 | 0.00004 |
| South America | 0.00000 | 0.00000 | 0.00000 | 0.00000 |
| China | 0.00040* | 0.00010 | 0.00019 | 0.00054 |
| Middle East | -0.00007 | 0.00012 | -0.00025 | 0.00019 |
| Tenure | -0.00007 | 0.00011 | -0.00028 | 0.00014 |
| Seen | -0.00010* | 0.00004 | -0.00018 | -0.00001 |
| Profile Views | 0.00001 | 0.00001 | -0.00001 | 0.00003 |
| Reputation | -0.00002 | 0.00001 | -0.00003 | 0.00000 |
| Gold Badges | 0.00229* | 0.00125 | 0.00004 | 0.00403 |
| Silver Badges | 0.00000 | 0.00003 | -0.00006 | 0.00006 |
| Bronze Badges | 0.00004 | 0.00008 | -0.00012 | 0.00020 |
| Number of Answers | -0.00002 | 0.00010 | -0.00021 | 0.00016 |
| Number of Questions | 0.00100 | 0.00060 | -0.00023 | 0.00211 |
| Reach | -0.00040 | 0.00023 | -0.00085 | 0.00003 |
| Cum Bronze Badge | | | | |
| website | 0.00000 | 0.00000 | -0.00001 | 0.00001 |
| USA | -0.00252* | 0.00074 | -0.00377 | -0.00117 |
| UK | 0.00010 | 0.00128 | -0.00171 | 0.00337 |
| Australia | -0.00007 | 0.00091 | -0.00179 | 0.00139 |
| India | 0.00033 | 0.00024 | -0.00010 | 0.00083 |
| Europe | 0.00000 | 0.00006 | -0.00012 | 0.00011 |
| Asia | -0.00002 | 0.00005 | -0.00010 | 0.00008 |
| South America | 0.00000 | 0.00000 | 0.00000 | 0.00000 |
| China | 0.00000 | 0.00000 | 0.00000 | 0.00000 |
| Middle East | 0.00000 | 0.00000 | 0.00000 | 0.00000 |
| Tenure | 0.00000 | 0.00000 | 0.00000 | 0.00000 |
| Seen | 0.00000 | 0.00000 | 0.00000 | 0.00000 |
| Profile Views | 0.00000 | 0.00000 | 0.00000 | 0.00000 |
| Reputation | 0.00000 | 0.00000 | 0.00000 | 0.00000 |
| Gold Badges | -0.000000* | 0.00000 | 0.00000 | 0.00000 |
| Silver Badges | 0.00000 | 0.00000 | 0.00000 | 0.00000 |
| Bronze Badges | 0.00000 | 0.00000 | 0.00000 | 0.00000 |
| Number of Answers | 0.00000 | 0.00000 | 0.00000 | 0.00000 |
| Number of Questions | 0.00000 | 0.00000 | 0.00000 | 0.00000 |
| Reach | 0.00000 | 0.00000 | 0.00000 | 0.00000 |

## 3.8. COUNTERFACTUAL ANALYSIS AND ITS MANAGERIAL IMPLICATIONS

An advantage of modeling consumers' choices from the utility primitive is the capability to run counterfactuals. One of the choices of a Gamification platform is to modify the threshold of earning badges. Given the heterogeneous short term and long term effects of different badges, a priori it might not be clear how changing the thresholds will affect the expected number of content contributions at aggregate level. Therefore, given the estimated parameters, I simulated the users' response to perturbation in the number of badges that they receive. As a measure, I used the expected total number of contributions, which as the sum of the predicted probability of the users' choices, is analogous to integrating the probability of choices across the population. In summary, to find the effect of each of the counterfactual scenarios of modifying the badges, I modified the related badge variable, and given the other entire variable and the parameters, I summed up the predicted choice probability of each of the users.

Table 3.14 summarizes the result of the counterfactual analysis of nine scenarios: shutting down or five percent increase or decrease of either silver and bronze badges, gold badges, or all the badges. First, shutting down the badges has increased the level of contribution by 3% for the duration of experiment. This result suggests that long term effect of distributing badges without expiry can negatively affect the contribution level of users. In addition, while increasing the number of silver and bronze badges by reducing the thresholds has negative impact on the expected number of contributions, increasing the gold badges by decreasing the threshold results has positive impact on the expected number of contributions. Therefore, the platform might be better off to increase the threshold for earning silver and bronze badges, but decrease the threshold for earning gold badges. If the platform wants to either increase or decrease the threshold across all the badges, then the counterfactual analysis suggests increasing the threshold, so that the platform extracts as many of users contributions as possible, before granting badges.

Table 3.14. Counterfactual Analysis Result

| Cases | Expected Number of Contributions | Absolute Change | Improvement Ratio |
|---|---|---|---|
| Real Case | 944,283 | - | - |
| Counterfactual 5% Increase in Silver and Bronze Badges | 943,075 | -1,208 | -0.13% |
| Counterfactual 5% Decrease in Silver and Bronze Badges | 945,557 | 1,274 | 0.13% |
| Counterfactual 5% Increase in Gold Badges | 945,226 | 942 | 0.10% |
| Counterfactual 5% Decrease in Gold Badges | 943,507 | -777 | -0.08% |
| Counterfactual 5% Increase in All Badges | 944,027 | -256 | -0.03% |
| Counterfactual 5% Decrease in All Badges | 944,819 | 536 | 0.06% |
| shut down silver and bronze badges | 979,687 | 35,404 | 3.75% |
| shut down gold badges | 944,775 | 492 | 0.05% |
| shut down all the badges | 975,344 | 31,061 | 3.29% |

## 3.9. CONCLUSION

In this paper, I developed a structural model that accounts for the effects of motivational factor of Gamification elements such as Badges and leaderboard on users' choice to contribute content.

To allow Gamification platforms to target their customers, I highlight the importance of controlling for user heterogeneity in the model using Hierarchical Dirichlet Process. First, using a large data set from Stack Overflow, I segment users' profile by a method that ensembles clustering assignments of LDA, mixture normal and k-mean methods. I showed heterogeneity in users' behavior by segmenting users into competitors, collaborators, achievers, explorers, and uninterested users. Then, by estimating the model over a sample of this data set, I showed that users' responses to various Gamification elements are heterogeneous. I showed that small sample size can return bias parameters' estimates. My results demonstrate that users with certain nationalities are sensitive to certain Gamification elements.

I further illustrated how the estimated model can be used to analyze a counterfactual scenario for Gamification platform's badge threshold modifications. This counterfactual analysis shows that, if the Gamification platform increases the threshold for earning silver and bronze badges, but decreases the threshold for earning gold badges, it can increase users' contribution. I believe that my modeling approach, proposed estimation method, and derived empirical insights in this paper can be of interest to both practitioners and scholars in academia.

# APPENDIX

## APPENDIX 1.A: DIRECT ACYCLIC GRAPH OF CONDITIONAL DISTRIBUTIONS

Probabilistic graphical approaches are popular in computer science, as they not only prove a visual tool to recognize conditional independence, but also they help saving space in representing probability distributions, and they facilitate probabilistic queries. Following represents the probabilistic graphical representation of the model I studied in this paper. Shaded circles represent the observed variables and un-shaded ones represent latent variables or parameters. The rectangles, called plate, represent the replication of variables with the number specified at their bottom right.

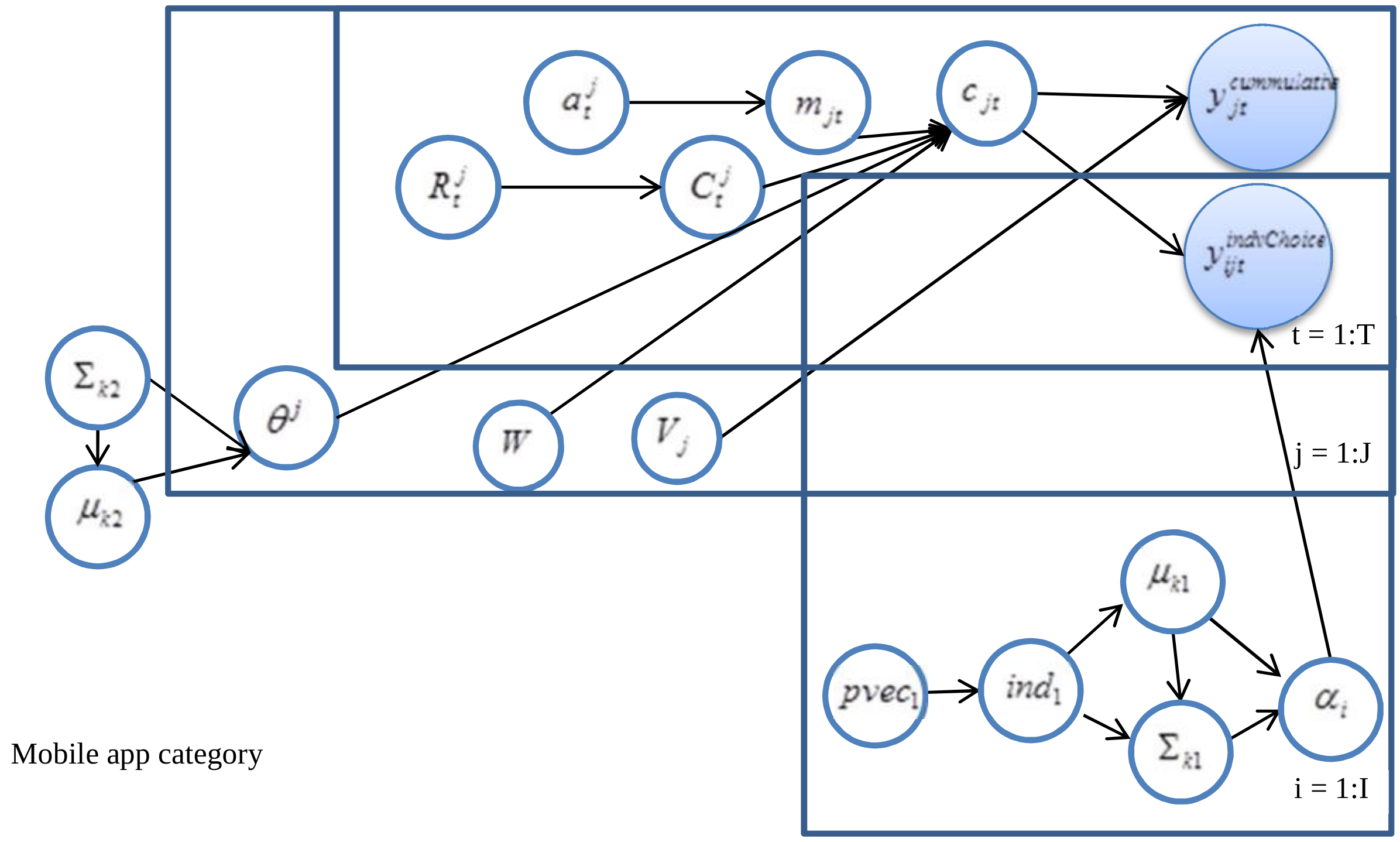

Figure 1.A.1. Probabilistic graphical model of customer mobile app choices under social influence

**APPENDIX 1.B: UNSCENTED KALMAN FILTER**

A recursive algorithm to update the latent state variable with Unscented Kalman Filter has the following steps. I refer the interested reader to Wan and van der Merve (2001).

Model has the following form, the first equation observation equation, and the second one state equation, with nonlinear functions H and F:

$$\begin{aligned} y_k &= H(x_k) + \eta_k, \eta_k \sim MVN(0, \mathrm{I}) \\ x_k &= F(x_{k-1}) + \upsilon_k, \upsilon_k \sim MVN(0, \Upsilon) \end{aligned} \quad \text{(B1)}$$

The estimation algorithm:

$$\begin{aligned} &\hat{x}_0 = E[x_0] \\ &P_0 = E[(x_0 - \hat{x}_0)(x_0 - \hat{x}_0)^{'}] \\ &k \in \{1, ..., \infty\} \\ &\lambda = \alpha^2 (L + \kappa) - L \\ &W_0^m = \frac{\lambda}{L + \lambda} \\ &W_0^c = \frac{\lambda}{L + \lambda} + (1 - \alpha^2 + \beta) \\ &W_i^m = W_i^c = \frac{1}{2(L + \lambda)}, i = 1, ..., 2L \\ &\beta = 2 \end{aligned} \quad \text{(B2)}$$

Drawing Sigma points:

$$\aleph_{k-1} = [\hat{x}_{k-1} \quad \hat{x}_{k-1} \pm \sqrt{(L + \lambda) P_{k-1}}] \text{ (B3)}$$

Updating Time:

$$\aleph_{k|k-1} = F[\aleph_{k|k-1}]$$
$$\hat{x}_k^- = \sum_{i=0}^{2L} W_i^m \aleph_{ik|k-1}$$
$$P_k^- = \sum_{i=0}^{2L} W_i^c [\aleph_{ik|k-1} - \hat{x}_k^-][\aleph_{ik|k-1} - \hat{x}_k^-]^T + I \quad \text{(B4)}$$
$$\Im_{k|k-1} = H[\aleph_{k|k-1}]$$
$$\hat{y}_k^- = \sum_{i=0}^{2L} W_i^m \Im_{ik|k-1}$$

Updating Measurement:

$$P_{y_k} = \sum_{i=0}^{2L} W_i^c [\Im_{k|k-1} - \hat{y}_k^-][\Im_{k|k-1} - \hat{y}_k^-]^T + Y$$
$$P_{x_k y_k} = \sum_{i=0}^{2L} W_i^c [\aleph_{ik|k-1} - \hat{x}_k^-][\Im_{k|k-1} - \hat{y}_k^-]^T$$
$$K = P_{x_k y_k} P_{y_k}^{-1} \quad \text{(B5)}$$
$$\hat{x}_k = \hat{x}_k^- + K(y_k - \hat{y}_k^-)$$
$$P_k = P_k^- - K P_{y_k} K^T$$

## APPENDIX 1.C: CONDITIONAL DISTRIBUTIONS FOR ESTIMATION OF THE MICRO CHOICE MODEL

Conditional distributions of the choice variable include the following:

$$A_i \mid y_{it}^j, s_{it}^j, \mu_i, \Sigma_i, \hat{c}_{jt}^{imm}, F_{jt} \qquad i = 1...I \quad \text{(C1)}$$

where this conditional distribution can be estimated by random walk metropolis hasting on the weighted likelihood.

The priors for normal mixture distribution of the individual and the category specific parameters used are:

$$\begin{aligned}
&\{(\mu_i, \Sigma_i)\} \mid A_i, \Delta, z_i, a, v, \vartheta, \alpha^d \\
&\alpha^d \mid I^* \\
&a \mid \{(\mu_i, \Sigma_i)^*\} \\
&v \mid \{(\mu_i, \Sigma_i)^*\}, \vartheta \\
&\vartheta \mid \{(\mu_i, \Sigma_i)^*\}, v
\end{aligned} \qquad \text{(C2)}$$

where the first conditional is the standard posterior Polya Urn representation for the mean and variance of individual specific random coefficient choice model parameters. $(\mu_i, \Sigma_i)^*$ denotes a set of unique $(\mu_i, \Sigma_i)$, which the DP process hyper-parameters depend only on (a posteriori). Given the $\{(\mu_i, \Sigma_i)^*\}$ set $\alpha^d$ and based measure parameters (i.e. $a, v, \vartheta$) are independent, a posteriori. The conditional posterior of the $G_0$ hyper-parameters (i.e. $a, v, \vartheta$), factors into two parts as $a$ is independent of $v, \vartheta$ given $\{(\mu_i, \Sigma_i)^*\}$. The form of this conditional posterior is:

$$p(a, v, \vartheta \mid \{(\mu_i, \Sigma_i)^*\}) \propto \prod_{j=1}^{I^*} \phi(\mu_j^* \mid 0, a^{-1}, \Sigma_j^*) IW(\Sigma_j^* \mid v, V = v\vartheta I_d) p(a, v, \vartheta) \quad \text{(C3)}$$

where $\phi(.|.,.)$ denotes the multivariate normal density. $IW(.|.,.)$ denotes Inverted-Wishart distribution. Finally, for Polya representation implementation the following conditional distribution is used:

$$(\mu_{i+1}, \Sigma_{i+1}) \mid \{(\mu_1, \Sigma_1), \ldots, (\mu_i, \Sigma_i)\} \sim \begin{cases} G_0(a, v, \vartheta) \text{ with prob } \dfrac{\alpha^d}{\alpha^d + i} \\ \delta_{(\mu_j, \Sigma_j)} \text{ with prob } \dfrac{1}{\alpha^d + i} \end{cases} \quad \text{(C4)}$$

I assessed the prior hyperparameters to provide proper but diffuse distributions, defined formally by:

$$\vec{a} = 0.00001, \bar{a} = 50, \vec{\vartheta} = 0.00001, \bar{\vartheta} = 600, \vec{v} = 0.00001, \bar{v} = 80 \qquad \text{(C5)}$$

Finally to complete the exposition, the posterior for the partition (segment) parameters has the following form:

$$\Sigma_k \mid \alpha_k, \Delta, z_k, v, V \sim IW(v + n_k, \\ v \times \vartheta \times I + (\alpha_k - \tilde{\mu}_k' - \Delta z_k)'(\alpha_k - \tilde{\mu}_k' - \Delta z_k) + a(\tilde{\mu}_k - \overline{\mu})(\tilde{\mu}_k - \overline{\mu}))$$

$$\mu_k \mid \alpha_k, \Sigma_k, \overline{\mu}, a, \Delta, z_k \sim N(\tilde{\mu}_k, \frac{\Sigma_k}{n_k + a})$$

$$\tilde{\mu}_k = \frac{n_k \overline{\alpha}_k + a\overline{\mu}_k}{n_k + a}, \overline{\alpha}_k = \frac{\sum_{i \in k} \alpha_i}{n_k}, \overline{\mu} = 0$$

# APPENDIX 1.D: CHOICE PARAMETER ESTIMATES FOR ALTERNATIVE MODELS

Table 1.D.1. PARAMETER ESTIMATES: Individual Choice effect (Local imitators)

| | **Estimate** | **Std. Dev.** | **2.5th** | **97.5th** |
|---|---|---|---|---|
| Category specific preference: | | | | |
| Device Tools $\alpha_1$ | -9.19* | 6.57 | -34.490 | -3.216 |
| eBooks $\alpha_2$ | -8.89* | 2.84 | -13.253 | -3.337 |
| Games $\alpha_3$ | -25.82* | 8.31 | -38.609 | -9.519 |
| Health/Diet/Fitness $\alpha_4$ | 1.17 | 1.83 | -6.588 | 2.295 |
| Humor/Jokes $\alpha_5$ | -0.41 | 7.11 | -31.222 | 1.509 |
| Internet/WAP $\alpha_6$ | -12.11* | 3.94 | -18.256 | -4.378 |
| Logic/Puzzle/Trivia $\alpha_7$ | -26.26* | 10.30 | -51.248 | -9.384 |
| Reference/Dictionaries $\alpha_8$ | -17.95* | 5.82 | -27.128 | -6.624 |
| Social Networks $\alpha_9$ | -3.23 | 1.33 | -5.043 | 0.184 |
| University $\alpha_{10}$ | -6.86* | 10.07 | -49.254 | -1.866 |
| States: | | | | |
| Individual download history State $\alpha_{11}$ | -17.79* | 18.37 | -93.057 | -5.689 |
| Latent imitation level $\alpha_{12}$ | 0.02* | 0.01 | 0.005 | 0.032 |
| App category characteristics (factors): | | | | |
| Popularity of app category $\alpha_{13}$ | 0.39 | 0.73 | -2.766 | 0.789 |
| Investment apps category $\alpha_{14}$ | -10.73* | 3.45 | -15.955 | -3.700 |
| Hedonic apps category $\alpha_{15}$ | 12.67 | 7.57 | -15.563 | 20.683 |

$* \ p<0.05$

Table 1.D.2. PARAMETER ESTIMATES: Individual Choice effect (Global imitators)

| | **Estimate** | **Std. Dev.** | **$2.5^{th}$** | **$97.5^{th}$** |
|---|---|---|---|---|
| Category specific preference: | | | | |
| Device Tools $\alpha_1$ | -2.13* | 0.26 | -2.79 | -1.89 |
| eBooks $\alpha_2$ | -0.6 | 0.84 | -0.68 | 0.52 |
| Games $\alpha_3$ | 0.22* | 0.37 | 0.17 | 1.18 |
| Health/Diet/Fitness $\alpha_4$ | 1.06 | 0.86 | -1.46 | 1.53 |
| Humor/Jokes $\alpha_5$ | -2.68* | 0.34 | -3.20 | -1.63 |
| Internet/WAP $\alpha_6$ | -0.53* | 0.28 | -1.76 | -0.44 |
| Logic/Puzzle/Trivia $\alpha_7$ | -1.72 | 0.75 | -2.07 | 1.08 |
| Reference/Dictionaries $\alpha_8$ | -1.45 | 0.46 | -1.76 | 0.64 |
| Social Networks $\alpha_9$ | -1.64* | 0.36 | -1.97 | -0.44 |
| University $\alpha_{10}$ | -2.05* | 0.37 | -2.44 | -1.61 |
| States: | | | | |
| Individual download history State $\alpha_{11}$ | -3.3* | 0.44 | -4.00 | -2.15 |
| Latent imitation level $\alpha_{12}$ | 0.01* | 0.00 | 0.00 | 0.01 |
| App category characteristics (factors): | | | | |
| Popularity of app category $\alpha_{13}$ | -0.08* | 0.15 | -0.38 | -0.06 |
| Investment apps category $\alpha_{14}$ | -0.42* | 1.09 | -1.24 | -0.40 |
| Hedonic apps category $\alpha_{15}$ | -0.31* | 0.49 | -2.29 | -0.22 |

Table 1.D.3. PARAMETER ESTIMATES: Individual Choice effect (Global Adopters)

| | **Estimate** | **Std. Dev.** | **2.5th** | **97.5th** |
|---|---|---|---|---|
| Category specific preference: | | | | |
| Device Tools $\alpha_1$ | -24.94* | 16.81 | -14.327 | -2.669 |
| eBooks $\alpha_2$ | -18.31* | 12.36 | -15.290 | -6.381 |
| Games $\alpha_3$ | -18.66* | 11.97 | -11.222 | -2.296 |
| Health/Diet/Fitness $\alpha_4$ | -13.37 | 8.16 | -5.939 | 2.982 |
| Humor/Jokes $\alpha_5$ | -11.26* | 6.43 | -22.097 | -9.715 |
| Internet/WAP $\alpha_6$ | -6.92* | 2.54 | -8.021 | -3.021 |
| Logic/Puzzle/Trivia $\alpha_7$ | -0.13* | 1.06 | -18.122 | -8.332 |
| Reference/Dictionaries $\alpha_8$ | -17.74* | 12.56 | -11.092 | -4.547 |
| Social Networks $\alpha_9$ | -26.77 | 17.29 | -15.530 | 0.076 |
| University $\alpha_{10}$ | -9.2* | 7.16 | -7.791 | -2.916 |
| States: | | | | |
| Individual download history State $\alpha_{11}$ | -35.93* | 22.88 | -34.350 | -13.821 |
| Latent imitation level $\alpha_{12}$ | 0.01* | 0.01 | 0.011 | 0.035 |
| App category characteristics (factors): | | | | |
| Popularity of app category $\alpha_{13}$ | -1.05 | 0.73 | -0.830 | 1.767 |
| Investment apps category $\alpha_{14}$ | 19.1 | 13.08 | -0.922 | 7.230 |
| Hedonic apps category $\alpha_{15}$ | -25.79 | 17.78 | -6.606 | 10.330 |

Table 1.D.4. PARAMETER ESTIMATES: Individual Choice effect (No social influence)

| | **Estimate** | **Std. Dev.** | **$2.5^{th}$** | **$97.5^{th}$** |
|---|---|---|---|---|
| Category specific preference: | | | | |
| Device Tools $\alpha_1$ | -20.51* | 6.14 | -28.162 | -2.250 |
| eBooks $\alpha_2$ | -2.43* | 0.39 | -2.647 | -1.372 |
| Games $\alpha_3$ | -11.89 | 5.26 | -19.040 | 0.482 |
| Health/Diet/Fitness $\alpha_4$ | -10.4* | 3.55 | -15.011 | -0.619 |
| Humor/Jokes $\alpha_5$ | -8.9 | 3.35 | -13.168 | 0.716 |
| Internet/WAP $\alpha_6$ | -21.99* | 7.19 | -31.258 | -2.083 |
| Logic/Puzzle/Trivia $\alpha_7$ | -16.36* | 4.60 | -21.876 | -1.765 |
| Reference/Dictionaries $\alpha_8$ | -7.59 | 1.80 | -8.841 | 0.232 |
| Social Networks $\alpha_9$ | -10.77* | 3.39 | -14.949 | -0.369 |
| University $\alpha_{10}$ | -2.44* | 0.46 | -2.661 | -0.457 |
| States: | | | | |
| Individual download history State $\alpha_{11}$ | -15.84* | 4.79 | -22.241 | -4.298 |
| Latent imitation level $\alpha_{12}$ | - | - | - | - |
| App category characteristics (factors): | | | | |
| Popularity of app category $\alpha_{13}$ | -1.77* | 0.68 | -2.708 | -0.254 |
| Investment apps category $\alpha_{14}$ | -7.51* | 2.81 | -11.425 | -1.362 |
| Hedonic apps category $\alpha_{15}$ | -9.29* | 3.67 | -14.263 | -0.893 |

Table 1.D.5. PARAMETER ESTIMATES: Individual Choice Hierarchical Model (Local imitators): Tenure explanation of the effects

| Parameter explained by Tenure | **Estimate** | **Std. Dev.** | **2.5$^{th}$** | **97.5$^{th}$** |
|---|---|---|---|---|
| Category specific preference: | | | | |
| Device Tools $\alpha_1$ | -0.0032* | 9.66E-05 | -0.0034 | -0.0030 |
| eBooks $\alpha_2$ | -0.0012* | 1.42E-04 | -0.0015 | -0.0010 |
| Games $\alpha_3$ | -0.0005* | 1.31E-04 | -0.0008 | -0.0002 |
| Health/Diet/Fitness $\alpha_4$ | -0.0023* | 1.37E-04 | -0.0026 | -0.0021 |
| Humor/Jokes $\alpha_5$ | 0.0006* | 7.44E-05 | 0.0004 | 0.0007 |
| Internet/WAP $\alpha_6$ | 0.0022* | 1.29E-04 | 0.0019 | 0.0023 |
| Logic/Puzzle/Trivia $\alpha_7$ | 0.0028* | 1.60E-04 | 0.0025 | 0.0031 |
| Reference/Dictionaries $\alpha_8$ | 0.0004* | 9.12E-05 | 0.0002 | 0.0006 |
| Social Networks $\alpha_9$ | 0.0034* | 1.46E-04 | 0.0031 | 0.0036 |
| University $\alpha_{10}$ | 0.0007* | 4.04E-05 | 0.0006 | 0.0007 |
| States: | | | | |
| Individual download history State $\alpha_{11}$ | -0.005* | 8.06E-05 | -0.0051 | -0.0048 |
| Latent imitation level $\alpha_{12}$ | 0.0001* | 5.88E-06 | 0.0000 | 0.0001 |
| App category characteristics (factors): | | | | |
| Popularity of app category $\alpha_{13}$ | 0.0001* | 1.13E-05 | 0.0001 | 0.0001 |
| Investment apps category $\alpha_{14}$ | 0.0025* | 6.35E-05 | 0.0024 | 0.0026 |
| Hedonic apps category $\alpha_{15}$ | -0.0012* | 1.09E-04 | -0.0014 | -0.0010 |

* $p<0.05$

Table 1.D.6. PARAMETER ESTIMATES: Individual Choice Hierarchical Model (Global imitators): Tenure explanation of the effects

| Parameter explained by Tenure | **Estimate** | **Std. Dev.** | **2.5th** | **97.5th** |
|---|---|---|---|---|
| Category specific preference: | | | | |
| Device Tools $\alpha_1$ | -0.0038* | 1.61E-04 | -0.0041 | -0.0035 |
| eBooks $\alpha_2$ | -0.0014* | 1.78E-04 | -0.0017 | -0.0011 |
| Games $\alpha_3$ | 0.0009* | 8.48E-05 | 0.0007 | 0.0010 |
| Health/Diet/Fitness $\alpha_4$ | 0.0046* | 4.85E-04 | 0.0040 | 0.0054 |
| Humor/Jokes $\alpha_5$ | -0.0061* | 3.34E-04 | -0.0065 | -0.0056 |
| Internet/WAP $\alpha_6$ | -0.0005* | 7.52E-05 | -0.0006 | -0.0004 |
| Logic/Puzzle/Trivia $\alpha_7$ | -0.0035* | 1.65E-04 | -0.0038 | -0.0032 |
| Reference/Dictionaries $\alpha_8$ | -0.0033* | 3.75E-04 | -0.0039 | -0.0028 |
| Social Networks $\alpha_9$ | -0.0034* | 2.16E-04 | -0.0037 | -0.0030 |
| University $\alpha_{10}$ | -0.0047* | 2.57E-04 | -0.0051 | -0.0043 |
| States: | | | | |
| Individual download history State $\alpha_{11}$ | -0.0086* | 5.63E-04 | -0.0095 | -0.0077 |
| Latent imitation level $\alpha_{12}$ | -0.0001* | 1.31E-05 | -0.0001 | -0.0001 |
| App category characteristics (factors): | | | | |
| Popularity of app category $\alpha_{13}$ | -0.0002* | 1.72E-05 | -0.0002 | -0.0002 |
| Investment apps category $\alpha_{14}$ | -0.0016* | 9.04E-05 | -0.0018 | -0.0014 |
| Hedonic apps category $\alpha_{15}$ | -0.0005* | 8.27E-05 | -0.0006 | -0.0004 |

* $p<0.05$

Table 1.D.7. PARAMETER ESTIMATES: Individual Choice Hierarchical Model (Global Adopters): Tenure explanation of the effects

| Parameter explained by Tenure | **Estimate** | **Std. Dev.** | **2.5th** | **97.5th** |
|---|---|---|---|---|
| Category specific preference: | | | | |
| Device Tools $\alpha_1$ | 0.0003* | 1.00E-04 | 1.36E-04 | 0.00046 |
| eBooks $\alpha_2$ | 0.00034* | 8.80E-05 | 1.46E-04 | 4.74E-04 |
| Games $\alpha_3$ | -0.00016 | 1.00E-04 | -3.15E-04 | 6.43E-05 |
| Health/Diet/Fitness $\alpha_4$ | 0.00028* | 7.29E-05 | 1.38E-04 | 0.000424 |
| Humor/Jokes $\alpha_5$ | 0.00027* | 9.25E-05 | 1.04E-04 | 0.000454 |
| Internet/WAP $\alpha_6$ | 0.00157* | 1.09E-04 | 1.35E-03 | 1.76E-03 |
| Logic/Puzzle/Trivia $\alpha_7$ | 0.00072* | 9.03E-05 | 5.92E-04 | 0.00091 |
| Reference/Dictionaries $\alpha_8$ | 0.00047* | 6.80E-05 | 3.06E-04 | 5.76E-04 |
| Social Networks $\alpha_9$ | -0.00006 | 1.00E-04 | -2.49E-04 | 9.45E-05 |
| University $\alpha_{10}$ | 0.00071* | 9.72E-05 | 5.37E-04 | 8.64E-04 |
| States: | | | | |
| Individual download history State $\alpha_{11}$ | 0.00119* | 1.98E-04 | 8.84E-04 | 0.001637 |
| Latent imitation level $\alpha_{12}$ | -0.00001 | 4.69E-06 | -1.43E-05 | 3.22E-06 |
| App category characteristics (factors): | | | | |
| Popularity of app category $\alpha_{13}$ | -0.00007* | 1.65E-05 | -9.67E-05 | -3.90E-05 |
| Investment apps category $\alpha_{14}$ | -0.00095* | 1.17E-04 | -1.14E-03 | -7.95E-04 |
| Hedonic apps category $\alpha_{15}$ | 0.0002* | 8.21E-05 | 7.29E-05 | 3.66E-04 |

* $p<0.05$

Table 1.D.8. PARAMETER ESTIMATES: Individual Choice Hierarchical Model (No social influence): Tenure explanation of the effects

| Parameter explained by Tenure | **Estimate** | **Std. Dev.** | **$2.5^{th}$** | **$97.5^{th}$** |
|---|---|---|---|---|
| Category specific preference: | | | | |
| Device Tools $\alpha_1$ | 0.00192* | 1.34E-04 | 1.63E-03 | 0.002171 |
| eBooks $\alpha_2$ | 0.00162* | 1.21E-04 | 1.33E-03 | 1.83E-03 |
| Games $\alpha_3$ | 0.00024 | 4.04E-04 | -4.20E-04 | 8.69E-04 |
| Health/Diet/Fitness $\alpha_4$ | -0.00004 | 1.86E-04 | -3.63E-04 | 0.000259 |
| Humor/Jokes $\alpha_5$ | 0.00019 | 1.53E-04 | -6.07E-05 | 0.000484 |
| Internet/WAP $\alpha_6$ | 0.00164* | 2.65E-04 | 1.15E-03 | 2.04E-03 |
| Logic/Puzzle/Trivia $\alpha_7$ | 0.00207* | 1.44E-04 | 1.87E-03 | 0.002432 |
| Reference/Dictionaries $\alpha_8$ | 0.00292* | 1.36E-04 | 2.58E-03 | 3.13E-03 |
| Social Networks $\alpha_9$ | 0.00128* | 1.39E-04 | 1.01E-03 | 0.001511 |
| University $\alpha_{10}$ | 0.00066* | 8.36E-05 | 4.77E-04 | 7.96E-04 |
| States: | | | | |
| Individual download history State $\alpha_{11}$ | 0.00084* | 1.19E-04 | 6.21E-04 | 1.08E-03 |
| Latent imitation level $\alpha_{12}$ | - | - | - | - |
| App category characteristics (factors): | | | | |
| Popularity of app category $\alpha_{13}$ | 0.000005 | 3.77E-05 | -6.62E-05 | 6.02E-05 |
| Investment apps category $\alpha_{14}$ | 0.00023 | 1.65E-04 | -1.11E-04 | 4.58E-04 |
| Hedonic apps category $\alpha_{15}$ | -0.00014 | 2.00E-04 | -4.80E-04 | 1.75E-04 |

* $p<0.05$

Table 1.D.9. PARAMETER ESTIMATES: Individual Choice effect (Local imitators)

| Total number of users: 1258 | **Positive Significant** | **Negative Significant** |
|---|---|---|
| Category specific preference: | | |
| Device Tools $\alpha_1$ | 0 | 1258 |
| eBooks $\alpha_2$ | 0 | 1258 |
| Games $\alpha_3$ | 0 | 1258 |
| Health/Diet/Fitness $\alpha_4$ | 1197 | 61 |
| Humor/Jokes $\alpha_5$ | 1197 | 61 |
| Internet/WAP $\alpha_6$ | 0 | 1258 |
| Logic/Puzzle/Trivia $\alpha_7$ | 0 | 1258 |
| Reference/Dictionaries $\alpha_8$ | 0 | 1258 |
| Social Networks $\alpha_9$ | 58 | 1197 |
| University $\alpha_{10}$ | 0 | 1258 |
| States: | | |
| Individual download history State $\alpha_{11}$ | 0 | 1258 |
| Latent imitation level $\alpha_{12}$ | 1217 | 0 |
| App category characteristics (factors): | | |
| Popularity of app category $\alpha_{13}$ | 1197 | 61 |
| Investment apps category $\alpha_{14}$ | 0 | 1258 |
| Hedonic apps category $\alpha_{15}$ | 1197 | 61 |

Table 1.D.10. PARAMETER ESTIMATES: Individual Choice effect
(Global imitators)

| Total number of users: 1258 | **Positive Significant** | **Negative Significant** |
|---|---|---|
| Category specific preference: | | |
| Device Tools $\alpha_1$ | 0 | 1258 |
| eBooks $\alpha_2$ | 42 | 1216 |
| Games $\alpha_3$ | 1250 | 8 |
| Health/Diet/Fitness $\alpha_4$ | 1208 | 50 |
| Humor/Jokes $\alpha_5$ | 0 | 1257 |
| Internet/WAP $\alpha_6$ | 0 | 1258 |
| Logic/Puzzle/Trivia $\alpha_7$ | 42 | 1216 |
| Reference/Dictionaries $\alpha_8$ | 42 | 1216 |
| Social Networks $\alpha_9$ | 8 | 1250 |
| University $\alpha_{10}$ | 8 | 1250 |
| States: | | |
| Individual download history State $\alpha_{11}$ | 0 | 1258 |
| Latent imitation level $\alpha_{12}$ | 1051 | 4 |
| App category characteristics (factors): | | |
| Popularity of app category $\alpha_{13}$ | 8 | 1250 |
| Investment apps category $\alpha_{14}$ | 8 | 1250 |
| Hedonic apps category $\alpha_{15}$ | 8 | 1250 |

Table 1.D.11. PARAMETER ESTIMATES: Individual Choice effect (Global Adopters)

| Total number of users: 1258 | **Positive Significant** | **Negative Significant** |
|---|---|---|
| Category specific preference: | | |
| Device Tools $\alpha_1$ | 0 | 1258 |
| eBooks $\alpha_2$ | 0 | 1258 |
| Games $\alpha_3$ | 0 | 1258 |
| Health/Diet/Fitness $\alpha_4$ | 0 | 1258 |
| Humor/Jokes $\alpha_5$ | 55 | 1203 |
| Internet/WAP $\alpha_6$ | 0 | 1258 |
| Logic/Puzzle/Trivia $\alpha_7$ | 438 | 545 |
| Reference/Dictionaries $\alpha_8$ | 0 | 1258 |
| Social Networks $\alpha_9$ | 0 | 1258 |
| University $\alpha_{10}$ | 0 | 1249 |
| States: | | |
| Individual download history State $\alpha_{11}$ | 0 | 1258 |
| Latent imitation level $\alpha_{12}$ | 607 | 257 |
| App category characteristics (factors): | | |
| Popularity of app category $\alpha_{13}$ | 55 | 1203 |
| Investment apps category $\alpha_{14}$ | 1203 | 55 |
| Hedonic apps category $\alpha_{15}$ | 55 | 1203 |

Table 1.D.12. PARAMETER ESTIMATES: Individual Choice effect (No social influence)

| Total number of users: 1258 | **Positive Significant** | **Negative Significant** |
|---|---|---|
| Category specific preference: | | |
| Device Tools $\alpha_1$ | 0 | 1258 |
| eBooks $\alpha_2$ | 0 | 1258 |
| Games $\alpha_3$ | 54 | 1103 |
| Health/Diet/Fitness $\alpha_4$ | 0 | 1258 |
| Humor/Jokes $\alpha_5$ | 54 | 1204 |
| Internet/WAP $\alpha_6$ | 0 | 1258 |
| Logic/Puzzle/Trivia $\alpha_7$ | 0 | 1258 |
| Reference/Dictionaries $\alpha_8$ | 54 | 1204 |
| Social Networks $\alpha_9$ | 0 | 1258 |
| University $\alpha_{10}$ | 0 | 1258 |
| States: | | |
| Individual download history State $\alpha_{11}$ | 2 | 1256 |
| Latent imitation level $\alpha_{12}$ | - | - |
| App category characteristics (factors): | | |
| Popularity of app category $\alpha_{13}$ | 2 | 1241 |
| Investment apps category $\alpha_{14}$ | 0 | 1243 |
| Hedonic apps category $\alpha_{15}$ | 2 | 1241 |

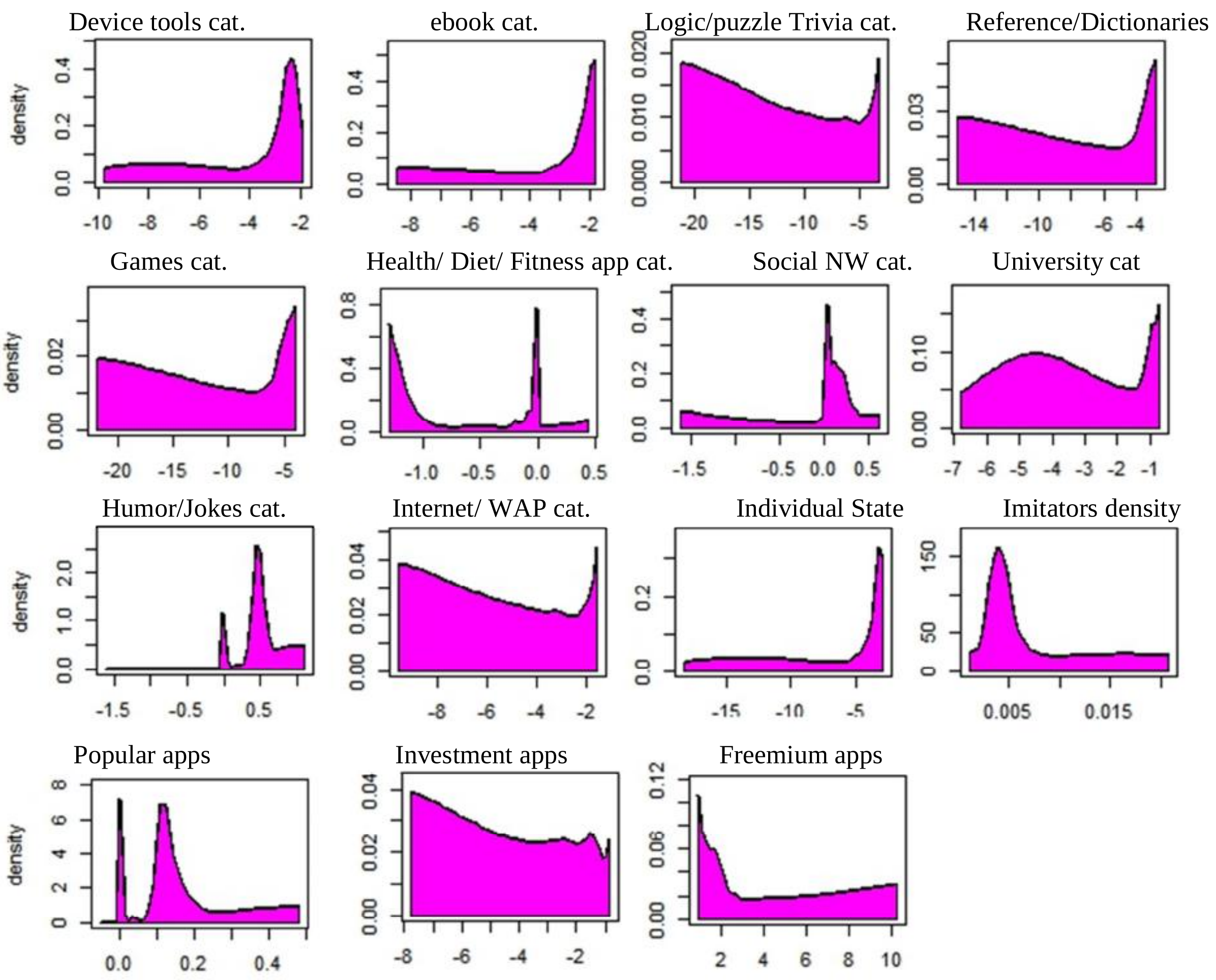

Figure 1.D.1. PARAMETER DISTRIBUTION: Heterogeneity in Individual Choice (Local Imitators)

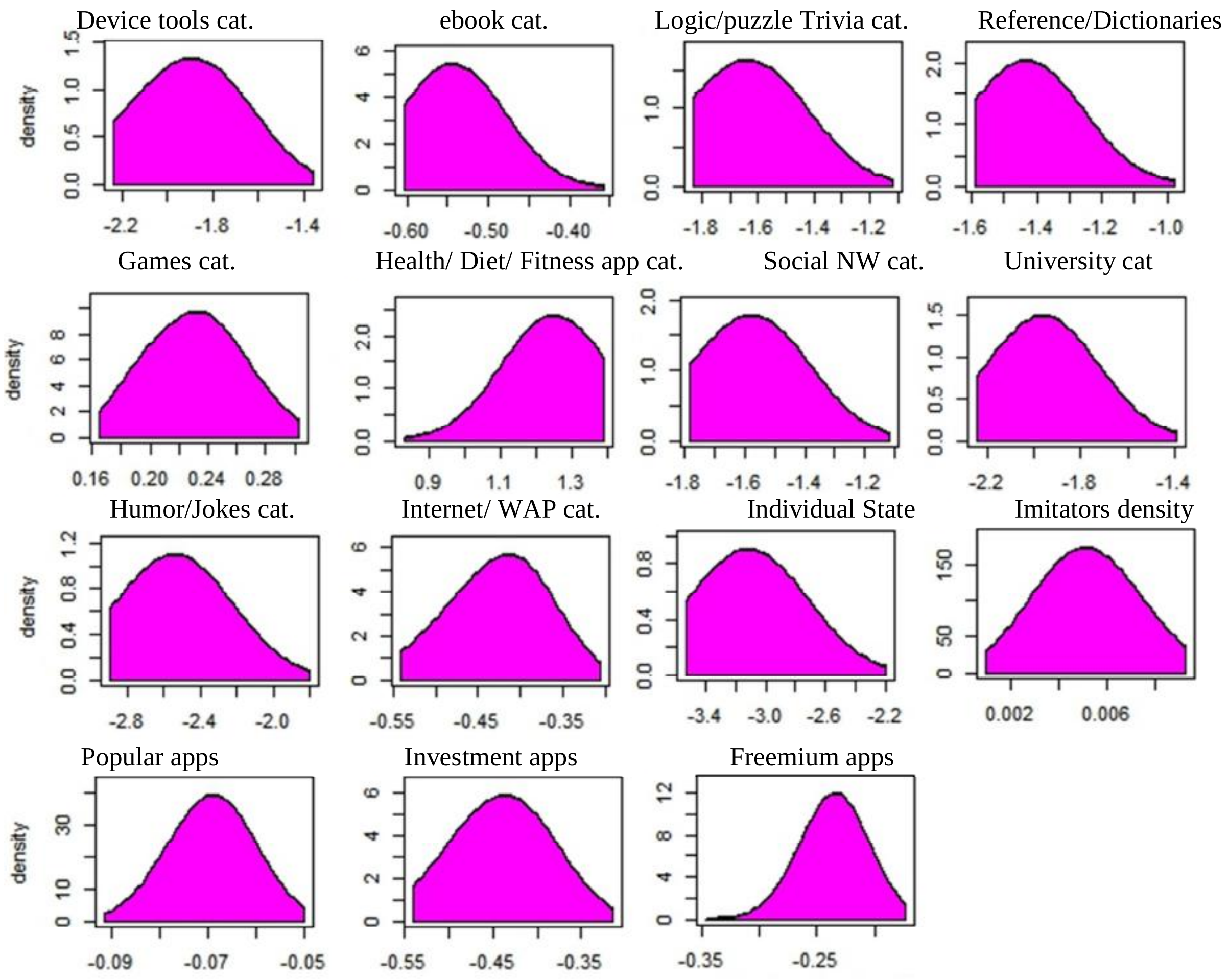

Figure 1.D.2. PARAMETER DISTRIBUTION: Heterogeneity in Individual Choice (Global Imitators)

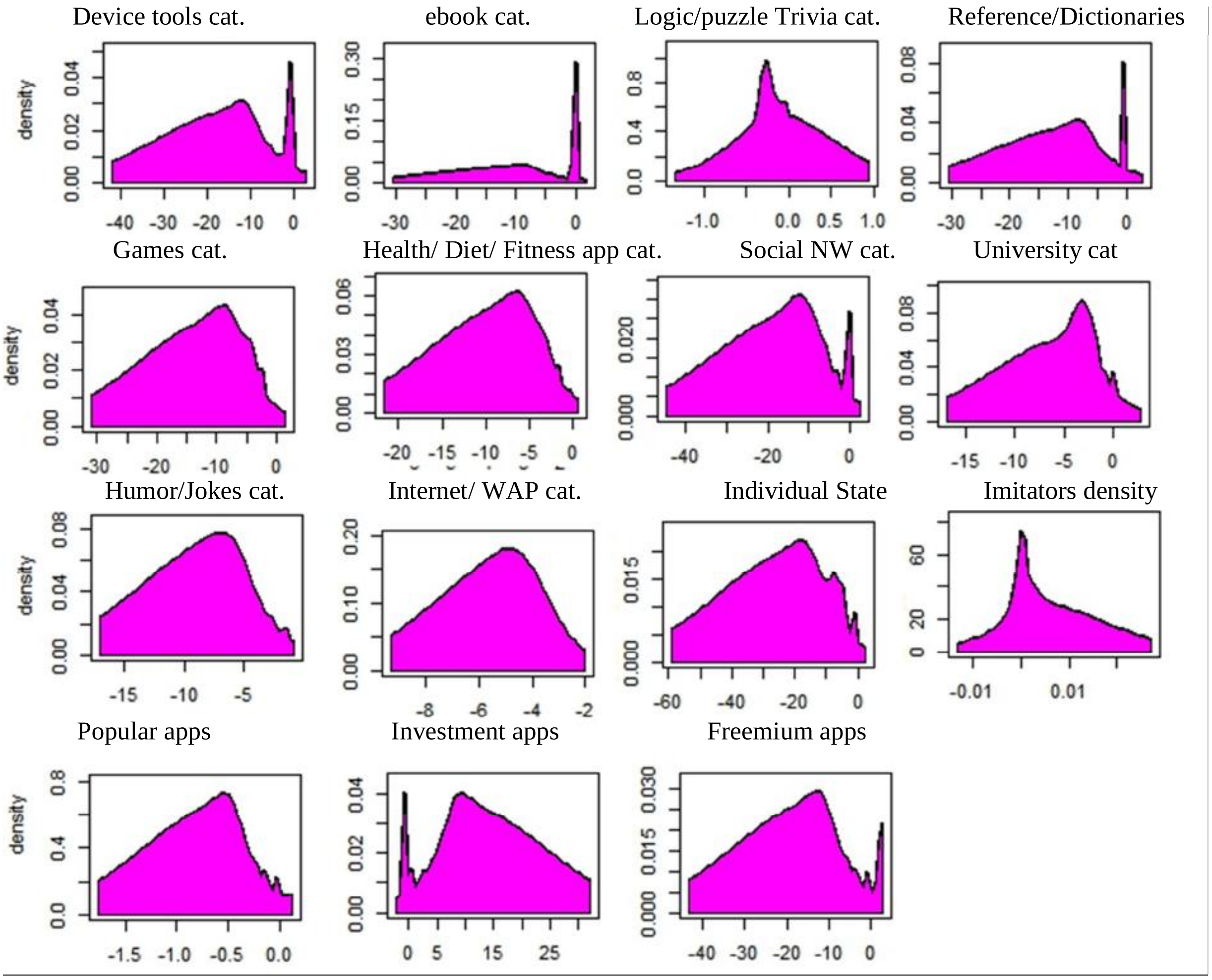

Figure 1.D.3. PARAMETER DISTRIBUTION: Heterogeneity in Individual Choice (Global Adopters)

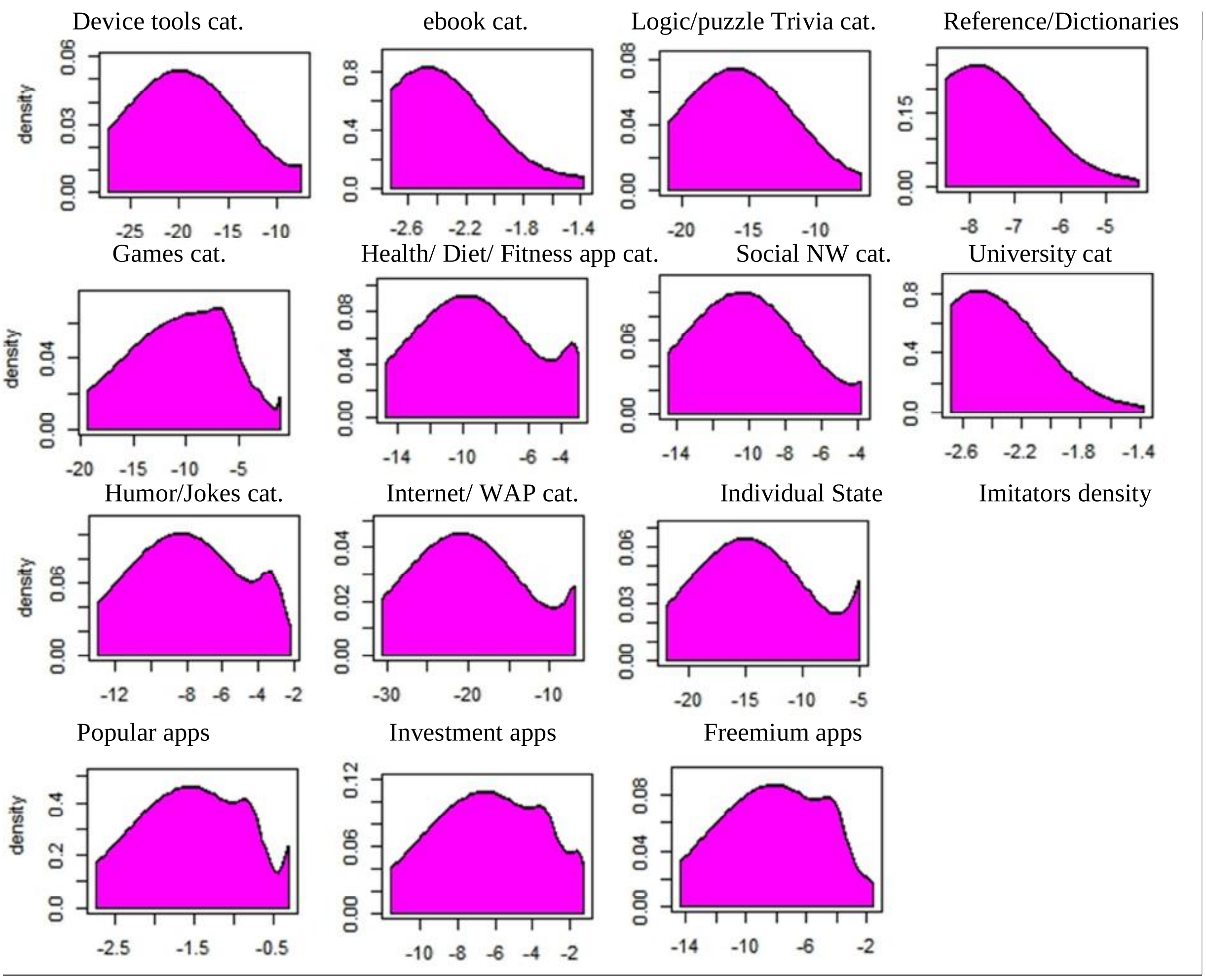

Figure 1.D.4. PARAMETER DISTRIBUTION: Heterogeneity in Individual Choice (No social influence)

# APPENDIX 2.A: LATENT DIRICHLET ALLOCATION

LDA is a three-level hierarchical Bayesian model, in which each item of a collection is modeled as a finite mixture over an underlying set of topics (Blei et al 2003). LDA is a generative approach; it use naïve conditional independence assumption, and it neglect the order of features by assuming exchangeability and using bag of words representation. These assumptions bring two main benefits to these approaches: simplicity, computational efficiency. Formally the LDA model assumes the following generative process for each item i in a collection C consisting of element (feature) e:

1. Choose N ~ Poisson ( $\xi$ ), where N is the number of elements e
2. Choose $\theta \sim Dir(\alpha)$, where $\theta$ is the probability that a given document has primitive topic
3. For each of the N features $i_n$:
    a. Choose a topic $z_n \sim Multinomial(\theta)$
    b. Choose a feature $i_n$ from $p(i_n \mid z_n, \beta)$, a multinomial probability conditioned on the topic

A k-dimensional Dirichlet random variable $\theta$ can take values in the (k-1)-simplex (a k-vector $\theta$ lies in the (k-1)-simplex if $\theta_i \geq 0, \sum_{i=1}^{k} \theta_i = 1$), and has the following probability density on this simplex:

$$p(\theta \mid \alpha) = \frac{\Gamma(\sum_{i=1}^{k} \alpha_i)}{\prod_{i=1}^{k} \Gamma(\alpha_i)} \theta_1^{\alpha_1 - 1} \backslash ...\theta_k^{\alpha_{k1} - 1}$$

I represented the Probability Graphical Model (PGM) of LDA in figure 1.4. As figure depicts, there are three levels to the LDA representation. The parameters $\alpha, \beta$ are collection level parameters, and they are sampled once. The variable $\theta_d$ has Dirichlet distribution, and it is document level variable, so it is sampled once per document. This variable simply defines the weight distribution of topics within the document. Finally variables $z_{d_n}$ and $w_{d_n}$ are feature level parameters and they are sampled once for each feature within each document. Variable $z_{d_n}$ defines the topic of n'ths word within document d, and variable $w_{d_n}$ defines the feature instance that appears at location n within document d. As I can see an LDA model is a type of conditionally independent hierarchical model, and it is often referred to as parametric empirical Bayes model. One of the advantages of an LDA model is that it is parsimonious, so unlike probabilistic Latent Semantic Indexing (pLSI) model, it does not suffer from over fitting.

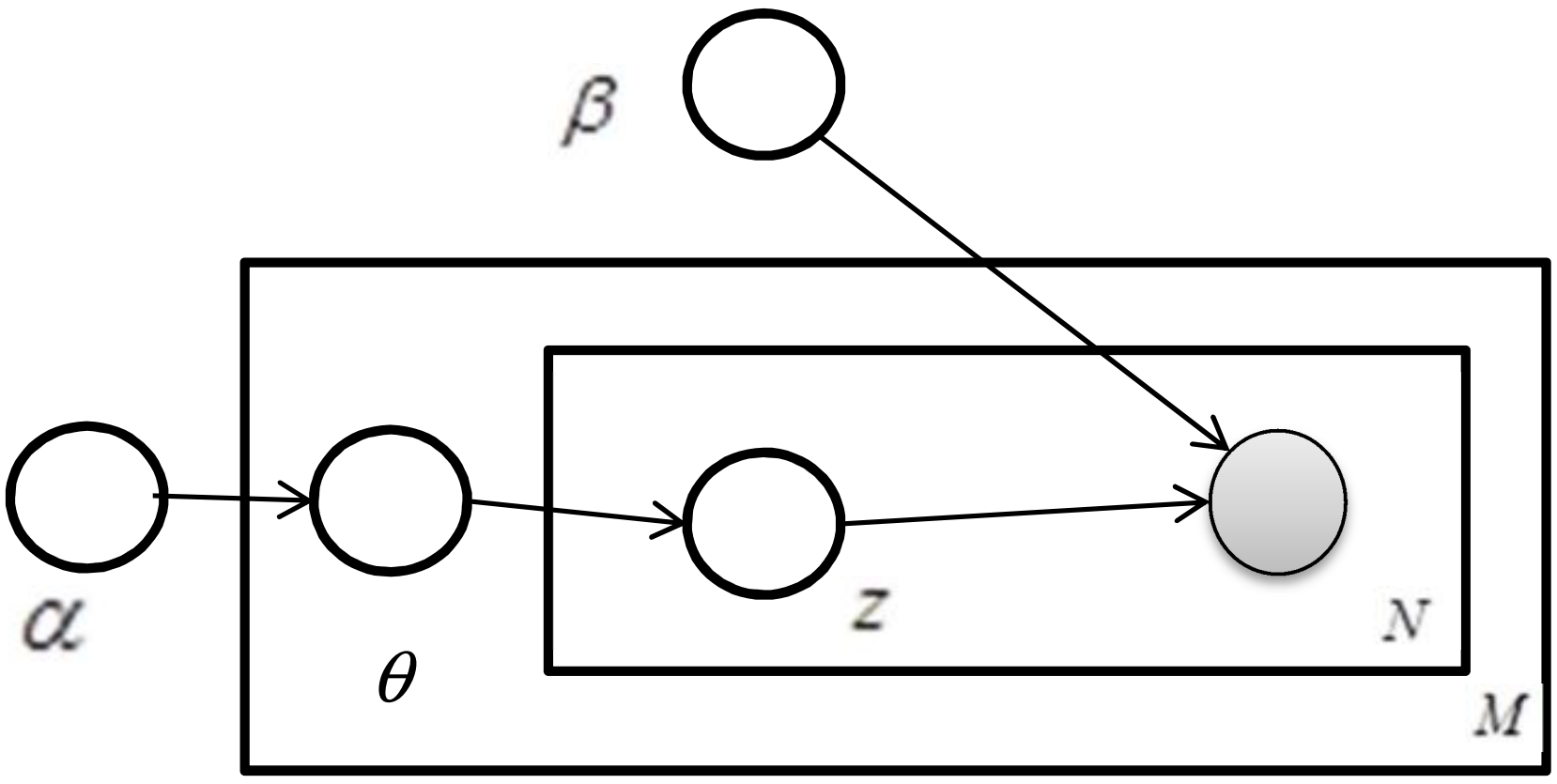

Figure 2.A.1. Graphical model representation of LDA

To estimate LDA model, I defined the likelihood of model in the following:

$$p(D \mid \alpha, \beta) = \prod_{d=1}^{M} \int p(\theta_d \mid \alpha) \left( \prod_{n=1}^{N_d} \sum_{z_{d_n}} p(z_{d_n} \mid \theta_d) p(w_{d_n} \mid z_{d_n}, \beta) d\theta_d \right.$$

The key inferential problem to solve for LDA is computing posterior distribution of topic hidden variables $\theta_d, z_d$, the first one with Dirichlet distribution, and the second one with multinomial distribution. To normalize the distribution of words given $\alpha, \beta$, I marginalized over the hidden variables as following:

$$p(D \mid \alpha, \beta) = \prod_{d=1}^{M} \frac{\Gamma(\sum_i \alpha_i)}{\prod_i \Gamma(\alpha_i)} \int \left( \prod_{i=1}^{k} \theta_i^{\alpha_i - 1} \right) \left( \prod_{n=1}^{N_d} \sum_{i=1}^{k} \prod_{j=1}^{V} (\theta_i \beta_{ij})^{w_n^j} d\theta \right.$$

Due to the coupling between $\theta$ and $\beta$ in the summation over latent topics this likelihood function is intractable. Therefore to estimate it Blei et al. (2003) suggests using variational inference method. Variational inference or variational Bayesian refers to a family of techniques for approximating intractable integrals arising in Bayesian inference and machine learning. These family of methods are an alternative to sampling methods, and they are basically used to analytically approximate the posterior probability of the unobservable variables, in order to do statistical inference over these variables. These methods also give a lower bound to the marginal log likelihood. This family of lower bounds is indexed by a set of variational parameters. To obtain tightest lower bound I used an optimization procedure to select the variational parameters. A simple way to obtain a tractable family of lower bounds is to consider simple modifications of the original graphical model, by removing dependencies and introducing new variational parameters instead. In the LDA model, I used the following variational distribution to approximate posterior distribution of unobserved variables given the observed data s follows:

$$q(\theta, z \mid \gamma, \phi) = q_1(\theta \mid \gamma)\prod_{n=1}^{N} q_2(z_n \mid \phi_n)$$

Where $q_1(.)$is a Dirichlet distribution with parameters $\gamma$ and $q_2(.)$is a multinomial distribution with parameters$\phi_n$. Variational parameters are result of solving the following optimization problem:

$$(\gamma^*, \phi^*) = \arg\min_{(\gamma,\phi)} D_{KL}(q(\theta, z \mid \gamma, \phi) \,\|\, p(\theta, z \mid w, \alpha, \beta))$$

where $D_{KL}$represents the Kullback-Leibler (KL) divergence between the variational distribution and the true joint posterior of latent parameters $p(\theta, z \mid w, \alpha, \beta)$. Formally, $D_{KL}$ is defined as follows:

$$D_{KL}(q(\theta,z\mid\gamma,\phi)\parallel p(\theta,z\mid w,\alpha,\beta) = \sum_{(\gamma,\phi)} q(\theta,z\mid\gamma,\phi)\log(\frac{q(\theta,z\mid\gamma,\phi)}{p(\theta,z\mid w,\alpha,\beta)})$$

As a result, I can write KL-divergence in the following format:

$$Logp(w\mid\alpha,\beta) = L(\gamma,\phi;\alpha,\beta) + D_{KL}(q(\theta,z\mid\gamma,\phi)\parallel p(\theta,z\mid w,\alpha,\beta))$$

where

$$L(\gamma,\phi;\alpha,\beta) = E_q[\log\ p(\theta,z,w\mid\alpha,\beta)] - E_q[\log q(\theta,z)]$$

This relation suggests that maximizing the lower bound $L(\gamma,\phi;\alpha,\beta)$ with respect to $\gamma$ and $\phi$ is equivalent to minimizing the KL divergence between the variational posterior probability and the true posterior probability. Expanding $L(\gamma,\phi;\alpha,\beta)$ using factorization of p and q gives the following:

$$\begin{aligned}L(\gamma,\phi;\alpha,\beta) &= E_q[\log p(\theta\mid\alpha)] + E_q[\log p(z\mid\theta)] + E_q[\log p(w\mid z,\beta)] - E_q[\log q(\theta)] - E_q[\log q(z)]\\ &= \log\Gamma(\sum_{j=1}^{k}\alpha_j) - \sum_{i=1}^{k}\log\Gamma(\alpha_i) + \sum_{i=1}^{k}(\alpha_i\ -1)(\Psi(\gamma_i) - \Psi(\sum_{j=1}^{k}\gamma_j) + \sum_{n=1}^{N}\sum_{i=1}^{k}\phi_{ni}(\Psi(\gamma_i) - \Psi(\sum_{j=1}^{k}\gamma_j))\\ &- \log\Gamma(\sum_{j=1}^{k}\gamma_j) - \sum_{i=1}^{k}\log\Gamma(\gamma_i) + \sum_{i=1}^{k}(\gamma_i\ -1)(\Psi(\gamma_i) - \Psi(\sum_{j=1}^{k}\gamma_j) + \sum_{n=1}^{N}\sum_{i=1}^{k}\phi_{ni}\log\phi_{ni}\end{aligned}$$

Where $\Gamma(.)$ is gamma function and $\Psi(.)$ is its derivative. They key for this derivation is the following equation: $E[\log\theta_i\mid\alpha] = \Psi(\alpha_i) - \Psi(\sum_{j=1}^{k}\alpha_j)$, which is direct derivative of general fact that the derivative of log normalization factor with respect to the natural parameter of an exponential distribution is equal to the expectation of sufficient statistics. Collecting terms that are only related to each of the variational parameters $\gamma$ and $\phi_{ni}$ from $L(\gamma,\phi;\alpha,\beta)$, and getting the derivative respectively give us an algorithm to solve the above optimization problem to find variational parameters. In particular, I can use a simple iterative fixed-point method and update two variational parameters by the following equations until convergance:

$$\phi_{ni} \propto \beta_{iw_n} \exp\{E_q[\log(\theta_i) \mid \gamma]\}$$
$$\gamma_i = \alpha_i + \sum_{n=1}^{n} \phi_{ni}$$

This optimization is document specific, so I viewed the Dirichlet parameter $\gamma^*(w)$ as providing a representation of a document in the topic simplex. In summary, I had the following variational inference algorithm for LDA (Blei et al 2003):

(1) Initialize $\phi_{ni}^{0} := 1/k$ for all i and n

(2) Initialize $\gamma_i := \alpha_i + N/k$ for all i and n

(3) **Repeat**

a. **For** n=1 **to** N

i. **For** i = 1 **to** k

1. $\phi_{ni}^{t+1} := \beta_{iw_n} \exp(\Psi(\gamma_i'))$

ii. Normalize $\phi_{ni}^{t+1}$ to sum to 1

b. $\gamma^{t+1} := \alpha + \sum_{n=1}^{N} \phi_n^{t+1}$

(4) **until** convergence

This algorithm has the order of $O(N^2k)$. Given the variational Bayesian method, I had tractable lower bound on the log likelihood, a bound which I can maximize with respect to $\alpha$ and $\beta$. I can thus find approximate empirical Bayes estimates for the LDA model via an alternating variational EM (VEM) procedure that maximizes a lower bound with respect to variational parameters $\gamma$ and $\phi$, and then, for fixed values of the variational parameters, maximizes the lower bound with respect to the model parameters $\alpha$ and $\beta$. The VEM algorithm is defined in the following:

1. (E-step) For each document, find the optimization value of the variational parameters $\{\gamma_d^*, \phi_d^* : d \in D\}$. This is done as described in the above variational inference algorithm.
2. (M-step) Maximize the resulting lower bound on the log likelihood with respect to the model parameters $\alpha$ and $\beta$. This corresponds to finding the maximum likelihood estimates with expected sufficient statistics for each document under the approximate posterior which is computed in the E-step. The update for the conditional multinomial parameter $\beta$ can be written out analytically as:

$$\beta_{ij} \propto \sum_{d=1}^{M} \sum_{n=1}^{N_d} \phi_{dni}^* w_{dn}^j$$

The last concern about LDA is to make sure that sparsity does not make the likelihood zero, an extended graphical model with prior on $\beta$, where $\beta$ is a k*V random matrix(k number of topics and V number of features, a row for each component), with independence identically Dirichlet distributed with parameter $\eta$ rows assumption. Now $\beta_i$ can be treated as a random variable to be endowed to the posterior distribution of hidden variables, giving us the following variational distribution with independence assumption:

$$q(\beta_{1:M}, z_{1:M}, \theta_{1:M} \mid \lambda, \gamma, \phi) = \prod_{i=1}^{k} Dir(\beta_i \mid \lambda_i) \prod_{d=1}^{M} q_d(\theta_d, z_d \mid \gamma_d, \phi_d)$$

To account for this modification, I only needed to change the variational inference algorithm by augmenting the following update of variational parameter $\lambda$ as follows:

$$\lambda_{ij} = \eta + \sum_{d=1}^{M} \sum_{n=1}^{N_d} \phi_{dni}^{*} w_{dn}^{j}$$

This equation finalizes the plot of VEM algorithm to estimate an LDA model. There is an alternative approach proposed by Phan et al. (2008) that uses Gibbs sampling to estimate an LDA model. This approach draws from the posterior distribution of p(z|w) by sampling as follows:

$$p(z_i = K \mid w, z_{-i}) \propto \frac{n_{-i,K}^{(j)} + \delta}{n_{-i,K}^{(.)} + V\delta} \frac{n_{-i,K}^{(d_i)} + \alpha}{n_{-i,.}^{(d_i)} + k\alpha}$$

where $z_{-i}$ is the vector of current topic memberships of all words without the i’th word $w_i$. The index j indicates that $w_i$ is equal to the j’th term in the vocabulary. $n_{-i,K}^{(j)}$ gives how often the j’th term of the vocabulary is currently assigned to topic K without the i'th word, and the dot implies the summation over all relevant index instances. $d_i$ indicates the document in the collection to which the word $w_i$ belongs to. In this Bayesian formulation $\delta$ and $\alpha$ are the prior parameters for the term distribution of topics $\beta$ and the topic distribution of documents $\theta$, respectively. The predictive distribution of the parameter $\theta$ and $\beta$ given w and z are given by:

$$\hat{\beta}_K^{(j)} = \frac{n_{-i,K}^{(j)} + \delta}{n_{-i,K}^{(.)} + V\delta}$$

$$\hat{\theta}_K^{(d)} = \frac{n_{-i,K}^{(d_i)} + \alpha}{n_{-i,.}^{(d_i)} + k\alpha}$$

The likelihood for the Gibbs sampling also has the following form:

$$\log(p(w \mid z)) = k \log(\frac{\Gamma(V\delta)}{\Gamma(\delta)^V}) + \sum_{K=1}^{k} \{[\sum_{j=1}^{V} \log(\Gamma(n_K^{(j)} + \delta))] - \log(\Gamma(n_K^{(.)} + V\delta))\}$$

**APPENDIX 2.B: K-MEANS CLUSTERING**

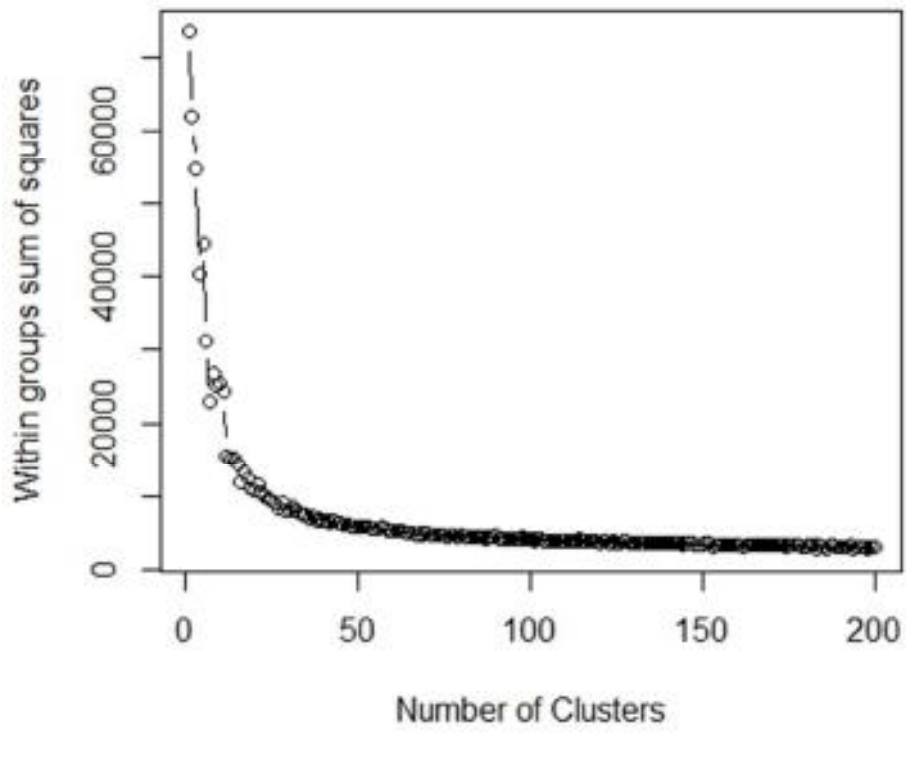

Figure 2.B.1. Within groups sum of square based on number of clusters in K-Means algorithm for bidders

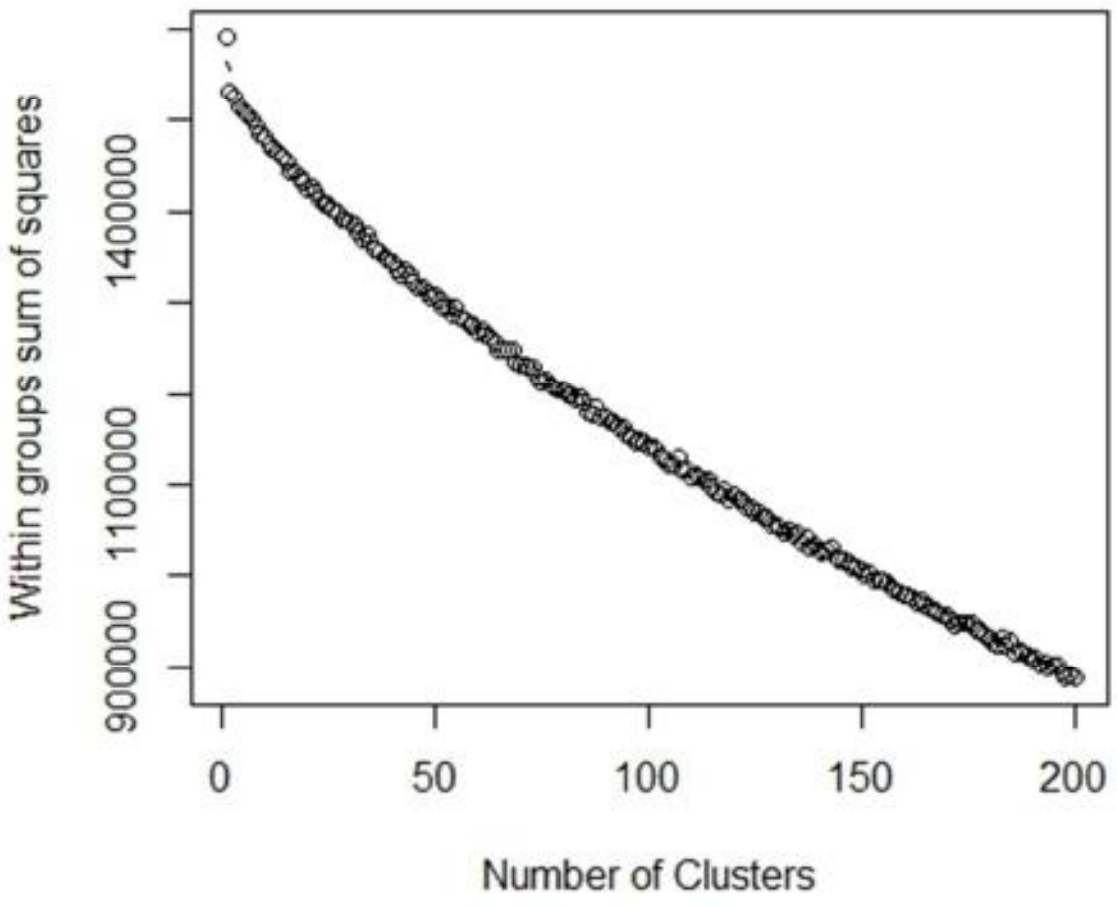

Figure 2.B.2. Within groups' sum of square based on number of clusters in K-Means algorithm for auctions

Table 2.B.1. Cluster center comparison between k-mean and mixture normal fuzzy clustering

| | Segment Size | Bidders Feedback mean | STD (Bidder's feedback) | Number of Bids on This item | STD(NBTI) | total number of bids in 30 days | STD (TNB30D) | Number of items bided on in 30 days | STD (NIB30D) | Bid activity with current Seller | STD(BACS) | Number of Cat bid on | STD(NCBO) |
|---|---|---|---|---|---|---|---|---|---|---|---|---|---|
| **k-means approach** | | | | | | | | | | | | | |
| mean | 261 | 5913 | 1531 | 9 | 3 | 867 | 307 | 456 | 134 | 25 | 11 | 2 | 1 |
| STD | 285 | 19587 | 2330 | 14 | 3 | 1569 | 408 | 989 | 169 | 28 | 8 | 2 | 1 |
| **Mixture normal Fuzzy clustering** | | | | | | | | | | | | | |
| mean | 261 | 3471 | 3635 | 9 | 7 | 680 | 631 | 258 | 267 | 24 | 19 | 2 | 1 |
| STD | 275 | 12400 | 8480 | 8 | 7 | 1073 | 1066 | 413 | 450 | 20 | 12 | 1 | 1 |

## APPENDIX 2.C: ESTIMATION PROCEDURE

We summarize the Monte Carlo (Generalized) Expectation Maximization algorithm using pseudo code.

[*Outline of the algorithm*]

Parameters to estimate:

$\Psi_j = (\gamma_j, \tau_j, \iota_j, \eta_j)$ : Auction specific, $\Theta_i = (\alpha_i, \beta_i, \delta_i, \rho_i)$ : Bidder specific

$\Sigma_j = (\sigma_{vj}, \sigma_{wj}, \sigma^1{}_{\zeta j}, \sigma^1{}_{\xi j}, \sigma^2{}_{\zeta j}, \sigma^2{}_{\xi j})$ : Variance of state space

Clustering step: eBay-specified auction cluster indices are denoted by $clus_j$. Bidders are clustered using mixture normal fuzzy clustering to extract $ind_i$ : index of membership of bidder i in bidder segment.

Generalized E-M algorithm:

Step 0: Initialize all parameters to estimate $\Psi$, $\Theta$, and $\Sigma$

E-Step:

- Compute weighted least square to estimate $(\hat{b}_{bidder}, \Sigma_k^{bidder}, \hat{b}_{auction}, \Sigma_k^{auction})$
- Compute prior over bidder and auction specific parameters
- Compute expected likelihood function using *Kalman forward filtering* and *backward smoothing* to estimate the distribution of state parameters. Then use *Monte Carlo Sampling* to integrate over latent state.

M-Step: Improve the expected likelihood function w.r.t $(\Psi, \Theta, \Sigma)$ using simulated annealing method and return to step 1a.

[*Details of the algorithm*]

*[Input data]:* a sequence of bid $b_{it}$ of individual $i = 1,...,I$ on t'th bid $t = 1,...,T$ within each auction $j = 1,...,J$, and a vector of cross sectional information about each bidder $d_i$ for each bidder, and a vector of cross sectional information about each auction $d_j$ for each auction.

*[Preprocessing]:*

1. eBay-specified auction cluster indices are denoted by $clus_j$.
2. Identify the segment of each of bidders by estimating mixture normal fuzzy clustering, specified in equation (15). With the following EM algorithm:
   [**E-step]**: Compute "expected" segment of all bidders for each segment by evaluating the Gaussian density of the bidder i's data for each segment:
   $$P(ind_i \mid \mu, \Sigma, \pi) = \frac{P(d_i \mid \mu_{ind_i}, \Sigma_{ind_i})\pi_{ind_i}}{\sum_{k=1}^{K} P(d_i \mid \mu_k, \Sigma_k)\pi_k}$$
   [**M-Step**]: Compute maximum likelihood of the model given the data's class membership distribution:
   $$\mu_k = \frac{\sum_{i=1}^{I} P(ind_i = k \mid \mu, \Sigma, \pi) d_i}{\sum_{i=1}^{I} P(ind_i \mid \mu, \Sigma, \pi)}$$
   $$\Sigma_k = \frac{\sum_{i=1}^{I} P(ind_i = k \mid \mu, \Sigma, \pi)[d_i - \mu_k][d_i - \mu_k]^T}{\sum_{i=1}^{I} P(ind_i \mid \mu, \Sigma, \pi)}$$
   $$\pi_k = \frac{\sum_{i=1}^{I} P(ind_i = k \mid \mu, \Sigma, \pi)}{I}$$
   The output of this algorithm after convergence is $ind_i$ which is the segment of bidder i.

3. Set initial value for the following parameters vectors:
   *[Auction specific parameters]*
   $\tau_j = (\tau_1,...,\tau_J), \gamma = (\gamma_1,...,\gamma_J), \iota = (\iota_1,...,\iota_J), \eta = (\eta_1,...,\eta_J)$
   Stacked in $\Psi_j = (\gamma_j, \tau_j, \iota_j, \eta_j)$, so $\Psi = (\Psi_1,...,\Psi_J)$
   *[Bidder specific parameters]*
   $\alpha = (\alpha_1,...,\alpha_I), \beta = (\beta_1,...,\beta_I), \delta = (\delta_1,...,\delta_I), \rho = (\rho_1,...,\rho_I)$
   Stacked in $\Theta_i = (\alpha_i, \beta_i, \delta_i, \rho_i)$ so $\Theta = (\Theta_1,...,\Theta_I)$
   *[Variance of the state space equations]*

$\sigma_v = (\sigma_{1v},...,\sigma_{Jv}), \sigma_w = (\sigma_{1w},...,\sigma_{Jw})$

$\sigma^1{}_\zeta = (\sigma^1{}_{1\zeta},...,\sigma^1{}_{J\zeta}), \sigma^1{}_\xi = (\sigma^1{}_{1\xi},...,\sigma^1{}_{J\xi})$

$\sigma^2{}_\zeta = (\sigma^2{}_{1\zeta},...,\sigma^2{}_{J\zeta}), \sigma^2{}_\xi = (\sigma^2{}_{1\xi},...,\sigma^2{}_{J\xi})$

Stacked in $\Sigma_j = (\sigma_{vj}, \sigma_{wj}, \sigma^1{}_{\zeta j}, \sigma^1{}_{\xi j}, \sigma^2{}_{\zeta j}, \sigma^2{}_{\xi j})$ so $\Sigma = (\Sigma_1,...,\Sigma_J)$

*[Main procedure to maximize a posteriori]*

1. Compute the prior on the auction specific parameters:

$$\hat{b}_{auction} = \left( \sum_{k=1}^{K} \left( \frac{d^T_{clust(j)=k} d_{clust(j)=k}}{\Sigma_k^{auction}} \right) \right)^{-1} \left( \sum_{k=1}^{K} \left( \frac{d^T_{clust(j)=k} \Psi_{clust(j)=k}}{\Sigma_k^{auction}} \right) \right)$$

Then prior is defined as follows:

$$P_{Norm}(\Psi_j^{clust(j)=k} \mid \hat{b}_{auction}, \Sigma_k^{auction}, d_j)$$

2. Compute prior on the bidder specific parameters:

$$\hat{b}_{bidders} = \left( \sum_{k=1}^{K} \left( \frac{d^T_{ind(i)=k} d_{ind(i)=k}}{\Sigma_k^{bidder}} \right) \right)^{-1} \left( \sum_{k=1}^{K} \left( \frac{d^T_{ind(i)=k} \Theta_{ind(i)=k}}{\Sigma_k^{bidder}} \right) \right)$$

Then prior is defined as follows:

$$P_{Norm}(\Theta_i^{ind(i)=k} \mid \hat{b}_{bidders}, \Sigma_k^{bidder}, d_i)$$

3. Compute the likelihood contribution of belief of bidder about the bids:

For j := 1, …, J do:

*[Kalman Filter on the evolution of bids in equation 4 and 5]*

For t := 1,…,T do:

*[Time updating (Prediction)]*

Project state ahead of a step ahead

$$\bar{\theta}^-_{jt} = \tau_j \bar{\theta}_{jt-1} + \gamma_j$$

Project the error covariance matrix a head

$$V_t(\theta_{jt})^- = \tau_j V_{t-1}(\theta_{jt-1})\tau_j' + \sigma_{jw}$$

*[Measurement update (Correction)]*

Compute the Kalman gain

$$K_{jt} = \frac{V(\theta_{jt-1})^-}{\left(V(\theta_{jt-1})^- + \sigma_{jv}\right)}$$

Compute estimate with measurement:

$$\bar{\theta}_{jt} = \bar{\theta}^-_{jt} + K_{jt}(b_{jt} - \bar{\theta}^-_{jt})$$

Update the error covariance:

$$V_t(\theta_{jt}) = (I - K_{jt})V_t(\theta_{jt})^-$$

EndFor

For t := T,…,0 do:

*[Backward Smoothing]*

Correction factor

$$C_{jt} = \frac{\tau_j V(\theta_{jt})}{V(\theta_{jt+1})}$$

Correct estimate with step ahead state prediction:

$$\bar{\theta}_{jt}^{+} = \bar{\theta}_{jt} + C_{jt}(\bar{\theta}_{jt+1}^{+} - \bar{\theta}_{jt+1})$$

Update the error covariance:

$$V_t(\theta_{jt})^{+} = V_t(\theta_{jt}) + C_{jt}\left(V_{t+1}(\theta_{jt+1})^{+} - V_{t+1}(\theta_{jt+1})\right)C_{jt}^{T}$$

EndFor

EndFor

*[Monte Carlo E-Step]*

From the time varying distribution of states draw S sample points

Compute the following likelihood contribution of the belief about bids based on the draws, by integrating out the latent state:

For j : = 1, …, J do:

$$\int P_{Norm}(b_{jt} \mid \theta_{jt}, \sigma_{jv}) \times P_{Norm}(\theta_{jt} \mid \theta_{jt-1}, \sigma_{jv}, \tau_j, \gamma_j) d\theta_{j1}, ..., \theta_{jT_j}$$

EndFor

4. Compute the likelihood contribution of belief of bidder about the number of bidders: Apply *Kalman Filter* and *backward smoothing* on the evolution of bids in equation (8) and (9). Then, apply the *Monte Carlo E-Step* (the pseudocode is similar to part (3), so I skip it here)

5. Compute the likelihood contribution of the evolution of valuation:
*[Invert the latent bids to recover a measure of valuation]*
For j : = 1, …, J do:

$$v_{it} = E_{\theta}\left[\frac{G_{t-1}(\theta_{it}) + \theta_{it} g_{t-1}(\theta_{it}) + \alpha_i G_{t-1}(\theta_{it}) - \alpha_i \theta_{it} g_{t-1}(\theta_{it}) + (\alpha_i + \beta_i)\theta_{it} g_{t-1}(\theta_{it})}{g_{t-1}(\theta_{it}) + \beta_i g_{t-1}(\theta_{it})}\right]$$

*An adaptive quadrature algorithm can be used to run the following integration:*

$$v_{it} = \int \left[ \frac{G_{t-1}(\theta_{it}) + \theta_{it} g_{t-1}(\theta_{it}) + \alpha_i G_{t-1}(\theta_{it}) - \alpha_i \theta_{it} g_{t-1}(\theta_{it}) + (\alpha_i + \beta_i)\theta_{it} g_{t-1}(\theta_{it})}{g_{t-1}(\theta_{it}) + \beta_i g_{t-1}(\theta_{it})} \right] f_{t-1}(\theta_{it}) d\theta_{it}$$

EndFor

Apply *Kalman Filter* and *backward smoothing* on the evolution of bids in equation (14). Then, apply the *Monte Carlo E-Step* (the pseudocode is similar to part (3), so I skip it here):

[Generalized M-Step]

Evaluate a posteriori of the parameters given the log of priors on auction and bidder specific parameters, and log of likelihood contribution of the belief about the bids and number of bidders in each auction, by summing them up, and optimize over the following vector of parameters $\Psi$, $\Theta$, and $\Sigma$. Due to the high number of parameters and multi-modality, I use simulated annealing with adaptive cooling for this step.

## APPENDIX 2.D: EXTRA TABLES FOR THE MAIN AND ALTERNATIVE MODEL (ONLINE COMPANION)

Table 2.D.1. Bidder's characteristics within each auction category

| Bidders segment | segment size | bidder feedback score (AVG) | SD (FDBK score) | Number of bids on this item (AVG) | SD (Num. Bidder) | total number of bids in 30 days (AVG) | SD (total number of bids in 30 days) | Number of items bided On (30 days) | SD(num. items Bided on) | Average num. bidding Activity with current Seller | SD(NBCS) | N categories bided on | SD (N cat. Bided on) |
|---|---|---|---|---|---|---|---|---|---|---|---|---|---|
| Jewelry and Watches | 1550 | 524 | 3077 | 5 | 10 | 275 | 710 | 127 | 341 | 21 | 28 | 2 | 1 |
| Collectibles | 859 | 863 | 4916 | 7 | 13 | 243 | 471 | 90 | 201 | 25 | 29 | 2 | 1 |
| Clothing, Shoes and Accessories | 453 | 342 | 1178 | 5 | 8 | 163 | 505 | 95 | 379 | 28 | 33 | 2 | 1 |
| Crafts | 558 | 536 | 1185 | 4 | 6 | 175 | 589 | 74 | 178 | 33 | 34 | 2 | 1 |
| Pottery and Glass | 607 | 967 | 4721 | 5 | 8 | 195 | 447 | 90 | 235 | 26 | 28 | 3 | 1 |
| Antiques | 546 | 643 | 1089 | 5 | 9 | 213 | 477 | 109 | 241 | 24 | 30 | 3 | 1 |
| Toys and Hobbies | 744 | 920 | 5589 | 5 | 9 | 159 | 357 | 64 | 170 | 26 | 30 | 2 | 1 |
| Stamps | 651 | 1188 | 1899 | 5 | 8 | 504 | 940 | 227 | 346 | 20 | 25 | 1 | 1 |
| Books | 482 | 651 | 1612 | 5 | 6 | 171 | 803 | 74 | 355 | 32 | 33 | 2 | 1 |
| Tickets and Experiences | 489 | 574 | 1224 | 3 | 4 | 56 | 119 | 33 | 85 | 40 | 36 | 2 | 1 |
| Art | 456 | 469 | 745 | 5 | 7 | 69 | 110 | 30 | 53 | 50 | 40 | 2 | 1 |
| Gift Cards and Coupons | 522 | 818 | 1784 | 4 | 6 | 373 | 1065 | 264 | 1001 | 17 | 27 | 2 | 1 |
| Music | 585 | 1047 | 3583 | 5 | 8 | 172 | 334 | 91 | 196 | 31 | 31 | 2 | 1 |
| Consumer Electronics | 734 | 425 | 2562 | 5 | 10 | 142 | 347 | 64 | 187 | 29 | 32 | 2 | 1 |
| DVDs and Movies | 602 | 635 | 1688 | 4 | 6 | 180 | 831 | 80 | 222 | 24 | 29 | 2 | 1 |
| Dolls and Bears | 679 | 1301 | 8033 | 5 | 9 | 219 | 370 | 90 | 155 | 17 | 24 | 2 | 1 |
| Entertainment Memorabilia | 506 | 626 | 1257 | 5 | 9 | 140 | 323 | 62 | 168 | 43 | 38 | 2 | 1 |
| Health and Beauty | 541 | 480 | 2098 | 5 | 9 | 100 | 191 | 44 | 94 | 35 | 34 | 2 | 1 |
| Video Games and Consoles | 682 | 567 | 4264 | 5 | 8 | 159 | 392 | 71 | 170 | 23 | 30 | 2 | 1 |

Table 2.D.2. Maximum A Posteriori of the model

| Element of the maximum a posteriori model selection criteria | Log Likelihood |
|---|---|
| Number of bidders evolution state space model | -788,079 |
| Bid evolution with each auction state space model | -92,739,612 |
| Valuation evolution state space model | -627,982 |
| prior on the auctions parameters | -16,854 |
| prior on the bidders parameters | -107,558 |

Table 2.D.3. Bidder's segment profile after mixture normal clustering

| Bidders 'segment Index | Segment Size | Bidders Feedback mean | STD (Bidder's feedback) | Number of Bids on This item | STD(NBTI) | total number of bids in 30 days | STD (TNB30D) | Number of items bided on in 30 days | STD (NIB30D) | Bid activity with current Seller | STD(BACS) | Number of categories Bided on Mean | STD(NCBO) |
|---|---|---|---|---|---|---|---|---|---|---|---|---|---|
| 1 | 215 | 70 | 68 | 15 | 17 | 211 | 151 | 48 | 38 | 19 | 24 | 2 | 1 |
| 2 | 23 | 630 | 606 | 7 | 8 | 3266 | 2604 | 856 | 715 | 9 | 20 | 2 | 1 |
| 3 | 89 | 335 | 379 | 21 | 19 | 940 | 679 | 163 | 138 | 12 | 17 | 2 | 1 |
| 4 | 963 | 588 | 1148 | 1 | 0 | 233 | 692 | 148 | 382 | 20 | 30 | 2 | 1 |
| 5 | 153 | 671 | 5 | 5 | 0 | 192 | 0 | 90 | 1 | 26 | 1 | 2 | 0 |
| 6 | 466 | 221 | 331 | 5 | 6 | 27 | 26 | 10 | 10 | 45 | 34 | 2 | 1 |
| 7 | 90 | 1864 | 3782 | 3 | 1 | 252 | 215 | 160 | 154 | 12 | 22 | 2 | 1 |
| 8 | 992 | 907 | 4109 | 1 | 0 | 214 | 545 | 136 | 357 | 19 | 29 | 2 | 1 |
| 9 | 535 | 452 | 444 | 4 | 2 | 101 | 88 | 48 | 46 | 21 | 17 | 2 | 1 |
| 10 | 284 | 75 | 77 | 15 | 14 | 107 | 182 | 14 | 19 | 56 | 33 | 2 | 1 |
| 11 | 42 | 1783 | 2171 | 26 | 18 | 186 | 121 | 52 | 33 | 22 | 18 | 3 | 2 |
| 12 | 83 | 978 | 1289 | 5 | 4 | 462 | 335 | 217 | 213 | 1 | 1 | 2 | 1 |
| 13 | 12 | 19165 | 29755 | 9 | 9 | 3314 | 2474 | 1146 | 842 | 3 | 3 | 4 | 3 |
| 14 | 52 | 449 | 598 | 21 | 14 | 1243 | 740 | 209 | 114 | 12 | 19 | 2 | 1 |
| 15 | 113 | 5046 | 18209 | 23 | 26 | 1778 | 1551 | 718 | 692 | 26 | 33 | 2 | 2 |
| 16 | 466 | 435 | 398 | 4 | 2 | 87 | 80 | 40 | 38 | 36 | 26 | 2 | 1 |
| 17 | 522 | 1146 | 1751 | 9 | 12 | 404 | 348 | 159 | 174 | 23 | 27 | 2 | 1 |
| 18 | 589 | 663 | 683 | 1 | 0 | 97 | 96 | 66 | 73 | 6 | 6 | 2 | 1 |
| 19 | 310 | 173 | 190 | 2 | 2 | 7 | 7 | 4 | 4 | 59 | 36 | 1 | 0 |
| 20 | 395 | 1426 | 1867 | 9 | 12 | 286 | 237 | 139 | 135 | 19 | 26 | 3 | 2 |
| 21 | 403 | 196 | 226 | 6 | 8 | 12 | 11 | 4 | 2 | 62 | 27 | 1 | 0 |
| 22 | 530 | 608 | 646 | 4 | 2 | 102 | 72 | 47 | 36 | 32 | 30 | 2 | 1 |
| 23 | 49 | 5504 | 11521 | 10 | 18 | 2369 | 2317 | 1283 | 1631 | 18 | 32 | 2 | 2 |
| 24 | 481 | 427 | 410 | 1 | 0 | 36 | 32 | 25 | 25 | 11 | 10 | 3 | 1 |
| 25 | 142 | 59 | 48 | 3 | 3 | 3 | 3 | 1 | 1 | 100 | 0 | 1 | 0 |
| 26 | 871 | 856 | 3857 | 1 | 0 | 81 | 162 | 56 | 124 | 19 | 29 | 2 | 1 |
| 27 | 242 | 518 | 322 | 5 | 2 | 201 | 153 | 83 | 52 | 20 | 11 | 2 | 1 |
| 28 | 569 | 122 | 156 | 7 | 10 | 29 | 34 | 9 | 9 | 46 | 33 | 2 | 1 |
| 29 | 62 | 567 | 1003 | 19 | 12 | 1123 | 911 | 225 | 201 | 4 | 4 | 2 | 1 |
| 30 | 83 | 1544 | 1884 | 13 | 15 | 162 | 170 | 40 | 29 | 14 | 13 | 3 | 1 |
| 31 | 102 | 340 | 348 | 3 | 1 | 420 | 546 | 194 | 265 | 2 | 2 | 3 | 2 |
| 32 | 7 | 14621 | 22217 | 2 | 0 | 4530 | 5814 | 1081 | 997 | 4 | 9 | 3 | 2 |
| 33 | 64 | 929 | 961 | 1 | 0 | 1939 | 2030 | 1631 | 2099 | 7 | 18 | 2 | 2 |
| 34 | 163 | 655 | 85 | 5 | 2 | 209 | 93 | 93 | 26 | 26 | 6 | 2 | 0 |
| 35 | 183 | 273 | 599 | 3 | 0 | 80 | 142 | 30 | 61 | 35 | 35 | 2 | 1 |
| 36 | 198 | 91 | 105 | 14 | 12 | 26 | 24 | 5 | 5 | 76 | 24 | 2 | 1 |
| 37 | 346 | 560 | 1422 | 2 | 0 | 108 | 256 | 57 | 161 | 31 | 36 | 2 | 1 |
| 38 | 971 | 714 | 1790 | 1 | 0 | 132 | 277 | 81 | 179 | 23 | 32 | 2 | 1 |
| 39 | 65 | 527 | 1202 | 5 | 1 | 29 | 18 | 12 | 8 | 32 | 20 | 2 | 1 |
| 40 | 14 | 2268 | 2423 | 10 | 5 | 301 | 147 | 115 | 58 | 17 | 19 | 3 | 1 |
| 41 | 73 | 175 | 174 | 16 | 14 | 808 | 612 | 159 | 84 | 3 | 3 | 2 | 1 |
| 42 | 73 | 350 | 449 | 7 | 4 | 143 | 80 | 45 | 23 | 7 | 4 | 3 | 1 |
| 43 | 93 | 48 | 45 | 12 | 8 | 134 | 118 | 30 | 29 | 32 | 24 | 2 | 1 |
| 44 | 41 | 198 | 264 | 35 | 17 | 177 | 97 | 35 | 35 | 27 | 11 | 2 | 1 |
| 45 | 19 | 1780 | 1464 | 18 | 10 | 822 | 577 | 212 | 133 | 15 | 13 | 2 | 1 |
| 46 | 5 | 84027 | 45365 | 3 | 1 | 831 | 893 | 644 | 703 | 2 | 2 | 3 | 3 |
| 47 | 3 | 8113 | 4001 | 11 | 8 | 3761 | 2896 | 1486 | 1391 | 39 | 40 | 3 | 1 |

Table 2.D.4. The winner regret $\alpha_i$ estimates across bidder's segments

| Bidders Segment | Segment Size | Estimate | STE | t-stat | p-value |
|---|---|---|---|---|---|
| Segment 1 | 215 | -1.36*** | 0.06 | -22.75 | <0.0001 |
| Segment 2 | 23 | -1.19*** | 0.19 | -6.32 | <0.0001 |
| Segment 3 | 89 | -1.31*** | 0.09 | -14.02 | <0.0001 |
| Segment 4 | 963 | -1.35*** | 0.03 | -49.76 | <0.0001 |
| Segment 5 | 153 | -1.33*** | 0.07 | -18.72 | <0.0001 |
| Segment 6 | 466 | -1.26*** | 0.04 | -33.82 | <0.0001 |
| Segment 7 | 90 | -1.19*** | 0.09 | -12.56 | <0.0001 |
| Segment 8 | 992 | -1.27*** | 0.03 | -45.81 | <0.0001 |
| Segment 9 | 535 | -1.33*** | 0.04 | -35.85 | <0.0001 |
| Segment 10 | 284 | -1.38*** | 0.05 | -25.37 | <0.0001 |
| Segment 11 | 42 | -1.67*** | 0.13 | -13.34 | <0.0001 |
| Segment 12 | 83 | -1.35*** | 0.09 | -14.82 | <0.0001 |
| Segment 13 | 12 | -1.13*** | 0.23 | -4.81 | <0.001 |
| Segment 14 | 52 | -1.16*** | 0.11 | -10.74 | <0.0001 |
| Segment 15 | 113 | -1.21*** | 0.08 | -15.86 | <0.0001 |
| Segment 16 | 466 | -1.35*** | 0.04 | -32.58 | <0.0001 |
| Segment 17 | 522 | -1.35*** | 0.04 | -38.33 | <0.0001 |
| Segment 18 | 589 | -1.32*** | 0.04 | -37.02 | <0.0001 |
| Segment 19 | 310 | -1.38*** | 0.05 | -27.72 | <0.0001 |
| Segment 20 | 395 | -1.28*** | 0.04 | -30.05 | <0.0001 |
| Segment 21 | 403 | -1.22*** | 0.04 | -29.13 | <0.0001 |
| Segment 22 | 530 | -1.30*** | 0.04 | -32.94 | <0.0001 |
| Segment 23 | 49 | -1.42*** | 0.12 | -12.18 | <0.0001 |
| Segment 24 | 481 | -1.36*** | 0.04 | -34.85 | <0.0001 |
| Segment 25 | 142 | -1.32*** | 0.07 | -17.62 | <0.0001 |
| Segment 26 | 871 | -1.28*** | 0.03 | -45.48 | <0.0001 |
| Segment 27 | 242 | -1.25*** | 0.06 | -22.10 | <0.0001 |
| Segment 28 | 569 | -1.30*** | 0.04 | -35.25 | <0.0001 |
| Segment 29 | 62 | -1.48*** | 0.11 | -12.93 | <0.0001 |
| Segment 30 | 83 | -1.46*** | 0.09 | -16.57 | <0.0001 |
| Segment 31 | 102 | -1.36*** | 0.09 | -14.97 | <0.0001 |
| Segment 32 | 7 | -0.87* | 0.38 | -2.30 | 0.027437 |
| Segment 33 | 64 | -1.43*** | 0.13 | -11.08 | <0.0001 |
| Segment 34 | 163 | -1.30*** | 0.07 | -19.08 | <0.0001 |
| Segment 35 | 183 | -1.28*** | 0.06 | -21.78 | <0.0001 |
| Segment 36 | 198 | -1.27*** | 0.06 | -20.19 | <0.0001 |
| Segment 37 | 346 | -1.31*** | 0.05 | -28.41 | <0.0001 |
| Segment 38 | 971 | -1.32*** | 0.03 | -47.30 | <0.0001 |
| Segment 39 | 65 | -1.34*** | 0.10 | -13.31 | <0.0001 |
| Segment 40 | 14 | -1.34*** | 0.19 | -7.05 | <0.0001 |
| Segment 41 | 73 | -1.33*** | 0.10 | -13.87 | <0.0001 |
| Segment 42 | 73 | -1.29*** | 0.08 | -15.56 | <0.0001 |
| Segment 43 | 93 | -1.28*** | 0.09 | -13.94 | <0.0001 |
| Segment 44 | 41 | -1.41*** | 0.13 | -10.50 | <0.0001 |
| Segment 45 | 19 | -1.33*** | 0.15 | -8.87 | <0.0001 |
| Segment 46 | 5 | -0.52* | 0.23 | -2.26 | 0.036761 |
| Segment 47 | 3 | -0.52 | 0.41 | -1.28 | 0.145362 |

* p<0.1, ** p<0.05, ***p<0.001

Table 2.D.5. The loser regret $\beta_i$ estimates across bidder's segments

| Bidders Segment | Segment Size | Estimate | STE | t-stat | p-value |
|---|---|---|---|---|---|
| Segment 1 | 215 | -1.34*** | 0.06 | -22.14 | <0.0001 |
| Segment 2 | 23 | -1.11*** | 0.19 | -5.69 | <0.0001 |
| Segment 3 | 89 | -1.39*** | 0.09 | -15.53 | <0.0001 |
| Segment 4 | 963 | -1.33*** | 0.03 | -47.22 | <0.0001 |
| Segment 5 | 153 | -1.37*** | 0.07 | -20.33 | <0.0001 |
| Segment 6 | 466 | -1.33*** | 0.04 | -32.76 | <0.0001 |
| Segment 7 | 90 | -1.15*** | 0.09 | -12.22 | <0.0001 |
| Segment 8 | 992 | -1.35*** | 0.03 | -48.69 | <0.0001 |
| Segment 9 | 535 | -1.31*** | 0.04 | -35.55 | <0.0001 |
| Segment 10 | 284 | -1.48*** | 0.05 | -29.18 | <0.0001 |
| Segment 11 | 42 | -1.55*** | 0.13 | -12.03 | <0.0001 |
| Segment 12 | 83 | -1.45*** | 0.09 | -15.86 | <0.0001 |
| Segment 13 | 12 | -1.46*** | 0.31 | -4.73 | <0.001 |
| Segment 14 | 52 | -1.29*** | 0.11 | -11.27 | <0.0001 |
| Segment 15 | 113 | -1.51*** | 0.08 | -18.92 | <0.0001 |
| Segment 16 | 466 | -1.26*** | 0.04 | -34.61 | <0.0001 |
| Segment 17 | 522 | -1.33*** | 0.04 | -36.52 | <0.0001 |
| Segment 18 | 589 | -1.28*** | 0.04 | -34.91 | <0.0001 |
| Segment 19 | 310 | -1.32*** | 0.05 | -26.63 | <0.0001 |
| Segment 20 | 395 | -1.36*** | 0.05 | -29.05 | <0.0001 |
| Segment 21 | 403 | -1.34*** | 0.05 | -29.24 | <0.0001 |
| Segment 22 | 530 | -1.24*** | 0.04 | -33.68 | <0.0001 |
| Segment 23 | 49 | -1.47*** | 0.14 | -10.88 | <0.0001 |
| Segment 24 | 481 | -1.36*** | 0.04 | -34.65 | <0.0001 |
| Segment 25 | 142 | -1.34*** | 0.07 | -19.36 | <0.0001 |
| Segment 26 | 871 | -1.36*** | 0.03 | -45.34 | <0.0001 |
| Segment 27 | 242 | -1.41*** | 0.05 | -26.47 | <0.0001 |
| Segment 28 | 569 | -1.29*** | 0.04 | -36.43 | <0.0001 |
| Segment 29 | 62 | -1.30*** | 0.11 | -11.96 | <0.0001 |
| Segment 30 | 83 | -1.31*** | 0.11 | -11.71 | <0.0001 |
| Segment 31 | 102 | -1.33*** | 0.08 | -16.93 | <0.0001 |
| Segment 32 | 7 | -1.70*** | 0.27 | -6.20 | <0.001 |
| Segment 33 | 64 | -1.51*** | 0.10 | -14.77 | <0.0001 |
| Segment 34 | 163 | -1.38*** | 0.07 | -18.55 | <0.0001 |
| Segment 35 | 183 | -1.21*** | 0.06 | -19.29 | <0.0001 |
| Segment 36 | 198 | -1.36*** | 0.06 | -22.33 | <0.0001 |
| Segment 37 | 346 | -1.33*** | 0.04 | -30.24 | <0.0001 |
| Segment 38 | 971 | -1.35*** | 0.03 | -47.39 | <0.0001 |
| Segment 39 | 65 | -1.32*** | 0.11 | -12.53 | <0.0001 |
| Segment 40 | 14 | -1.27*** | 0.26 | -4.92 | <0.001 |
| Segment 41 | 73 | -1.30*** | 0.10 | -13.57 | <0.0001 |
| Segment 42 | 73 | -1.40*** | 0.10 | -14.16 | <0.0001 |
| Segment 43 | 93 | -1.32*** | 0.10 | -13.39 | <0.0001 |
| Segment 44 | 41 | -1.51*** | 0.11 | -13.92 | <0.0001 |
| Segment 45 | 19 | -1.30*** | 0.18 | -7.41 | <0.0001 |
| Segment 46 | 5 | -0.79 | 0.49 | -1.63 | 0.082383 |
| Segment 47 | 3 | -1.23** | 0.32 | -3.85 | <0.05 |

* p<0.1, ** p<0.05, ***p<0.001

Table 2.D.6. The update of valuation parameters $\delta_i$ and learning parameter $\rho_i$ estimates across bidder's segments

| Bidders Segment | Segment Size | Valuation revelation $\delta_i$ | STE ($\delta_i$) | Learning $\rho_i$ | STE ($\rho_i$) |
|---|---|---|---|---|---|
| Segment 1 | 215 | 1.21*** | 0.05 | 0.15** | 0.06 |
| Segment 2 | 23 | 0.79*** | 0.09 | 0.21 | 0.15 |
| Segment 3 | 89 | 1.23*** | 0.08 | 0.25*** | 0.10 |
| Segment 4 | 963 | 1.22*** | 0.03 | 0.26*** | 0.03 |
| Segment 5 | 153 | 1.30*** | 0.07 | 0.16** | 0.07 |
| Segment 6 | 466 | 1.20*** | 0.04 | 0.25*** | 0.04 |
| Segment 7 | 90 | 1.25*** | 0.08 | 0.28*** | 0.09 |
| Segment 8 | 992 | 1.25*** | 0.02 | 0.27*** | 0.03 |
| Segment 9 | 535 | 1.25*** | 0.03 | 0.18*** | 0.04 |
| Segment 10 | 284 | 1.22*** | 0.05 | 0.28*** | 0.05 |
| Segment 11 | 42 | 1.33*** | 0.11 | 0.27** | 0.13 |
| Segment 12 | 83 | 1.33*** | 0.08 | 0.07 | 0.09 |
| Segment 13 | 12 | 1.37*** | 0.23 | 0.48* | 0.26 |
| Segment 14 | 52 | 1.15*** | 0.10 | 0.39*** | 0.12 |
| Segment 15 | 113 | 1.22*** | 0.07 | 0.32*** | 0.08 |
| Segment 16 | 466 | 1.22*** | 0.04 | 0.24*** | 0.04 |
| Segment 17 | 522 | 1.20*** | 0.03 | 0.28*** | 0.04 |
| Segment 18 | 589 | 1.26*** | 0.03 | 0.27*** | 0.04 |
| Segment 19 | 310 | 1.22*** | 0.04 | 0.23*** | 0.05 |
| Segment 20 | 395 | 1.20*** | 0.04 | 0.24*** | 0.04 |
| Segment 21 | 403 | 1.24*** | 0.04 | 0.26*** | 0.04 |
| Segment 22 | 530 | 1.20*** | 0.03 | 0.16*** | 0.04 |
| Segment 23 | 49 | 1.26*** | 0.12 | 0.21* | 0.12 |
| Segment 24 | 481 | 1.21*** | 0.03 | 0.23*** | 0.04 |
| Segment 25 | 142 | 1.29*** | 0.06 | 0.12* | 0.06 |
| Segment 26 | 871 | 1.27*** | 0.03 | 0.24*** | 0.03 |
| Segment 27 | 242 | 1.18*** | 0.04 | 0.40*** | 0.06 |
| Segment 28 | 569 | 1.23*** | 0.03 | 0.29*** | 0.04 |
| Segment 29 | 62 | 1.33*** | 0.09 | 0.05 | 0.14 |
| Segment 30 | 83 | 1.40*** | 0.09 | 0.29*** | 0.10 |
| Segment 31 | 102 | 1.13*** | 0.08 | 0.39*** | 0.09 |
| Segment 32 | 7 | 1.04*** | 0.09 | 0.19 | 0.20 |
| Segment 33 | 64 | 1.25*** | 0.10 | 0.25** | 0.10 |
| Segment 34 | 163 | 1.21*** | 0.06 | 0.33*** | 0.07 |
| Segment 35 | 183 | 1.27*** | 0.06 | 0.26*** | 0.07 |
| Segment 36 | 198 | 1.24*** | 0.06 | 0.14** | 0.06 |
| Segment 37 | 346 | 1.22*** | 0.04 | 0.25*** | 0.05 |
| Segment 38 | 971 | 1.19*** | 0.02 | 0.25*** | 0.03 |
| Segment 39 | 65 | 1.13*** | 0.09 | 0.28*** | 0.09 |
| Segment 40 | 14 | 1.28*** | 0.24 | 0.50* | 0.25 |
| Segment 41 | 73 | 1.25*** | 0.08 | 0.30*** | 0.10 |
| Segment 42 | 73 | 1.17*** | 0.09 | 0.39*** | 0.10 |
| Segment 43 | 93 | 1.20*** | 0.07 | 0.25*** | 0.09 |
| Segment 44 | 41 | 1.10*** | 0.12 | 0.33** | 0.14 |
| Segment 45 | 19 | 1.42*** | 0.17 | 0.28* | 0.15 |
| Segment 46 | 5 | 1.07*** | 0.24 | 0.81*** | 0.18 |
| Segment 47 | 3 | 1.17** | 0.27 | -0.08 | 0.26 |

* p<0.1, ** p<0.05, ***p<0.001

Table 2.D.7.The winner regret $\alpha_i$ and the loser regret $\beta_i$ estimates across auction categories

| Auction Category | number of bidders | winner regret | STE (winner regret) | t-stat (winner regret) | p-value (winner regret) | Loser regret | STE (Loser regret) | t-stat (Loser regret) | p-value (Loser regret) |
|---|---|---|---|---|---|---|---|---|---|
| Jewelry and Watches | 1550 | -1.32*** | 0.02 | -60.59 | <0.0001 | -1.32*** | 0.02 | -60.33 | <0.0001 |
| Collectibles | 859 | -1.35*** | 0.03 | -46.96 | <0.0001 | -1.38*** | 0.03 | -47.07 | <0.0001 |
| Clothing, Shoes and Accessories | 453 | -1.24*** | 0.04 | -29.94 | <0.0001 | -1.40*** | 0.04 | -35.06 | <0.0001 |
| Crafts | 558 | -1.33*** | 0.04 | -37.27 | <0.0001 | -1.33*** | 0.04 | -34.27 | <0.0001 |
| Pottery and Glass | 607 | -1.38*** | 0.04 | -38.98 | <0.0001 | -1.35*** | 0.04 | -38.37 | <0.0001 |
| Antiques | 546 | -1.28*** | 0.04 | -33.61 | <0.0001 | -1.30*** | 0.04 | -33.74 | <0.0001 |
| Toys and Hobbies | 744 | -1.27*** | 0.03 | -40.54 | <0.0001 | -1.35*** | 0.03 | -41.96 | <0.0001 |
| Stamps | 651 | -1.38*** | 0.03 | -41.66 | <0.0001 | -1.32*** | 0.03 | -41.22 | <0.0001 |
| Books | 482 | -1.37*** | 0.04 | -35.47 | <0.0001 | -1.31*** | 0.04 | -33.14 | <0.0001 |
| Tickets and Experiences | 489 | -1.26*** | 0.04 | -32.47 | <0.0001 | -1.28*** | 0.04 | -34.57 | <0.0001 |
| Art | 456 | -1.25*** | 0.04 | -31.10 | <0.0001 | -1.28*** | 0.04 | -31.00 | <0.0001 |
| Gift Cards and Coupons | 522 | -1.32*** | 0.04 | -32.59 | <0.0001 | -1.30*** | 0.04 | -31.89 | <0.0001 |
| Music | 585 | -1.32*** | 0.04 | -36.80 | <0.0001 | -1.34*** | 0.04 | -35.76 | <0.0001 |
| Consumer Electronics | 734 | -1.30*** | 0.03 | -42.21 | <0.0001 | -1.36*** | 0.03 | -42.92 | <0.0001 |
| DVDs and Movies | 602 | -1.28*** | 0.04 | -35.48 | <0.0001 | -1.37*** | 0.03 | -39.65 | <0.0001 |
| Dolls and Bears | 679 | -1.29*** | 0.03 | -40.39 | <0.0001 | -1.37*** | 0.03 | -41.79 | <0.0001 |
| Entertainment Memorabilia | 506 | -1.37*** | 0.04 | -36.04 | <0.0001 | -1.31*** | 0.04 | -34.57 | <0.0001 |
| Health and Beauty | 541 | -1.31*** | 0.04 | -35.79 | <0.0001 | -1.36*** | 0.04 | -35.91 | <0.0001 |
| Video Games and Consoles | 682 | -1.29*** | 0.03 | -38.58 | <0.0001 | -1.32*** | 0.03 | -39.19 | <0.0001 |

* p<0.1, **p<0.05, ***p<0.01

Table 2.D.8. The update of valuation parameters $\delta_i$ and learning parameter $\rho_i$ estimates across bidder's segments

| Bidders segment | number of bidders | Learning value from bid parameter ($\rho_i$) | STE ($\rho_i$) | Valuation revelation parameter ($\delta_i$) | STE ($\delta_i$) |
|---|---|---|---|---|---|
| Jewelry and Watches | 1550 | 0.25*** | 0.02 | 1.22*** | 0.02 |
| Collectibles | 859 | 0.26*** | 0.03 | 1.25*** | 0.03 |
| Clothing, Shoes and Accessories | 453 | 0.20*** | 0.04 | 1.23*** | 0.04 |
| Crafts | 558 | 0.25*** | 0.04 | 1.21*** | 0.03 |
| Pottery and Glass | 607 | 0.26*** | 0.03 | 1.26*** | 0.03 |
| Antiques | 546 | 0.24*** | 0.04 | 1.20*** | 0.03 |
| Toys and Hobbies | 744 | 0.25*** | 0.03 | 1.24*** | 0.03 |
| Stamps | 651 | 0.25*** | 0.03 | 1.24*** | 0.03 |
| Books | 482 | 0.22*** | 0.04 | 1.25*** | 0.03 |
| Tickets and Experiences | 489 | 0.22*** | 0.04 | 1.22*** | 0.03 |
| Art | 456 | 0.18*** | 0.04 | 1.19*** | 0.04 |
| Gift Cards and Coupons | 522 | 0.29*** | 0.04 | 1.26*** | 0.03 |
| Music | 585 | 0.23*** | 0.03 | 1.19*** | 0.03 |
| Consumer Electronics | 734 | 0.25*** | 0.03 | 1.23*** | 0.03 |
| DVDs and Movies | 602 | 0.29*** | 0.04 | 1.23*** | 0.03 |
| Dolls and Bears | 679 | 0.32*** | 0.03 | 1.28*** | 0.03 |
| Entertainment Memorabilia | 506 | 0.24*** | 0.04 | 1.17*** | 0.03 |
| Health and Beauty | 541 | 0.24*** | 0.04 | 1.24*** | 0.03 |
| Video Games and Consoles | 682 | 0.24*** | 0.03 | 1.19*** | 0.03 |

* $p<0.1$, ** $p<0.05$, *** $p<0.01$

Table 2.D.9. The growth of bids and their drift parameters, $\tau_i$ and $\gamma_i$, and the rush of bidders at the end of auction rate and average entrance rate in each period, $\eta_j$ and $\iota_j$, estimates across auction segments

| Auction Cluster | Auction Cluster Size | growth of bids ($\tau_i$) | STE ($\tau_i$) | Drift of bids ($\gamma_i$) | STE ($\gamma_i$) | Last minute flood ($\eta_j$) | STE ($\eta_j$) | Mean entrance rate ($\iota_j$) | STE ($\iota_j$) |
|---|---|---|---|---|---|---|---|---|---|
| Jewelry and Watches | 150 | 1.65*** | 0.07 | 5.44*** | 0.08 | 1.01*** | 0.06 | 1.84*** | 0.08 |
| Collectibles | 104 | 1.63*** | 0.09 | 5.63*** | 0.10 | 0.93*** | 0.06 | 1.90*** | 0.09 |
| Clothing, Shoes and Accessories | 85 | 1.86*** | 0.09 | 5.56*** | 0.10 | 0.93*** | 0.07 | 2.11*** | 0.08 |
| Crafts | 79 | 1.81*** | 0.11 | 5.48*** | 0.10 | 0.97*** | 0.08 | 2.03*** | 0.11 |
| Pottery and Glass | 75 | 1.50*** | 0.10 | 5.47*** | 0.10 | 1.10*** | 0.10 | 2.16*** | 0.11 |
| Antiques | 69 | 1.68*** | 0.12 | 5.60*** | 0.11 | 0.95*** | 0.11 | 1.89*** | 0.13 |
| Toys and Hobbies | 94 | 1.73*** | 0.10 | 5.61*** | 0.09 | 0.90*** | 0.09 | 2.21*** | 0.10 |
| Stamps | 73 | 2.00*** | 0.12 | 5.64*** | 0.12 | 1.04*** | 0.12 | 2.08*** | 0.13 |
| Books | 85 | 1.83*** | 0.12 | 5.50*** | 0.11 | 1.13*** | 0.12 | 2.12*** | 0.12 |
| Tickets and Experiences | 92 | 1.74*** | 0.13 | 5.58*** | 0.12 | 0.95*** | 0.12 | 2.10*** | 0.13 |
| Art | 71 | 1.79*** | 0.16 | 5.52*** | 0.12 | 1.05*** | 0.16 | 2.04*** | 0.17 |
| Gift Cards and Coupons | 86 | 1.77*** | 0.15 | 5.36*** | 0.12 | 1.06*** | 0.15 | 2.14*** | 0.15 |
| Music | 87 | 1.86*** | 0.16 | 5.65*** | 0.12 | 1.15*** | 0.16 | 2.03*** | 0.16 |
| Consumer Electronics | 84 | 1.76*** | 0.17 | 5.37*** | 0.15 | 1.13*** | 0.17 | 2.13*** | 0.17 |
| DVDs and Movies | 88 | 1.82*** | 0.17 | 5.75*** | 0.14 | 1.11*** | 0.17 | 2.26*** | 0.18 |
| Dolls and Bears | 85 | 1.66*** | 0.19 | 5.81*** | 0.16 | 1.26*** | 0.19 | 2.08*** | 0.19 |
| Entertainment Memorabilia | 89 | 1.86*** | 0.19 | 5.63*** | 0.15 | 1.12*** | 0.19 | 2.06*** | 0.20 |
| Health and Beauty | 75 | 1.84*** | 0.24 | 5.77*** | 0.20 | 1.15*** | 0.24 | 2.26*** | 0.24 |
| Video Games and consoles | 94 | 1.86*** | 0.20 | 5.65*** | 0.17 | 1.04*** | 0.20 | 2.21*** | 0.20 |

* p<0.1, **p<0.05, ***p<0.01

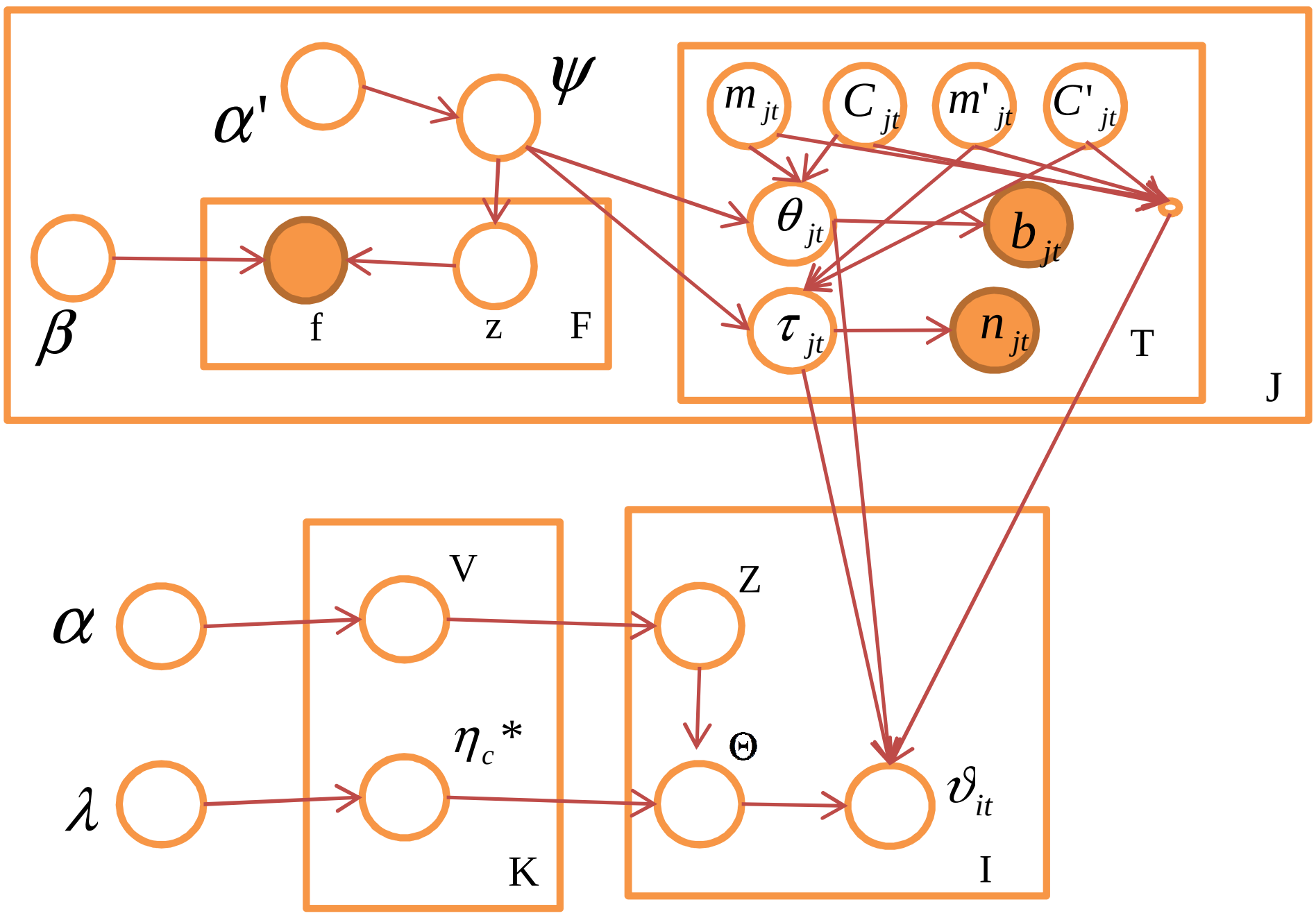

Figure 2.D.1. The probabilistic graphical plate model of the main model

### 2.D.2.1. ESTIMATION RESULTS OF THE MODEL WITH LDA-ESTIMATED AUCTION CLUSTERS

Table 2.D.10. Summary statistics for the bidder specific parameter estimations

| | within each auction category (19) | | | | within each bidder segment (47) | | | |
|---|---|---|---|---|---|---|---|---|
| Parameter | min | max | Mean | SD | min | max | Mean | SD |
| avg. winner regret | -1.37 | -1.27 | -1.33 | 0.03 | -1.9 | -1.09 | -1.35 | 0.13 |
| se winner regret | 0.02 | 0.05 | 0.04 | 0.01 | 0.03 | 0.49 | 0.11 | 0.10 |
| avg. loser regret | -1.38 | -1.25 | -1.32 | 0.03 | -1.98 | -1.08 | -1.35 | 0.17 |
| se loser regret | 0.03 | 0.05 | 0.04 | 0.005 | 0.03 | 0.5 | 0.11 | 0.10 |
| avg. valuation param. | 1.2 | 1.3 | 1.26 | 0.03 | 0.85 | 1.89 | 1.27 | 0.14 |
| se valuation param. | 0.02 | 0.04 | 0.04 | 0.005 | 0.03 | 0.54 | 0.10 | 0.10 |
| avg. learning param. | 0.16 | 0.33 | 0.25 | 0.04 | 0 | 1.33 | 0.27 | 0.19 |
| se learning param. | 0.02 | 0.04 | 0.03 | 0.01 | 0.03 | 0.78 | 0.12 | 0.14 |

Table 2.D.11. Relation between the winner regret $\alpha_i$, the loser regret $\beta_i$, the update of valuation parameters $\delta_i$ and learning parameter $\rho_i$ estimates across forty seven bidder segments

| | *winner regret* | *loser regret* | *valuation revelation* | *learning* |
|---|---|---|---|---|
| winner regret | 1 | | | |
| | 0.25 | | | |
| loser regret | | 1 | | |
| | -0.14 | 0.45 | | |
| valuation revelation | | | 1 | |
| | -0.38 | 0.03 | 0.61 | |
| learning | | | | 1 |

Table 2.D.12. Relation between the winner regret $\alpha_i$, the loser regret $\beta_i$, the update of valuation parameters $\delta_i$ and learning parameter $\rho_i$ estimates across forty seven bidder segments

| Regressand | Regressor | Estimate | SE | Lower 95% | Upper 95% |
|---|---|---|---|---|---|
| Winner regret | | | | | |
| | Intercept | -1.10** | 0.15 | -1.40 | -0.80 |
| | loser regret | 0.18* | 0.11 | -0.03 | -0.41 |
| Winner Regret | | | | | |
| | Intercept | -1.28** | 0.03 | -1.35 | -1.22 |
| | Learning | -0.25** | 0.09 | -0.45 | -0.07 |
| Loser Regret | | | | | |
| | Intercept | -2.03** | 0.20 | -2.44 | 1.62 |
| | Valuation revelation | 0.53** | 0.16 | 0.21 | 0.86 |

** Two tail 0.95% confidence interval significance   * One tail 0.95% confidence interval significance

Table 2.D.13. Explaining winner regret $\alpha_i$, the loser regret $\beta_i$, the update of valuation parameters $\delta_i$ and the learning parameter $\rho_i$ estimates across 47 bidder segments

| Regressand | Regressor | Estimate | SE | Lower 95% | Upper 95% |
|---|---|---|---|---|---|
| Winner Regret $(Adjusted\text{-}R^2 = 0.26)$ | | | | | |
| | Intercept | -1.3535* | 0.0163 | -1.3864 | -1.3206 |
| | Segment Size | 0.0000 | 0.0001 | -0.0002 | 0.0001 |
| | Bidders Feedback mean | -0.0001 | 0.0001 | -0.0004 | 0.0002 |
| | Number of Bids on This item | -0.0005 | 0.0026 | -0.0057 | 0.0046 |
| | total number of bids in 30 days | -0.0075* | 0.0035 | -0.0145 | -0.0004 |
| | Number of items bid on in 30 days | 0.0000 | 0.0001 | -0.0002 | 0.0002 |
| | Bid activity with current Seller | -0.0016 | 0.0011 | -0.0038 | 0.0006 |
| | Number of categories Bid on Mean | -0.0396 | 0.0439 | -0.1284 | 0.0491 |
| Loser Regret $(Adjusted\text{-}R^2 = 0.26)$ | | | | | |
| | Intercept | -1.3509* | 0.0221 | -1.3956 | -1.3061 |
| | Segment Size | 0.0000 | 0.0001 | -0.0002 | 0.0002 |
| | Bidders Feedback mean | 0.0000 | 0.0002 | -0.0004 | 0.0004 |
| | Number of Bids on This item | 0.0037 | 0.0035 | -0.0033 | 0.0107 |
| | total number of bids in 30 days | -0.0122* | 0.0047 | -0.0218 | -0.0026 |
| | Number of items bid on in 30 days | 0.0002 | 0.0001 | -0.0001 | 0.0004 |
| | Bid activity with current Seller | -0.0018 | 0.0015 | -0.0048 | 0.0011 |
| | Number of categories Bid on Mean | -0.1336* | 0.0597 | -0.2544 | -0.0129 |
| Learning value from bids $(Adjusted\text{-}R^2 = 0.30)$ | | | | | |
| | Intercept | 1.2688* | 0.0160 | 1.2364 | 1.3013 |
| | Segment Size | 0.0000 | 0.0001 | -0.0002 | 0.0001 |
| | Bidders Feedback mean | 0.0007* | 0.0001 | 0.0004 | 0.0010 |
| | Number of Bids on This item | 0.0038 | 0.0025 | -0.0013 | 0.0089 |
| | total number of bids in 30 days | -0.0108* | 0.0034 | -0.0178 | -0.0039 |
| | Number of items bid on in 30 days | 0.0002 | 0.0001 | -0.0000 | 0.0003 |
| | Bid activity with current Seller | -0.0002 | 0.0011 | -0.0024 | 0.0019 |
| | Number of categories Bid on Mean | -0.0105 | 0.0433 | -0.0980 | 0.0770 |
| Valuation update $(Adjusted\text{-}R^2 = 0.76)$ | | | | | |
| | Intercept | 0.2707* | 0.0166 | 0.0000 | 0.2372 |
| | Segment Size | 0.0001 | 0.0001 | 0.2501 | -0.0001 |
| | Bidders Feedback mean | 0.0013* | 0.0001 | 0.0000 | 0.0010 |
| | Number of Bids on This item | 0.0032 | 0.0026 | 0.2314 | -0.0021 |
| | total number of bids in 30 days | 0.0021 | 0.0036 | 0.5510 | -0.0051 |
| | Number of items bid on in 30 days | 0.0000 | 0.0001 | 0.8363 | -0.0002 |
| | Bid activity with current Seller | 0.0006 | 0.0011 | 0.5586 | -0.0016 |
| | Number of categories Bid on Mean | -0.0227 | 0.0447 | 0.6151 | -0.1132 |

* Two tail 0.95% confidence interval significance

Table 2.D.14. Summary statistics for the auction specific parameter estimations

| | within each auction category (19) | | | | within each auction cluster (50) | | | |
|---|---|---|---|---|---|---|---|---|
| Parameter | min | max | Mean | SD | min | max | Mean | SD |
| avg. growth of bids | 1.61 | 2.03 | 1.79 | 0.12 | 1.67 | 9.87 | 3.20 | 1.53 |
| se growth of bids | 0.07 | 0.24 | 0.15 | 0.05 | 0.04 | 7.33 | 1.46 | 1.38 |
| avg. drift of bids | 5.24 | 5.82 | 5.58 | 0.15 | 5.15 | 12.72 | 6.79 | 1.42 |
| se drift of bids | 0.08 | 0.2 | 0.13 | 0.03 | 0.05 | 6.81 | 1.28 | 1.28 |
| avg. last minute flood | 0.93 | 1.33 | 1.10 | 0.10 | 0.92 | 9.33 | 2.52 | 1.56 |
| se last minute flood | 0.06 | 0.24 | 0.14 | 0.05 | 0.04 | 7.43 | 1.49 | 1.40 |
| avg. mean entrance rate | 1.92 | 2.31 | 2.08 | 0.13 | 1.39 | 10.42 | 3.43 | 1.58 |
| se mean entrance rate | 0.08 | 0.25 | 0.15 | 0.04 | 0.05 | 7.23 | 1.45 | 1.37 |

Table 2.D.15. Counterfactual analysis of shutting down only winner and both winner/loser regret

| Auction Category | Number of Auctions | Average improvement of shutting down winner regret | Average improvement of shutting down both winner and loser |
|---|---|---|---|
| Jewelry and Watches | 149 | 32% | 29% |
| Collectibles | 103 | 26% | 23% |
| Clothing, Shoes and Accessories | 84 | 21% | 39% |
| Crafts | 78 | 50% | 42% |
| Pottery and Glass | 74 | 18% | 28% |
| Antiques | 68 | 45% | 49% |
| Toys and Hobbies | 93 | 31% | 30% |
| Stamps | 72 | 52% | 28% |
| Books | 84 | 50% | 42% |
| Tickets and Experiences | 91 | 22% | 6% |
| Art | 70 | 31% | 28% |
| Gift Cards and Coupons | 85 | 40% | 21% |
| Music | 86 | 32% | 39% |
| Consumer Electronics | 83 | 34% | 35% |
| DVDs and Movies | 87 | 47% | 47% |
| Dolls and Bears | 84 | 29% | 29% |
| Entertainment Memorabilia | 88 | 25% | 6% |
| Health and Beauty | 74 | 38% | 19% |
| Video Games and Consoles | 93 | 21% | 21% |
| Total improvement | | 26% | 23% |
| Average improvement across all auctions | | 34% | 29% |

Table 2.D.16. Auction's Cluster profile

| Auction Cluster index | Cluster size | number of bidders mean | STD (number of bidders) | number of bids | STD (number of bids) | mean duration (Days) | STD (duration in Days) | Dominant Auction Categories |
|---|---|---|---|---|---|---|---|---|
| 1 | 14 | 10 | 4 | 44 | 11 | 4 | 2 | Jewelry, collectible |
| 2 | 12 | 9 | 5 | 44 | 15 | 5 | 2 | Clothing, antique |
| 3 | 46 | 9 | 5 | 42 | 14 | 5 | 2 | Jewelry, collectible, pottery |
| 4 | 13 | 10 | 4 | 45 | 16 | 5 | 2 | Stamps and books |
| 5 | 14 | 10 | 5 | 44 | 18 | 5 | 2 | Jewelry |
| 6 | 9 | 6 | 4 | 25 | 19 | 5 | 1 | Jewelry |
| 7 | 454 | 9 | 4 | 63 | 19 | 5 | 2 | Toys and hobbies |
| 8 | 19 | 8 | 3 | 33 | 9 | 5 | 2 | Collectible |
| 9 | 10 | 9 | 5 | 41 | 19 | 5 | 1 | Stamps |
| 10 | 27 | 8 | 3 | 44 | 18 | 5 | 2 | DVD and entertainment |
| 11 | 5 | 1 | 0 | 1 | 0 | 1 | 0 | Pottery |
| 12 | 214 | 13 | 5 | 44 | 15 | 4 | 2 | Video Games and Consoles |
| 13 | 32 | 10 | 5 | 47 | 16 | 5 | 2 | Pottery |
| 14 | 14 | 8 | 3 | 48 | 20 | 5 | 2 | Dolls and Bears |
| 15 | 39 | 8 | 3 | 44 | 17 | 5 | 2 | Healthy and beauty |
| 16 | 11 | 6 | 4 | 50 | 11 | 6 | 1 | Jewelry |
| 17 | 48 | 12 | 6 | 50 | 19 | 5 | 2 | Music and DVDs |
| 18 | 15 | 8 | 4 | 38 | 16 | 5 | 2 | Video Games and Consoles |
| 19 | 37 | 9 | 4 | 48 | 15 | 5 | 1 | Clothing |
| 20 | 22 | 8 | 5 | 47 | 21 | 5 | 2 | Stamps, entertainment, jewelry |
| 21 | 29 | 10 | 4 | 50 | 18 | 5 | 1 | Entertainment and music |
| 22 | 24 | 8 | 4 | 40 | 19 | 5 | 2 | Clothing and jewelry |
| 23 | 29 | 10 | 4 | 55 | 18 | 4 | 2 | DVD, consumer electronics |
| 24 | 12 | 7 | 3 | 36 | 9 | 5 | 2 | Toy, music, gift |
| 25 | 19 | 10 | 3 | 47 | 13 | 5 | 1 | Clothing and gift |
| 26 | 24 | 10 | 4 | 44 | 15 | 5 | 2 | Crafts |
| 27 | 38 | 10 | 5 | 51 | 20 | 5 | 2 | Art |
| 28 | 19 | 9 | 4 | 41 | 13 | 5 | 1 | Dolls and Bears |
| 29 | 30 | 9 | 4 | 43 | 18 | 5 | 2 | Pottery |
| 30 | 23 | 8 | 4 | 38 | 14 | 5 | 2 | Craft and book |
| 31 | 16 | 7 | 3 | 40 | 14 | 5 | 1 | Antique |
| 32 | 8 | 8 | 4 | 43 | 17 | 5 | 2 | Craft, toys, book |
| 33 | 13 | 7 | 4 | 32 | 13 | 5 | 2 | Helath and beauty |
| 34 | 38 | 9 | 4 | 43 | 14 | 5 | 1 | Art |
| 35 | 20 | 7 | 3 | 34 | 12 | 5 | 2 | Clothing and art |
| 36 | 23 | 8 | 3 | 42 | 10 | 5 | 2 | Stamps |
| 37 | 8 | 10 | 5 | 36 | 13 | 4 | 2 | Antique, collectible |
| 38 | 30 | 8 | 3 | 44 | 20 | 5 | 1 | Gift Cards and Coupons |
| 39 | 10 | 10 | 3 | 38 | 12 | 4 | 2 | Video game, dolls |
| 40 | 7 | 11 | 6 | 44 | 12 | 5 | 2 | Video game, gift card, pottery |
| 41 | 17 | 7 | 2 | 50 | 16 | 5 | 2 | Clothing |
| 42 | 16 | 10 | 4 | 45 | 15 | 5 | 2 | Pottery, dolls |
| 43 | 16 | 8 | 3 | 43 | 16 | 5 | 2 | Stamps, video games |
| 44 | 20 | 9 | 5 | 42 | 19 | 4 | 2 | Stamps, toys |
| 45 | 17 | 8 | 4 | 48 | 18 | 5 | 2 | Books entertainment |
| 46 | 10 | 11 | 6 | 43 | 12 | 5 | 2 | Gift cards |
| 47 | 17 | 9 | 5 | 44 | 12 | 5 | 2 | Books |
| 48 | 42 | 9 | 5 | 42 | 19 | 5 | 2 | Consumer electronics |
| 49 | 11 | 11 | 4 | 45 | 16 | 4 | 2 | Ticket experience |
| 50 | 5 | 8 | 1 | 48 | 7 | 6 | 0 | Books, music, DVD |

Table 2.D.17. Maximum A Posteriori of the model

| Element of the maximum a posteriori model selection criteria | Log Likelihood |
|---|---|
| Number of bidders evolution state space model | -2,055,997 |
| Bid evolution with each auction state space model | -96,099,682 |
| Valuation evolution state space model | -991,098 |
| prior on the auctions parameters | -17,261 |
| prior on the bidders parameters | -112,190 |

Table 2.D.18. The winner regret $\alpha_i$ estimates across bidder's segments

| Bidders Segment | Segment Size | Estimate | STE | t-stat | p-value |
|---|---|---|---|---|---|
| Segment 1 | 215 | -1.30*** | 0.07 | -19.28 | <0.0001 |
| Segment 2 | 23 | -1.68*** | 0.18 | -9.19 | <0.0001 |
| Segment 3 | 89 | -1.28*** | 0.10 | -13.02 | <0.0001 |
| Segment 4 | 963 | -1.29*** | 0.03 | -42.10 | <0.0001 |
| Segment 5 | 153 | -1.36*** | 0.09 | -15.53 | <0.0001 |
| Segment 6 | 466 | -1.31*** | 0.05 | -26.69 | <0.0001 |
| Segment 7 | 90 | -1.21*** | 0.11 | -11.38 | <0.0001 |
| Segment 8 | 992 | -1.36*** | 0.03 | -45.07 | <0.0001 |
| Segment 9 | 535 | -1.24*** | 0.04 | -30.68 | <0.0001 |
| Segment 10 | 284 | -1.23*** | 0.06 | -21.24 | <0.0001 |
| Segment 11 | 42 | -1.11*** | 0.14 | -7.87 | <0.0001 |
| Segment 12 | 83 | -1.33*** | 0.10 | -13.70 | <0.0001 |
| Segment 13 | 12 | -1.55*** | 0.31 | -4.99 | <0.001 |
| Segment 14 | 52 | -1.41*** | 0.14 | -10.04 | <0.0001 |
| Segment 15 | 113 | -1.53*** | 0.09 | -16.17 | <0.0001 |
| Segment 16 | 466 | -1.43*** | 0.05 | -31.14 | <0.0001 |
| Segment 17 | 522 | -1.21*** | 0.05 | -25.80 | <0.0001 |
| Segment 18 | 589 | -1.28*** | 0.04 | -32.03 | <0.0001 |
| Segment 19 | 310 | -1.32*** | 0.06 | -23.60 | <0.0001 |
| Segment 20 | 395 | -1.32*** | 0.05 | -26.16 | <0.0001 |
| Segment 21 | 403 | -1.34*** | 0.05 | -27.61 | <0.0001 |
| Segment 22 | 530 | -1.36*** | 0.04 | -33.75 | <0.0001 |
| Segment 23 | 49 | -1.32*** | 0.11 | -11.76 | <0.0001 |
| Segment 24 | 481 | -1.33*** | 0.04 | -30.30 | <0.0001 |
| Segment 25 | 142 | -1.35*** | 0.09 | -15.08 | <0.0001 |
| Segment 26 | 871 | -1.35*** | 0.03 | -40.01 | <0.0001 |
| Segment 27 | 242 | -1.32*** | 0.06 | -20.62 | <0.0001 |
| Segment 28 | 569 | -1.36*** | 0.04 | -33.70 | <0.0001 |
| Segment 29 | 62 | -1.33*** | 0.14 | -9.79 | <0.0001 |
| Segment 30 | 83 | -1.54*** | 0.10 | -15.96 | <0.0001 |
| Segment 31 | 102 | -1.33*** | 0.10 | -12.96 | <0.0001 |
| Segment 32 | 7 | -1.37** | 0.40 | -3.41 | <0.05 |
| Segment 33 | 64 | -1.29*** | 0.11 | -12.25 | <0.0001 |
| Segment 34 | 163 | -1.36*** | 0.07 | -18.81 | <0.0001 |
| Segment 35 | 183 | -1.37*** | 0.08 | -17.58 | <0.0001 |
| Segment 36 | 198 | -1.34*** | 0.06 | -21.06 | <0.0001 |
| Segment 37 | 346 | -1.33*** | 0.05 | -26.05 | <0.0001 |
| Segment 38 | 971 | -1.33*** | 0.03 | -42.88 | <0.0001 |
| Segment 39 | 65 | -1.09*** | 0.12 | -8.97 | <0.0001 |
| Segment 40 | 14 | -1.30*** | 0.28 | -4.64 | <0.001 |
| Segment 41 | 73 | -1.41*** | 0.12 | -11.29 | <0.0001 |
| Segment 42 | 73 | -1.44*** | 0.11 | -12.63 | <0.0001 |
| Segment 43 | 93 | -1.32*** | 0.10 | -13.58 | <0.0001 |
| Segment 44 | 41 | -1.36*** | 0.13 | -10.59 | <0.0001 |
| Segment 45 | 19 | -1.24*** | 0.18 | -6.97 | <0.0001 |
| Segment 46 | 5 | -1.45** | 0.40 | -3.60 | <0.05 |
| Segment 47 | 3 | -1.90** | 0.49 | -3.88 | <0.05 |

* p<0.1, **p<0.05, ***p<0.01

Table 2.D.19. The loser regret $\beta_i$ estimates across bidder's segments

| Bidders Segment | Segment Size | Estimate | STE | t-stat | p-value |
|---|---|---|---|---|---|
| Segment 1 | 215 | -1.29*** | 0.07 | -18.31 | <0.0001 |
| Segment 2 | 23 | -1.15*** | 0.28 | -4.12 | <0.001 |
| Segment 3 | 89 | -1.21*** | 0.09 | -13.11 | <0.0001 |
| Segment 4 | 963 | -1.36*** | 0.03 | -43.41 | <0.0001 |
| Segment 5 | 153 | -1.32*** | 0.08 | -16.87 | <0.0001 |
| Segment 6 | 466 | -1.33*** | 0.05 | -29.32 | <0.0001 |
| Segment 7 | 90 | -1.27*** | 0.10 | -12.46 | <0.0001 |
| Segment 8 | 992 | -1.34*** | 0.03 | -43.52 | <0.0001 |
| Segment 9 | 535 | -1.26*** | 0.04 | -29.25 | <0.0001 |
| Segment 10 | 284 | -1.29*** | 0.06 | -21.81 | <0.0001 |
| Segment 11 | 42 | -1.14*** | 0.14 | -8.35 | <0.0001 |
| Segment 12 | 83 | -1.38*** | 0.11 | -12.98 | <0.0001 |
| Segment 13 | 12 | -1.37*** | 0.32 | -4.30 | <0.01 |
| Segment 14 | 52 | -1.50*** | 0.13 | -11.52 | <0.0001 |
| Segment 15 | 113 | -1.40*** | 0.10 | -14.72 | <0.0001 |
| Segment 16 | 466 | -1.29*** | 0.05 | -28.33 | <0.0001 |
| Segment 17 | 522 | -1.39*** | 0.04 | -31.98 | <0.0001 |
| Segment 18 | 589 | -1.39*** | 0.04 | -34.08 | <0.0001 |
| Segment 19 | 310 | -1.25*** | 0.06 | -22.04 | <0.0001 |
| Segment 20 | 395 | -1.38*** | 0.05 | -30.21 | <0.0001 |
| Segment 21 | 403 | -1.25*** | 0.05 | -27.67 | <0.0001 |
| Segment 22 | 530 | -1.27*** | 0.04 | -30.60 | <0.0001 |
| Segment 23 | 49 | -1.19*** | 0.15 | -8.13 | <0.0001 |
| Segment 24 | 481 | -1.34*** | 0.04 | -33.17 | <0.0001 |
| Segment 25 | 142 | -1.34*** | 0.09 | -15.08 | <0.0001 |
| Segment 26 | 871 | -1.34*** | 0.03 | -39.38 | <0.0001 |
| Segment 27 | 242 | -1.25*** | 0.06 | -20.25 | <0.0001 |
| Segment 28 | 569 | -1.29*** | 0.04 | -30.67 | <0.0001 |
| Segment 29 | 62 | -1.55*** | 0.11 | -14.03 | <0.0001 |
| Segment 30 | 83 | -1.40*** | 0.12 | -11.81 | <0.0001 |
| Segment 31 | 102 | -1.32*** | 0.09 | -14.93 | <0.0001 |
| Segment 32 | 7 | -1.98*** | 0.50 | -3.93 | <0.01 |
| Segment 33 | 64 | -1.29*** | 0.13 | -10.21 | <0.0001 |
| Segment 34 | 163 | -1.27*** | 0.08 | -16.42 | <0.0001 |
| Segment 35 | 183 | -1.32*** | 0.07 | -17.79 | <0.0001 |
| Segment 36 | 198 | -1.35*** | 0.07 | -18.79 | <0.0001 |
| Segment 37 | 346 | -1.30*** | 0.05 | -23.93 | <0.0001 |
| Segment 38 | 971 | -1.35*** | 0.03 | -44.24 | <0.0001 |
| Segment 39 | 65 | -1.18*** | 0.11 | -10.54 | <0.0001 |
| Segment 40 | 14 | -1.98*** | 0.24 | -8.34 | <0.0001 |
| Segment 41 | 73 | -1.29*** | 0.13 | -10.03 | <0.0001 |
| Segment 42 | 73 | -1.30*** | 0.09 | -14.57 | <0.0001 |
| Segment 43 | 93 | -1.50*** | 0.10 | -15.09 | <0.0001 |
| Segment 44 | 41 | -1.08*** | 0.14 | -7.86 | <0.0001 |
| Segment 45 | 19 | -1.42*** | 0.18 | -7.77 | <0.0001 |
| Segment 46 | 5 | -1.33*** | 0.17 | -7.59 | <0.001 |
| Segment 47 | 3 | -1.73** | 0.50 | -3.46 | <0.05 |

* p<0.1, **p<0.05, ***p<0.01

Table 2.D.20. The update of valuation parameters $\delta_i$ and learning parameter $\rho_i$ estimates across bidder's segments

| Bidders Segment | Segment Size | Valuation revelation $\delta_i$ | STE ($\delta_i$) | Learning $\rho_i$ | STE ($\rho_i$) |
|---|---|---|---|---|---|
| Segment 1 | 215 | 1.27*** | 0.06 | 0.28*** | 0.07 |
| Segment 2 | 23 | 1.53*** | 0.19 | 0.53** | 0.23 |
| Segment 3 | 89 | 1.22*** | 0.07 | 0.22** | 0.11 |
| Segment 4 | 963 | 1.22*** | 0.03 | 0.22*** | 0.03 |
| Segment 5 | 153 | 1.28*** | 0.06 | 0.11 | 0.08 |
| Segment 6 | 466 | 1.26*** | 0.04 | 0.26*** | 0.05 |
| Segment 7 | 90 | 1.16*** | 0.09 | 0.20* | 0.11 |
| Segment 8 | 992 | 1.29*** | 0.03 | 0.25*** | 0.03 |
| Segment 9 | 535 | 1.16*** | 0.03 | 0.20*** | 0.04 |
| Segment 10 | 284 | 1.29*** | 0.05 | 0.26*** | 0.06 |
| Segment 11 | 42 | 1.38*** | 0.16 | 0.27* | 0.15 |
| Segment 12 | 83 | 1.35*** | 0.08 | 0.15 | 0.10 |
| Segment 13 | 12 | 1.21*** | 0.22 | 0.38* | 0.21 |
| Segment 14 | 52 | 1.04*** | 0.11 | 0.49*** | 0.14 |
| Segment 15 | 113 | 1.23*** | 0.07 | 0.23** | 0.09 |
| Segment 16 | 466 | 1.24*** | 0.04 | 0.33*** | 0.05 |
| Segment 17 | 522 | 1.29*** | 0.04 | 0.25*** | 0.04 |
| Segment 18 | 589 | 1.22*** | 0.04 | 0.26*** | 0.04 |
| Segment 19 | 310 | 1.26*** | 0.05 | 0.19*** | 0.05 |
| Segment 20 | 395 | 1.30*** | 0.04 | 0.29*** | 0.05 |
| Segment 21 | 403 | 1.23*** | 0.04 | 0.32*** | 0.05 |
| Segment 22 | 530 | 1.22*** | 0.03 | 0.29*** | 0.04 |
| Segment 23 | 49 | 1.31*** | 0.12 | 0.26* | 0.14 |
| Segment 24 | 481 | 1.27*** | 0.04 | 0.19*** | 0.04 |
| Segment 25 | 142 | 1.22*** | 0.06 | 0.33*** | 0.08 |
| Segment 26 | 871 | 1.28*** | 0.03 | 0.26*** | 0.03 |
| Segment 27 | 242 | 1.23*** | 0.06 | 0.23*** | 0.06 |
| Segment 28 | 569 | 1.28*** | 0.03 | 0.29*** | 0.04 |
| Segment 29 | 62 | 1.07*** | 0.09 | 0.00 | 0.12 |
| Segment 30 | 83 | 1.40*** | 0.09 | 0.27** | 0.11 |
| Segment 31 | 102 | 1.28*** | 0.07 | 0.22** | 0.09 |
| Segment 32 | 7 | 0.85** | 0.27 | 0.20 | 0.47 |
| Segment 33 | 64 | 1.25*** | 0.10 | 0.14 | 0.13 |
| Segment 34 | 163 | 1.15*** | 0.07 | 0.22*** | 0.07 |
| Segment 35 | 183 | 1.18*** | 0.06 | 0.11 | 0.07 |
| Segment 36 | 198 | 1.33*** | 0.06 | 0.05 | 0.07 |
| Segment 37 | 346 | 1.35*** | 0.04 | 0.27*** | 0.05 |
| Segment 38 | 971 | 1.24*** | 0.03 | 0.25*** | 0.03 |
| Segment 39 | 65 | 1.29*** | 0.09 | 0.25** | 0.12 |
| Segment 40 | 14 | 1.24*** | 0.17 | 0.32 | 0.30 |
| Segment 41 | 73 | 1.27*** | 0.10 | 0.27** | 0.12 |
| Segment 42 | 73 | 1.42*** | 0.09 | 0.15 | 0.12 |
| Segment 43 | 93 | 1.34*** | 0.08 | 0.09 | 0.09 |
| Segment 44 | 41 | 1.51*** | 0.14 | 0.42** | 0.17 |
| Segment 45 | 19 | 1.13*** | 0.19 | 0.05 | 0.18 |
| Segment 46 | 5 | 1.89** | 0.54 | 1.33** | 0.49 |
| Segment 47 | 3 | 1.20* | 0.40 | 0.55 | 0.78 |

* p<0.1, **p<0.05, ***p<0.01

Table 2.D.21. The growth of bids and their drift parameters, $\tau_i$ and $\gamma_i$, and the rush of bidders at the end of auction rate and average entrance rate in each period, $\eta_j$ and $\iota_j$, estimates across auction segments

| Auction Cluster | Auction Cluster Size | growth of bids ( $\tau_i$ ) | STE ( $\tau_i$ ) | Drift of bids ( $\gamma_i$ ) | STE ( $\gamma_i$ ) | Last minute flood ( $\eta_j$ ) | STE( $\eta_j$ ) | Mean entrance rate ( $\iota_j$ ) | STE ( $\iota_j$ ) |
|---|---|---|---|---|---|---|---|---|---|
| 1 | 15 | 2.28*** | 0.20 | 5.47*** | 0.36 | 1.27*** | 0.28 | 1.88*** | 0.25 |
| 2 | 13 | 1.93*** | 0.23 | 5.15*** | 0.33 | 1.16*** | 0.20 | 1.93*** | 0.32 |
| 3 | 47 | 1.73*** | 0.15 | 5.32*** | 0.14 | 1.22*** | 0.12 | 2.07*** | 0.16 |
| 4 | 14 | 1.84*** | 0.28 | 5.43*** | 0.27 | 0.96*** | 0.27 | 1.39*** | 0.28 |
| 5 | 15 | 1.89*** | 0.37 | 5.46*** | 0.21 | 0.92*** | 0.30 | 2.01*** | 0.30 |
| 6 | 10 | 1.74*** | 0.49 | 5.50*** | 0.21 | 1.13* | 0.53 | 2.75*** | 0.37 |
| 7 | 455 | 1.67*** | 0.04 | 5.54*** | 0.05 | 0.96*** | 0.04 | 2.03*** | 0.05 |
| 8 | 20 | 1.89*** | 0.40 | 5.29*** | 0.28 | 1.23*** | 0.37 | 2.13*** | 0.37 |
| 9 | 11 | 2.25*** | 0.67 | 5.78*** | 0.43 | 1.35* | 0.74 | 2.45*** | 0.72 |
| 10 | 28 | 2.15*** | 0.33 | 5.82*** | 0.29 | 1.29*** | 0.34 | 2.53*** | 0.32 |
| 11 | 6 | 3.82** | 1.34 | 6.34*** | 0.91 | 2.48 | 1.57 | 3.82** | 1.34 |
| 12 | 215 | 1.68*** | 0.08 | 5.59*** | 0.08 | 1.10*** | 0.07 | 2.16*** | 0.09 |
| 13 | 33 | 2.08*** | 0.37 | 5.39*** | 0.29 | 1.76*** | 0.38 | 2.43*** | 0.35 |
| 14 | 15 | 2.39** | 0.84 | 6.49*** | 0.57 | 2.22** | 0.84 | 2.66*** | 0.81 |
| 15 | 40 | 2.05*** | 0.36 | 5.91*** | 0.27 | 1.53*** | 0.36 | 1.95*** | 0.37 |
| 16 | 12 | 2.89** | 1.16 | 6.33*** | 0.90 | 2.08 | 1.22 | 3.18** | 1.15 |
| 17 | 49 | 1.84*** | 0.33 | 5.68*** | 0.28 | 1.40*** | 0.34 | 2.29*** | 0.34 |
| 18 | 16 | 3.10*** | 1.00 | 7.17*** | 0.74 | 2.43** | 1.04 | 2.87** | 1.00 |
| 19 | 38 | 2.15*** | 0.47 | 5.73*** | 0.38 | 1.53*** | 0.48 | 2.28*** | 0.47 |
| 20 | 23 | 2.34*** | 0.81 | 6.12*** | 0.64 | 1.76** | 0.83 | 2.64*** | 0.80 |
| 21 | 30 | 2.43*** | 0.66 | 6.05*** | 0.54 | 1.62** | 0.67 | 2.77*** | 0.66 |
| 22 | 25 | 2.49*** | 0.82 | 6.29*** | 0.66 | 1.73** | 0.84 | 2.82*** | 0.80 |
| 23 | 30 | 2.36*** | 0.72 | 6.09*** | 0.61 | 1.51** | 0.74 | 2.64*** | 0.71 |
| 24 | 13 | 3.31* | 1.67 | 6.70*** | 1.40 | 2.78 | 1.71 | 3.48* | 1.65 |
| 25 | 20 | 2.72** | 1.15 | 6.31*** | 0.98 | 2.43* | 1.17 | 2.72** | 1.16 |
| 26 | 25 | 2.77*** | 0.96 | 6.20*** | 0.83 | 2.03* | 0.99 | 2.81*** | 0.97 |
| 27 | 39 | 2.49*** | 0.65 | 5.87*** | 0.58 | 1.57** | 0.67 | 2.55*** | 0.65 |
| 28 | 20 | 3.06** | 1.29 | 6.66*** | 1.11 | 2.08 | 1.33 | 3.57** | 1.28 |
| 29 | 31 | 2.49*** | 0.89 | 6.24*** | 0.76 | 2.06** | 0.89 | 2.99*** | 0.87 |
| 30 | 24 | 2.78** | 1.17 | 6.60*** | 1.01 | 2.23* | 1.19 | 2.90** | 1.18 |
| 31 | 17 | 3.54** | 1.67 | 7.20*** | 1.46 | 2.94 | 1.71 | 3.72** | 1.66 |
| 32 | 9 | 5.32 | 3.16 | 8.36** | 2.81 | 4.31 | 3.27 | 5.30 | 3.16 |
| 33 | 14 | 3.66 | 2.19 | 7.64*** | 1.89 | 3.18 | 2.21 | 3.68 | 2.19 |
| 34 | 39 | 2.43*** | 0.83 | 6.13*** | 0.74 | 1.76** | 0.85 | 2.82*** | 0.82 |
| 35 | 21 | 3.29** | 1.56 | 7.06*** | 1.38 | 2.55 | 1.59 | 3.50** | 1.54 |
| 36 | 24 | 2.97** | 1.42 | 6.93*** | 1.25 | 2.19 | 1.44 | 3.23** | 1.41 |
| 37 | 9 | 5.70 | 3.70 | 8.99** | 3.31 | 5.46 | 3.72 | 6.31 | 3.63 |
| 38 | 31 | 3.06** | 1.16 | 6.85*** | 1.04 | 2.28* | 1.18 | 2.98** | 1.16 |
| 39 | 11 | 5.00 | 3.25 | 8.59** | 2.91 | 4.54 | 3.30 | 5.02 | 3.25 |
| 40 | 8 | 6.52 | 4.48 | 9.71** | 4.06 | 5.84 | 4.57 | 6.91 | 4.43 |
| 41 | 18 | 3.85* | 2.13 | 7.66*** | 1.92 | 3.17 | 2.17 | 4.05* | 2.12 |
| 42 | 17 | 4.05* | 2.32 | 7.94*** | 2.07 | 3.51 | 2.34 | 4.17* | 2.30 |
| 43 | 17 | 3.97 | 2.38 | 7.41*** | 2.17 | 3.35 | 2.41 | 4.47* | 2.35 |
| 44 | 21 | 3.80* | 1.97 | 7.38*** | 1.80 | 3.03 | 2.00 | 4.39** | 1.94 |
| 45 | 18 | 4.16* | 2.34 | 7.79*** | 2.13 | 3.56 | 2.38 | 4.30* | 2.34 |
| 46 | 11 | 5.75 | 3.84 | 9.18** | 3.52 | 5.05 | 3.91 | 6.19 | 3.81 |
| 47 | 18 | 4.04 | 2.46 | 8.12*** | 2.23 | 3.31 | 2.50 | 4.68* | 2.43 |
| 48 | 43 | 2.87** | 1.07 | 6.28*** | 0.99 | 1.90* | 1.09 | 3.13*** | 1.07 |
| 49 | 12 | 5.69 | 3.78 | 8.96** | 3.49 | 5.13 | 3.83 | 5.59 | 3.79 |
| 50 | 6 | 9.87 | 7.33 | 12.72 | 6.81 | 9.33 | 7.43 | 10.42 | 7.23 |

* p<0.1, **p<0.05, ***p<0.01

Table 2.D.22. The winner regret $\alpha_i$ and the loser regret $\beta_i$ estimates across auction categories

| Auction Category | number of bidders | winner regret | STE (winner regret) | t-stat (winner regret) | p-value (winner regret) | Loser regret | STE (Loser regret) | t-stat (Loser regret) | p-value (Loser regret) |
|---|---|---|---|---|---|---|---|---|---|
| Jewelry and Watches | 1550 | -1.30*** | 0.02 | -52.40 | <0.0001 | -1.31*** | 0.03 | -51.92 | <0.0001 |
| Collectibles | 859 | -1.27*** | 0.04 | -36.14 | <0.0001 | -1.33*** | 0.03 | -39.61 | <0.0001 |
| Clothing, Shoes and Accessories | 453 | -1.28*** | 0.04 | -30.07 | <0.0001 | -1.31*** | 0.05 | -28.91 | <0.0001 |
| Crafts | 558 | -1.31*** | 0.04 | -29.89 | <0.0001 | -1.38*** | 0.04 | -34.35 | <0.0001 |
| Pottery and Glass | 607 | -1.29*** | 0.04 | -32.21 | <0.0001 | -1.27*** | 0.04 | -32.27 | <0.0001 |
| Antiques | 546 | -1.37*** | 0.04 | -32.08 | <0.0001 | -1.34*** | 0.04 | -33.99 | <0.0001 |
| Toys and Hobbies | 744 | -1.31*** | 0.03 | -38.17 | <0.0001 | -1.32*** | 0.04 | -36.28 | <0.0001 |
| Stamps | 651 | -1.36*** | 0.04 | -36.26 | <0.0001 | -1.32*** | 0.04 | -34.17 | <0.0001 |
| Books | 482 | -1.32*** | 0.04 | -29.72 | <0.0001 | -1.25*** | 0.04 | -29.83 | <0.0001 |
| Tickets and Experiences | 489 | -1.37*** | 0.04 | -31.65 | <0.0001 | -1.37*** | 0.04 | -30.73 | <0.0001 |
| Art | 456 | -1.29*** | 0.05 | -26.94 | <0.0001 | -1.32*** | 0.05 | -28.36 | <0.0001 |
| Gift Cards and Coupons | 522 | -1.34*** | 0.04 | -29.96 | <0.0001 | -1.34*** | 0.04 | -30.39 | <0.0001 |
| Music | 585 | -1.36*** | 0.04 | -34.28 | <0.0001 | -1.35*** | 0.04 | -36.23 | <0.0001 |
| Consumer Electronics | 734 | -1.35*** | 0.03 | -39.15 | <0.0001 | -1.35*** | 0.04 | -35.97 | <0.0001 |
| DVDs and Movies | 602 | -1.36*** | 0.04 | -33.58 | <0.0001 | -1.32*** | 0.04 | -34.18 | <0.0001 |
| Dolls and Bears | 679 | -1.30*** | 0.04 | -32.33 | <0.0001 | -1.31*** | 0.04 | -35.22 | <0.0001 |
| Entertainment Memorabilia | 506 | -1.34*** | 0.04 | -32.42 | <0.0001 | -1.36*** | 0.04 | -31.30 | <0.0001 |
| Health and Beauty | 541 | -1.37*** | 0.04 | -34.49 | <0.0001 | -1.30*** | 0.04 | -30.04 | <0.0001 |
| Video Games and Consoles | 682 | -1.35*** | 0.04 | -37.11 | <0.0001 | -1.32*** | 0.04 | -35.68 | <0.0001 |

* p<0.1, **p<0.05, ***p<0.01

Table 2.D.23. The update of valuation parameters $\delta_i$ and learning parameter $\rho_i$ estimates across bidder's segments

| Bidders segment | number of bidders | Learning value from bid parameter ($\rho_i$) | STE ($\rho_i$) | Valuation revelation parameter ($\delta_i$) | STE ($\delta_i$) |
|---|---|---|---|---|---|
| Jewelry and Watches | 1550 | 1.24*** | 0.02 | 0.25*** | 0.02 |
| Collectibles | 859 | 1.20*** | 0.03 | 0.23*** | 0.03 |
| Clothing, Shoes and Accessories | 453 | 1.28*** | 0.04 | 0.23*** | 0.04 |
| Crafts | 558 | 1.21*** | 0.03 | 0.21*** | 0.04 |
| Pottery and Glass | 607 | 1.28*** | 0.03 | 0.24*** | 0.04 |
| Antiques | 546 | 1.26*** | 0.04 | 0.25*** | 0.04 |
| Toys and Hobbies | 744 | 1.29*** | 0.03 | 0.28*** | 0.04 |
| Stamps | 651 | 1.27*** | 0.03 | 0.29*** | 0.04 |
| Books | 482 | 1.20*** | 0.04 | 0.33*** | 0.04 |
| Tickets and Experiences | 489 | 1.25*** | 0.04 | 0.20*** | 0.04 |
| Art | 456 | 1.28*** | 0.04 | 0.24*** | 0.04 |
| Gift Cards and Coupons | 522 | 1.24*** | 0.04 | 0.26*** | 0.04 |
| Music | 585 | 1.27*** | 0.03 | 0.26*** | 0.04 |
| Consumer Electronics | 734 | 1.28*** | 0.03 | 0.27*** | 0.04 |
| DVDs and Movies | 602 | 1.27*** | 0.03 | 0.29*** | 0.04 |
| Dolls and Bears | 679 | 1.29*** | 0.03 | 0.16*** | 0.04 |
| Entertainment Memorabilia | 506 | 1.21*** | 0.04 | 0.25*** | 0.04 |
| Health and Beauty | 541 | 1.30*** | 0.04 | 0.26*** | 0.04 |
| Video Games and Consoles | 682 | 1.26*** | 0.03 | 0.21*** | 0.04 |

* p<0.1, **p<0.05, ***p<0.01

Table 2.D.24. The growth of bids and their drift parameters, $\tau_i$ and $\gamma_i$, and the rush of bidders at the end of auction rate and average entrance rate in each period, $\eta_j$ and $\iota_j$, estimates across auction segments

| Auction Cluster | Auction Cluster Size | growth of bids ($\tau_i$) | STE ($\tau_i$) | Drift of bids ($\gamma_i$) | STE ($\gamma_i$) | Last minute flood ($\eta_j$) | STE ($\eta_j$) | Mean entrance rate ($\iota_j$) | STE ($\iota_j$) |
|---|---|---|---|---|---|---|---|---|---|
| Jewelry and Watches | 150 | 1.62*** | 0.07 | 5.47*** | 0.08 | 0.93*** | 0.06 | 1.99*** | 0.08 |
| Collectibles | 104 | 1.79*** | 0.09 | 5.24*** | 0.10 | 1.03*** | 0.08 | 1.99*** | 0.10 |
| Clothing, Shoes and Accessories | 85 | 1.62*** | 0.10 | 5.46*** | 0.12 | 1.07*** | 0.09 | 1.94*** | 0.11 |
| Crafts | 79 | 1.84*** | 0.10 | 5.58*** | 0.12 | 0.99*** | 0.08 | 2.10*** | 0.13 |
| Pottery and Glass | 75 | 1.78*** | 0.12 | 5.40*** | 0.10 | 1.01*** | 0.10 | 2.09*** | 0.12 |
| Antiques | 69 | 1.77*** | 0.11 | 5.40*** | 0.12 | 1.05*** | 0.11 | 1.92*** | 0.12 |
| Toys and Hobbies | 94 | 1.70*** | 0.11 | 5.55*** | 0.10 | 1.05*** | 0.09 | 1.95*** | 0.14 |
| Stamps | 73 | 1.77*** | 0.14 | 5.63*** | 0.11 | 1.07*** | 0.12 | 2.13*** | 0.14 |
| Books | 85 | 1.61*** | 0.12 | 5.75*** | 0.12 | 1.02*** | 0.12 | 2.08*** | 0.13 |
| Tickets and Experiences | 92 | 1.64*** | 0.13 | 5.60*** | 0.12 | 1.17*** | 0.12 | 1.92*** | 0.15 |
| Art | 71 | 1.87*** | 0.17 | 5.54*** | 0.15 | 1.05*** | 0.16 | 2.05*** | 0.17 |
| Gift Cards and Coupons | 86 | 2.03*** | 0.15 | 5.63*** | 0.13 | 1.21*** | 0.15 | 2.00*** | 0.16 |
| Music | 87 | 1.65*** | 0.16 | 5.53*** | 0.14 | 1.07*** | 0.16 | 2.20*** | 0.16 |
| Consumer Electronics | 84 | 1.84*** | 0.17 | 5.50*** | 0.15 | 1.04*** | 0.17 | 2.27*** | 0.18 |
| DVDs and Movies | 88 | 1.80*** | 0.18 | 5.68*** | 0.14 | 1.24*** | 0.18 | 2.31*** | 0.18 |
| Dolls and Bears | 85 | 1.92*** | 0.20 | 5.65*** | 0.17 | 1.23*** | 0.19 | 2.27*** | 0.20 |
| Entertainment Memorabilia | 89 | 1.87*** | 0.20 | 5.82*** | 0.17 | 1.24*** | 0.19 | 2.15*** | 0.20 |
| Health and Beauty | 75 | 2.00*** | 0.24 | 5.73*** | 0.20 | 1.33*** | 0.24 | 1.93*** | 0.25 |
| Video Games and consoles | 94 | 1.83*** | 0.21 | 5.81*** | 0.17 | 1.12*** | 0.21 | 2.24*** | 0.21 |

* p<0.1, **p<0.05, ***p<0.01

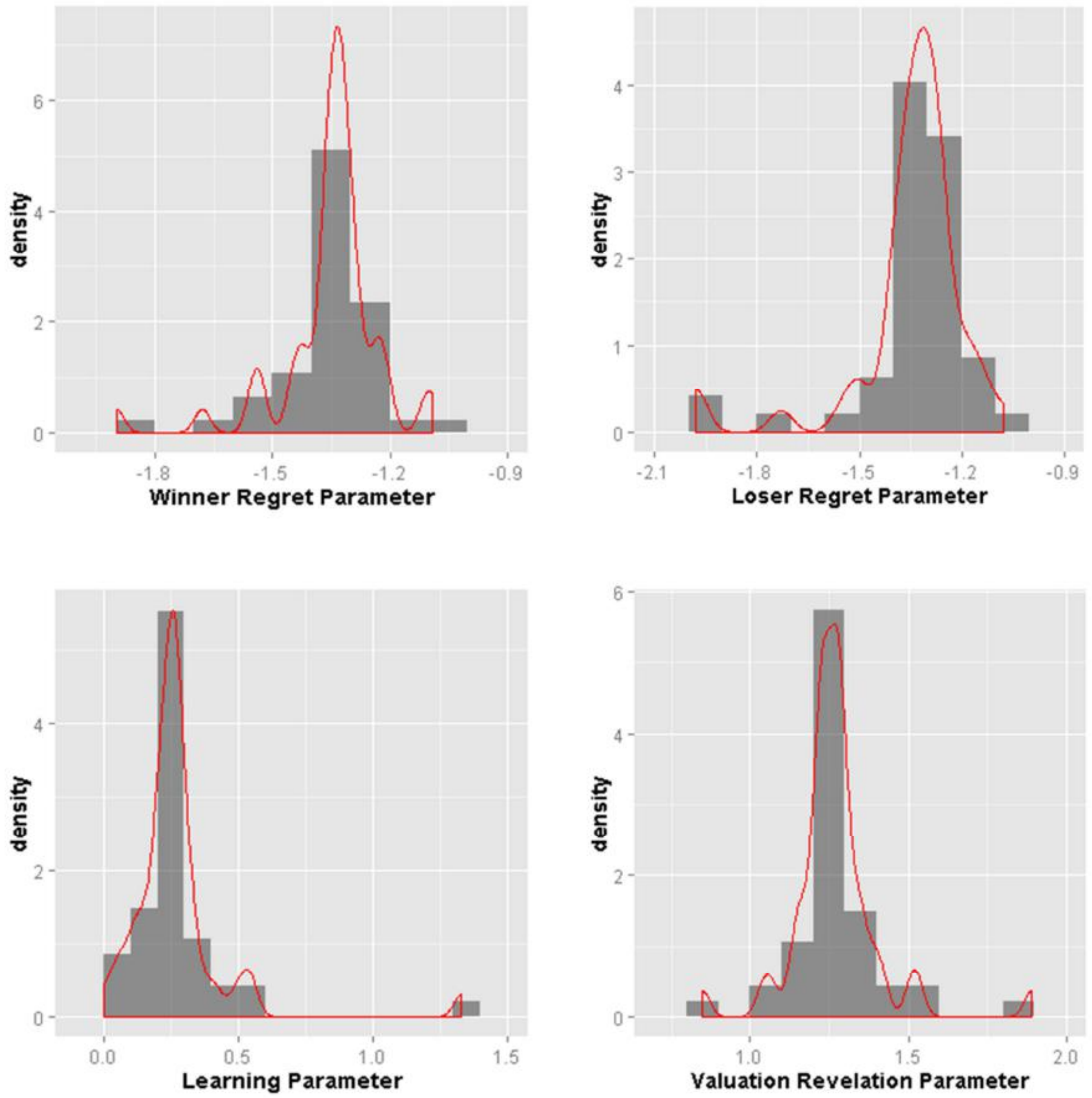

Figure 2.D.2. Histogram of regret and valuation evolution parameters across bidder segments

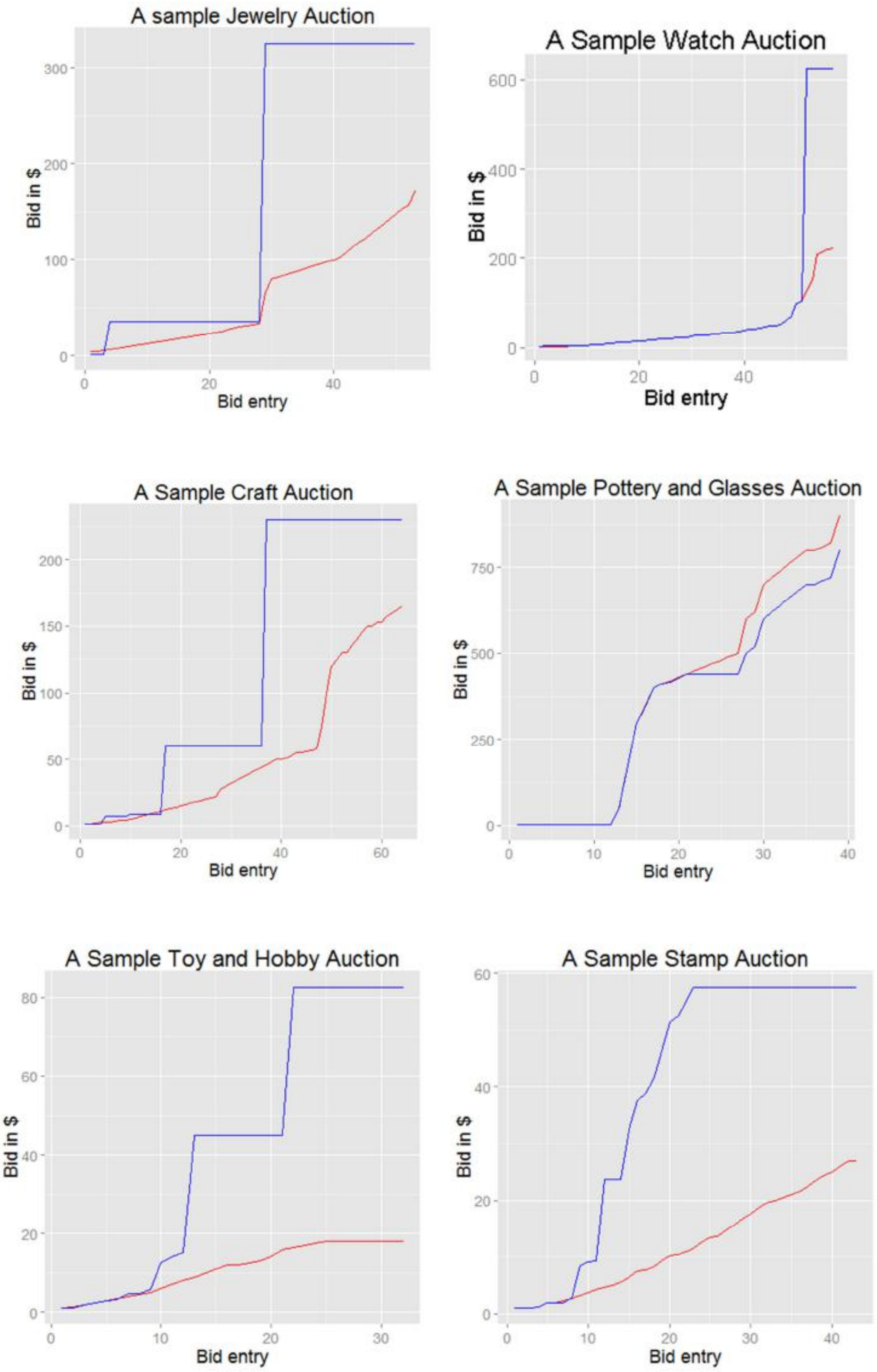

Figure 2.D.3. Counterfactual analysis of shutting down winner regret (blue line the optimal bidding when regret is shut down, and red line the observed)

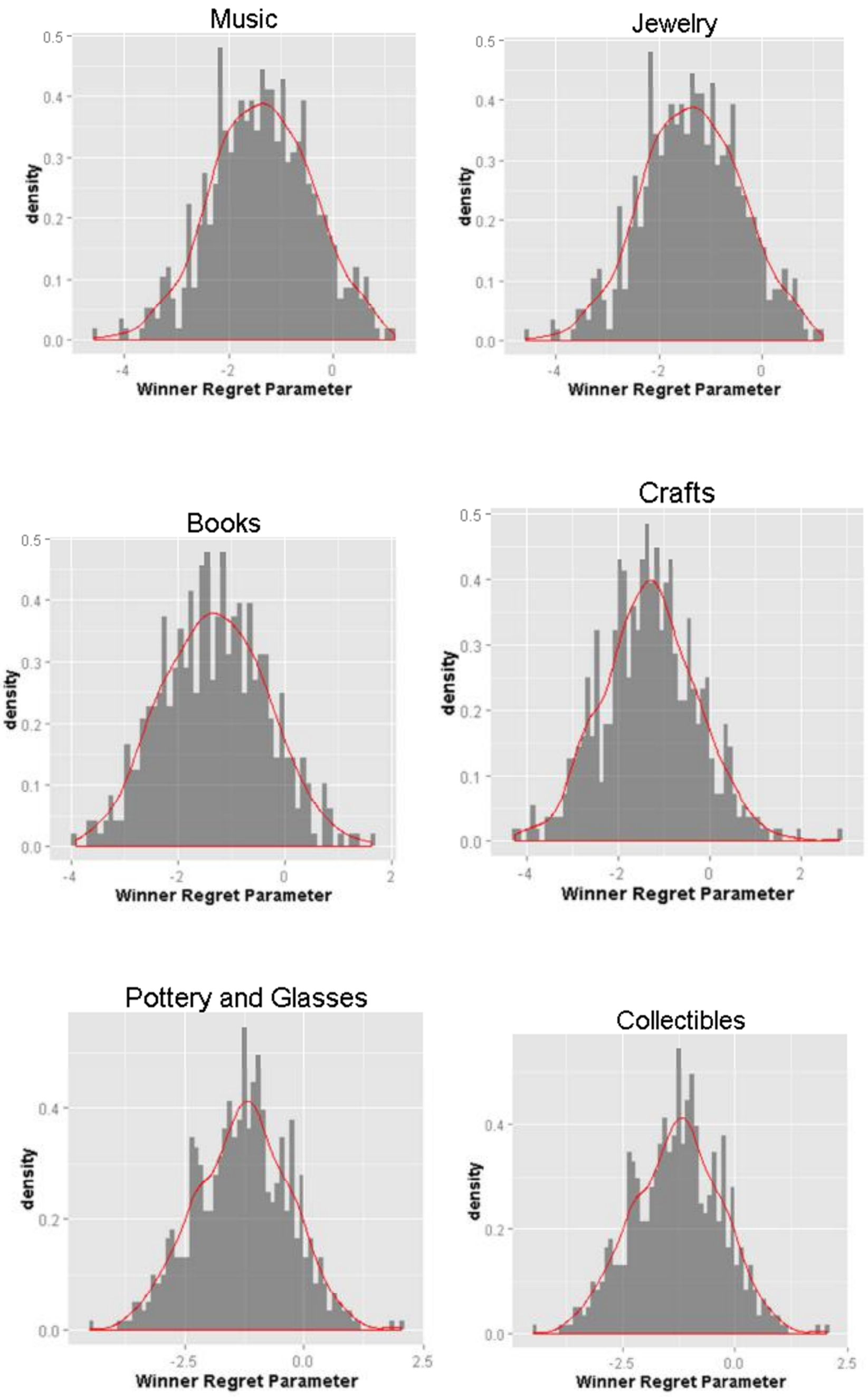

Figure 2.D.4. Histogram of winner regret parameter distribution across item categories

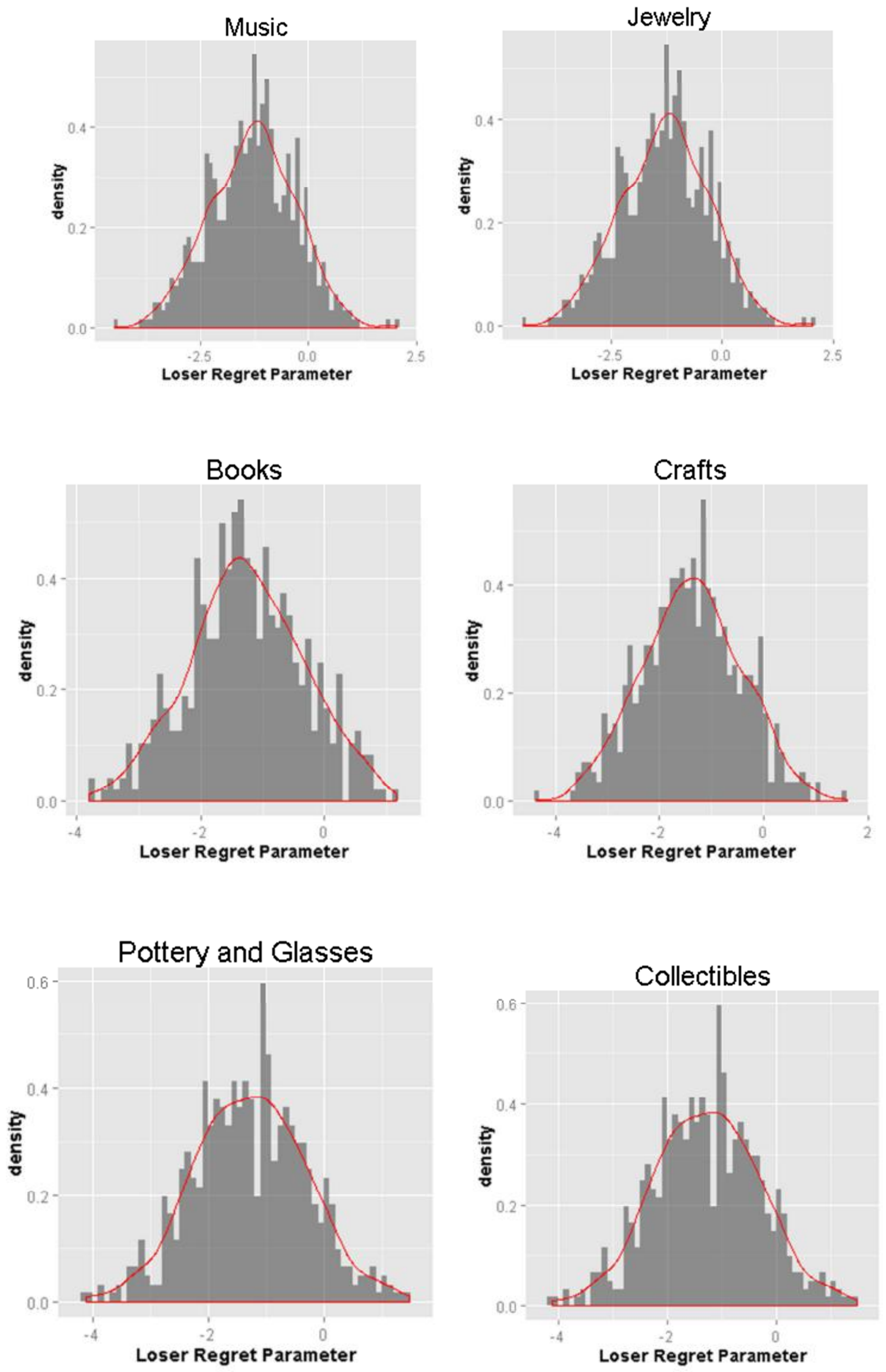

Figure 2.D.5. Histogram of loser regret parameter distribution across item categories

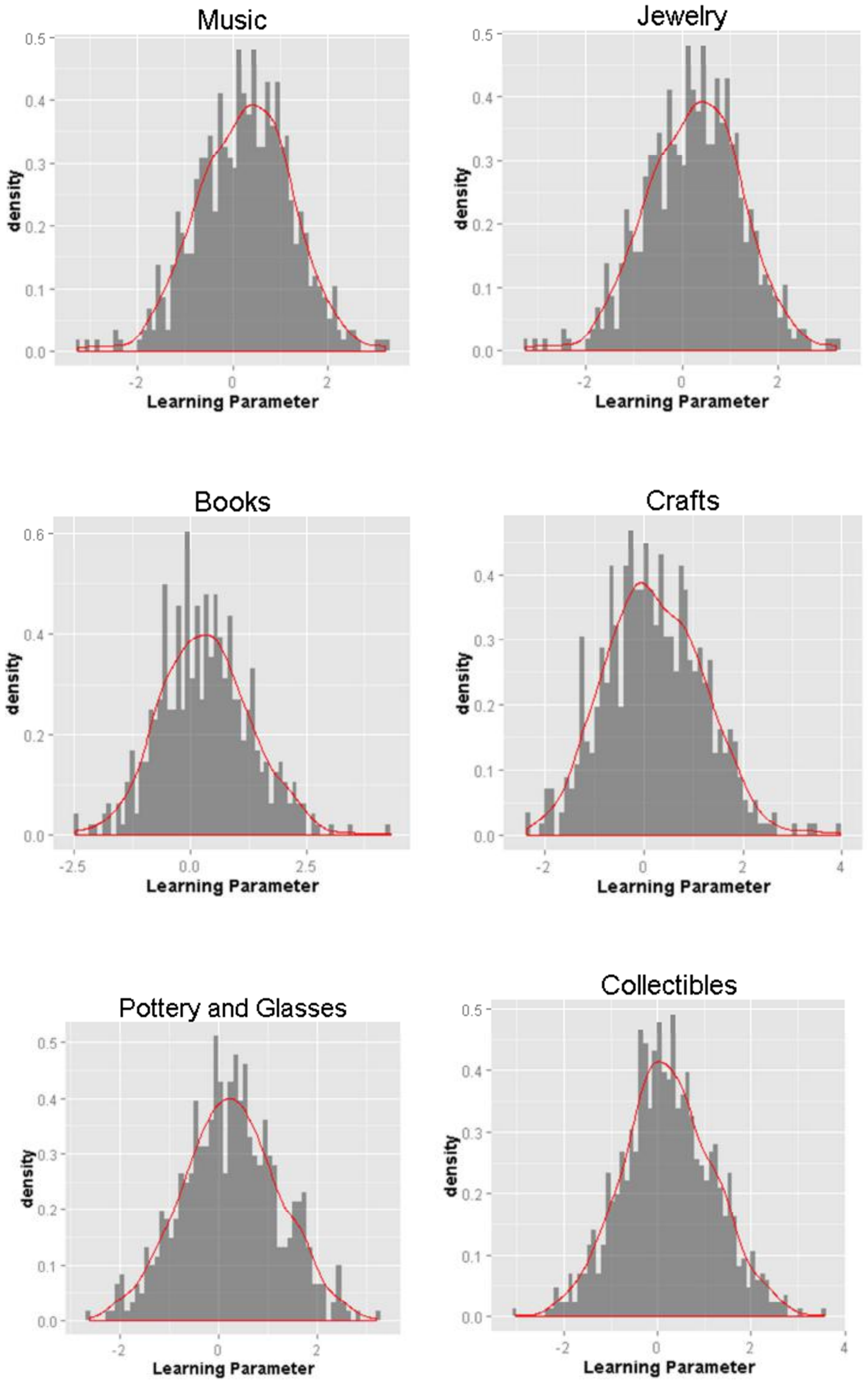

Figure 2.D.6. Histogram of distribution of learning parameter distribution across item categories

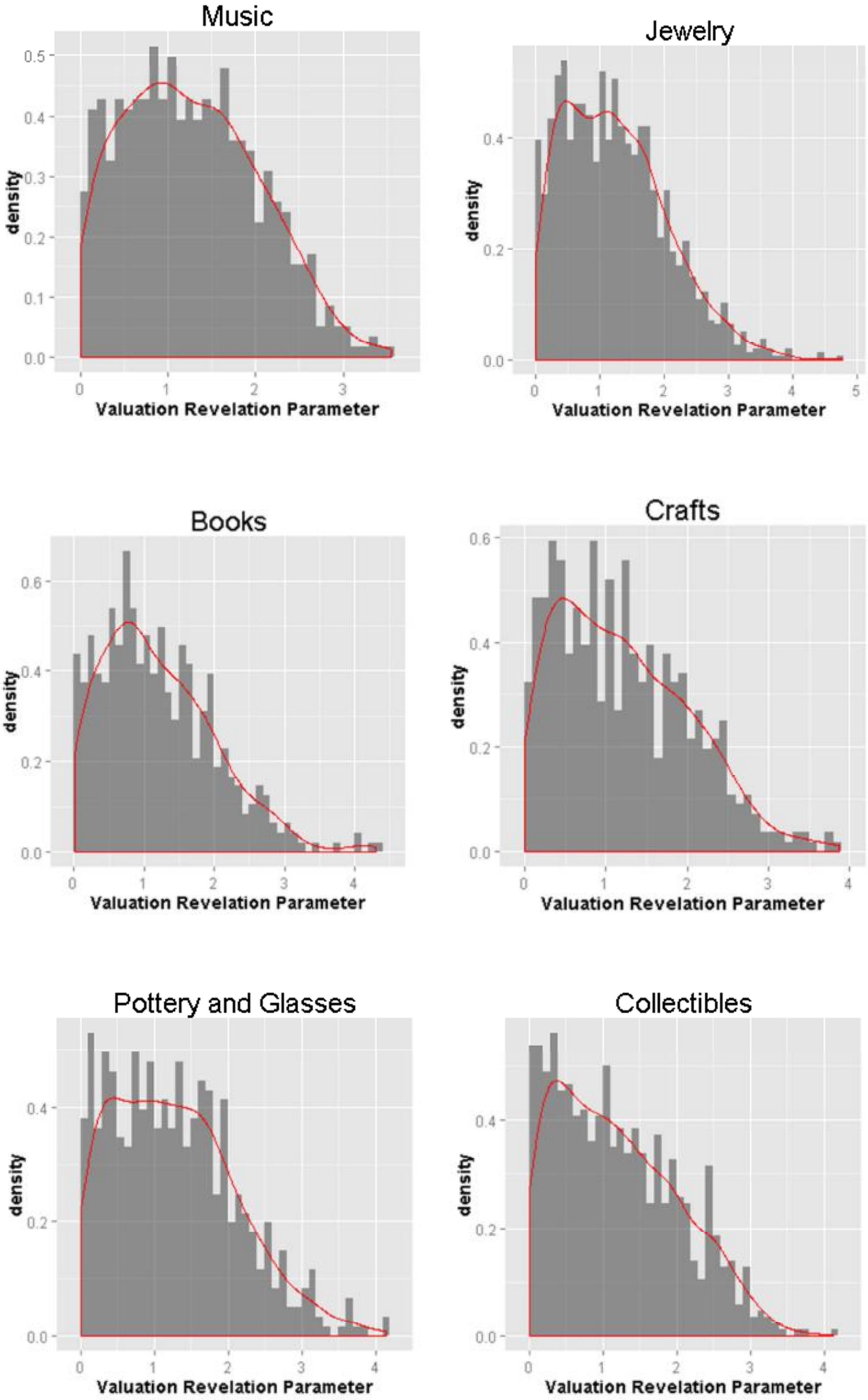

Figure 2.D.7. Histogram of valuation revalation parameter distribution across item categories

## 2.D.3. K-MEANs BIDDER CLUSTERS

Table 2.D.25. Bidder's segment profile (based on k-means approach)

| Bidders 'segment Index | Segment Size | Bidders Feedback mean | STD (Bidder's feedback) | Number of Bids on This item | STD(NBTI) | total number of bids in 30 days | STD (TNB30D) | Number of items bided on in 30 days | STD (NIB30D) | Bid activity with current Seller | STD(BACS) | Number of categories Bided on Mean | STD(NCBO) |
|---|---|---|---|---|---|---|---|---|---|---|---|---|---|
| 1 | 252 | 319 | 695 | 4 | 2 | 33 | 75 | 18 | 54 | 70 | 7 | 1 | 0 |
| 2 | 940 | 654 | 242 | 5 | 1 | 189 | 55 | 88 | 25 | 26 | 3 | 2 | 0 |
| 3 | 185 | 233 | 389 | 14 | 4 | 215 | 174 | 48 | 42 | 13 | 9 | 2 | 0 |
| 4 | 27 | 728 | 623 | 15 | 14 | 3660 | 696 | 876 | 325 | 18 | 28 | 3 | 1 |
| 5 | 51 | 1539 | 1578 | 1 | 1 | 2218 | 543 | 1643 | 287 | 5 | 14 | 2 | 1 |
| 6 | 542 | 573 | 678 | 2 | 2 | 93 | 93 | 50 | 51 | 6 | 6 | 4 | 0 |
| 7 | 384 | 402 | 709 | 3 | 2 | 43 | 94 | 20 | 44 | 65 | 13 | 2 | 0 |
| 8 | 128 | 101 | 157 | 23 | 4 | 97 | 191 | 11 | 26 | 86 | 14 | 1 | 1 |
| 9 | 70 | 539 | 951 | 2 | 3 | 122 | 127 | 57 | 57 | 70 | 20 | 5 | 1 |
| 10 | 77 | 2944 | 2913 | 2 | 2 | 1232 | 300 | 918 | 198 | 4 | 8 | 1 | 1 |
| 11 | 89 | 660 | 1178 | 18 | 5 | 326 | 277 | 73 | 69 | 14 | 13 | 5 | 1 |
| 12 | 170 | 346 | 746 | 11 | 3 | 152 | 153 | 41 | 43 | 15 | 9 | 1 | 0 |
| 13 | 156 | 281 | 611 | 28 | 5 | 295 | 249 | 56 | 54 | 22 | 13 | 1 | 0 |
| 14 | 994 | 533 | 662 | 2 | 1 | 83 | 79 | 44 | 41 | 6 | 5 | 3 | 0 |
| 15 | 389 | 258 | 432 | 3 | 2 | 27 | 54 | 14 | 28 | 34 | 5 | 1 | 0 |
| 16 | 102 | 203 | 480 | 42 | 7 | 143 | 184 | 17 | 39 | 72 | 22 | 1 | 1 |
| 17 | 428 | 849 | 987 | 2 | 2 | 145 | 168 | 83 | 98 | 6 | 8 | 5 | 1 |
| 18 | 569 | 274 | 378 | 2 | 1 | 30 | 48 | 16 | 24 | 19 | 4 | 1 | 0 |
| 19 | 66 | 862 | 994 | 2 | 2 | 1341 | 455 | 638 | 237 | 6 | 13 | 4 | 1 |
| 20 | 3 | 63923 | 4161 | 2 | 0 | 2033 | 492 | 1631 | 332 | 0 | 0 | 8 | 1 |
| 21 | 283 | 749 | 930 | 2 | 2 | 345 | 132 | 189 | 76 | 3 | 4 | 2 | 0 |
| 22 | 1024 | 207 | 462 | 2 | 1 | 21 | 85 | 8 | 25 | 99 | 2 | 1 | 0 |
| 23 | 18 | 4219 | 3906 | 1 | 1 | 6547 | 796 | 5777 | 810 | 0 | 0 | 0 | 1 |
| 24 | 24 | 1610 | 1138 | 12 | 12 | 7469 | 2529 | 1533 | 583 | 9 | 21 | 2 | 1 |
| 25 | 245 | 334 | 372 | 3 | 3 | 71 | 116 | 37 | 69 | 34 | 9 | 3 | 0 |
| 26 | 102 | 384 | 449 | 19 | 7 | 1483 | 455 | 240 | 111 | 9 | 13 | 2 | 1 |
| 27 | 354 | 305 | 500 | 2 | 2 | 29 | 68 | 17 | 43 | 51 | 4 | 1 | 0 |
| 28 | 534 | 331 | 475 | 2 | 2 | 25 | 27 | 13 | 13 | 15 | 4 | 2 | 0 |
| 29 | 112 | 6040 | 1989 | 2 | 2 | 135 | 137 | 91 | 88 | 9 | 13 | 1 | 0 |
| 30 | 337 | 342 | 545 | 2 | 2 | 24 | 41 | 13 | 20 | 33 | 7 | 2 | 0 |
| 31 | 170 | 748 | 811 | 2 | 2 | 539 | 217 | 296 | 121 | 3 | 5 | 3 | 0 |
| 32 | 91 | 357 | 762 | 48 | 6 | 781 | 584 | 105 | 125 | 16 | 12 | 2 | 1 |
| 33 | 959 | 345 | 495 | 1 | 1 | 49 | 41 | 26 | 21 | 5 | 3 | 1 | 0 |
| 34 | 910 | 441 | 594 | 2 | 1 | 71 | 55 | 38 | 29 | 4 | 3 | 2 | 0 |
| 35 | 217 | 579 | 741 | 2 | 2 | 496 | 161 | 268 | 78 | 3 | 6 | 1 | 0 |
| 36 | 8 | 118039 | 13206 | 2 | 1 | 1973 | 810 | 1517 | 624 | 1 | 2 | 7 | 3 |
| 37 | 145 | 185 | 417 | 16 | 4 | 65 | 86 | 13 | 19 | 49 | 12 | 2 | 1 |
| 38 | 358 | 931 | 1016 | 2 | 1 | 220 | 80 | 132 | 43 | 3 | 5 | 1 | 0 |
| 39 | 7 | 35175 | 8251 | 3 | 1 | 103 | 76 | 74 | 55 | 20 | 32 | 2 | 1 |
| 40 | 29 | 16632 | 2988 | 2 | 2 | 240 | 316 | 164 | 191 | 17 | 32 | 2 | 1 |
| 41 | 28 | 477 | 786 | 2 | 2 | 1330 | 580 | 805 | 278 | 82 | 18 | 1 | 0 |
| 42 | 21 | 5739 | 6559 | 2 | 2 | 3898 | 631 | 3013 | 368 | 1 | 2 | 3 | 3 |
| 43 | 91 | 606 | 1181 | 35 | 7 | 357 | 244 | 58 | 51 | 25 | 18 | 4 | 1 |
| 44 | 205 | 1030 | 1122 | 2 | 2 | 848 | 254 | 447 | 142 | 7 | 12 | 1 | 0 |
| 45 | 73 | 5572 | 2184 | 1 | 1 | 180 | 176 | 128 | 130 | 7 | 11 | 3 | 1 |
| 46 | 238 | 133 | 349 | 9 | 3 | 62 | 165 | 13 | 37 | 94 | 8 | 1 | 0 |
| 47 | 41 | 161 | 182 | 71 | 10 | 908 | 1084 | 79 | 86 | 36 | 30 | 2 | 1 |

## APPENDIX 3.A: CONDITIONAL DISTRIBUTIONS FOR ESTIMATION OF THE GAMIFICTION CHOICE MODEL

Conditional distributions of the choice variable include the following:

$$\begin{aligned}\Lambda_i \mid y_{it}^{j}, cont_{it-1}, rcv_{it-1}, crep_{iw-1}, rep_{iw-1}, \\ rnk_{iw-1}, \Delta rnk_{iw-1}, bdg_{it-1}, cbdg_{it-1}, \mu_i, \Sigma_i \qquad i = 1...I\end{aligned} \qquad \text{(A1)}$$

where this conditional distribution can be estimated by random walk metropolis hasting on the weighted likelihood.

The priors for normal mixture distribution of the individual and the category specific parameters used are:

$$\begin{aligned}&\{(\mu_i, \Sigma_i)\} \mid \Lambda_i, \Delta, z_i, a, v, \vartheta, \alpha^d \\ &\alpha^d \mid I^* \\ &a \mid \{(\mu_i, \Sigma_i)^*\} \\ &v \mid \{(\mu_i, \Sigma_i)^*\}, \vartheta \\ &\vartheta \mid \{(\mu_i, \Sigma_i)^*\}, v\end{aligned} \qquad \text{(A2)}$$

where the first conditional is the standard posterior Polya Urn representation for the mean and variance of individual specific random coefficient choice model parameters. $(\mu_i, \Sigma_i)^*$ denotes a set of unique $(\mu_i, \Sigma_i)$, which the DP process hyper-parameters depend only on (a posteriori). Given the $\{(\mu_i, \Sigma_i)^*\}$ set $\alpha^d$ and based measure parameters (i.e. $a, v, \vartheta$) are independent, a posteriori. The conditional posterior of the $G_0$ hyper-parameters (i.e. $a, v, \vartheta$), factors into two parts as $a$ is independent of $v, \vartheta$ given $\{(\mu_i, \Sigma_i)^*\}$. The form of this conditional posterior is:

$$p(a,v,\vartheta \mid \{(\mu_i,\Sigma_i)^*\}) \propto \prod_{j=1}^{I^*} \phi(\mu_j^* \mid 0, a^{-1}, \Sigma_j^*) IW(\Sigma_j^* \mid v, V = v\vartheta I_d)\, p(a,v,\vartheta) \quad \text{(A3)}$$

where $\phi(.\mid.,.)$ denotes the multivariate normal density. $IW(.\mid.,.)$ denotes Inverted-Wishart distribution. Finally, for Polya representation implementation the following conditional distribution is used:

$$(\mu_{i+1},\Sigma_{i+1}) \mid \{(\mu_1,\Sigma_1),...,(\mu_i,\Sigma_i)\} \sim \begin{cases} G_0(a,v,\vartheta) \text{ with prob } \dfrac{\alpha^d}{\alpha^d + i} \\ \delta_{(\mu_j,\Sigma_j)} \text{ with prob } \dfrac{1}{\alpha^d + i} \end{cases} \quad \text{(C4)}$$

I assessed the prior hyperparameters to provide proper but diffuse distributions, defined formally by:

$$\vec{a} = 0.01, \bar{a} = 2, \vec{\vartheta} = 0.1, \bar{\vartheta} = 3, \vec{v} = 0.1, \bar{v} = 4 \qquad \text{(C5)}$$

Finally to complete the exposition, the posterior for the partition (segment) parameters has the following form:

$$\Sigma_k \mid \alpha_k, \Delta, z_k, v, V \sim IW(v + n_k, \; v \times \vartheta \times I + (\alpha_k - \tilde{\mu}_k' - \Delta z_k)'(\alpha_k - \tilde{\mu}_k' - \Delta z_k) + a(\tilde{\mu}_k - \bar{\mu})(\tilde{\mu}_k - \bar{\mu}))$$

$$\mu_k \mid \alpha_k, \Sigma_k, \bar{\mu}, a, \Delta, z_k \sim N(\tilde{\mu}_k, \frac{\Sigma_k}{n_k + a})$$

$$\tilde{\mu}_k = \frac{n_k \bar{\alpha}_k + a\bar{\mu}_k}{n_k + a}, \bar{\alpha}_k = \frac{\sum_{i \in k} \alpha_i}{n_k}, \bar{\mu} = 0$$

**APPENDIX 3.B:** EXTRA TABLES FOR THE ALTERNATIVE MODEL

Table 3.B.1. PARAMETER ESTIMATES: Individual Choice effect (10K sample size with model that explains parameters with fixed variables at Hierarchy)

| | **Estimate** | **Std. Dev.** | **2.5th** | **97.5th** |
|---|---|---|---|---|
| Fixed Effect | 50.881 | 160.168 | 50.144 | 53.493 |
| States: | | | | |
| Previous contribution | -0.004 | 0.042 | -0.097 | 0.095 |
| Reciprocity (contribution received) | -0.065 | 0.201 | -0.489 | 0.256 |
| Leader Board: | | | | |
| Cum Reputation | 0.101 | 0.190 | -0.335 | 0.455 |
| Reputation | -6.340 | 18.781 | -7.643 | -4.691 |
| Rank | -0.053 | 0.419 | -0.926 | 0.797 |
| Rank Change | 0.000 | 0.001 | -0.001 | 0.001 |
| Badges | | | | |
| Gold Badge | -0.198 | 0.826 | -1.074 | 0.350 |
| Silver Bade | -0.084 | 0.443 | -0.973 | 0.526 |
| Bronze Badge | -0.002 | 0.370 | -0.711 | 0.679 |
| Cum Gold Badge | 0.014 | 0.197 | -0.387 | 0.344 |
| Cum Silver Badge | -0.015 | 0.157 | -0.317 | 0.284 |
| Cum Bronze Badge | -0.010 | 0.125 | -0.252 | 0.235 |

Table 3.B.2. PARAMETER ESTIMATES: Individual Choice effect
(10K sample size with model that explains parameters with fixed variables at Hierarchy)

| | **Positive Significant** | **Negative Significant** | **% positive** | **% Negative** |
|---|---|---|---|---|
| Fixed Effect | 9711 | 60 | 97% | 1% |
| States: | | | | |
| Previous contribution | 864 | 902 | 9% | 9% |
| Reciprocity (contribution received) | 450 | 1250 | 5% | 13% |
| Leader Board: | | | | |
| Cum Reputation | 1124 | 565 | 11% | 6% |
| Reputation | 60 | 9706 | 1% | 97% |
| Rank | 552 | 860 | 6% | 9% |
| Rank Change | 1316 | 1407 | 13% | 14% |
| Badges | | | | |
| Gold Badge | 33 | 318 | 0% | 3% |
| Silver Bade | 69 | 91 | 1% | 1% |
| Bronze Badge | 208 | 201 | 2% | 2% |
| Cum Gold Badge | 300 | 298 | 3% | 3% |
| Cum Silver Badge | 837 | 945 | 8% | 9% |
| Cum Bronze Badge | 968 | 986 | 10% | 10% |

Table 3.B.3. PARAMETER ESTIMATES: Individual Choice effect (5K sample size with model that explains parameters with all variables at Hierarchy)

| | **Estimate** | **Std. Dev.** | **2.5th** | **97.5th** |
|---|---|---|---|---|
| Fixed Effect | -0.405 | 0.938 | -2.826 | 0.830 |
| States: | | | | |
| Previous contribution | -0.008 | 0.065 | -0.100 | 0.101 |
| Reciprocity (contribution received) | -0.085 | 0.370 | -0.429 | 0.217 |
| Leader Board: | | | | |
| Cum Reputation | 0.111 | 0.198 | -0.318 | 0.455 |
| Reputation | -0.441 | 0.874 | -1.706 | 1.014 |
| Rank | -0.032 | 0.374 | -0.805 | 0.750 |
| Rank Change | 0.000 | 0.001 | -0.002 | 0.002 |
| Badges | | | | |
| Gold Badge | 0.078 | 1.237 | -1.291 | 1.955 |
| Silver Bade | -0.155 | 0.519 | -0.818 | 0.566 |
| Bronze Badge | -0.011 | 0.615 | -0.862 | 1.174 |
| Cum Gold Badge | 0.040 | 0.269 | -0.388 | 0.460 |
| Cum Silver Badge | -0.011 | 0.187 | -0.290 | 0.277 |
| Cum Bronze Badge | -0.010 | 0.152 | -0.237 | 0.217 |

Table 3.B. 4. PARAMETER ESTIMATES: Individual Choice effect
(5K sample size with model that explains parameters with all variables at Hierarchy)

| | **Positive Significant** | **Negative Significant** | **% positive** | **% Negative** |
|---|---|---|---|---|
| Fixed Effect | 94 | 725 | 2% | 15% |
| States: | | | | |
| Previous contribution | 428 | 498 | 9% | 10% |
| Reciprocity (contribution received) | 205 | 624 | 4% | 12% |
| Leader Board: | | | | |
| Cum Reputation | 541 | 261 | 11% | 5% |
| Reputation | 167 | 362 | 3% | 7% |
| Rank | 308 | 360 | 6% | 7% |
| Rank Change | 665 | 716 | 13% | 14% |
| Badges | | | | |
| Gold Badge | 643 | 569 | 13% | 11% |
| Silver Bade | 53 | 58 | 1% | 1% |
| Bronze Badge | 147 | 92 | 3% | 2% |
| Cum Gold Badge | 158 | 163 | 3% | 3% |
| Cum Silver Badge | 429 | 468 | 9% | 9% |
| Cum Bronze Badge | 491 | 476 | 10% | 10% |

Table 3.B.5. PARAMETER ESTIMATES: Individual Choice effect (5K sample size with model that explains parameters with full variables at Hierarchy)

| | **Estimate** | **Std. Dev.** | **$2.5^{th}$** | **$97.5^{th}$** |
|---|---|---|---|---|
| Fixed Effect | -0.094 | 0.818 | -1.414 | 2.455 |
| States: | | | | |
| Previous contribution | -0.010 | 0.079 | -0.107 | 0.110 |
| Reciprocity (contribution received) | -0.089 | 0.449 | -0.481 | 0.232 |
| Leader Board: | | | | |
| Cum Reputation | 0.111 | 0.193 | -0.312 | 0.444 |
| Reputation | -0.447 | 0.622 | -1.478 | 0.940 |
| Rank | -0.023 | 0.362 | -0.803 | 0.735 |
| Rank Change | 0.000 | 0.001 | -0.002 | 0.002 |
| Badges | | | | |
| Gold Badge | -0.089 | 0.641 | -1.730 | 0.844 |
| Silver Bade | -0.045 | 0.312 | -0.592 | 0.446 |
| Bronze Badge | -0.049 | 0.390 | -0.833 | 1.005 |
| Cum Gold Badge | 0.022 | 0.230 | -0.418 | 0.451 |
| Cum Silver Badge | -0.012 | 0.189 | -0.294 | 0.279 |
| Cum Bronze Badge | -0.011 | 0.152 | -0.249 | 0.222 |

Table 3.B.6. PARAMETER ESTIMATES: Individual Choice effect
(5K sample size with model that explains parameters with fixed variables at Hierarchy)

| | **Positive Significant** | **Negative Significant** | **% positive** | **% Negative** |
|---|---|---|---|---|
| Fixed Effect | 356 | 536 | 7% | 11% |
| States: | | | | |
| Previous contribution | 427 | 513 | 9% | 10% |
| Reciprocity (contribution received) | 208 | 622 | 4% | 12% |
| Leader Board: | | | | |
| Cum Reputation | 532 | 251 | 11% | 5% |
| Reputation | 147 | 390 | 3% | 8% |
| Rank | 301 | 363 | 6% | 7% |
| Rank Change | 684 | 707 | 14% | 14% |
| Badges | | | | |
| Gold Badge | 116 | 310 | 2% | 6% |
| Silver Bade | 49 | 54 | 1% | 1% |
| Bronze Badge | 144 | 92 | 3% | 2% |
| Cum Gold Badge | 154 | 164 | 3% | 3% |
| Cum Silver Badge | 412 | 459 | 8% | 9% |
| Cum Bronze Badge | 490 | 478 | 10% | 10% |

Table 3.B.7.PARAMETER ESTIMATES: Individual Choice effect (1K size for k-mean stratified sample with model that explains parameters with fixed variables at Hierarchy)

| | **Estimate** | **Std. Dev.** | **$2.5^{th}$** | **$97.5^{th}$** |
|---|---|---|---|---|
| Fixed Effect | -0.003 | 2.045 | -3.356 | 2.772 |
| States: | | | | |
| Previous contribution | -0.034 | 0.089 | -0.236 | 0.165 |
| Reciprocity (contribution received) | -0.031 | 0.213 | -0.557 | 0.403 |
| Leader Board: | | | | |
| Cum Reputation | 0.095 | 0.233 | -0.384 | 0.540 |
| Reputation | -0.271 | 0.869 | -1.937 | 1.391 |
| Rank | 0.004 | 0.386 | -0.818 | 0.834 |
| Rank Change | 0.000 | 0.009 | -0.003 | 0.004 |
| Badges | | | | |
| Gold Badge | -0.134 | 1.138 | -2.132 | 2.822 |
| Silver Bade | -0.448 | 1.019 | -1.756 | 1.265 |
| Bronze Badge | -0.193 | 2.120 | -5.708 | 0.877 |
| Cum Gold Badge | 0.021 | 0.494 | -0.584 | 0.432 |
| Cum Silver Badge | -0.020 | 0.170 | -0.394 | 0.330 |
| Cum Bronze Badge | -0.011 | 0.188 | -0.288 | 0.330 |

Table 3.B.8. PARAMETER ESTIMATES: Individual Choice effect (1K size for k-mean stratified sample with model that explains parameters with fixed variables at Hierarchy)

| | **Positive Significant** | **Negative Significant** | **% positive** | **% Negative** |
|---|---|---|---|---|
| Fixed Effect | 215 | 284 | 4% | 6% |
| States: | | | | |
| Previous contribution | 97 | 134 | 10% | 13% |
| Reciprocity (contribution received) | 62 | 121 | 6% | 12% |
| Leader Board: | | | | |
| Cum Reputation | 122 | 73 | 12% | 7% |
| Reputation | 78 | 106 | 8% | 11% |
| Rank | 66 | 68 | 7% | 7% |
| Rank Change | 122 | 123 | 12% | 12% |
| Badges | | | | |
| Gold Badge | 128 | 213 | 13% | 21% |
| Silver Bade | 26 | 98 | 3% | 10% |
| Bronze Badge | 28 | 57 | 3% | 6% |
| Cum Gold Badge | 39 | 34 | 4% | 3% |
| Cum Silver Badge | 77 | 98 | 8% | 10% |
| Cum Bronze Badge | 126 | 96 | 13% | 10% |

Table 3.B.9. PARAMETER ESTIMATES: Individual Choice effect (1K size for LDA stratified sample with model that explains parameters with fixed variables at Hierarchy)

| | **Estimate** | **Std. Dev.** | **2.5th** | **97.5th** |
|---|---|---|---|---|
| Fixed Effect | 14.635 | 2.287 | 12.119 | 19.938 |
| Times: | | | | |
| Previous contribution | -0.015 | 0.079 | -0.174 | 0.174 |
| Reciprocity (contribution received) | -0.110 | 0.208 | -0.613 | 0.272 |
| Leader Board: | | | | |
| Cum Reputation | 0.069 | 0.183 | -0.318 | 0.435 |
| Reputation | -1.989 | 0.847 | -3.334 | -0.197 |
| Rank | -0.092 | 0.357 | -0.853 | 0.654 |
| Rank Change | 0.000 | 0.003 | -0.005 | 0.003 |
| Badges | | | | |
| Gold Badge | -0.314 | 1.366 | -2.758 | 2.276 |
| Silver Bade | -0.136 | 1.076 | -2.265 | 1.370 |
| Bronze Badge | 0.047 | 0.535 | -1.181 | 1.053 |
| Cum Gold Badge | 0.044 | 0.262 | -0.472 | 0.526 |
| Cum Silver Badge | -0.034 | 0.178 | -0.413 | 0.329 |
| Cum Bronze Badge | -0.017 | 0.137 | -0.266 | 0.236 |

Table 3.B.10. PARAMETER ESTIMATES: Individual Choice effect (1K size for LDA stratified sample with model that explains parameters with fixed variables at Hierarchy)

| | **Positive Significant** | **Negative Significant** | **% positive** | **% Negative** |
|---|---|---|---|---|
| Fixed Effect | 1000 | 0 | 20% | 0% |
| States: | | | | |
| Previous contribution | 114 | 116 | 11% | 12% |
| Reciprocity (contribution received) | 42 | 156 | 4% | 16% |
| Leader Board: | | | | |
| Cum Reputation | 99 | 49 | 10% | 5% |
| Reputation | 5 | 936 | 1% | 94% |
| Rank | 60 | 81 | 6% | 8% |
| Rank Change | 128 | 145 | 13% | 15% |
| Badges | | | | |
| Gold Badge | 162 | 320 | 16% | 32% |
| Silver Bade | 46 | 69 | 5% | 7% |
| Bronze Badge | 25 | 19 | 3% | 2% |
| Cum Gold Badge | 38 | 32 | 4% | 3% |
| Cum Silver Badge | 84 | 114 | 8% | 11% |
| Cum Bronze Badge | 94 | 93 | 9% | 9% |

Table 3.B.11. PARAMETER ESTIMATES: Individual Choice effect (1K size for Uniform stratified sample with model that explains parameters with fixed variables at Hierarchy)

| | **Estimate** | **Std. Dev.** | **$2.5^{th}$** | **$97.5^{th}$** |
|---|---|---|---|---|
| Fixed Effect | -0.260 | 1.280 | -2.227 | 3.432 |
| States: | | | | |
| Previous contribution | -0.027 | 0.080 | -0.177 | 0.153 |
| Reciprocity (contribution received) | -0.109 | 0.272 | -0.690 | 0.318 |
| Leader Board: | | | | |
| Cum Reputation | 0.091 | 0.182 | -0.319 | 0.424 |
| Reputation | -0.083 | 1.060 | -1.772 | 1.937 |
| Rank | -0.073 | 0.415 | -0.995 | 0.863 |
| Rank Change | 0.000 | 0.005 | -0.004 | 0.004 |
| Badges | | | | |
| Gold Badge | -0.351 | 1.971 | -3.725 | 2.775 |
| Silver Bade | -0.533 | 1.056 | -2.180 | 1.218 |
| Bronze Badge | 0.060 | 0.892 | -1.307 | 1.414 |
| Cum Gold Badge | -0.015 | 0.275 | -0.503 | 0.580 |
| Cum Silver Badge | -0.027 | 0.163 | -0.313 | 0.271 |
| Cum Bronze Badge | -0.003 | 0.150 | -0.288 | 0.305 |

Table 3.B.12. PARAMETER ESTIMATES: Individual Choice effect (1K size for Uniform stratified sample with model that explains parameters with fixed variables at Hierarchy)

| | **Positive Significant** | **Negative Significant** | **% positive** | **% Negative** |
|---|---|---|---|---|
| Fixed Effect | 82 | 89 | 2% | 2% |
| States: | | | | |
| Previous contribution | 93 | 118 | 9% | 12% |
| Reciprocity (contribution received) | 52 | 152 | 5% | 15% |
| Leader Board: | | | | |
| Cum Reputation | 100 | 52 | 10% | 5% |
| Reputation | 126 | 103 | 13% | 10% |
| Rank | 69 | 92 | 7% | 9% |
| Rank Change | 137 | 139 | 14% | 14% |
| Badges | | | | |
| Gold Badge | 158 | 323 | 16% | 32% |
| Silver Bade | 14 | 145 | 1% | 15% |
| Bronze Badge | 21 | 23 | 2% | 2% |
| Cum Gold Badge | 49 | 45 | 5% | 5% |
| Cum Silver Badge | 82 | 98 | 8% | 10% |
| Cum Bronze Badge | 133 | 111 | 13% | 11% |

Table 3.B.13. PARAMETER ESTIMATES: Individual Choice effect (1K size for mixed-normal stratified sample with model that explains parameters with fixed variables at Hierarchy)

| | **Estimate** | **Std. Dev.** | **2.5$^{th}$** | **97.5$^{th}$** |
|---|---|---|---|---|
| Fixed Effect | 0.065 | 1.353 | -2.403 | 3.649 |
| States: | | | | |
| Previous contribution | -0.023 | 0.076 | -0.153 | 0.103 |
| Reciprocity (contribution received) | -0.055 | 0.279 | -0.437 | 0.237 |
| Leader Board: | | | | |
| Cum Reputation | 0.119 | 0.206 | -0.294 | 0.506 |
| Reputation | -0.456 | 0.621 | -1.603 | 0.814 |
| Rank | -0.048 | 0.282 | -0.676 | 0.494 |
| Rank Change | 0.000 | 0.005 | -0.006 | 0.004 |
| Badges | | | | |
| Gold Badge | -0.085 | 0.933 | -1.858 | 2.451 |
| Silver Bade | -0.108 | 0.895 | -2.220 | 1.019 |
| Bronze Badge | -0.029 | 0.607 | -0.800 | 0.798 |
| Cum Gold Badge | 0.019 | 0.319 | -0.445 | 0.397 |
| Cum Silver Badge | -0.007 | 0.196 | -0.321 | 0.339 |
| Cum Bronze Badge | -0.023 | 0.149 | -0.312 | 0.259 |

Table 3.B.14. PARAMETER ESTIMATES: Individual Choice effect (1K size for mixed-normal stratified sample with model that explains parameters with fixed variables at Hierarchy)

| | **Positive Significant** | **Negative Significant** | **% positive** | **% Negative** |
|---|---|---|---|---|
| Fixed Effect | 195 | 137 | 4% | 3% |
| States: | | | | |
| Previous contribution | 88 | 105 | 9% | 11% |
| Reciprocity (contribution received) | 41 | 124 | 4% | 12% |
| Leader Board: | | | | |
| Cum Reputation | 114 | 44 | 11% | 4% |
| Reputation | 31 | 152 | 3% | 15% |
| Rank | 44 | 58 | 4% | 6% |
| Rank Change | 121 | 125 | 12% | 13% |
| Badges | | | | |
| Gold Badge | 81 | 108 | 8% | 11% |
| Silver Bade | 25 | 42 | 3% | 4% |
| Bronze Badge | 25 | 21 | 3% | 2% |
| Cum Gold Badge | 40 | 38 | 4% | 4% |
| Cum Silver Badge | 97 | 83 | 10% | 8% |
| Cum Bronze Badge | 108 | 109 | 11% | 11% |

Table 3.B.15. PARAMETER ESTIMATES: Individual Choice effect (1K size for mixed-normal stratified sample with model that explains parameters with full variables at Hierarchy)

| | **Estimate** | **Std. Dev.** | **$2.5^{th}$** | **$97.5^{th}$** |
|---|---|---|---|---|
| Fixed Effect | -0.294 | 1.505 | -3.123 | 2.165 |
| States: | | | | |
| Previous contribution | -0.037 | 0.103 | -0.195 | 0.143 |
| Reciprocity (contribution received) | -0.119 | 0.498 | -0.927 | 0.356 |
| Leader Board: | | | | |
| Cum Reputation | 0.134 | 0.273 | -0.373 | 0.533 |
| Reputation | -0.579 | 1.416 | -2.484 | 2.422 |
| Rank | 0.031 | 0.583 | -1.019 | 1.080 |
| Rank Change | -0.001 | 0.026 | -0.005 | 0.005 |
| Badges | | | | |
| Gold Badge | 0.046 | 1.667 | -5.272 | 1.733 |
| Silver Bade | -0.094 | 1.547 | -2.989 | 3.156 |
| Bronze Badge | 0.115 | 1.429 | -1.215 | 1.436 |
| Cum Gold Badge | -0.025 | 0.176 | -0.398 | 0.300 |
| Cum Silver Badge | -0.019 | 0.182 | -0.399 | 0.324 |
| Cum Bronze Badge | -0.006 | 0.160 | -0.320 | 0.329 |

Table 3.B.16. PARAMETER ESTIMATES: Individual Choice effect (1K size for mixed-normal stratified sample with model that explains parameters with full variables at Hierarchy)

| | **Positive Significant** | **Negative Significant** | **% positive** | **% Negative** |
|---|---|---|---|---|
| Fixed Effect | 155 | 288 | 3% | 6% |
| States: | | | | |
| Previous contribution | 92 | 145 | 9% | 15% |
| Reciprocity (contribution received) | 57 | 135 | 6% | 14% |
| Leader Board: | | | | |
| Cum Reputation | 131 | 57 | 13% | 6% |
| Reputation | 78 | 220 | 8% | 22% |
| Rank | 85 | 88 | 9% | 9% |
| Rank Change | 134 | 157 | 13% | 16% |
| Badges | | | | |
| Gold Badge | 135 | 89 | 14% | 9% |
| Silver Bade | 122 | 129 | 12% | 13% |
| Bronze Badge | 32 | 27 | 3% | 3% |
| Cum Gold Badge | 24 | 20 | 2% | 2% |
| Cum Silver Badge | 73 | 115 | 7% | 12% |
| Cum Bronze Badge | 129 | 80 | 13% | 8% |

Table 3.B.17. PARAMETER ESTIMATES: Individual Choice effect (1K size for k-mean stratified sample with model that explains parameters with full variables at Hierarchy)

| | **Estimate** | **Std. Dev.** | **2.5$^{th}$** | **97.5$^{th}$** |
|---|---|---|---|---|
| Fixed Effect | -0.419 | 1.661 | -4.120 | 2.602 |
| States: | | | | |
| Previous contribution | -0.024 | 0.150 | -0.199 | 0.147 |
| Reciprocity (contribution received) | -0.168 | 0.397 | -0.734 | 0.286 |
| Leader Board: | | | | |
| Cum Reputation | 0.118 | 0.199 | -0.330 | 0.472 |
| Reputation | -0.419 | 0.893 | -2.209 | 1.093 |
| Rank | -0.049 | 0.376 | -0.777 | 0.742 |
| Rank Change | 0.000 | 0.008 | -0.006 | 0.004 |
| Badges | | | | |
| Gold Badge | -0.131 | 1.950 | -3.361 | 3.930 |
| Silver Bade | 0.052 | 1.412 | -2.208 | 1.787 |
| Bronze Badge | 0.068 | 0.847 | -1.282 | 1.994 |
| Cum Gold Badge | -0.021 | 0.727 | -0.440 | 0.546 |
| Cum Silver Badge | -0.021 | 0.216 | -0.364 | 0.317 |
| Cum Bronze Badge | -0.006 | 0.167 | -0.266 | 0.274 |

Table 3.B.18. PARAMETER ESTIMATES: Individual Choice effect (1K size for k-mean stratified sample with model that explains parameters with full variables at Hierarchy)

| | **Positive Significant** | **Negative Significant** | **% positive** | **% Negative** |
|---|---|---|---|---|
| Fixed Effect | 202 | 380 | 4% | 8% |
| States: | | | | |
| Previous contribution | 99 | 109 | 10% | 11% |
| Reciprocity (contribution received) | 48 | 135 | 5% | 14% |
| Leader Board: | | | | |
| Cum Reputation | 114 | 53 | 11% | 5% |
| Reputation | 56 | 218 | 6% | 22% |
| Rank | 70 | 72 | 7% | 7% |
| Rank Change | 141 | 140 | 14% | 14% |
| Badges | | | | |
| Gold Badge | 222 | 292 | 22% | 29% |
| Silver Bade | 131 | 75 | 13% | 8% |
| Bronze Badge | 52 | 32 | 5% | 3% |
| Cum Gold Badge | 47 | 36 | 5% | 4% |
| Cum Silver Badge | 97 | 82 | 10% | 8% |
| Cum Bronze Badge | 103 | 84 | 10% | 8% |

Table 3.B.19. PARAMETER ESTIMATES: Individual Choice effect (1K size for LDA stratified sample with model that explains parameters with full variables at Hierarchy)

| | **Estimate** | **Std. Dev.** | **2.5$^{th}$** | **97.5$^{th}$** |
|---|---|---|---|---|
| Fixed Effect | 13.886 | 2.334 | 9.135 | 17.403 |
| States: | | | | |
| Previous contribution | -0.034 | 0.082 | -0.197 | 0.134 |
| Reciprocity (contribution received) | -0.104 | 0.249 | -0.695 | 0.410 |
| Leader Board: | | | | |
| Cum Reputation | 0.130 | 0.203 | -0.287 | 0.522 |
| Reputation | -1.850 | 0.930 | -3.612 | -0.218 |
| Rank | -0.093 | 0.448 | -1.088 | 0.860 |
| Rank Change | 0.000 | 0.002 | -0.004 | 0.005 |
| Badges | | | | |
| Gold Badge | -0.874 | 1.402 | -3.808 | 2.347 |
| Silver Bade | 0.034 | 1.194 | -2.064 | 3.101 |
| Bronze Badge | 0.130 | 0.548 | -0.800 | 1.268 |
| Cum Gold Badge | 0.000 | 0.289 | -0.493 | 0.600 |
| Cum Silver Badge | -0.035 | 0.176 | -0.403 | 0.364 |
| Cum Bronze Badge | -0.008 | 0.140 | -0.289 | 0.280 |

Table 3.B.20.PARAMETER ESTIMATES: Individual Choice effect (1K size for LDA stratified sample with model that explains parameters with full variables at Hierarchy)

| | **Positive Significant** | **Negative Significant** | **% positive** | **% Negative** |
|---|---|---|---|---|
| Fixed Effect | 997 | 1 | 20% | 0% |
| States: | | | | |
| Previous contribution | 82 | 107 | 8% | 11% |
| Reciprocity (contribution received) | 43 | 131 | 4% | 13% |
| Leader Board: | | | | |
| Cum Reputation | 115 | 39 | 12% | 4% |
| Reputation | 6 | 850 | 1% | 85% |
| Rank | 58 | 70 | 6% | 7% |
| Rank Change | 131 | 140 | 13% | 14% |
| Badges | | | | |
| Gold Badge | 70 | 555 | 7% | 56% |
| Silver Bade | 163 | 144 | 16% | 14% |
| Bronze Badge | 39 | 16 | 4% | 2% |
| Cum Gold Badge | 33 | 30 | 3% | 3% |
| Cum Silver Badge | 89 | 94 | 9% | 9% |
| Cum Bronze Badge | 106 | 89 | 11% | 9% |

Table 3.B.21. PARAMETER ESTIMATES: Individual Choice effect (1K size for Uniform stratified sample with model that explains parameters with full variables at Hierarchy)

| | **Estimate** | **Std. Dev.** | **2.5th** | **97.5th** |
|---|---|---|---|---|
| Fixed Effect | 0.459 | 2.628 | -4.001 | 4.872 |
| States: | | | | |
| Previous contribution | -0.021 | 0.080 | -0.190 | 0.156 |
| Reciprocity (contribution received) | -0.104 | 0.244 | -0.662 | 0.294 |
| Leader Board: | | | | |
| Cum Reputation | 0.124 | 0.203 | -0.327 | 0.489 |
| Reputation | -0.651 | 1.382 | -2.805 | 1.752 |
| Rank | 0.062 | 0.534 | -0.922 | 0.892 |
| Rank Change | 0.000 | 0.002 | -0.004 | 0.004 |
| Badges | | | | |
| Gold Badge | 0.012 | 2.723 | -4.257 | 5.609 |
| Silver Bade | -0.167 | 1.287 | -4.334 | 3.800 |
| Bronze Badge | -0.014 | 0.966 | -2.003 | 1.228 |
| Cum Gold Badge | 0.051 | 0.211 | -0.385 | 0.426 |
| Cum Silver Badge | -0.023 | 0.149 | -0.325 | 0.269 |
| Cum Bronze Badge | 0.004 | 0.120 | -0.230 | 0.274 |

Table 3.B.22. PARAMETER ESTIMATES: Individual Choice effect (1K size for Uniform stratified sample with model that explains parameters with full variables at Hierarchy)

| | **Positive Significant** | **Negative Significant** | **% positive** | **% Negative** |
|---|---|---|---|---|
| Fixed Effect | 358 | 155 | 7% | 3% |
| States: | | | | |
| Previous contribution | 85 | 119 | 9% | 12% |
| Reciprocity (contribution received) | 44 | 132 | 4% | 13% |
| Leader Board: | | | | |
| Cum Reputation | 120 | 48 | 12% | 5% |
| Reputation | 76 | 243 | 8% | 24% |
| Rank | 72 | 70 | 7% | 7% |
| Rank Change | 135 | 140 | 14% | 14% |
| Badges | | | | |
| Gold Badge | 257 | 332 | 26% | 33% |
| Silver Bade | 53 | 91 | 5% | 9% |
| Bronze Badge | 47 | 57 | 5% | 6% |
| Cum Gold Badge | 34 | 34 | 3% | 3% |
| Cum Silver Badge | 69 | 95 | 7% | 10% |
| Cum Bronze Badge | 112 | 84 | 11% | 8% |

## VITA

Meisam Hejazi Nia was born in Tehran, Iran. Meisam attended Allame Helli (exceptional talents) high school, during 1999-2003. He received the Bronze medal for national mathematics Olympiad contents in 2002. He completed his undergraduate coursework as top 1% at Amirkabir University of Tehran (Polytechnic University of Tehran) double majoring in computer engineering and information technology engineering. He earned his master degree in software system engineering from the same institute, and an MBA from Sharif University of Technology. He published his research on various information systems (e.g., commercial distributed processing) and marketing (mobile marketing) issues in IJMM, JETWI, and JACS journals. He has 3 years of experience as information system specialist and manager in an international textile company, and 2 years of experience as product specialist and manager in marketing department of an international telecom operator (MTN) in Tehran. He entered the Ph.D. program in Management Science (Marketing) at the Naveen Jindal School of Management of The University of Texas at Dallas in August 2012. During the course of his doctoral studies, Meisam presented his work in Marketing Science, Marketing Dynamics, International Industrial Organization, NYU Digital Big Data, AMA and POMS conferences. Prior to the completion of his studies he also held “senior e-commerce Data Scientist” position at Saber Airline Solutions in Southlake, Texas.